\documentclass[a4paper,11pt]{book} 

\usepackage{amstext,amsmath,amssymb}

\usepackage[dvips]{graphicx}

\usepackage{fancyhdr}
\numberwithin{equation}{chapter}%AMS - numbers equations by section

\usepackage[active]{srcltx}

\usepackage{mymacros}

\newcommand{\FP}{Fokker--Planck}
\newcommand{\LRT}{linear-response theory}
\newcommand{\QME}{quantum master equation}

\begin{document}

%%%%%%%%%%%%%%%%%%%%%%%%%%%%%%%%%%%%%%%%%%%%%%%%%%%%%%%%%%%%
%%%%%%%%%%%%%%%%%   TITLE  &  ABSTRACT   %%%%%%%%%%%%%%%%%%%
%%%%%%%%%%%%%%%%%   TITLE  &  ABSTRACT   %%%%%%%%%%%%%%%%%%%
%%%%%%%%%%%%%%%%%   TITLE  &  ABSTRACT   %%%%%%%%%%%%%%%%%%%
%%%%%%%%%%%%%%%%%   TITLE  &  ABSTRACT   %%%%%%%%%%%%%%%%%%%
%%%%%%%%%%%%%%%%%   TITLE  &  ABSTRACT   %%%%%%%%%%%%%%%%%%%
%%%%%%%%%%%%%%%%%%%%%%%%%%%%%%%%%%%%%%%%%%%%%%%%%%%%%%%%%%%%

\title{
Quantum and statistical mechanics in open systems: \\
theory and examples
}

\author{David Zueco
\protect\\[1.ex]\parbox{0.9\textwidth}{\centering\small
Departamento de F\'{\i}sica de la Materia Condensada e 
Instituto de Ciencia de Materiales de Arag\'on,
C.S.I.C. -- Universidad de Zaragoza, E-50009 Zaragoza, Spain
}%
}

%\date{\today}
\date{2007}

\maketitle %normal

\thispagestyle{empty}

\tableofcontents

\newpage

\thispagestyle{empty}

%%%%%%%%%%%%%%%%%%%%%%%%%%%%%%%%%%%%%%%%%%%%%%%%%%%%%%%%%%%%
%%%%%%%%%%%%%%%%%%%%%%%%%%%%%%%%%%%%%%%%%%%%%%%%%%%%%%%%%%%%
%%%%%%%%%%%% nolismens %%%%%%%%%%%%%%%%%%%%%%%%%%%%%%%%%%%%%
%%%%%%%%%%%%%%%%%%%%%%%%%%%%%%%%%%%%%%%%%%%%%%%%%%%%%%%%%%%%

 \chapter*{acknowledgments}
 \addcontentsline{toc}{chapter}{acknowledgments}
 \chaptermark{acknowledgments}

This English translation of my thesis has been done by my former boss and phD advisor, Jos\'e Luis Garc\'{\i}a-Palacios.  If you find anything useful in the next pages you should acknowledge him rather than me.  Gracias Jose.

\vspace{3.ex}
\subsection*{Acknowledgments in the spanish version}

Todo lo bueno que se pueda extraer de las siguientes p\'aginas se debe a las siguientes personas.

A mis padres y hermana, Luisa, Juanjo y Laura.  Ellos han hecho lo que soy y
sobre todo han aguantado mi parte más fea.  Esta tesis es vuestra.
A mis otros hermanos Diego y Javi y a los que vinieron después In\'es, Pedro y
Pablo. 
A mis abuelos.

A mi director, Jos\'e Luis Garc\'{\i}a-Palacios, la mayor\'{\i}a de las ideas 
que aqu\'{\i} se exponen son suyas.  A los ``chain-gang'' Fernando, Juanjo, Pedro y
Mario, ellos me dieron la primera oportunidad en la investigaci\'on y siempre me
dejaron la libertad de poder elegir.
A los dem\'as miembros del departamento de F\'{\i}sica de la Materia Condensada, en
especial a Fernando Luis (siempre fu\'e un placer hablar de f\'{\i}sica contigo) y a
  Luis Mart\'{\i}n-Moreno, a ti Luis te debo muchas cosas, gracias.
A los profesores Jos\'e Luis Alonso, por sus \'animos y apoyo y a Javier Sesma por
todos los libros que me prest\'o.  A mi primo y profesor Juan Pablo Mart\'{\i}nez,
por empezar a ayudarme desde el primer d\'{\i}a que pis\'e la facultad.

A pesar de ser como soy tengo unos amigos cojonudos que siempre me han ayudado
y con los que estoy y estaré\'e siempre en deuda.  Gracias al Chato, Ram\'on,
Alberto, L\'opez, Blas,
Colibr\'{\i}, Chotillo, Sanvi, Sudakilla, Yisush, Diego, Buggie, Rom\'an, Piojo y Fran
por convertirme en el Dr. Pavos.  Aupa Chirikov!.

No me puedo olvidar de mis suegros y ``cu\~{n}ao'' quienes no s\'olo no prohibieron
a su hija y hermana salir con un tipo como yo sino que siempre me han tratado
con cariño.

Y a t\'{\i} Mar\'{\i}a, mi cari, mi mujer, mi peque\~{n}a.  A t\'{\i}, es imposible dibujar con
palabras todo lo que has hecho por m\'{\i}.  As\'{\i} que gracias por todo y algo m\'as,
pero sobre todo por ser la demostraci\'on de mi teorema.

%%%%%%%%%%%%%%%%%%%%%%%%%%%%%%%%%%%%%%%%%%%%%%%%%%%%%%%%%%%%
%%%%%%%%%%%%%%%%%%%%%%%%%%%%%%%%%%%%%%%%%%%%%%%%%%%%%%%%%%%%
%%%%%%%%%%%% articulos %%%%%%%%%%%%%%%%%%%%%%%%%%%%%%%%%%%%%
%%%%%%%%%%%%%%%%%%%%%%%%%%%%%%%%%%%%%%%%%%%%%%%%%%%%%%%%%%%%

 \chapter*{articles}
 \addcontentsline{toc}{chapter}{articles}
 \chaptermark{articles}

This thesis is based on the following articles

\begin{itemize}

\item [{\Large \bf I} ] 
{{\bf Caldeira--Leggett quantum master equation in Wigner phase space:  \\
continued-fraction solution  and application to 
 Brownian motion 
\\
in periodic potentials}. 
\\
J.L. Garcia-Palacios and  D. Zueco
\\
\underline{\sc Journal of Physics A: Mathematical and General} 
{\bf 37} pp 10735-10770 (2004)  
\\({\tt cond-mat/0407454})}

\item [{\Large \bf II}] {{\bf Quantum ratchets at high temperatures}. 
\\
 D. Zueco and J.L. Garcia-Palacios
\\
\underline{\sc Physica E} {\bf 29}   pp 435-441 (2005) 
({\tt cond-mat/04122566})
}

\item [{\Large \bf III}] {{\bf  Longitudinal relaxation of quantum
      superparamagnets}. 
\\
 D. Zueco and J.L. Garcia-Palacios
\\
\underline{\sc Physical Review B} {\bf 73} pp 104448 (2006)   
({\tt cond-mat/0509627})
}

\item [{\Large \bf IV}] {{\bf Solving spin quantum-master equations with
      matrix continued-fraction methods:   application to superparamagnets}.
\\ 
J.L. Garcia-Palacios and  D. Zueco
\\
\underline{\sc Journal of Physics A: Mathematical and General} 
{\bf 39} pp 13243-13284 (2006)   \\ ({\tt cond-mat/0603730})}

\item [{\Large \bf V}] {{\bf Non-linear response of single-molecule magnets: 
\\
field-tuned      quantum-to-classical crossovers}
\\
R. Lopez-Ruiz, F. Luis, A. Millan, C. Rillo, D. Spec and
      J.L. Garcia-Palacios
\\
\underline{\sc Physical Review B}
{\bf 75}, pp 012402 (2007) ({\tt cond-mat/0606091})

}

\item [{\Large \bf VI}] {{\bf Bopp operators and phase-space spin dynamics: 
\\
application to       rotational quantum Brownian motion.}
\\
D. Zueco and I. Calvo 
\\
\underline{\sc Journal of Physics A: Mathematical and General} 
{\bf 40} pp 4635--4648 (2007) ({\tt quant-ph/0611194})
}

%% \item [{\Large \bf VII}] {{\bf Equilibrium properties of open quantum systems:
%% \\
%%  dissipative      corrections to the  Brillouin magnetizations.} 
%% \\
%% D. Zueco and  J.L. Garcia-Palacios
%% \\
%% {\bf en preparation}
%% }

\end{itemize}

%%%%%%%%%%%%%%%%%%%%%%%%%%%%%%%%%%%%%%%%%%%%%%%%%%%%%%%%%%%%
%%%%%%%%%%%%%%%%%%%%%%%%%%%%%%%%%%%%%%%%%%%%%%%%%%%%%%%%%%%%
%%%%%%%%%%%%% introduction              %%%%%%%%%%%%%%%%%%%%
%%%%%%%%%%%%%%%%%%%%%%%%%%%%%%%%%%%%%%%%%%%%%%%%%%%%%%%%%%%%
%%%%%%%%%%%%%%%%%%%%%%%%%%%%%%%%%%%%%%%%%%%%%%%%%%%%%%%%%%%%

\chapter {introduction}
\label{chap:intro}

In physics the ``system'' is the small part of the universe on which
we have some control.
In other words, it is the part of ``interest'' that we measure and/or
manipulate.
Knowing the system Hamiltonian, $\HS$, we can calculate its time
evolution and find the future state of the system, as well as telling
its past (how it reached the current state).
The system of interest can be, for instance, an electron, an atom, a
piece of metal, a cup of coffee, or the solar system.

Everyday experience teaches us that any system interacts with its
environment.
For example, we all know that a coffee thermos in a cold winter
morning will eventually cool down, no matter how sophisticated the
flask is.
Nobody is astonished by this: {\em we cannot fully isolate the coffee from
the outside}.
Admittedly, in many cases the effects of the environment can be
neglected, and the dynamics is then determined by the system
Hamiltonian alone.
In other cases, however, we must take into account that our system is
not isolated, but coupled to its surroundings \cite{weiss,zwanzig}.

Either because we do not know the evolution of the environment (on
which we have not control), or because we only look at the system
dynamics, we do not have complete info on the whole (system plus bath).
Think about classical mechanics, for example; one point in the system's
phase space may correspond to several configurations of the total system.
If we cannot fully determine the current state, we are unable to
reconstruct the past or predict the future evolution from $\HS$ alone.
On the other hand, the total phase space of system plus environment is
very large.
As a result the Poincare recurrence time (to return to the initial
state) becomes infinite for most practical purposes: we never find the
coffee warming up after some time ({\em irreversibility}).

In this context the formalism of open systems is important in that we
need not change the current theoretical frameworks to account for
irreversibility.
In addition, this formalism provides some clues for the difficulty in
finding quantum superpositions in everyday macroscopic life.
The theory saves face.

In the following pages we study the effects of the environment on
the dynamics of the system of interest.
Although we start surveying the classical limit, our work will focus
on the quantum domain (non-relativistic).
A constant throughout the manuscript will be to relate the results
obtained with their classical limits (appropriately defined).

Although the formalism of the first chapters is general, the theory
will be applied later on to systems with one degree of freedom.
Examples from solid-state physics where the calculations can be
applied are the phase difference across a Josephson junction, or
collective excitations in lattices, like solitons or fluxons
\cite{weiss, wallraff}.
In these systems openness is important, while they can be approximately
described by a few collective degrees of freedom.
We will also tackle spin systems, like quantum superparamagnets or
ensembles of $1/2$-spins (two-level systems)
\cite{mildri01,vorbrakra04}.

As for the spirit of this manuscript, we have chosen not to repeat here
the details of the calculations included in the articles we have
written in the last three years.
The reader interested in those details can consult our publications,
which are cited as PAPER~I, II,\dots
In any case the thesis is quite self-contained, and can be followed
without recourse to the articles (at least that was our aim).

The text is organized as follows.
We begin reviewing the classical limit in Chap.~\ref{chap:clasico}.
We introduce the bath-of-oscillators formalism for the environment,
obtaining {\em reduced\/} dynamical equations (Langevin \& \FP\
equations).
Then we quantize and proceed to derive dynamical equations for the
reduced density matrix (\QME s; Chap.~\ref{chap:dinamica}).
Chapter~\ref{chap:equilibrio} is devoted to the thermal equilibrium
properties of open quantum systems.
Then in Chapter~\ref{chap:phase-space} we introduce the phase-space
formalism and apply it to \QME s, giving a natural link to the classical
equations.
This first part is general, and we close it by introducing in
Chap.~\ref{chap:metodos} the tools we will employ later on in explicit
calculations: \LRT\ \& the continued-fraction technique to solve
master equations.
The last two chapters (the longest) are devoted to the application of
the general formalism to two problems of interest.
In chapter \ref{chap:aplicationesI} we study the Bloch electron (a
particle in a periodic potential) coupled to a dissipative bath.
Finally, in Chap.~\ref{chap:aplicationesII} we study the equilibrium
and dynamical properties of quantum superparamagnets.

%%%%%%%%%%%%%%%%%%%%%%%%%%%%%%%%%%%%%%%%%%%%%%%%%%%%%%%%%%%%
%%%%%%%%%%%%%%%%%%%%%%%%%%%%%%%%%%%%%%%%%%%%%%%%%%%%%%%%%%%%
%%%%%%%%%%%%%%%%%%%%%%%%%%%%%%%%%%%%%%%%%%%%%%%%%%%%%%%%%%%%
%%%%%%%%%%%%% resumen clasico           %%%%%%%%%%%%%%%%%%%%
%%%%%%%%%%%%%%%%%%%%%%%%%%%%%%%%%%%%%%%%%%%%%%%%%%%%%%%%%%%%
%%%%%%%%%%%%%%%%%%%%%%%%%%%%%%%%%%%%%%%%%%%%%%%%%%%%%%%%%%%%
%%%%%%%%%%%%%%%%%%%%%%%%%%%%%%%%%%%%%%%%%%%%%%%%%%%%%%%%%%%%

\chapter [irreversibility and quantization]
{classical theory: \\ irreversibility and quantization}
\label{chap:clasico}

In this chapter we place the problem in a historical context,
beginning with the 19th century discovery of Brownian motion.
Then we introduce the bath-of-oscillators formalism and address in
this frame the problems of irreversibility and quantization.
We conclude introducing the specific systems we will study in this thesis.
%%

%%%%%%%%%%%%%%%%%%%%%%%%%%%%%%%%%%%%%%%%%%%%%%%%%%%%%%%%%%%%%%%
%%%%%%%%%%%%%%%%%%%%%%%%%%%%%%%%%%%%%%%%%%%%%%%%%%%%%%%%%%%%%%%
%%%%%%%%%%%%%%%%%% historia.

\section [history of Brownian motion]
{history of Brownian motion \cite[Chap.~1]{gardiner}}
\label{sec:int-his}

\begin{figure}
\centerline{\resizebox{7.cm}{!}{%
\includegraphics[angle = -0]{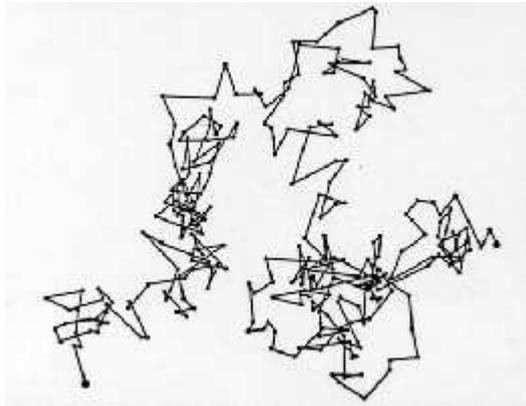}
}
}
\caption{
Possible trajectory observed by Brown.
%
%{\tt http://www.bromot.picts}
%
}
\label{fig:brow-mov}
\end{figure}

In 1827 Father Brown observes under his microscope the irregular
motions executed by pollen grains when suspended in a fluid
(figure \ref{fig:brow-mov}).
%
%___________________________________________________
%___________________________________________________
%___________________________________________________
%___________________________________________________
%___________________________________________________
%___________________________________________________
Such a motion disagreed with physical laws known at the time, where
particles followed smooth and predictable trajectories according to
Newton laws (we left aside chaos, unknown then).
Due to his background, he was a botanist, Brown thought that the
animated motion observed might be a manifestation of life.
Soon he would be discarding such an explanation.

It wasn't until 1905 when Einstein, in his {\it annus mirabilis},
explains in a satisfactory way the phenomenon, rightly called Brownian
motion.%
\footnote{
In 1906 Smoluchowsky's paper is published, putting forward the same
explanation of Brownian motion.
We would like to dispel an urban legend attributing the name to the
irregular and unpredictable motions on stage of the late James Brown.
} % END OF FOOTNOTE
Einstein's approach is based on two main points:
(i) the irregular motion is caused by the impact of the fluid's
molecules on the colloidal grain
(ii) the motion of those molecules is so intricate that the grain
motion can only be described statistically.
From these assumptions an equation for $\W(\q,t)$ could be derived,
giving the probability of the pollen grain being around $\q$ at time
$t$ (a sort of Liouville equation, but in configuration space).
%__________________________
\begin{equation}
\label{ein1905}
\frac{\partial \W (\q,t)}{\partial t}
=
\Desv
\frac {\partial^2 \W (\q,t)}{\partial x^2}
\; .
\end{equation}
%__________________________
This is a diffusion equation where $\Desv$ is related with the
standard deviation of the grain position 
$\langle \q (t)^2 \rangle - \langle \q (0)^2 \rangle = 2 \Desv\, t$.

A few years later Langevin proposed an explanation, a model, which
using his own words was ``infinitely more simple'' than that of
Einstein \& Smoluchowsky.
He assumed that the force due to the fluid can be decomposed
into:
(i') a systematic part (viscous force) and
(ii') a fluctuating force, accounting for the unpredictability of the
collisions.
Incorporating these forces into Newton's equation, Langevin wrote:
%_______________________________________________
\begin{equation}
\label{lan1908}
m \frac {\dif^2 \q}{\dif t^2}
=
-  m \dampp
\frac{\dif \q}{\dif t}
+
\xi (t)
\end{equation}
%_______________________________________________
Here $\dampp$ is the viscous damping parameter and $\xi(t)$ the
fluctuating force, on which he only assumed that it can take random
values both positive and negative.
This was the first example of a stochastic differential equation
\cite{gardiner, vankampen}.

Except for the short time regime (inertial), the treatments of
Einstein and Langevin (more complete) lead to equivalent results.
Both equation (\ref{ein1905}) and (\ref{lan1908}) describe the pollen
grain dynamics.
The Liouville and Newton time evolutions, respectively, get modified
due to the interaction with the fluid, in such a way that the
effective equations for the grain display irreversible dynamics.
%%

%%%%%%%%%%%%%%%%%%%%%%%%%%%%%%%%%%%%%%%%%%%%%%%%%%%%%%%%%%%%%%%%%%%%%%%
\section{problems with irreversibility and quantization}
\label{sec:probII}

At this point we could raise the following question.
Assuming that, though certainly complex, both the pollen grain and the
fluid molecules obey Newton's {\em reversible\/} laws, how comes one can
arrive at {\em irreversible\/} equations? 
This is the irreversibility paradox \cite{zwanzig}:
\begin{itemize}
\item
{\it Loschmidt objection\/}: 
equations (\ref{ein1905}) and (\ref{lan1908}) are {\em not\/}
invariant under time-reversal.
\end{itemize}
To this, a second complain was added: %
\footnote{
These objections were first raised against Boltzmann's kinetic
equation \cite{schwabl}.
} % END OF FOOTNOTE
\begin{itemize}
\item
{\it Zermelo's paradox\/}: 
If we start from a configuration with many pollen grains at the center
of the fluid, we know that after some time they will spread throughout
the volume, according to the diffusion equation (\ref{ein1905}).
That is, we will never find that at some later time all grains return
to the center, the initial configuration.
This is at odds with Poincare's theorem, which asserts that the
trajectories of a bounded system (in phase space) will pass
arbitrarily close to the initial state, after a time called the
{\em recurrence time}.
\end{itemize}
%%%

Finally, along with the irreversibility/recurrence issues, we find the
problem of quantization.
How could we quantize an equation that is irreversible, or stochastic?
In general phenomenological quantization schemes (say, the counterpart
of plugging $-\dampp\,\dot \q$ in Newton equation) can lead to
violations of basic issues (like the very normalization of the
state, or commutation relations).%
\footnote{
Several attempts of direct quantization of dissipative equations are
critically discussed in \cite{calleg83pa} and in \cite[Sec. 2.2]{weiss}.
} % END OF FOOTNOTE
With the formalism introduced in next section %, hamiltonian embebding,
we will try to answer both questions.

%%%%%%%%%%%%%%%%%%%%%%%%%%%%%%%%%%%%%%%%%%%%%%%%%%%%%%%%%%%%%%%%%%%%%%%%%%%%%%%
%%%%%%%%%%%%%%%%%%%%%%%%%%%%%%%%%%%%%%%%%%%%%%%%%%%%%%%%%%%%%%%%%%%%%%%%%%%%%%%
%%%%%%%%%%%% bath of oscillators model

\section{bath of oscillators formalism}
\label{sec:int-for}

Here we discuss the rigorization of the diffusion and Langevin
equations.
This will allow us to address the issues of irreversibility and
quantization.

As physicists we feel comfortable when we see a Hamiltonian.
Therefore we formally write:
%____________________________________________
\begin{equation}
\HTOT = \HS + \HI + \HB
\; .
\end{equation}
%____________________________________________
%
This total Hamiltonian is made of the {\em system\/} Hamiltonian
$\HS$, the environment $\HB$ (the {\em bath}), and the interaction
between them $\HI$.
Naturally, we will have to give content to each part of $\HTOT$.
This looks like a though task though.
On top of the problem of modelling the system itself (think of nuclear
physics), now we have to model everything in its surroundings.
Fortunately, in many cases of interest one has the following
simplifying conditions:
\begin{itemize}
\item[(i)]
The environment ``feels'' only weakly the presence of the system.
Loosely speaking, the bath is big, a macroscopic object (but not
necessarily classical); the liquid carrier in Brown's set up.
The fluid's properties change very little when a pollen grain, or a
handful, are suspended on it.
Naturally, the converse does not hold.
The grain is indeed affected by the bath.

\item[(ii)] 
The environment dynamics is faster that the characteristic
time scales of the system.
For example, due to their microscopic nature the fluid molecules move
around so fast that they look like random to us.

\end{itemize}

Under these conditions the bath can be modelized as a set of harmonic
oscillators linearly coupled to the system \cite{calleg83, mohsmi80}:
%___________________________________________
\begin{equation}
\label{clmodel}
\HTOT
=
\HS (\q, \p)
+
\frac{1}{2}
\sum_\bindex
\bigg[
\pb^2
+
\omega_\bindex^2
\left (
\qb 
+
\frac {u_ \bindex}{\omega_\bindex^2}
\F (\q,\p)
\right )
^2
\bigg]
\end{equation}
%___________________________________________
Here $(\q,\p)$ are the system canonical variables, not necessarily
coordinates and velocities.
Note that we can extract the interaction as $\sim \qb \F(\q,\p)$.
Thus, the coupling is linear in the bath coordinates $\qb$, whereas it can
be arbitrary in the system variables $\F(\q,\p)$. %
\footnote{
The most general Hamiltonian linear in the bath would be
%___________________________________________
\begin{equation}
\label{clmodelgral}
\HTOT
=
\HS (\q, \p)
+
\frac{1}{2}
\sum_\bindex
\Big (
\pb
+
v_ \bindex
\G_ \bindex (\q,\p)
\Big )
^2
+
\omega_\bindex^2
\Big (
\qb
+
\frac {u_ \bindex}{\omega_\bindex^2}
\F_\bindex (\q,\p)
\Big )
^2
\end{equation}
%___________________________________________
where one accounts for a possible coupling with the oscillator
momenta, and for different $\F$ and $\G$ for each mode.
However, to illustrate the formalism is enough with (\ref{clmodel}).
The general (\ref{clmodelgral}) brings little new, while it makes the
calculations cumbersome.
} % END OF FOOTNOTE

When extracting the coupling $\sim \qb \F(\q,\p)$, one also gets a
term $\propto \F^2(\q,\p)$ which only depends on the system variables.
This {\em counter-term\/} cancels the renormalization of $\HS$
produced by a purely linear $ \sim \qb \F$ (generally speaking, an
arbitrary coupling induces fluctuations, dissipation {\em and\/}
renormalizations).
So what we are doing is assuming that $\HS$ is the Hamiltonian to
which we have access experimentally, where all relevant
renormalizations have already been included (see \cite{calleg83} and
what follows).

A further case where Hamiltonian (\ref{clmodel}) can be used is:
\begin{itemize}
\item [(iii)] The bath oscillators provides an ``exact'' description of
the environment.
Think about the phonons in a solid, or the normal modes of the
electromagnetic field, the photons, or less evident cases like
electron-hole excitations.
\end{itemize}

Well, once we have introduced our working Hamiltonian, we would like
to obtain dynamical equations for the (sub)system degrees of freedom.
To this end we consider a dynamical variable $A(\q,\p)$ depending only
on the system, and write Hamilton's equations for it:
$\dif A /\dif t = \{A, \HTOT \}$
with $\{\, , \, \}$ the Poisson bracket and the total
Hamiltonian~(\ref{clmodel}).
Writing similarly the equations for the bath variables, $\qb$ and
$\pb$, we have \cite{gar1999}
%
%_______________________________________________
\begin{eqnarray}
\label{hama}
\frac {\dif A}{\dif t}
&=&
\{A, \HS \}
%_{{\rm P}}
+
\sum_\bindex \frac{u_\bindex^2}{2 \omega_\bindex^2} \{A,\F^2 \}
%_{{\rm P}}
+
\sum_\bindex u_\bindex \qb
\{ A, \F \}
%_{{\rm P}}
\\
\label{hamosc}
\frac {\dif \qb}{\dif t}
&=&
\pb
,
\qquad
\qquad
\frac{\dif \pb}{\dif t}
=
-\omega^2_\bindex \qb -u_\bindex \F
\; .
\end{eqnarray}
%_______________________________________________
%
The first line includes the contribution of the system dynamics in the
absence of interaction (first term), the part from the renormalization
terms (middle), and the coupling to the bath coordinates (last).
The second line is simply the equations of motion for the bath
oscillators (as forced by the system).

Here we start to appreciate the advantages of model~(\ref{clmodel}),
as we know how to solve Eqs.~(\ref{hamosc}).
They are just scores of non-interacting and forced oscillators,
with $\F(\q,\p)$ playing the role of a driving.
This was made possible by the coupling being linear in the bath
coordinates.
Now, from undergraduate physics we know that the solution of a
harmonic oscillator forced by $F(\tau)$ reads:
%_________________________________
\begin{equation}
\label{solosc}
\qb =
\qb^{\rm h}
-
\frac {u_\bindex}{\omega_\bindex}
\int_{t_0}^{t}
\dif \tau
\sin [ \omega_\bindex (t-\tau)]
\F(\tau)
\;.
\end{equation}
%_________________________________
where
$\qb^{\rm h}
=
\qb (t_0) 
\cos [ \omega_\bindex (t-t_0)]
+
\pb (t_0)/\omega_\bindex
\sin [ \omega_\bindex (t-t_0)]
$
is the general solution of the homogeneous equation (free oscillators,
without $F$), while the last term is a particular solution of the
inhomogeneous equation.

Plugging now the solution~(\ref{solosc}) in the system
equation~(\ref{hama}), and integrating it by parts, one finds
($\beta\equiv1/\kT$)
%_________________________________
\begin{equation}
\label{Alan}
\frac {\dif A}{\dif t}
=
\{A, \HS \}
%_{{\rm P}}
+
\{A, \F \}
%_{{\rm P}}
\bigg [
\f (t)
+
\beta
\int_{t_0}^{t}
\dif \tau
\corr (t - \tau)
\frac {\dif \F}{\dif t}
(\tau)
\bigg ]
\end{equation}
%_________________________________
%
We see that the renormalization term in (\ref{hama}) was cancelled by
the term
$-\sum_\bindex \frac{u_\bindex^2}{2 \omega_\bindex^2} \{A,\F^2\}$
arising from the integration by parts.
This was indeed the reason behind including $\sim \F^2$ in the
starting Hamiltonian~(\ref{clmodel}).
It ensures that the renormalization (if physical) was already included
in $\HS$, and not counted twice; then $\HS$ is the Hamiltonian (the
levels) to which we have access in experiments.
As for the other two terms in (\ref{Alan}), we are going to see they
describe fluctuations and relaxation with
%_________________________________
\begin{equation}
\label{fcorr}
\f (t)
=
\sum_\bindex 
u_\bindex
\qb^{\rm h}
\; ;
\qquad
\corr(t - \tau) = \kT \sum_\bindex 
\frac {u_\bindex^2}{\omega_\bindex^2}
\cos [ \omega_\bindex (t - \tau) ]
\end{equation}
%_________________________________
%
We have extracted $\beta$ from $\corr(t - \tau) $ so that it matches
the correlator of $\f (t)$ [Eq.~(\ref{meanf}) below], in
correspondence with the quantum definition
(chapter~\ref{chap:dinamica}).

\begin{figure}
\centerline{\resizebox{9.cm}{!}{%
\includegraphics[angle = -0]{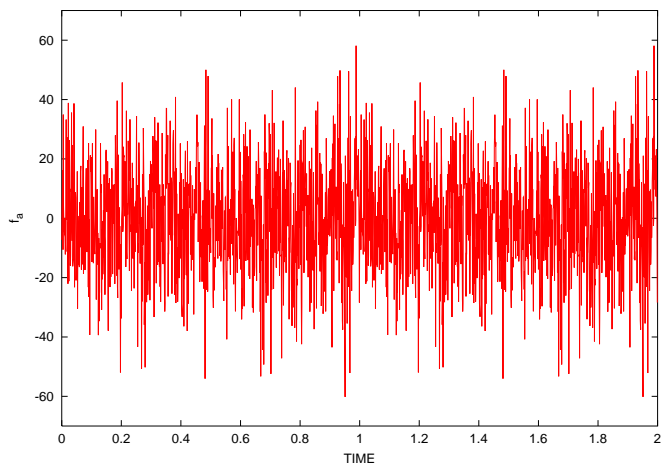}
}
}
\caption{
$\f(t)$ obtained from (\ref{fcorr}) summing over 1000 oscillators with
  initial conditions drawn from a Gaussian distribution.
The force really looks random.
}
\label{fig:noise}
\end{figure}

From the equations above we begin to grasp the structure of a Langevin
equation like (\ref{lan1908}).
We would have the ``random force'' $\f(t)$, which is just the {\em
free\/} evolution of the bath oscillators (see Fig.~\ref{fig:noise}),
while the memory integral in Eq.~(\ref{Alan}) would correspond to the
viscous damping.
The later originates from the inhomogeneous solution
in~(\ref{solosc}), and brings back the dynamics at previous times (a
sort of back-reaction on the system of its previous action on the
bath).
As for the fluctuations, recall that we do not have full control on the
bath, whereas $f(t)$ includes the initial conditions
$\qb(t_0),\pb(t_0)$ for each oscillator.
It seems reasonable to draw them from an equilibrium state at the
initial time.%
\footnote {
This does not mean that the bath is at equilibrium for $t > t_0$, since
the system induces dynamics on the bath through the interaction; see
the $\qb$ evolution~(\ref{solosc}).
} % END OF FOONOTE
%
%This will fix the statistical properties to $\f(t)$.
%
Being harmonic oscillators, the distribution of positions and momenta
would be Gaussian, so we just need to specify the first two moments of
$\f(t)$
%______________________________
\begin{eqnarray}
\label{meanf}
\langle \f (t) \rangle 
&=&
0
\\
\label{corrclas}
\langle \f (t) \f (\tau)
\rangle
&=&
\corr (t - \tau)
\;.
%\qquad\corr \equiv \frac {1}{\beta} \overline \corr
\end{eqnarray}
%______________________________ 
Here we see the link mentioned between the integral kernel and the
force correlator.
The latter can be written in terms of the bath {\em spectral density\/}:
%________________________________
\begin{equation}
\label{jomega}
J(\omega)
=
\pi
\sum_\bindex
\frac {u_\bindex^2}{2 \omega_\bindex}
\delta (\omega - \omega_\alpha)
\end{equation}
%________________________________
in the following way \cite{ing02}
%_______________________________
\begin{equation}
\label{corrcl}
%\beta
\corr (\tau)
=
2
\kT
\int_{0}^{\infty}
\frac {\dif \omega}{\pi}
\frac{J (\omega)}{\omega}
\cos \omega \tau
\; .
\end{equation}
%________________________________
%
Now we are ready to give a couple of examples to give content and fix ideas.
%
%___________________________________________________
%___________________________________________________
%___________________________________________________
%___________________________________________________
%___________________________________________________
%___________________________________________________

%%%%%%%%%%%%%%%%%%%%%%%%%%%%%%%%%%%%%%%%%%%%%%%%%%%%%%%%%%%%%%%%%%%%%%%%55
\subsection {examples}
\label{subs:ejcl}

\subsubsection {\it Brownian particle}

Let us consider the Hamiltonian of mechanics  
$\HS = \p^2/2m + V(\q)$,
and $\F(\q,\p) = -\q$
(i.e., the coupling is linear in the system coordinate as well).
Assume now that the correlator $\corr (\tau)$ is a function strongly
peaked around $\tau=0$, expressing that the system loses memory
quickly ({\em Markovian\/} limit).
Quantitatively
$\corr(\tau) = 2 m\kT \dampp\,\delta (\tau)$,
corresponding to
$J(\omega) = m \dampp\,\omega$,
with $\dampp$ a measure of the coupling strength.%
\footnote{
The spectral density~(\ref{corrcl}) is customarily assumed of the form  
$J(\omega) \sim \omega^\alpha$.
If $\alpha = 1$ the bath is called {\em Ohmic\/} (Markovian case) and
the cases $\alpha \gtrless 1$ are called super/sub-Ohmic.
The choice $\alpha = 1$ if often done for simplicity; it is hard to
solve non-local (integro-differential) equations!
But if required by the physics of the problem, one cannot escape
working with $\alpha\neq1$, as we will do later on.
\label{foot:jomega}
} % END OF FOOTNOTE 
Setting first $A = \q$ and then $A = \p$ in the equation of
motion~(\ref{Alan}) one gets
%
%________________________________
\begin{equation}
\label{lanrig}
\frac {\dif \q}{\dif t}
=
\frac {\partial \HS}{\partial \p}
\; ;
\qquad
\frac {\dif \p}{\dif t}
=
-
\frac {\partial \HS}{\partial \q}
+
\f (t)
-
m \dampp\,\frac {\dif \q}{\dif t}
\end{equation}
%________________________________
This equation reduces to the one proposed by Langevin when $\HS =
\p^2/2m$ (that is, for the free particle).

A stochastic equation with Gaussian  $\f(t)$ and $\delta$-correlation 
$\langle f(t) f(\tau)\rangle \propto \delta(t-\tau)$,
can be converted into an equation for the probability distribution
$\W$
(\FP\ equation; see e.g.~\cite{zwanzig,vankampen}).
For the mechanical problem (\ref{lanrig}) one obtains the
Klein--Kramers equation~\cite{risken}:
%_____________________________________
\begin{equation}
\label{wkk}
%%%%%%%%%%%%%%%%%%
\partial_{t}
\W (\q, \p, t)
=
\Big[
-\p/m \,
\partial_\q
+
V' 
\partial_\p
+
\dampp
\partial_\p
\big(\p+ m\,\kT \partial_\p \big)
\Big ]
\W(\q, \p, t)
\end{equation}
%_____________________________________
%
This \FP\ equation for a particle in a potential is a
generalization of the diffusion equation (\ref{ein1905}). %
\footnote{
Actually, Einstein Eq.~(\ref{ein1905}) follows when neglecting
inertial effects, the so called overdamped regime.
And it was for a free particle; in a potential the overdamped equation
for the distribution $ \W(\q,t) = \int \dif p \W(\q, \p; t)$ is called
the Smoluchowsky equation \cite{risken}
%_____________________________________
\begin{equation}
\dampp\,\partial_t \W (\q, t)
=
\big(
\kT
\,
\partial_{\q \q}^2
-
\partial_\q
V
\big )
\W(\q, t)
\;.
\end{equation}
%_____________________________________
} % END OF FOOTNOTE 

\enlargethispage*{1.cm}

\subsubsection{\it classical Brownian spin/dipole}

Fortunately the equation of motion (\ref{Alan}) is valid for any
Poisson bracket, so we can handle problems more general than the
Brownian particle.
Given that in this thesis we will study simple spin Hamiltonians, we
are going to discuss first the classical case $\HS=\HS(m_x,m_y,m_z)$
with the angular-momentum Poisson algebra
%_____________________________
\begin{equation}
\label{poiss}
\bm (\theta,\phi) 
= 
( \sin \theta \cos \phi, \sin \theta \sin \phi, \cos \theta)
\; ;
\qquad
\qquad
\{
m_i
, m_j
\}
=
\gyr\,
\epsilon_{ijk} m_k
\end{equation}
%______________________________
Here $\theta$ and $\phi$ are the polar and azimuthal angles on the
sphere.
The constant $\gyr$ is the {\it gyromagnetic ratio\/} related with the
spin number by $\hbar S= \mu_{\rm B}/\gyr$, with $\mu_{\rm B}$ Bohr's
magnetron.

In this case we set $A = m_i$ in (\ref{Alan}) and we get (Ohmic bath) %
\footnote{
Note the useful relation:
%________________________________
\begin{equation}
\{ m_i, f(\bm) \}  
= 
-  \gyr 
\Big (\bm \times
\frac {\partial f(\bm)}{\partial \bm}
\Big)_i
\end{equation}
%________________________________
valid for any function of the spin variables $m_i$.
It follows from the Poisson bracket (\ref{poiss}) and the
definition of the vector product $\times$.
} % END OF FOOTNOTE
%___________________________________
\begin{equation}
\label{lanspin}
\frac{1}{\gyr}
\frac {\dif \bm}{\dif t}
=
\bm  
\times
\Big [
\bBef
+
{\bf {\xi}}(t)
\Big ]
-
\frac {1}{S}
\bm
\times \hat {\tL}
\Big (
\bm
\times
\bBef
\Big )
\; ;
\qquad
\bBef :=
-
\frac {\partial \HS}{\partial \bm}
\;.
\end{equation}
%__________________________________
%
This equation has the structure of the phenomenological equation
proposed by Landau--Lifshitz in 1935.
Here $(\partial /\partial \bm )_i \equiv \partial /\partial m_i$ is
the gradient in $\bm$-space, while the noise and the damping include:
%___________________________________________

\begin{equation}
{\bf \xi}(t)
=
\f(t) \frac {\partial F}{\partial \bm}
\; ;
\qquad
\qquad
\tL_{jk}
=
\damps
\frac  {\partial F}{\partial m_j}
\frac  {\partial F}{\partial m_k}
\; .
\end{equation}
%__________________________________________
We have used an Ohmic $J(\omega) = \damps\,\omega$, with $\damps$ the
coupling strength;
the analogue of the damping $\dampp$ in particle problems
(footnote~\ref{foot:jomega}).
We will use two different letters for particles and spins following
the standard notation in the literature (besides, $\dampp$ has
dimensions of frequency $1/t$ whereas $\damps$ is dimensionless).

As we did for the Brownian particle, we can also write a \FP\ equation
associated to the spin Langevin equation~(\ref{lanspin})
%______________________________________________
\begin{equation}
\label{clfp}
\frac{1}{\gyr}
\frac {\partial \W}{\partial t}
=
-
\frac{\partial}{\partial \bm} 
\cdot
\bigg \{
(\bm \times \bBef)
-
\frac {1}{S}
\bm
\times
\hat \tL
\Big [
\bm
\times
\Big (
\bBef
-
\kT
\frac{\partial}{\partial \bm}
\Big )
\Big ]
\bigg \}
\W
\; .
\end{equation}
%___________________________________________
where now $(\partial/\partial \bm) \cdot$ is the divergence operator
in $\bm$-space.
Let us illustrate this with a simple case:
coupling $F \sim \sum S_i$ linear in the spin variables plus axially
symmetric $\HS(\theta)$, as in
$\HS \propto - K \,\cos^2 \theta - B_z \cos \theta$.
Under these conditions $\partial_{\phi} \bBef = 0$ and we can
restrict ourselves to axially symmetric solutions
$\partial_{\phi}\W=0$.
Then, introducing $z = \cos \theta$ and $2 \tau_{\rm N} = \beta/(\gyr
\damps)$, equation (\ref{clfp}) reduces to \cite{gar2000}:
%_______________________________
\begin{equation}
\label{clfptheta}
%\frac{\beta}{\damps g}
2 \tau_{\rm N}
\frac {\partial W}{\partial t}
=
\frac {\partial}{\partial z}
\left [
(1 - z^2) 
\left (
\frac {\partial}{\partial z}
+
\beta \frac {\partial \HS}{\partial z}
\right )
\right ]
\W
\end{equation}
%________________________________
This is nothing but Debye's equation for dipolar relaxation (possibly
the first \FP\ equation describing {\em rotational\/} Brownian motion).

%%%%%%%%%%%%%%%%%%%%%%%%%%%%%%%%%%%%%%%%%%%%%%%%%%%%%%%%%%%%%%%%%%%%
%%%%%%%%%%%%%%%%%%%%%%%%%%%%%%%%%%%%%%%%%%%%%%%%%%%%%%%%%%%%%%%%%%%%
%%%%%%%%%%%%%%%%%%%%%%%%%%%%%%%%%%%%%%%%%%%%%%%%%%%%%%%%%%%%%%%%%%%%

\section {revisiting the paradoxes of irreversibility}
\label{sec:irrepar}

Now we are in a position permitting to address the two objections of
section~\ref{sec:probII}:
\begin{itemize}
\item
First, let us tackle the time-reversal objection due to Loschmidt.
It is clear that the Hamilton equations~(\ref{hama}) and
(\ref{hamosc}) are reversible.
However, the generalized Langevin equation (\ref{Alan}) is completely
equivalent to them.
What happens is that one thinks of reversing the velocities of the
system only.
Therefore, we have a ``practical'' irreversibility: although the whole
system is reversible, we cannot change the sign of the velocities of
all degrees of freedom, including those of the bath (on which in
practice we have little control).
\item
Second, Zermelo's paradox.
The fact that both the effective equations (and the experiments) show
that the pollen grains are never brought back together, is just a
consequence of the total phase space being very large.
Again, we cannot overlook the bath.
Indeed one can estimate the recurrence time and find that it diverges
as we let the number of bath oscillators tend to infinity
\cite{schwabl, greiner-CM2}.
It is only for finite ``baths'' that such time remains finite.
But in practical cases the recurrence time can be several times the
estimated age of the universe.
For practical purposes we can be sure that the grains will never
return to the initial configuration.
\end{itemize}
%%

%%%%%%%%%%%%%%%%%%%%%%%%%%%%%%%%%%%%%%%%%%%%%%%%%%%%%%%%%%%%%%%%%%
%%%%%%%%%%%%%%%%%%%%%%%%%%%%%%%%%%%%%%%%%%%%%%%%%%%%%%%%%%%%%%%%%%
%%%%%%%%%%%%%%%%%%%%%% quantization

\section {quantization}
\label{sec:cuan}

We have just seen how to handle the irreversibility problem (get along
with it, would be more correct).
Let us move on to the issue of quantizing a system with Langevin or
\FP\ dynamics.
The idea is to consider the Hamiltonian (\ref{clmodel}) as the problem
to be quantized.
Then we would just need to transform the variables $(\q,\p)$ \&
$(\qb,\pb)$ into Hilbertian operators and proceed by the book.

Actually, the main derivation of Sec.~\ref{sec:int-for} can be
followed analogously in the quantum case; ``simply'' replace Poisson
brackets $\{\;,\;\}$ by commutators $[\;,\;]$, with $A$ an operator in
the Heisenberg picture.
This is how one derives {\em quantum Langevin equations\/}
\cite{forlewoco88}.
However, these are not as useful/generic as their classical
counterparts; now both $\q$ and the force $\f(t)$ are operators, and
calculations to-the-end exist only in simple cases \cite{haning05,
sto2003}
On top of this, there are no simple recipes (as those in the classical
case), to convert a Langevin equation into an equation for the
distribution (\FP\ like).

To bypass these problems one usually resorts to quantum master
equations.
Recall that the bath is not under control, so it is out of question any
preparation of pure states (of system plus bath).
Then one needs the density matrix formulation (more general) to
compute averages $\langle A \rangle$, our main interest.
If $A$ acts only on the Hilbert space of our system, we can introduce
the {\em reduced density matrix\/}  $\dm$.
Then using the properties of partial tracing, we get $\langle A
\rangle$ just as a trace over the system:
%__________________________________
\begin{equation}
\label{tracepartial}
\langle A \rangle
= 
\Tr_{{\rm s +b}} 
\big (
A \dm_{\rm tot}
\big )
=
\Tr_{{\rm s}}
\big (
A
\dm
\big )
\; ;
\qquad
\qquad
\dm
:=
\Tr_{\rm b} \big ( \dm_{\rm tot} \big )
\;.
\end{equation}
%__________________________________
%
Here ${\rm s + b}$ indicates the original trace over the whole
system+bath using the total density matrix $\dm_{\rm tot}$.

Much as in the classical case the distribution $\W(\q,\p)$ involved
the system variables only, now we have $\dm$ playing this role.
In this sense the equations of motion for $\dm$ will be the analogue of
the \FP\ equations.
Those equations (next chapter) generalize the Von Neumann equation for
the closed evolution:
%_________________
\begin{equation}
\frac {\dif \dm} {\dif t}
=
-
\frac{\iu}{\hbar} 
[\HS, \dm]
\nonumber
\;.
\end{equation}
%__________________
This simply parallels the fact that the Hamiltonian part of the \FP\
was the Liouville equation
%______________________
\begin{equation}
\partial_t \W
=
\{ \HS, \W \}
\; .
\nonumber
\end{equation}
%______________________

%%%%%%%%%%%%%%%%%%%%%%%%%%%%%%%%%%%%%%%%%%%%%%%%%%%%%%%%%%%%%%%
%%%%%%%%%%%%%%%%%%%%%%%%%%%%%%%%%%%%%%%%%%%%%%%%%%%%%%%%%%%%%%%
%%%%%%%%%% the systems we are going to treat

\section {systems studied in this thesis}
\label{sec:sis}

%_____________________________
%_____________________________
%_____________________________
\begin{figure}
\includegraphics[width=14.em,angle=-90]{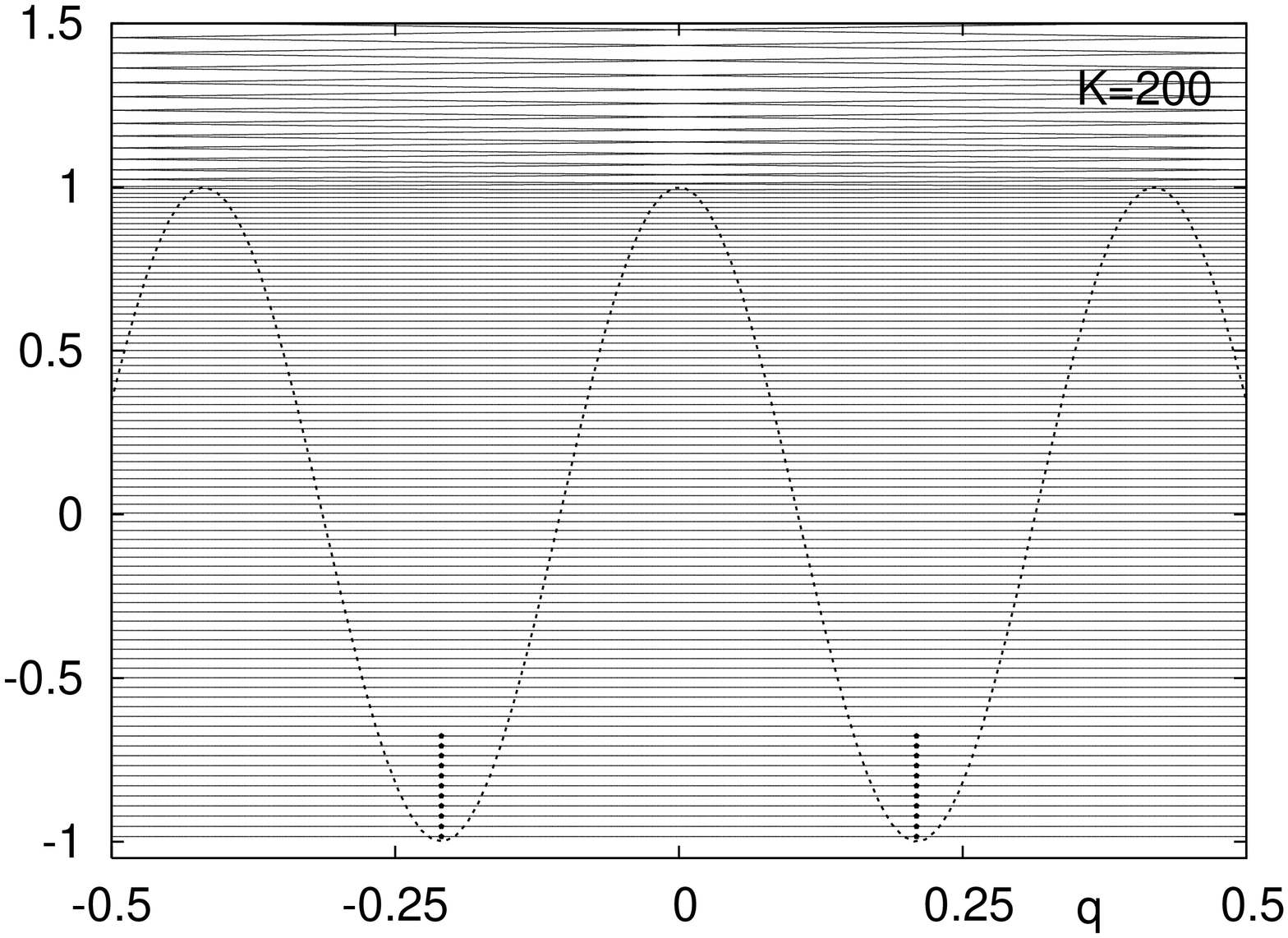}
\includegraphics[width=14.em,angle=-90]{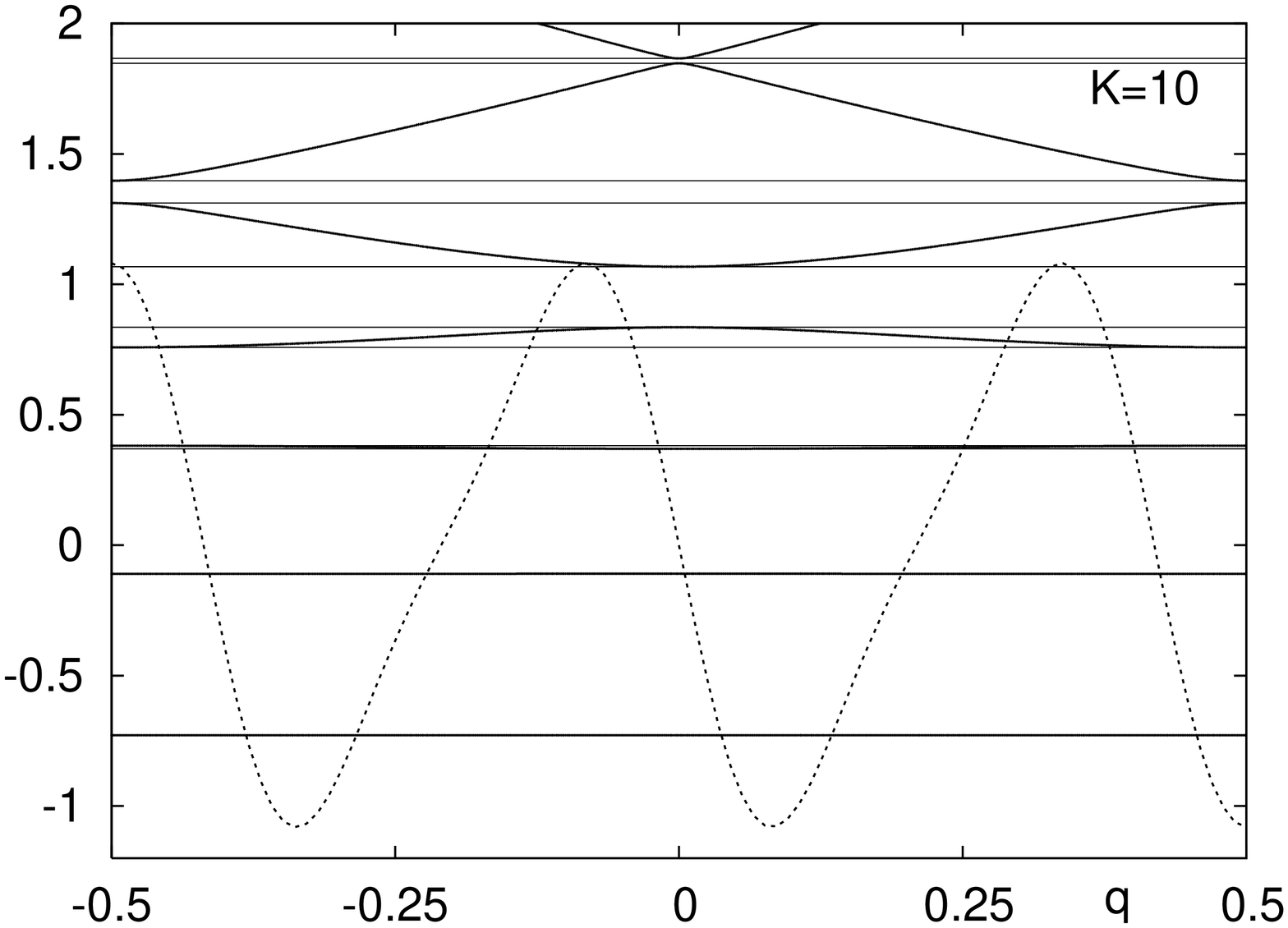}
\caption{
Left: energy bands for the cosine potential~(\ref{cosine}) with
$\kondobar = 200$ (the quantumness parameter $\sim S_0/\hbar$).
Right: ratchet potential (\ref{Vrat}) with $\rat$=0.44 and bands
plotted for $\kondobar = 10$ (see the text).
}
\label{fig:bands}
\end{figure}
%_____________________________
%_____________________________
%_____________________________

We are going to deal with both particle problems and spin Hamiltonians
(translational and rotational problems, respectively).
In the final applications we will consider Hamiltonians with one
degree of freedom (but which could describe {\em effectively\/}
composite systems like a Josephson junction or a molecular magnet).

\subsection {particle systems}
We will consider the standard Hamiltonian
%_________________________________
\begin{equation}
\label{Hpar}
\HS
=
\frac {\p^2}{2m}
+
V(\q)
\end{equation}
%_________________________________
with $[\q, \p] = \iu \hbar$. 
In particular we will address the problem of a quantum particle
running in a periodic potential $V(\q) = V(\q + L)$ plus a constant
force (this is related with the Wannier-Stark problem
\cite{glukolkor02}).
We will study the dynamics in a cosine potential
(Fig.~\ref{fig:bands}):
%________________________________
\begin{equation}
\label{cosine}
V (\q)
=  
V_0 
\cos \q
\end{equation}
%________________________________
and in ratchet potentials (lacking inversion symmetry)
%_____________________________-
\begin{equation}
\label{Vrat}
V(\q)
=
-
V_0
\left [
\sin \q
+
(\rat/2) \sin (2 \q)
\right ]
\;.
\end{equation}
%______________________________

In  figure~\ref{fig:bands} we have plotted the energy bands for these
potentials (see $\jpai$, Sec. 7.3).
We have introduced a dimensionless parameter
%
%_______________________________
\begin{equation}
\label{kondos}
\kondobar =
2 \pi \frac {S_0}{\hbar}
\; ;
\end{equation}
%______________________________
with $S_0$ some relevant/characteristic action; for instance
$S_0 = E_0/\omega_0$ 
with $E_0$ the barrier height and $\omega_0$ the oscillation frequency
at the bottom of the wells.
The parameter $\kondobar$ tells us how classical (or quantum) the
system is.
Letting $\kondobar\to\infty$ gives the classical limit; then the bands tend to
a continuum (for $\kondobar = 200$ the levels of Fig.~\ref{fig:bands}
are already very close).

\subsection {spin systems}
\label{sub:ourspin}

%__________________________________________________________
%__________________________________________________________
%__________________________________________________________
\begin{figure}[t]
\centerline{\resizebox{6.cm}{!}{%
\includegraphics[angle = -0]{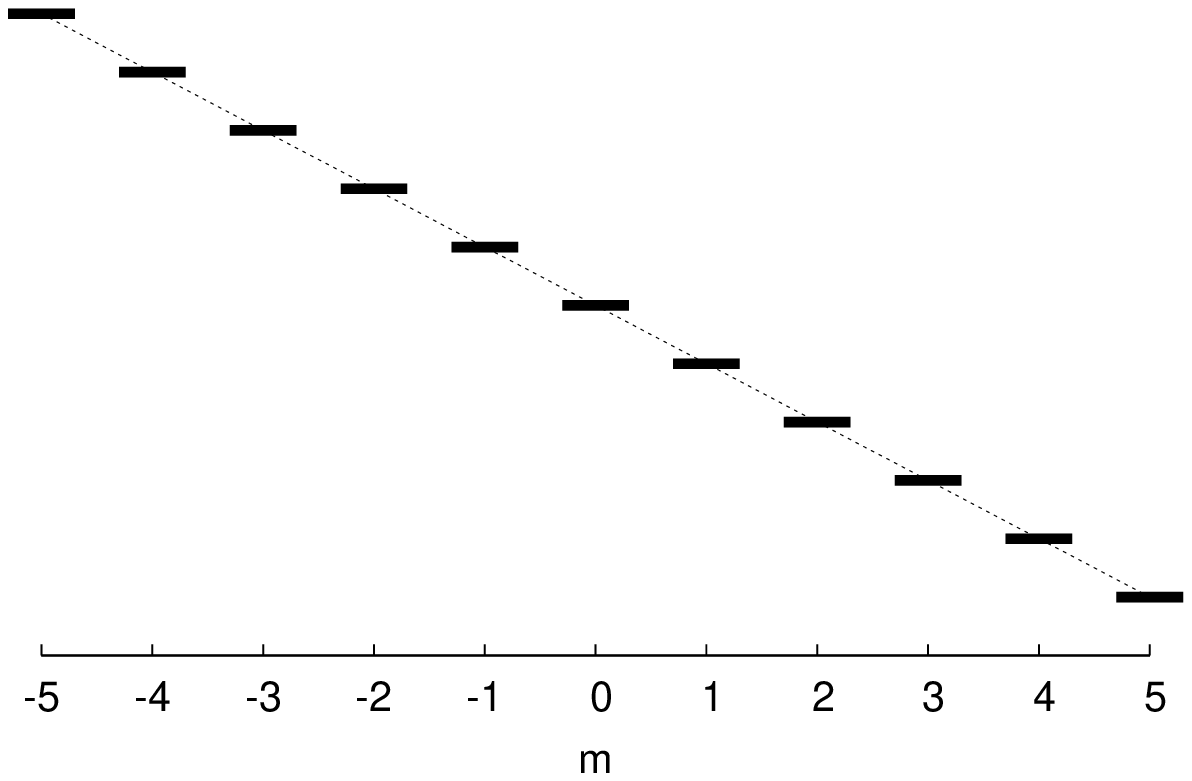}
}
\resizebox{6.cm}{!}{%
\includegraphics[angle = -0]{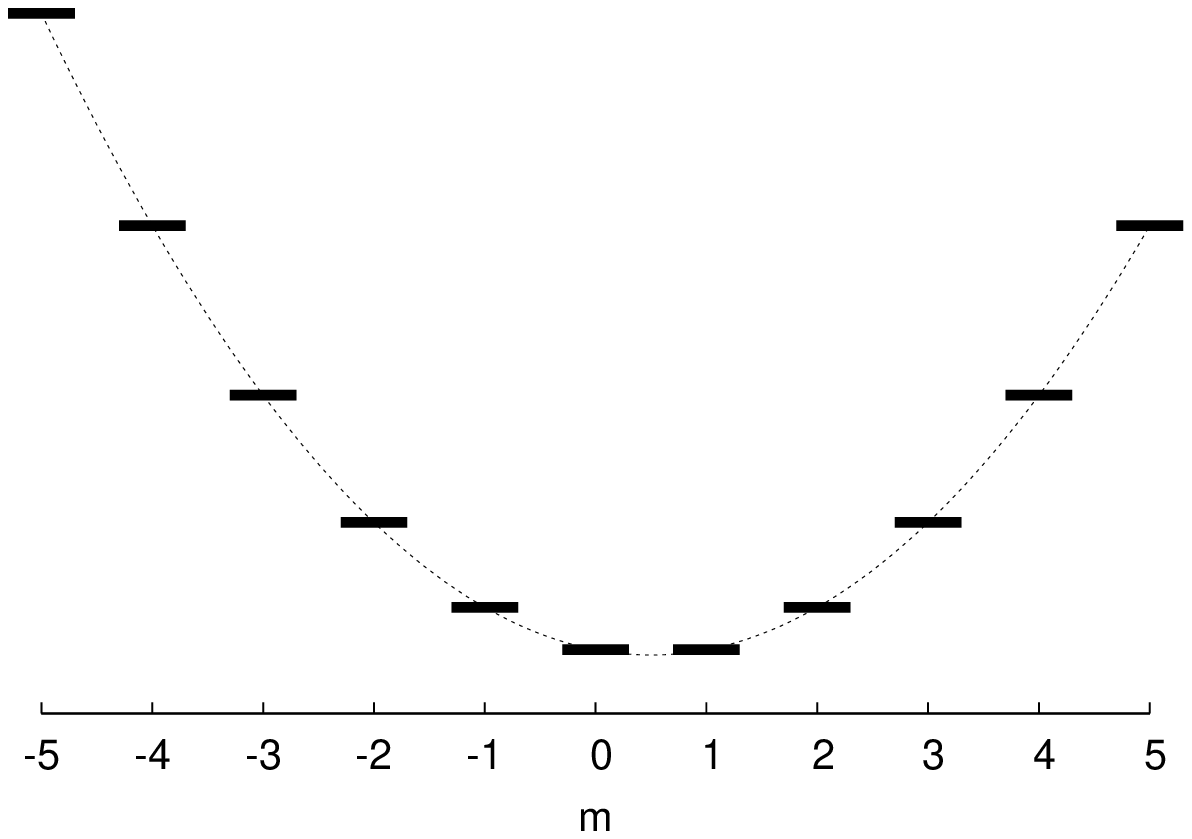}
}
}
\centerline{\resizebox{6.cm}{!}{%
\includegraphics[angle = -0]{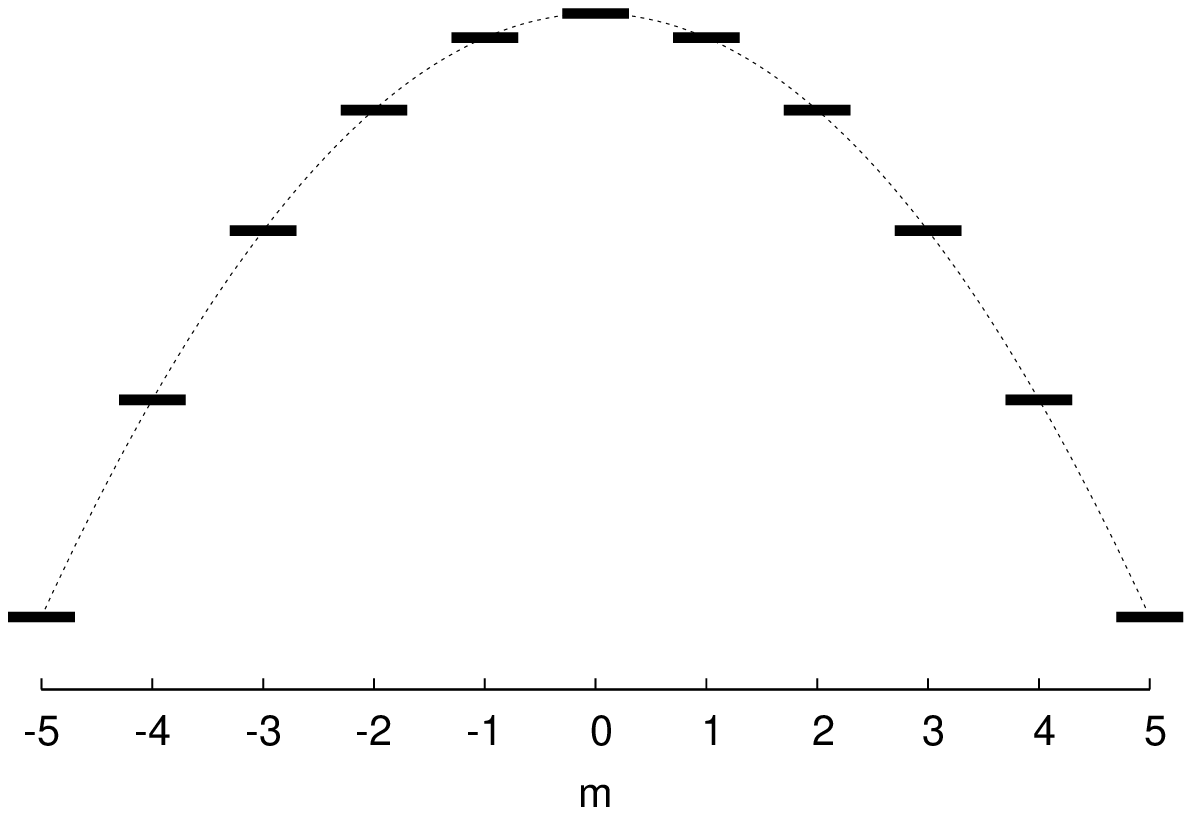}
}
\resizebox{6.cm}{!}{%
\includegraphics[angle = -0]{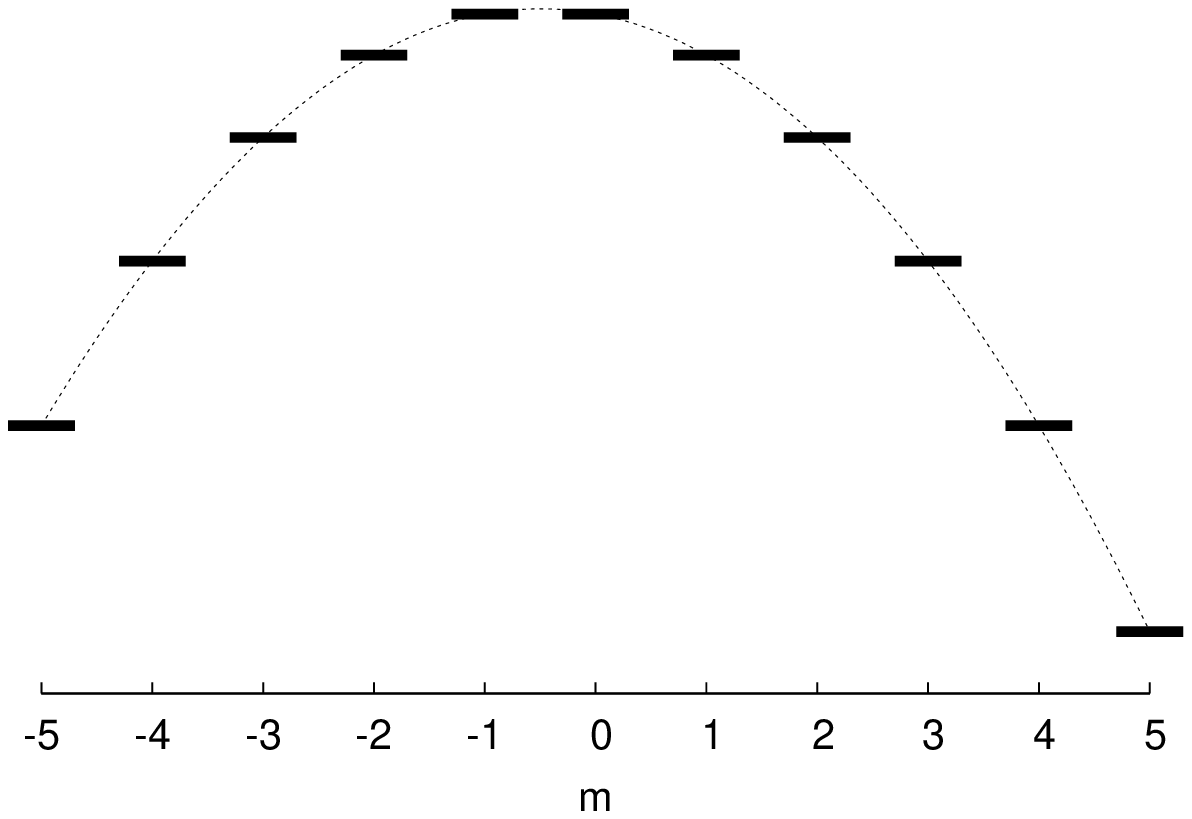}
}
}
\caption{
Energy levels of the spin Hamiltonian~(\ref{spinham}) for $S=5$ and
different anisotropy and field parameters.
Top left: isotropic spin ($\ani=0$).
Top right: spin with $\ani <0$ at $\Bz = \vl D \vl$;
the spectrum has a single well structure.
Bottom: double minima structure, $\ani >0$, in the absence of applied
field (left) and at the first crossing field $\Bz = \ani$ (right).
}
\label{fig:els}
\end{figure}
%___________________________________________________
%___________________________________________________
%___________________________________________________
%
In parallel with mechanical problems, we will consider simple spin
Hamiltonians (Fig.~\ref{fig:els})
%____________________________
\begin{equation}
\label{spinham}
\HS
=
-\ani S_z^2
-
\Bz S_z
%{\bf B}
\; .
\end{equation}
%_____________________________
Here 
$[S_i, S_k] = \iu \epsilon_{ijk} S_k$, 
with 
$S^2 \vl m \rangle = S(S+1) \vl m \rangle$ 
and $S_z \vl m \rangle = \vl m \rangle$.
The above is an effective spin Hamiltonian \cite{white}.
The term $-\Bz S_z$ is the Zeeman coupling to the external field,
while $\ani$ is the anisotropy constant, of spin-orbit origin.
When $D = 0$, Hamiltonian (\ref{spinham}) reduces to the often
studied isotropic spin.
The cases $D \gtrless 0$ correspond to anisotropic spins, with ``easy
axes'' and ``easy plane'' anisotropy, respectively.
Later, in chapter~\ref{chap:aplicationesII} we will focus on $\ani \geq
0$ in~(\ref{spinham}), which is the minimal model for superparamagnets
\cite{blupra2004}.

The classical limit will be taken letting 
$S = (\mu_{\rm  B}/\gyr)/\hbar$ 
tend to infinity.
Note the analogy between $\kondobar$ for periodic potentials and $S$
here.
When taking the limit, it will be convenient to ``normalize'' the
operators as
$\HS = -\ani S^2 (S_z/S)^2 - {\Bz }S  (S_z/S)$.
Then we let $S$ grow keeping constant $DS^2$ and $\Bz S$, the amount
of anisotropy and Zeeman energies.
Mathematically, more and more levels are introduced, towards a
continuum, keeping constant the barriers' height.

%%%%%%%%%%%%%%%%%%%%%%%%%%%%%%%%%%%%%%%%%%%%%%%%%%%%%%%%%%%%%%%%%%%%%%%%%%%
%%%%%%%%%%%%%%%%%%%%%%%%%%%%%%%%%%%%%%%%%%%%%%%%%%%%%%%%%%%%%%%%%%%%%%%%%%%
%%%%%%%%%%%%%%%%%%%%%%%%%%%%%%%%%%%%%%%%%%%%%%%%%%%%%%%%%%%%%%%%%%%%%%%%%%%
%%%%%%%%%%%%%%%%%%%%%%%%%%%%%%%%%%%%%%%%%%%%%%%%%%%%%%%%%%%%%%%%%%%%%%%%%%%
%%%%%%%%%%%%%%%%%%%%%%%%%%%%%%%%%%%%%%%%%%%%%%%%%%%%%%%%%%%%%%%%%%%%%%%%%%%
%%%%%%%%%%%%%%%%  CHAPTER DYNAMICS
%%%%%%%%%%%%%%%%%%%%%%%%%%%%%%%%%%%%%%%%%%%%%%%%%%%%%%%%%%%%%%%%%%%%%%%%%%%
%%%%%%%%%%%%%%%%%%%%%%%%%%%%%%%%%%%%%%%%%%%%%%%%%%%%%%%%%%%%%%%%%%%%%%%%%%%
%%%%%%%%%%%%%%%%%%%%%%%%%%%%%%%%%%%%%%%%%%%%%%%%%%%%%%%%%%%%%%%%%%%%%%%%%%%
%%%%%%%%%%%%%%%%%%%%%%%%%%%%%%%%%%%%%%%%%%%%%%%%%%%%%%%%%%%%%%%%%%%%%%%%%%%

\chapter {dynamics of open quantum systems}
\label{chap:dinamica}

In this chapter we address the dynamics of open quantum systems.
We first discuss the notion of quantum decoherence, through a solvable
example.
To handle more complex systems one can resort to quantum master
equations, the quantum analogue of the \FP\ equations of
the previous chapter.
Here we introduce those equations and give some examples.
We discuss the approximations involved in their derivation, as well as
other invoked in solving master equations.

%%%%%%%%%%%%%%%%%%%%%%%%%%%%%%%%%%%%%%%%%%%%%%%%%%%%%%%%%%%%%%%%%%%%%%%%%
%%%%%%%%%%%%%%%%%%%%%%%%%%%%%%%%%%%%%%%%%%%%%%%%%%%%%%%%%%%%%%%%%%%%%%%%%
%%%%%%%%%%%%%%%%%%%%%%%%%%%%%%%%%%%%%%%%%%%%%%%%%%%%%%%%%%%%%%%%%%%%%%%%%

\section{quantum decoherence}
\label{sec:decoherence}

In the previous chapter we discussed how irreversibility could emerge
from reversible equations.
Now we would like to see how the classical world we are familiar with
arises out of the microscopic quantum world it is made of.
A proposed explanation for such emergence resorts to the coupling with
the environment, in a {\em irreversible\/} process called decoherence
(see \cite{zur91} for an introduction to the trendy topic).
It can be seen as a consequence of having the dynamics of very many
degrees of freedom but only looking at some reduced dynamics.
The notion can be introduced with a simple example, due to Zurek
\cite{zur82}, which is fully solvable.

The question one aims to answer is: if everything around us, including
ourselves, is made of atoms obeying the laws of quantum mechanics, how
comes we do not find superpositions of states in the macroscopic
world?
The most famous/shocking of those superpositions is due to
Schr\"odinger:
%
%_____________________
\begin{equation}
\label{cat}
\vl \Psi \rangle
=
\frac{1}{\sqrt {2}}
\vl {\rm alive} \; {\rm cat} \rangle
+
\frac{1}{\sqrt {2}}
\vl {\rm dead} \; {\rm cat} \rangle
\end{equation}
%_____________________
%
%\newpage\noindent
Or others equally exotic like
$
%_____________________
%\begin{equation}
\vl \Psi \rangle
\propto
%\frac{1}{\sqrt {2}}
\vl {\rm here}  \rangle
+
%\frac{1}{\sqrt {2}}
\vl {\rm there} \rangle
%\end{equation}
%_____________________
$,
i.e., the possibility of having delocalized objects.
To address these points let's go through the promised simple example.%
\footnote {
It is worth recalling that the superposition (\ref{cat}) is different
from a statistical mixture (as Schr\"odinger himself noted).
In the density-matrix language, a statistical mixture and a
superposition are represented by
%_____________________
\begin{eqnarray}
\varrho^{\rm mixed}
=
\left (
\begin{array}{cc}
1/2 & 0
\\
0 & 1/2
\end{array}
\right)
\; ;
\qquad
\varrho^{\rm \Psi}
=
\left (
\begin{array}{cc}
1/2 & 1/2
\\
1/2 & 1/2
\end{array}
\right)
\; .
\end{eqnarray}
} % END OF FOOTNOTE   

As usual, we will consider that the system interacts with many degrees
of freedom, so that the total quantum Hamiltonian is
$\HTOT = \HS + \HI + \HB$.
For simplicity, we assume that both system and bath have discrete and
non-degenerate spectra.
In the basis of eigenstates of $\HS$ and $\HB$ one can write
%_________________________________
\begin{equation}
\HS
=
\sum_i \delta_i \vl s_i \rangle \langle s_i \vl
\; ;
\qquad
\HB
=
\sum_\bindex \epsilon_\bindex \vl e_\bindex \rangle \langle e_\bindex \vl
\; .
\end{equation}
%_________________________________
Here $\vl e_\bindex \rangle$ are the eigenstates of the bath
oscillators, and $\vl s_j \rangle$ those of the system.
In this basis, the $\HI$ in the bath-of-oscillators
model~(\ref{clmodel}) can be written as
\[
\sum_{i \bindex} \gamma_{i \bindex} \vl s_i \rangle \langle s_i \vl
\otimes \vl e_\bindex \rangle \langle e_\bindex \vl + \sum_{i i'
\bindex \bindex'} \sigma_{i i'\bindex \bindex'} \vl s_i \rangle
\langle s_{i'} \vl \otimes \vl e_\bindex \rangle \langle e_{\bindex'}
\vl
\]
To proceed we assume that
$\sigma_{i i'\bindex \bindex'} \ll \gamma_{i \bindex}$,
meaning that decoherence times are much shorter than relaxation processes.
Then
%______________________________
\begin{equation}
\HI
\cong
\sum_{i \bindex} \gamma_{i \bindex} 
\Big (
\vl s_i \rangle \langle s_i \vl \otimes \vl e_\bindex
\rangle \langle e_\bindex \vl
\Big )
\;,
\end{equation}
%______________________________
which allows to solve analytically the problem.
If 
$\sigma_{i i'\bindex \bindex'} \not \ll \gamma_{i  \bindex}$ 
we would have to resort to a quantum master equation (next section),
and solve it numerically, most likely.

Let us assume now that our system is in a superposition
like~(\ref{cat}) and let us study the dynamical evolution.
Since we have to specify the state of the bath too, we assume for
simplicity that it is a pure state (though everything can be repeated
for the realistic case of a mixed bath).
Then, the initial state reads:
%________________________________
\begin{equation}
\label{Psicero}
\vl \Psi (t=0) \rangle
=
\sum_j a_j \vl s_j \rangle
\otimes
\sum_\bindex b_\bindex \vl e_\bindex \rangle
\end{equation}
%________________________________
%
Now with the $\HI$ considered, the time evolution at time $t$ results in the
{\em entangled\/} state:%
\footnote{
The state~(\ref{Psicero}) is ``separable'', that is, it can be
expressed as the tensor product of a state of the system's Hilbert
space times a state in the bath's Hilbert space.
On the contrary, (\ref{entangled}) is a linear combination of
separable states, and it is no longer possible to assign state
vectors to the system and bath separately.
These states are called {\it entangled}.
} % END OF FOOTNOTE
%_________________________________
\begin{equation}
\label{entangled}
\vl \Psi (t) \rangle 
= 
\sum a_j b_\bindex\,
\e^{-\frac{\iu}{\hbar} t(\delta_j + \epsilon_\bindex + \gamma_{j \bindex})} 
\vl s_j \rangle
\otimes 
\vl e_\bindex \rangle
\end{equation}
%_________________________________
Calculating now the reduced density matrix~(\ref{tracepartial}), that
is,
$\varrho = \Tr_{\rm b} (\vl \Psi (t) \rangle \langle \Psi (t) \vl )$
one finds in the $\{ \vl s_j \rangle \}$ basis
%____________________________________
\begin{eqnarray}
\varrho_{jj} (t)
&=&
\vl a_j \vl^2
\\
\varrho_{ij}(t)
&=&
a_i a^*_j\,
\e^{- \frac {\iu}{\hbar} t(\delta_i - \delta_j)}
z_{ij}(t)
\; ;
\qquad
z_{ij}(t)
=
\sum_\bindex
\vl b_\bindex \vl^2 \e^{-\iu \omega_\bindex^{ij} t}
\end{eqnarray}
%____________________________________
%
Here
$a_i a^*_j \e^{-\frac{\iu}{\hbar} (\delta_i - \delta_j)t}$
is just the coherent rotation due to the evolution under $\HS$, while 
$z_{ij}(t)$
is due to the interaction with the environment
[$\omega_\bindex^{ij}
= 
(\gamma_{i\bindex} - \gamma_{j\bindex})/\hbar$].

\begin{figure}
\centerline{\resizebox{7.cm}{!}{%
\includegraphics[angle = -0]{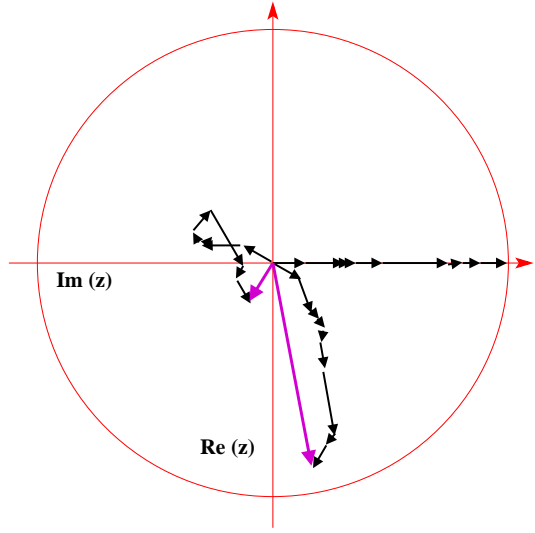}
}
}
\caption{
Several realizations of a random walk in 2 dimensions:
$z= \sum \vl b_\alpha \vl \e^{\iu \omega_\alpha t}$.
The total sum is constrained as $\sum \vl b_\alpha \vl^2 = 1$ (see text).
}
\label{fig:randomwalk}
\end{figure}

In general $ \omega_\bindex^{ij} \neq 0$ and hence $z_{ij}(t)$ results
to be the sum of many oscillating factors with different frequencies.
Then at long enough times
$\langle z_{ij}(t) \rangle \to 0$, 
so that 
$ \e^{-\iu \omega_\bindex^{ij} t }$
can be seen as a random phase.
Therefore 
$\vl b_\bindex \vl^2 \e^{-\iu \omega_\bindex^{ij} t}$
becomes a random vector in the complex plane.
The problem can thus be mapped onto the random walk in 2D, where 
each move is
 $\vl b_\bindex \vl^2 \e^{-\iu 
  \omega_\bindex^{ij} t}$
and
$\sum \vl b_\bindex \vl^2 = 1$
is kept fixed (see Fig.~\ref{fig:randomwalk}).
%
%___________________________________________________
%___________________________________________________
%___________________________________________________
%___________________________________________________
%___________________________________________________
%___________________________________________________
%
Then $z_{ij}(t)$ is a random variable with a probability distribution
given by $P(z) \propto \e^{-N z^2}$, where $N$ is the number of bath
oscillators.
Therefore at sufficiently long times, and considering the bath to be
macroscopic ($N \to \infty$), one writes
%___________________
\begin{equation}
z_{ij}(t)  \cong 0
\; .
\end{equation}
%_________________
%
Or in other words, the off-diagonal elements drop to zero and the
original superposition $\sum_j \alpha_j \vl s_j \rangle$ is ``destroyed''.
In the sense that the reduced density matrix is indistinguishable from a
statistical mixure with $\vl a_i \vl^2$ as diagonal elements.

The time  $\td$, from which we can take $z_{ij}(t) \cong 0$, is called
the decoherence time.
The loose of coherence is a consequence of looking only at the system.
In fact, everything follows from the unitary evolution of the total
quantum system; that is, the coherence is still out there, but now is
diffused/shared between a large number of degrees of freedom.
In a sense, this decoherence follows from the ``non-locality'' of
quantum mechanics, as the state~(\ref{entangled}) is fully entangled
between system and bath.

Finally, is worth to recall that every density matrix is always
diagonalizable (as it is Hermitian).
Or conversely, given a diagonal density matrix, it can be seen as
non-diagonal by some other basis.
Then, the concept of decoherence seems to be ``basis dependent''.
What happens, decoherentists say, is that the interaction with the
bath ``determines'' in which basis $\dm$ will be diagonal; the so
called {\it pointer basis} \cite{zur91}.
It seems that Nature chose us to be localized objects, as well as
being alive or death.
To give some numbers, the decoherence time for a superposition of dust
grains with a diameter $10^{-3}$\,cm, set $1$\,cm apart is estimated
to be around $10^{-36}$ seconds \cite{kiejoo98}.
That is, for most practical purposes, and for most macroscopic
objects, quantum superpositions cannot be observed.%
\footnote{
The decoherence approach does not exclude the existence of those
superpositions.
It only asserts that they vanish at too short a time to be observable
\cite{joo99}.
For a discussion of the state of the play in the search of quantum
superpositions in macroscopic objects, see Leggett's insightful
review \cite{leg2002}.
On the other hand, decoherence does not seem to solve the
``measurement problem''.
It is true that it provides a diagonal $\dm$, with its elements giving
the outcome probabilities of each eigenstate.
But it does not furnish the mechanism by which, is a given
measurement, the system would ``collapse'' in the corresponding
subspace.
}

%%%%%%%%%%%%%%%%%%%%%%%%%%%%%%%%%%%%%%%%%%%%%%%%%%%%%%%%%%%%%%%%%%%%%%%%%%%%%
%%%%%%%%%%%%%%%%%%%%%%%%%%%%%%%%%%%%%%%%%%%%%%%%%%%%%%%%%%%%%%%%%%%%%%%%%%%%%
%%%%%%%  equations maestras

\section{derivation of master equations}
\label{sec:dedqme}

It will not be always possible to express $\dm(t)$ analytically, as in
the previous section.
Our goal here is to derive differential equations for $\dm$.
In contrast to the classical case [Eq.~(\ref{Alan})], we do not know
how to obtain in general an exact equation for the
model~(\ref{clmodel}).
Therefore, we will be happy with an approximate equation.
The main assumption involved will be to consider that the system-bath
interaction is {\em weak}.

The most usual techniques in deriving master equations are those of
projection operators \cite{zwanzig, brepet}, path integral
\cite{calleg83pa, kleinert,kargra97,ank2003qme}, or cumulant
expansions \cite{vankampen,dattpur}.
For our purposes, it will suffice with a simpler derivation, based on plain
quantum-mechanical perturbation theory.
The approximations invoked will be clearly seen as we proceed.

We will start working with $\dm_{\rm tot}$; we would trace later on to
obtain the effective dynamics for $\dm$ [Eq.~(\ref{tracepartial})].
The evolution of $\dm_{\rm tot}$ is given by
%_____________________________________-
\begin{equation}
\label{dmtotev}
\dm_{\rm tot} (t)
=
U(t;\tcero)
\dm_{\rm tot}(\tcero)
U^{+} (t;\tcero)
\; ;
\qquad
U(t; \tcero)
=
\e^{-\frac{\iu}{\hbar}  \HTOT (t -t_0)}
\end{equation}
%_____________________________________
where we have assumed that $\HTOT$ is not time dependent.
This is not essential though; we could consider that $\HTOT$ depends
on $t$, and then the evolution operator would be a time-ordered exponential
$U(t;\tcero) = {\cal T} [\exp \{ - \iu/\hbar \int \dif t' \HTOT(t')\} ]$.

In what follows it will be convenient to split $\V = \HI$ from $\HO =
\HS + \HB$, and assume that the {\bf system-bath interaction is weak}.
That is, we will treat $\V = \HI$ as a perturbation (we shall quantify
later what we mean by ``weak''; be patient!).
We want to calculate $U(t;\tcero)$ perturbatively in $\V=\HI$;
to that end, we will use the following Kubo-type identity
\cite[p.~148]{kuboII}: 
%________________________________________
\begin{equation}
\label{kuboreal}
\e^{-\frac{\iu}{\hbar}  (\HO + \V) (t-\tcero)}
=
\e^{-\frac{\iu}{\hbar}   \HO  (t- \tcero)}
\left [
1
-
\frac{\iu}{\hbar}
\int^{t-\tcero}_0  \dif s  \; \e^{\frac{\iu}{\hbar}  \HO \, s } 
\, \V \, 
\e^{- \frac{\iu}{\hbar}  (\HO + \V) s} 
\right ]
\end{equation}
%________________________________________
%
This equation%
\footnote{
The proof of (\ref{kuboreal}) follows multiplying both sides by
$\e^{+\frac{\iu}{\hbar} \HO (t-t_0)}$
and differentiating with respect to $t$.
} % END OF FOOTNOTE
can be iterated by expressing
$\e^{ -  (\iu/\hbar)  (\HO + \V) s }$
in the integrand in just the same way, and so on.
To second order in $\V$ one has:
%_____________________________
\begin{equation}
\label{U2nd}
U(t; \tcero)
\cong
\e^{-\frac{\iu}{\hbar} \HO (t-t_0)}
\left [
1
-
\frac{\iu}{\hbar}
\int^{t-\tcero}_0  \dif s
\;
\V(s)
-
\frac{1}{\hbar^2}
\int^{t -\tcero}_0  \dif s
\int^{t-\tcero}_{s}  \dif u
\;
\V(u) \V(s)
\right ]
\end{equation}
%_____________________________
where $V(s)$ is the Heisenberg evolution under $\HO$ (interaction
picture), that is,
$\V(s)=\e^{+(\iu/\hbar)  \HO\,s} \V \e^{-(\iu/\hbar)\HO\,s}$.
Taking the Hermitian conjugate of (\ref{U2nd}) we get a similar
expression for $U^{+}(t;\tcero)$.
Both results can be inserted now in (\ref{dmtotev}), giving the formal
time evolution of $\dm_{\rm tot}(t) = U \dm_{\rm tot}(t_0) U^+$.

As we seek for a differential equation, we can differentiate $\dm_{\rm
  tot}$ with respect to $t$.
Keeping only terms to order $\V^2$, one gets:
%_______________________________
\begin{equation}
\label{dmtot2nd}
\frac{\dif \dm_{\rm tot}}{\dif t}
=
-\frac{\iu}{\hbar}
[\HO, \dm_{\rm tot}]
- \frac{\iu}{\hbar}[\V, \widetilde {\dm}_{\rm tot} ]
+
\frac{1}{\hbar^2}
\int_{0}^{t-\tcero}
\dif s
\;
\Big (
\V [\widetilde {\dm}_{\rm tot}, \V(s - t)] + [\V(s-t),\widetilde {\dm}_{\rm
    tot}] \V  
\Big )
\end{equation}
%_______________________________
where the operators without argument are assumed in the Schr\"odinger
picture, whereas $\V(s-t)$ is the formal Heisenberg evolution under
$\HO$.
The tilde in $\widetilde {\dm}_{\rm tot}$ means unperturbed
evolution under $\HO$, that is, 
$\widetilde {\dm}_{\rm tot} 
= 
\e^{-(\iu/\hbar) \HO(t-\tcero)} \dm(t_0) \e^{+(\iu/\hbar) \HO(t-\tcero)}$.
Keep in mind that $\dm_{\rm tot} \neq \widetilde {\dm}_{\rm tot}$.

Now we have to trace over the bath to get the evolution equation for
$\dm$, Eq.~(\ref{tracepartial}).
It is convenient to write the system-bath coupling as ($\F=\F^+$, $\B = \B^+$)
%____________________________
\begin{equation}
\label{HIfactorised}
\HI
=
\F \otimes \B
\end{equation}
%____________________________
with $\F$ acting only on the system's Hilbert space and $\B$ over the
bath's.%
\footnote{
The more general form  $\sum_j \F_j \otimes \B_j$ is easily handled
adding summations here and there.
} % END OF FOOTNOTE
Now we will assume that {\bf at the initial time $\tcero$ the density
  matrix  $\dm_{\rm tot}$ is given by the following tensor product}
  (decoupled initial conditions):
%
%_____________________________________-
\begin{equation}
\label{dmfactorised}
\dm_{\rm tot} (\tcero)
=
\dm_{\rm s}
\otimes
\dm_{\rm b}
\; .
\end{equation}
%_____________________________________-
%
Besides, $\dm_{\rm b}$ would be the equilibrium density matrix of the
bath in the absence of the system, i.e.,
 $\dm_{\rm b} 
= 
\e^{-\beta \HB}/\ZB $ with $\ZB 
= 
\Tr_{\rm b} (\e^{-\beta \HB})$.
That is, the bath is considered at thermal equilibrium before
``meeting'' the system [see paragraph before (\ref{meanf}) and
(\ref{corrclas})].
The factorized form~(\ref{dmfactorised}) is difficult of justify, but
it is an assumption required to proceed further with the derivation.
To be more reassured, think that in this work we are interested in the
long time behavior, where the system would have ``forgotten'' any
initial conditions.

After all this preparations, let us trace over the bath's degrees of
freedom, $\Tr_{\rm b}(\cdots)$.
Thanks to (\ref{dmfactorised}), we know that 
$\widetilde {\dm}_{\rm tot} (t) 
=
\dm_{\rm s}(t) \otimes \dm_{\rm b}(t)$
because $\HS$ and $\HB$ act on different Hilbert spaces ($[\HS, \HB] =
0$).
Besides, 
$\dm_{\rm b} \propto
\e^{-\beta \HB}\to \dif \dm_{\rm b}/\dif t = 0$
and
$\dm_{\rm s}(t) 
= 
\e^{-(\iu/\hbar) \HS (t-\tcero)} 
\dm_{\rm s}(\tcero) 
\e^{+(\iu/\hbar) \HS (t-\tcero)}$.
Therefore, for the first order term of (\ref{dmtot2nd}) we have
%__________________________
\begin{equation}
\frac{\iu}{\hbar} 
\Tr_{\rm b}
\left (
 [\V, \widetilde {\dm}_{\rm tot}(t)]
\right )
=
\frac{\iu}{\hbar} 
\langle \B \rangle_{\rm b}
[\F, \widetilde{\dm}_{\rm s}(t)]
\end{equation}
%__________________________
where 
$\langle \cdots \rangle_{\rm b} \equiv \Tr_{\rm b}(\cdots \dm_{\rm b})$.
In most cases of interest the bath is such that 
$\langle \B \rangle_{\rm b} = 0$,
and the above term drops.
For the first term in the integrand of (\ref{dmtot2nd}) we have:
%____________________________________
\begin{eqnarray}
\label{traceex}
\Tr_{\rm b}
\left (
\V [\widetilde {\dm}_{\rm tot}, \V(s - t)]
\right )
=
&+&
\F \widetilde {\dm}_{\rm s}(t) 
\F (s-t)
\Tr_{\rm b} \left ( 
\B {\dm}_{\rm b}
\B (s-t) 
\right )
\\ \nonumber
&-&
\;
\F \F(s-t) \widetilde {\dm}_{\rm s}(t) 
\Tr_{\rm b} \left ( 
\B 
\B (s-t)
{\dm}_{\rm b}
\right )
\\ \nonumber
=
&+&
\F \widetilde {\dm}_{\rm s}(t) 
\F (s-t)
\corr (s-t)
-
\F \F(s-t) \widetilde {\dm}_{\rm s}(t)
\corr (t-s)
\end{eqnarray}
%____________________________________
where
%_____________________________________
\begin{equation}
\label{corrdef}
\corr (\tau)
=
%\Tr_{\rm b}
%\left (
\langle
\B (\tau) \B
\rangle_{\rm b}
%\right )
\end{equation}
%_____________________________________
is the correlator of the bath operator.
This shows a clear analogy with the classical case [see
  Eq.~(\ref{corrcl})], so we have used the same letter to denote it.

Notice that to derive (\ref{traceex}), the
factorization~(\ref{dmfactorised}) has been essential;
then 
$\widetilde {\dm}_{\rm
tot} (t) = \dm_{\rm s}(t) \otimes \dm_{\rm b}(t)$.
As the coupling $\HI$ is factorized too, Eq.~(\ref{HIfactorised}), we
were able to take the parts acting on the system out of the trace
action.
Proceeding in the same way with the term 
$[\V(s-t),\widetilde {\dm}_{\rm tot}] \V $
in (\ref{dmtot2nd}), we finally arrive at
%________________________________
\begin{equation}
\label{dmegral}
\frac{\dif \dm}{\dif t}
=
-
\frac{\iu}{\hbar}
[\HS, \dm]
%-
%\frac{\iu}{\hbar}
%\langle V \rangle_{\rm b}
%[\F, \dm]
+
\R
\;,
\end{equation}
%______________________________
with the following {\em relaxation term}
%____________________________
\begin{equation}
\label{Rsch}
\R
=
-\frac{1}{\hbar^2}
\int_{\tcero}^t
\dif s
\,
\Big \{
\corr(t-s)
[\F, \F(s-t) \dm]
-
\corr(s-t)
[\F, \dm \F(s-t)]
\Big \}
\end{equation}
%________________________________
%
Note that we have replaced $\widetilde {\dm}_{\rm s}$ by $\dm$ inside
$\R$, as they differ in second order terms, while $\R$ is already of
second order in $V$.

Now we can be more specific on what we mean by weak interaction.
The equation obtained, Eq.~(\ref{dmegral}), was based on the
expansion~(\ref{U2nd}).
There we assumed that $\V$ was a perturbation on $\HO$, so that it
made sense to retain only the first few terms in the expansion.
Well, but (\ref{U2nd}) would only be valid at short times,
$\V t /\hbar \ll 1$,
because letting $\vl t-\tcero \vl \to \infty$ the integral could
 become very large and break down the perturbation theory
 \cite{galindo}.
So, is it that the master equation only makes sense at short times?
This would reduce considerably its applicability (let alone the
description at short times being biased by the decoupled initial
condition).
Fortunately, the fact is that we implicitly assume the existence of
correlation time $\tc$ obeying
%____________________________
\begin{equation}
\corr (\tc) \cong 0, \qquad {\rm para} \quad t > \tc
\, .
\end{equation}
%___________________________
%
Therefore, it is this $\tc$ what limits the validity of the expansion
as it bounds the integral term.
Following Van Kampen \cite[Chap.~XVII]{vankampen} one introduces some
$\dampp$ as a measure of the coupling strength (in the appropriate
units $[\dampp] = [t^{-1}]$), and asserts that the master
equation~(\ref{dmegral}) is valid if
%_____________________________
\begin{equation}
\label{validity:range}
\dampp\,\tc \ll 1
\; ,
\end{equation}
%_____________________________
quantifying in this way what we mean by weak-coupling.

%%%%%%%%%%%%%%%%%%%%%%%%%%%%%%%%%%%%%%%%%%%%%%%%%%%%%%%%%%%%%%%%%%%%%%%
%%%%%%%%%%%%%%%%%%%%%%%%%%%%%%%%%%%%%%%%%%%%%%%%%%%%%%%%%%%%%%%%%%%%%%%

\section{Heisenberg evolution}

In much the same way as we obtained an equation governing the
evolution of the reduced density matrix, we can obtain the
corresponding equation for any operator $A$ within the Heisenberg
picture.

The Heisenberg evolution for $A$ is given by 
$A(t) = U^+(t, \tcero) A(\tcero) U(t, \tcero)$.
We can use the same approximate expressions~(\ref{U2nd}) for $U$ and
$U^+$ and then differentiate with respect to $t$.
Keeping terms to second order in $\V$, we would get an equation $\dif
A/\dif t = (\cdots)$ analogue to (\ref{dmtot2nd}).
Assuming then
(i) the factorized initial conditions~(\ref{dmfactorised}),
(ii) that the operator $A$ only acts on the system's variables
$A = A \otimes \mathbb {I}_{\rm b}$
($\mathbb {I}_{\rm b}$ is the identity in the Hilbert space of the
bath),
(iii) taking into account that $ A (t) = \Tr_{\rm b} ( A (t) \dm_b )$,
and finally (iv) taking the partial trace
$\Tr_{\rm b}(\cdots \dm_{\rm b})$,
one arrives at
%____________________________________-
\begin{equation}
\label{dmH}
\frac {\dif   A} {\dif t}
=
\frac{\iu}{\hbar}
[\HS,  A] 
+
\R
\end{equation}
%____________________________________
with the following relaxation term
%__________________________________
\begin{equation}
\R
=
-\frac{1}{\hbar^2}
\int_{t_0}^{t}
\, \dif s
\,
\Big \{
\corr(s-t)
\F (s)
[\F, A]
-
\corr(t-s)
[\F, A] \F(s)
\Big \}
\; .
\end{equation}
%__________________________________
Here the operators without time argument are understood to be
evaluated at $t$.
Note that $\R$ in the Heisenberg equation is structurally different
(simpler if you want) from the $\R$ obtained for the density matrix
$\dm$, Eq.~(\ref{Rsch}).
Remember that now we are in the Heisenberg picture, so that 
$\dm_{\rm tot}(t) =
 \dm_{\rm tot}(0) = \dm_{\rm s} \otimes \dm_{\rm b}$;
we also exploited this to trace over the bath above.
Finally, from the $A$ solving the above equation one can get the
desired average as
%_______________
\begin{equation}
\langle A \rangle = \Tr_{\rm
  s} ( A \dm_s) \; .
\end{equation}
%________________

%%%%%%%%%%%%%%%%%%%%%%%%%%%%%%%%%%%%%%%%%%%%%%%%%%%%%%%%%%%%%%%%%%%%%%%
%%%%%%%%%%%%%%%%%%%%%%%%%%%%%%%%%%%%%%%%%%%%%%%%%%%%%%%%%%%%%%%%%%%%%%%

\section{application to the bath-of-oscillators model}
\label{sec:dmeclmodel}

Let us now particularize the master equation~(\ref{dmegral}) to the
bath of oscillators model~(\ref{clmodel}).
Note first that the Hamiltonian~(\ref{clmodel}) can be rewritten as:
%________________________________-
\begin{equation}
\label{clmodelre}
\HTOT
=
\underbrace{\quad \widetilde \HS \quad}_{\substack {\HS + 
\case F \sum_\bindex \vl u_\bindex
  \vl^2/\omega_\bindex^2 } }
+
\F \otimes 
\underbrace{\sum_\bindex u_\bindex \qb}_ {\substack {\B}}
+
\underbrace{
\frac{1}{2} 
\sum_\bindex
(\pb^2 + \omega_\bindex^2 \qb^2)
}
_
{\substack{ \HB}}
\end{equation}
%__________________________________
%
where the counter-term 
$\case \F \sum_\bindex \vl u_\bindex \vl^2/\omega_\bindex^2$
has been absorbed into $\widetilde \HS$ [see the discussion after
(\ref{clmodel}) and (\ref{Alan})].
In this form, the model has the structure $\HI = \F \otimes \B$ for
which we derived the master equation~(\ref{dmegral}).
(The case of Heisenberg evolution~(\ref{dmH}) is analogous.)

In the quantum case it is convenient to rewrite~(\ref{clmodelre}) as:
%_________________________________
\begin{equation}
\label{clmodelct}
\HTOT
=
\widetilde {\HS}
+
\F
\otimes
\sum_\bindex
\left (
\ccbath_\bindex
\bosonb^+
+
\ccbath^*_\bindex
\bosonb
\right )
+
\sum_\bindex
\hbar
\omega_\bindex 
\bosonb^+
\bosonb
%-
%\left (
%\sum_\bindex
%\frac {\vl \cbath_\bindex \vl^2}
%{\omega \bindex}
%\right ) 
%\F^2
\end{equation}
%_________________________________
where $\bosonb^+, \bosonb$ are bosonic operators (creation and
destruction) with the new constants
$\ccbath_\bindex = \sqrt {\hbar/(2 \omega_\bindex)}\, u_\bindex$.
We recognize the last term in (\ref{clmodelct}) as the standard
quantum oscillator Hamiltonian $\hbar\omega\bosonb^+\bosonb$.

Let us now particularize the equations of the previous section.
The correlator~(\ref{corrdef}) follows from the time evolution of the
bosonic operators
$\bosonb \propto \e^{\iu \omega_{\bindex} t}$
as
%____________________________________
\[
\corr (\tau)
=
\sum_{\bindex}
|\cbath_{\bindex}|^{2}
\big[
n_{\bindex}\,
\e^{+\iu \omega_{\bindex} \tau}
+
(n_{\bindex}+1)\,
\e^{-\iu \omega_{\bindex}\tau}
\big]
\]
%____________________________________
%
where $n_{\bindex} = 1/(\e^{\beta \hbar \omega_{\bindex}} -1)$ is the Bose
distribution.
Introducing now the spectral density~(\ref{jomega}) as in the
classical case:
%___________________________
\begin{equation}
J(\omega) = 
\pi
\sum
\frac {u_\bindex^2}{2 \omega_\bindex}
\delta (\omega - \omega
_\bindex)
=
\pi \sum \frac {\vl \ccbath_\bindex \vl^2}{\hbar} 
 \delta (\omega - \omega
_\bindex)
\end{equation}
%_____________________________
we can write the correlator as
%_______________________________________________
\begin{equation}
\label{corrc}
\corr (\tau)
=
\hbar \int_0^\infty
\frac {\dif \omega}{\pi}
J(\omega)
\left [
\coth \left ( \case \beta \hbar \omega  \right )
\cos (\omega \tau)
-
\iu
\sin (\omega \tau)
\right ]
\; .
\end{equation}
%_______________________________________________
%
This renders the classical result~(\ref{corrcl}) in the limit $\hbar
\to 0$, letting
$\coth ( \beta \hbar \omega/2) \to 2 /(\beta \hbar \omega)$.

Now we can write the master equation for the bosonic bath model
(\ref{clmodel}) to second order in perturbation theory
%________________________________
\begin{equation}
\label{dmebom}
\frac{\dif \dm}{ \dif t}
=
-
\frac{\iu}{\hbar}
[\HS, \dm]
%-
%\frac{\iu}{\hbar}
%\langle V \rangle_{\rm b}
%[\F, \dm]
+
\R
\end{equation}
%______________________________
with [cf. Eq.~(\ref{Rsch})]
%____________________________
\begin{equation}
\label{Rbom}
\R
=
-\frac{1}{\hbar^2}
\left \{
 \iu \gamma(0)
[\F^2, \dm]
+
\int_{\tcero}^t
\dif s \, 
\corr(t-s)
[\F, \F(s-t) \dm]
-
\corr(s-t)
[\F, \dm \F(s-t)]
\right \}
\; .
\end{equation}
%________________________________
%
Here we have added to $\R$ the term coming from the counter-term in
(\ref{clmodelre}), with
$\gamma (0) = \hbar \sum_\bindex \cbath_\bindex^2/2 \omega_\bindex^2$
which can be expressed in terms of the so-called damping kernel
$\gamma(\tau) = \hbar \int_0^\infty \dif \omega J(\omega)/\omega \cos
(\omega \tau)$.
That term cancels with $\iu \gamma(0) [\F^2, \dm]$ arising after
integration by parts in $\R$.
Therefore, as discussed in the classical case, all renormalizations on
the system have been accounted for from the outset.

Let us close with a few comments.
Equation (\ref{dmebom}) is sometimes criticized for not being of
Lindblad type, and hence for not ensuring positivity in the time evolution
of $\dm$. %
\footnote { 
A Lindblad type equation can be written as (h.c. $\equiv$ Hermitian conjugate)
%____________________________________
\begin{equation}
\dif \dm/ \dif t
=
-{\iu \over \hbar}
[\HS,\dm]
-
{1\over\hbar}
\sum_{n,m}h_{n,m}
\big(\dm L_m L_n+L_m L_n\dm-2L_n\dm L_m\Big)
+
\mathrm{h.c.}
\end{equation}
%____________________________________
This is the more general time-local equation preserving the positivity
of $\dm$.
Its derivation is not based on any microscopic model, but in the
theory of dynamical semigroups \cite[Sec. 3.2]{brepet}, and is popular
in mathematical-physics quarters.
}
In appendix~D.~3 of $\jpaii$ we addressed this point from a physical
rather than a rigorous point of view.
Let us summarize our position.
Equation~(\ref{dmebom}) was derived under several approximations, the
most important being $ \dampp\,\tau_c\ll 1$;
outside its validity range the equation would violate positivity and
what not.
Besides, the derivation assumed decoupled initial conditions at
$\tcero=-\infty$ [see Eq.~(\ref{dmfactorised})].
Therefore, taking arbitrary initial conditions at $t=0$ (which are the
only cases of reported violations of positivity), it seems natural
that the approach may produce nonsense.
%%
%
%___________________________________________________
%___________________________________________________
%___________________________________________________
\begin{figure}
\centerline{\resizebox{9.cm}{!}{%
\includegraphics[angle = -0]{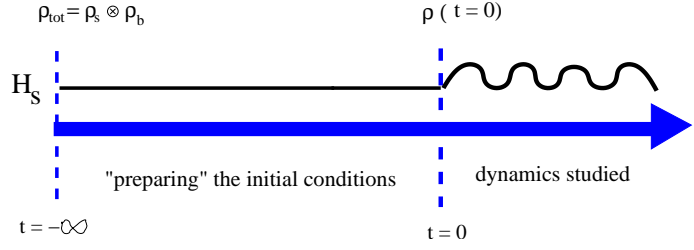}
}
}
\caption{
Scheme of the protocol discussed in the text to work consistently with
the master equations.
We assume that bath and system first met a long long time ago: long
enough so as the decoupled initial conditions no longer matter.
Then, from $t=0$ onwards we manipulate $\HS$ with fields, etc., to
study the response of the system.
} 
\label{fig:timeline}
\end{figure}
%___________________________________________________
%___________________________________________________
%___________________________________________________

So then, can we only study the evolutions under those conditions?
Yes and no.
Imagine we want to study the evolution from an equilibrium state at
$t=0$ of system plus bath.
If the total system is in some sense ``ergodic'' it does not matter the
initial condition we used at $t_0 = -\infty$ (for instance,
factorized, or other);
at $t=0$ we will have our Boltzmann initial condition.
From that time onwards, we can manipulate the system with fields,
etc. (that is, modifying via $\HS$) and study the corresponding
response (see Fig.~\ref{fig:timeline}).
And all this because we are respecting a continuous process coming all
the way from $t_0 = -\infty$. %
\footnote{
Now we can understand why if we use arbitrary initial conditions at
$t=0$ with the constant coefficients of the asymptotic range ($t>0$),
we can get any kind of funny result.
The $\dm (t_0)$ allowed would only be those compatible with the
original initial conditions at $t_0=-\infty$ plus the joint
system-bath evolution.
} % END OF FOOTNOTE
Therefore, if kept within its range of validity (weak coupling), the
equation would preserve positivity, hermiticity, etc.  to the
approximation considered.
\footnote{ 
Note that in the derivation we used a time independent $\HS$, while
the manipulation protocol discussed is based on modifying $\HS$.
There is no problem about this: as such time dependence results in a
time dependence of some of the equation coefficients, but the basic
structure remains the same \cite{kohdithan97}.
} % END OF FOOTNOTE
%%

%%%%%%%%%%%%%%%%%%%%%%%%%%%%%%%%%%%%%%%%%%%%%%%%%%%%%%%%%%%%%%%%%%%%%%%
\section {a few examples}

Within the oscillator bath model, there are some cases where we can write
the master equation~(\ref{dmebom}) explicitly.

%%%%%%%%%%%%%%%%%%%%%%%%%%%%%%%%%%%%%%%%%%%%%%%%%%%%%%%%%%%%%%%%%%%%%%%%%

\subsection{when the eigenstates of $\HS$ are known}
\label{sub:conocemos}

If we know the eigenvalues $\epsilon_n$ and eigenvectors $\vl n
\rangle$ of the system Hamiltonian, 
$\HS \vl n \rangle = \epsilon_n \vl n\rangle$,
we can write explicitly the equation for the components of the density
matrix $\dm_{nm}$ in that basis.
To this end, we sandwich both sides of (\ref{dmegral}) as
$\langle n \vl \partial_t  \dm \vl m \rangle$, and employ:
%______________________
\begin{equation}
\langle n \vl \F (\tau) \vl m \rangle
=
\e^{-\frac{\iu}{\hbar} \Delta_{nm} \tau} \F_{nm}
\end{equation}
%______________________
where $\Delta_{nm} = \epsilon_n -\epsilon_m$.
After some algebra (done with care in appendix~D.1 of $\jpaii$), one obtains
%_________________________________
\begin{eqnarray}
\label{dmebomnm}
\frac {\dif  \dm_{nm}}{\dif t}
&=&
- \frac{\iu}{\hbar} \Delta_{nm} \dm_{nm}
+
\sum_{n'm'}
\R_{nn'mm'}
\dm_{n'm'}
\end{eqnarray}
%___________________________________
with the explicit relaxation term
%__________________________________
\begin{eqnarray}
%%%%%%%%%%%%%%%%%%
\label{Rdmebomnm}
\sum_{n'm'}
\R_{nn'mm'}
\dm_{n'm'}
=
\sum_{n'm'}
\big[
&-&
\delta_{m m'}
\big(
\tsum_{l}
\Wu_{l \vl n'}^{*}
\F_{n l}
\F_{l n'}
\big)
\nonumber \\[-2.ex]
%%%%%%%%%%%%%%%%%%
&+&
\big(
\Wu_{n \vl n'}
+
\Wu_{m\vl m'}^*
\big)
\,
\F_{n'n}
\F_{mm'}
\\
%%%%%%%%%%%%%%%%%%
& 
-&
\delta_{ n n'}
\big(
\tsum_{l}
\Wu_{l \vl m'}
\F_{m' l}
\F_{l m}
\big)
\quad
\big]
\;
\dm_{n'm'}
\nonumber
%%%%%%%%%%%%%%%%%%
\;.
\end{eqnarray}
%_________________________________
Here the relaxation rates are
%____________________________
\begin{equation}
\label{wuri}
\Wu_{n \vl m}
:=
\Wu (\Delta_{nm})
=
\iu \gamma(0)
+
\int_0^{-\tcero}
\dif \tau
\e^{-\frac{\iu}{\hbar} \Delta_{nm}\tau}
\corr (\tau)
=
\W (\Delta_{nm})  + \iu \se  (\Delta_{nm})
\end{equation}
%____________________________
with the part
$\gamma(0) = \sum_\bindex \cbath_\bindex^2/\omega_\bindex $
coming from the counter-term [see after Eq.~(\ref{Rbom})].
When the correlator $\corr (\tau)$ is given by Eq.~(\ref{corrc}), one
can prove the important relations
%_________________________________-
\begin{equation}
\label{detailed}
\W (\Delta)
=
\e^{-\beta \Delta}
\W (-\Delta)
\;,
\qquad
\qquad
\se (0) = 0
\end{equation}
%________________________________
%
The first is the {\em detailed balance\/} condition.
The second is a consequence of having accounted for the
renormalization (and will be exploited later on).
The equation obtained is of the Redfield type, like those introduced for
the study of magnetic resonance \cite{blum}.
%%

%%%%%%%%%%%%%%%%%%%%%%%%%%%%%%%%%%%%%%%%%%%%%%%%%%%
\subsection {RWA}
\label{sub:RWA}

The ``rotating wave approximation'' is common in many branches of
physics.
Roughly speaking, the  RWA  consists in dropping the terms which
 ``rotate'' much faster than the characteristic frequencies of the
 problem addressed.
As an  illustration imagine we have terms like
 $\e^{\iu (\omega_1
  + \omega_2)t}$
and
 $\e^{\iu (\omega_1
  - \omega_2)t}$
(with $\omega_1 \, , \omega_2 >0$).
Then one argues that 
 $\e^{\iu
(\omega_1 + \omega_2)t}$ 
oscillates much faster than 
$\e^{\iu (\omega_1 - \omega_2)t}$
and can be replaced by its average (zero).
Let us see how the RWA is implemented in our problems.

Let us jump into the  ``rotating frame'':
$\overline \dm_{nm}(t) = \dm_{nm} \e^{- \iu/\hbar \Delta_{nm} t}$
where the density-matrix equation~(\ref{dmebomnm}) reads:
%___________________
\begin{equation}
\frac {\dif \overline {\dm}_{nm}(t)}{\dif t}
=
\sum_{n'm'} \R_{nn'mm'}
\overline {\dm}_{n'm'} \e^{\frac {\iu}{\hbar} (\Delta_{nm} - \Delta_{n'm'}) t}
\; .
\end{equation}
%___________________
%
The crudest approximation assumes that the terms
$\e^{\iu/\hbar (\Delta_{nm} - \Delta_{n'm'})t}$
with
$\Delta_{nm} - \Delta_{n'm'} \neq 0$
revolve in time much faster than the characteristic relaxation scales
of the system, so they are averaged to zero.
That is, we discard {\em all\/} terms with
 $\Delta_{nm} - \Delta_{n'm'} \neq 0$.
If we have a system spectrum without degenerations or regularities
(see below), that is, where all $\Delta$'s are different, we have
$\Delta_{nm} - \Delta_{n'm'} = 0$,
only if $n = n'$ \& $m = m'$ or if $n = m$ \& $n'= m'$ \cite{blum}.
Therefore, in such a RWA the master equation splits into diagonal and
off-diagonal elements as follows
%______________________-
\begin{eqnarray}
\label{dmeRWA}
\frac {\dif \dm_{nn}}{\dif t}
&=& \sum_{ n'}\R_{nn'nn'} \dm_{n'n'}
\\ \nonumber
\frac {\dif  \dm_{nm}}{\dif t}
&=&
- \frac{\iu}{\hbar} \Delta_{nm} \dm_{nm}
+
\R_{nnmm}
\dm_{nm} 
\; .
\end{eqnarray}
%______________________
where we have returned to the unrotated basis.
\enlargethispage*{1.cm}

It is not difficult to see that $\R_{nn'nn'}$ is real:
$\R_{nn'nn'} = 2 \vl \F_{nn'}\vl^2 W(\Delta_{nn'})$.
Besides, $\R_{nnmm} = \R_{mmnn}^*$ (as $\dm$ is Hermitian).
But in the RWA one only retains the real parts of $\R_{nnmm}$,
arguing that the imaginary parts reflect an extra renormalization.
Notice that the equations for the off-diagonal terms are now
uncoupled, and their solution is simply
%_________________________________-
\begin{equation}
\label{nondiaRWA}
\dm_{nm}(t)
=
\dm_{nm}(0)\,
\e^{\left (-\frac{\iu}{\hbar} \Delta_{nm} + \R_{nnmm}\right ) t}
\end{equation}
%________________________________
%
Therefore, we just need to solve the equation for the diagonal terms.
Then analytical advances are sometimes possible, and in the worst
case, it will be computationally ``cheaper''.
Furthermore, equation~(\ref{dmeRWA}) is of Lindblad type \cite{brepet}
(by some chance), and the equilibrium solution
$\partial_t \dm = 0$
results to be the canonical equilibrium $\propto \e^{-\beta \HS}$
(see next chapter), what makes it more ``user friendly''.

As for our discussion of decoherence, given that the $W (\Delta_{nm})
>0$ (see $\jpaii$), the $\R_{nnmm}$ give directly the inverses of the
decoherence times.
In this case the basis where the density matrix becomes diagonal, the
pointer basis, is that of eigenstates of $\HS$ (this is related with the
equilibrium solution being $\propto \e^{-\beta \HS}$).
Finally, the equation for the diagonal terms has the same form as the
Pauli master equation, but there the off-diagonal dynamics is
neglected.

The RWA is many times difficult to justify (for a illustration of the
limitations see \cite{raujohshn04}).
For instance, we could have systems with 
$\Delta_{nm} - \Delta_{n', m'}$
very small.
Or it could also happen that 
$\Delta_{nm} - \Delta_{n', m'}= 0$
in more cases than those considered, for example, if there is a
degeneracy, or if $\HS$ has an equispaced spectrum
$\Delta_{n n \pm 1} ={\rm cte}$.
In the latter case, we would have 
$\Delta_{nm} - \Delta_{n', m'}= 0$
also when $n-m = n'- m'$ for any shift $p$ in the indexes.
Then, we are not allowed to drop the $\R_{n n + p, m m +p}$ terms in
(\ref{dmebomnm}) and we write
%_____________________________
\begin{equation}
\label{RWAmej}
\frac {\dif
\dm_{nm}}{\dif t}
= 
- \frac{\iu}{\hbar}
\Delta_{nm}
\dm_{nm}
+
\sum_p
\R_{n n+p, m m+p} 
\; \dm_{n+p, m+p}
\end{equation}
%_____________________________
%
We will refer to this equation as ``improved RWA'', and it will be
invoked in the last chapter.
%%

%%%%%%%%%%%%%%%%%%%%%%%%%%%%%%%%%%%%%%%%%%%%%%%%%%%%%%%%%%%%%%%%%%%%
\subsection {semiclassical or high temperature equation}
\label{sub:cldme}

This is another instance where the relaxation term can be written in a
simple form.
Let us start having a look at the formula~(\ref{corrc}) for the
correlator.
Although the frequency integral extends to infinity, we have 
$J(\omega > \cutoff) = 0$
with $\cutoff$ the bath cutoff.%
\footnote{
Remember that in the classical case we wrote $J(\omega) \sim
\omega^\alpha$ (see the footnote in Sec.~\ref{subs:ejcl}).
But this is an idealized situation; in fact any physical $J(\omega)$
should go to zero at high enough frequencies \cite[Sec. 3.1]{weiss}.
Therefore, one inserts a regularizator, exponential for example
\begin{equation}
J(\omega) \propto \omega^\alpha \e^{-\omega/\cutoff}
\end{equation}
and at frequencies higher than the  cutoff $\cutoff$ the spectral
density goes to zero.
} % END OF FOOTNOTE
Then, at high enough temperatures, 
$\beta \hbar \cutoff \leq 1$
one can approximate 
$\coth ( \beta \hbar \omega/2) \cong 2 /(\beta \hbar \omega)$.

Now let us consider a low frequency part
$J(\omega) \sim m \dampp\,\omega$, 
which corresponds to the {\it Markovian\/} case discussed in the
classical limit (see Sec.~\ref{subs:ejcl}).
Under these conditions, the correlator~(\ref{corrcl}) is left as
%_______________________________-
\begin{equation}
\corr (\tau)
=
2 m \dampp\,\kT  \delta (\tau)
+
\iu m \hbar \dampp\,\frac {\dif}{\dif \tau}
\delta (\tau)
\end{equation}
%________________________________
Therefore the relaxation term reduces to
%____________________________
\begin{equation}
\label{RhT}
\R =
-\frac{m \dampp\,\kT}{\hbar^2}
\left ( [ \F, \F \dm] + {\rm h.c.} \right )
+ \frac {m \dampp}{2 \hbar^2}
\left ( [\F, [\HS, \F] \dm ] + {\rm h.c.} \right )
\end{equation}
%___________________________
This is the semiclassical or high-temperature equation.
It has the advantage of not having an integral in $\R$ (time local)
and hence is easier to handle.

If we now particularize the above  $\R$ to the case of a mechanical
Hamiltonian
$ \HS = \p^2/2m + V(\q)$, 
with the particle linearly coupled to the bath coordinates, $\F = -
\q$, we get
%____________________________________
\begin{equation}
\label{cleq}
\frac {\dif  \dm }{\dif t}
=
-\frac {\iu}{\hbar}
[\HS, \dm]
-\frac {m \dampp\,\kT}{\hbar^2}
[\q, [\q, \dm]] -
\iu \frac {\dampp}{2  \hbar}
[\q, [ \p, \dm ]_{+}]
\; ,
\end{equation}
%___________________________________
with $[A, B]_{+} = AB + BA$ the anticommutator.
This is nothing but the ``famous'' {\it Caldeira--Leggett\/} equation,
originally derived with the path integral formalism \cite{calleg83pa}.

Let us close with a few remarks on the validity range as compared
to the original equation (\ref{dmebom}).
First, we used $J(\omega) = m \dampp\,\omega$, with $\dampp$ having
units of $t^{-1}$ and being related with the coupling strength [see
Eq.~(\ref{jomega})].
On the other hand, for such $J(\omega)$ the bath correlation time is
$\tc = \beta \hbar$ (see \cite[Sec.~3.6]{brepet}) and hence the
validity range from Eq.~(\ref{validity:range}) would be 
$ \beta \hbar \dampp \ll 1$.
But on top of this restriction, {\it Caldeira--Leggett} put the $\beta
\hbar \cutoff \leq 1$ condition.
Then, one fears that  {\it a priori} the validity range would be quite
restrictive.
However, in the case of the harmonic oscillator,
$\HS = \p^2/2 m + \case m \who^2 \q^2$,
exact quantum master equations have been obtained
\cite{hupazzha92,kargra97}.
And for this case Unruh and Zurek \cite{unrzur89} checked empirically
that, except at short times, the restrictions on the high-temperature
equation could effectively be relaxed to
$\cutoff/\who \ll \e^{\kT/\hbar \dampp}$.
Although limited to the harmonic oscillator, this is anyway good news
on the applicability range of the Caldeira--Leggett master equation.
%%

%%%%%%%%%%%%%%%%%%%%%%%%%%%%%%%%%%%%%%%%%%%%%%%%%%%%%%%%%%%%%%%%%
\section {summary}

We are aware that this chapter has been terse.
Let us summarize the main points.

First we have discussed the concept of quantum decoherence.
The decoherence program is an example of how quantum theory can save
face.
In a way similar to how reversible dynamics could account for the
irreversibility found in nature, by invoking the interaction with the
environment, quantum mechanics manages to explain the
``non-observation'' of certain odd superpositions in the macroscopic
world (for the good of cats and other creatures).
The reason, we are told, is the extremely fast lost of coherence when
we only look at our system, through the notion of partial tracing
[Eq.~(\ref{tracepartial})].

Then we discussed the description of the dynamics of open systems by
means of a master equation for the reduced density matrix $\dm$.
Those master equations are valid in the range $\dampp\,\tc \ll 1$ with
$\dampp$ quantifying the system-bath coupling strength and $\tc$ the
bath correlation time.
The prime difficulty with such equations is in dealing with the time
integral in the relaxation term $\R$ [Eq.~(\ref{Rsch})].
We saw how this can be handled without problems when the spectrum of
$\HS$ is known, getting a Redfield-type equation [Eq.~(\ref{Rdmebomnm})].

Finally, we discussed a couple of approximations widely used in the
specialized literature, like the rotating wave approximation
[Eq.~(\ref{dmeRWA})] and the high-temperature approximation {\em \`a
la\/} Caldeira--Leggett [Eq.~(\ref{cleq})].
Both equations have validity ranges, in principle, more restrictive
than $\dampp\,\tc \ll 1$, although when they can be used result in a
great simplification.

%%%%%%%%%%%%%%%%%%%%%%%%%%%%%%%%%%%%%%%%%%%%%%%%%%%%%%%%%%%%
%%%%%%%%%%%%%%%%%%%%%%%%%%%%%%%%%%%%%%%%%%%%%%%%%%%%%%%%%%%%
%%%%%%%%%%%%%%%%%%%%%%%%%%%%%%%%%%%%%%%%%%%%%%%%%%%%%%%%%%%%
%%%%%%%%%%%%%%%%%%%%%%%%%%%%%%%%%%%%%%%%%%%%%%%%%%%%%%%%%%%%
%%%%%%%%%%%%%%%%%%  equilibrium
%%%%%%%%%%%%%%%%%%%%%%%%%%%%%%%%%%%%%%%%%%%%%%%%%%%%%%%%%%%%
%%%%%%%%%%%%%%%%%%%%%%%%%%%%%%%%%%%%%%%%%%%%%%%%%%%%%%%%%%%%
%%%%%%%%%%%%%%%%%%%%%%%%%%%%%%%%%%%%%%%%%%%%%%%%%%%%%%%%%%%%

\chapter{equilibrium properties}
\label{chap:equilibrio}

Up to this point we were not concerned with the equilibrium properties
in open systems.
The equations discussed, classical or quantal, describe an
irreversible dynamics in which the system in contact with the
reservoir exchanges energy (chapters \ref{chap:clasico} and
\ref{chap:dinamica}).
Waiting long enough the system will equilibrate to the bath,
reaching a stationary regime.
Which is the state reached is the question that we will try to answer
in this chapter.

We will start revisiting the classical case.
Then we introduce the thermodynamic perturbation theory formalism, and
expand the partition function in powers of the system-bath coupling,
getting explicitly the partition function $\Zr$ to second order.
After this, we will check that the master equations derived in the
previous chapter are thermodynamically consistent.
That is, that they have as stationary solution the equilibrium distribution
obtained here to the same order in the bath coupling.
We conclude discussing how to compute equilibrium quantities in open
systems, and illustrate this with the example of the harmonic
oscillator.

%%%%%%%%%%%%%%%%%%%%%%%%%%%%%%%%%%%%%%%%%%%%%%%%%%%%%%%%%%%%
\section{the classical limit}
\label{sec:cllim}

The best way to address the equilibrium properties starting from the
dynamics is to resort to the \FP\ equations.
The necessary condition for $W_{\rm eq}$ to be a stationary solution
is  
$ \partial_t \W_{\rm eq}= {\cal L}_{\rm cl} \W_{\rm eq} = 0$,
where ${\cal L}_{\rm cl}$ is a compact writing of the \FP\ differential
operator.
If we focus on the particle case [Eq.~(\ref{wkk})], we can easily
check that ${\cal L}_{\rm cl} \e^{-\beta \HS} = 0$. %
\footnote {
The first two terms in (\ref{wkk}), arising from the Poisson bracket,
give zero when applied to $\e^{-\beta \HS}$.
Letting the other two terms (diffusion and relaxation) act on
$\e^{-\beta \HS}$ one finds they cancel by using
$(\beta/m) \partial_p [p \, \exp(-\beta \p^2/2m) ]
= 
- \partial^2_\p [\exp(-\beta \p^2/2m)]$.
For spin problems 
$ {\cal L}_{\rm cl} \W_{\rm eq} = 0$ 
follows analogously from Eq.~(\ref{clfp}), see \cite{gar2000}.
}
Furthermore, it happens that any initial condition, in the limit $t
\to \infty$, tends to the same solution \cite{risken}.
Therefore, the system relaxes to the equilibrium Gibbs--Boltzmann
  distribution $\Weq \propto \e^{-\beta \HS}$.
The dynamics of classical open systems places us in a comfortable
theoretical framework where the \FP\ equations capture the approach to
equilibrium.

Let us reexamine this from another point of view.
Although in an effective/reduced way, the \FP\ or Langevin equations
describe the total dynamics of system plus bath~(\ref{clmodel}) (see
the discussion on irreversibility in Sec.~\ref{sec:irrepar}).
But the total system is a macroscopic entity, and hence we can apply
the Gibbsian hypothesis.
That is, at sufficiently long times the averages can be computed with
the Gibbs distribution for the whole system
%______________________________
\begin{equation}
\W_{\rm tot} 
= 
\frac{\e^{-\beta \HTOT}}{\ZTOT}
\; ;
\qquad
\qquad
\ZTOT
= 
\int \dif \p \dif \q \int \prod \dif \pb \prod \dif \qb \e^{-\beta \HTOT}
\end{equation}
%______________________________
%
As we are discussing the reduced distribution, we integrate out the
bath, getting
%__________________________
\begin{equation}
\label{Wredcl}
\Weq = 
\int \prod \dif \pb \prod \dif \qb \frac {\e^{-\beta \HTOT}}{\ZTOT}
=
\frac {\e^{-\beta \HS}}{\ZS}
\end{equation}
%___________________________
or in other words, $\ZTOT = \ZB \ZS$, with 
$\ZB = \int \prod \dif \pb \prod \dif \qb \e^{-\beta \HB}$
and 
$\ZS = \int \dif \p \dif \q \, \e^{-\beta \HS}$.

\begin{figure}
\centerline{\resizebox{6.7cm}{!}{%
\includegraphics[angle = -0]{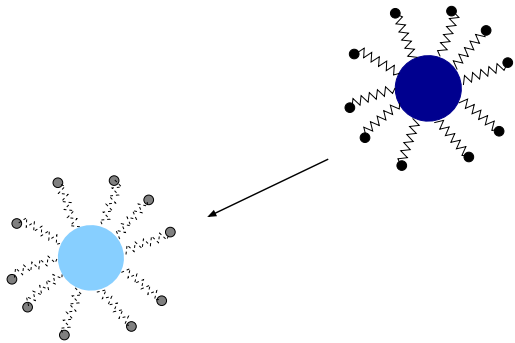}
}
}
\caption{
Pictorial representation of the change of variables 
$\qb - \cbath_\bindex \q/\omega_\bindex^2 \to \qb$
where, for simplicity, we have used $\F = -\q$ [see after
Eq.~(\ref{change})].
Physically, what we do is to shift everything by a distance $ \propto
\q$.
As the integrals extend from $-\infty$ to $+\infty$ their limits do not
change.
Compare with the analogy of the Bohr-Van Leeuwen particle in a
magnetic field (final comment in Sec.~\ref{sub:pert}).
}
\label{fig:change}
\end{figure}

This result has relied on the integrals over the bath being Gaussian
(see Fig.~\ref{fig:change})
%___________________________
\begin{equation}
\label{change}
\Weq = \frac{\e^{-\beta \HS}}{\ZTOT} \int \prod \dif \pb \prod \dif \qb
\e^{-\beta /2\sum \pb^2 + \omega_\bindex^2 (\qb + \cbath_\bi
  \F/\omega_\bindex^2)^2}
\; ,
\end{equation}
%___________________________
so that the change of variables
 $\qb + \cbath_\bindex \F/\omega_\bindex^2 \to \qb$
gives
%__________________________
%\begin{equation}
%\nonumber
$
\Weq = 
\e^{-\beta \HS}/\ZS.
$
%\end{equation}
%___________________________
This is the reason behing (\ref{Wredcl}) giving the reduced
distribution equal to that of the system without couping.
Now we understand that $\e^{-\beta \HS}/\ZS$ is stationary solution of
(\ref{wkk}), since the \FP\ equation is nothing but the evolution of
the total system with the bath variables integrated out.
%%
%___________________________________________________
%___________________________________________________
%___________________________________________________
%___________________________________________________
%___________________________________________________
%___________________________________________________

We remark that, although it results natural to accept that the system
relaxes to the distribution $\e^{-\beta \HS}$, from the point of view
of (\ref{Wredcl}) this is a ``happy'' coincidence.
We mean that the result is  ``model dependent'' and for other models
(coupling nonlinear in the bath coordinates, or other kinds of baths),
the reduced distribution could be different from  $\e^{-\beta \HS}$.
Otherwise, every system with several degrees of freedom would factor
into the product of reduced distributions for each degree of freedom.

In fact, this result is so particular that, as we will soon see, does not
hold for the same Hamiltonian in the quantum case, and the system
relaxes to a distribution different from the canonical one.
What is the meaning behind this?
In the books of statistical mechanics one assumes that the body is
macroscopic and that the interaction with the bath is somehow mediated
by its ``surface''.
As the surface $L^2$ to volume $L^3$ ratio goes to zero in large
systems $\propto1/L$, the interaction with the reservoir $\HI$ is put
aside, leaving the Gibbs distribution \cite{balescu}.
But for the open systems we are treating here we cannot neglect $\HI$.
System plus bath, on the other hand, constitute a macroscopic object.
And here it is where we can assume Gibbs.
But then nothing ensures that the reduced distribution is going to be
$ \propto \e^{-\beta \HS}$.
We insist again, the result (\ref{Wredcl}) is quite particular.
%%

%%%%%%%%%%%%%%%%%%%%%%%%%%%%%%%%%%%%%%%%%%%%%%%%%%%%%%%%%%%%%%%%%%%%%%%%%%
\section{the quantum case}
\label{sec:cuaneq}

We now address the quantum problem.
We will try to answer to the following questions
%_____
\begin{itemize}
\item
Is the reduced distribution the canonical one for the system without
coupling?
%__________________________
\begin{equation}
\dm_{\rm eq} := \Tr_{\rm b} \left ( \e^{-\beta \HTOT} \right ) 
\stackrel{?}{\propto} \e^{-\beta \HS}
\end{equation}
%___________________________
To which we have already advanced the answer (previous paragraph),
and, if it is not so, how is that distribution?
\item
Is the reduced density matrix a stationary solution of the master
equations discussed in chapter~\ref{chap:dinamica}? 
That is
%___________________________
\begin{equation}
{\cal L}_{\rm q} \, \dm_{\rm eq} \stackrel{?}{=} 0
\; .
\end{equation}
%___________________________
with ${\cal L}_{\rm q}$ a compact way of writing the master
operator in~(\ref{dmebom}).
\end{itemize}

%%%%%%%%%%%%%%%%%%%%%%%%%%%%%%%%%%%%%%%%%%%%%%%%%%%%%%%%%%%
\subsection {thermodynamic perturbation theory}
\label{sub:pert}

In principle we do not have a practical expression for
$\e^{-\beta\HTOT}$ (much as in the dynamics we did not know an exact
differential equation for the reduced density matrix).
Therefore, we will assume again that the system-bath coupling is weak.
This will allow us to perform the equilibrium calculation to the same
perturbative order as we did in chapter \ref{chap:dinamica} for the
master equation.
Besides, with this calculation we will gain some insight in the way
the first corrections to the uncoupled distribution emerge.

In order to obtain an approximate expression for $\e^{-\beta \HTOT}$
we will start again from the Kubo identity~(\ref{kuboreal}).
As in Sec.~\ref{sec:dedqme} we will identify $\HO \equiv \HS + \HB$
and $\HI \equiv \V $ and expand to the second order:
%_______________________________________
%________________________________________________
\begin{equation}
\label{rho2nd}
\e^{-\beta (\HO + \V)}
\cong
\e^{-\beta \HO}
\left [
1
-
\int^\beta_0  \dif \sigma \,
\V (-\iu \hbar \sigma)
+
\int^\beta_0  \dif \sigma \,
%\e^{\lambda \HO} \V \e^{ - \lambda \HO}
\int^\sigma_0  \dif \theta\, 
%\e^{\sigma \HO} \V \e^{ - \sigma \HO}
\V (-\iu \hbar \sigma)
\V (-\iu \hbar \theta)
\right ]
\; .
\end{equation}
%_______________________________________________________
%
%_______________________________________
This is nothing but the evolution~(\ref{U2nd}) in imaginary time, with
$V(-\iu \hbar \sigma) = \e^{\sigma \HS} V \e^{-\sigma \HS}$.
But note that here the expansion is in powers of $\beta V$.

\renewcommand{\req}{R}

We now trace with respect to the reservoir and apply the result to the
bosonic bath model (as in section~\ref{sec:dmeclmodel}), getting
%__________________________
\begin{equation}
\label{dmeq2nd}
\dm_{\rm eq} =
\underbrace{\frac {\e^{-\beta \HS}}{\ZS}}_ {\substack {\dm^{(0)}}}
-
\underbrace{\frac {\Zdos}{\ZS^2} \e^{-\beta \HS} + 
\req}_{\substack {\dm^{(2)}}}
\end{equation}
%___________________________
Here $\ZS = \Tr_{\rm s} ( \e^{-\beta \HS})$ and the ``rest'' $\req$
%___________________________
\begin{equation}
\req
=
\frac {\e^{-\beta \HS}}{\ZS}
\left [
\int_0^\beta
\dif \sigma\,
\int_0^\sigma
\dif \theta \,
\F (-\iu \hbar \sigma)
\F(-\iu \hbar (\sigma - \theta))
\corr (-\iu \hbar \theta)
+\gamma (0) \int_0^\beta  \dif \sigma\, \F^2(-\iu \hbar \sigma)
\right ]
\; ,
\end{equation}
%___________________________
involves the correlator~(\ref{corrdef}) in imaginary time
$\corr(- \iu \sigma) = \langle \B (- \iu \sigma) \B \rangle_{\rm b}$.
We already defined
$\gamma(0) = \hbar \sum_\bindex \cbath_\bindex^2/2 \omega_\bindex^2$
after (\ref{wuri}), which is a byproduct of the counter-term in
(\ref{clmodel}).

The result above is already properly normalized, $\Tr (\dm ) = 1$, and
$\Zdos$ is essentially the second order correction to
 $\ZTOT = \Tr (\e^{-\beta \HTOT})$:
%___________________________________________
\begin{equation}
\label{zgral}
\frac {\ZTOT}{\ZB}
=
\ZS
+
\Zdos
\; ;
\qquad
\Zdos 
=
-
\frac {1}{2}
\beta
\sum_{nm}
|F_{nm}|^2
\e^{-\beta \epsilon_n}
\left [
\int^\beta_0
\dif \sigma\,
\e^{ \sigma \Delta_ {nm}}
\corr (- \iu \sigma )
-
\gamma(0)
\right ]
\; .
\end{equation}
%___________________________________________
%
The superindexes $(2)$ tell that this result is valid to second order.
%
% with $\corr(- \iu \sigma) = \langle
%\B (- \iu \sigma) \B
%\rangle_{\rm b}$ [Eq.~(\ref{corrdef})].
%
The first order terms disappear as they did in the dynamics since in
the bosonic bath (\ref{clmodelct}) we have
$\langle E \rangle_{\rm b} = 0$.

We have thus answered the {\bf first question}, showing that the reduced
distribution, contrary to the classical counterpart, does not agree
with the canonical result at zero coupling.

We close this section with a few more remarks on the result obtained.
First notice that
%__________________________
\begin{equation}
\label{zse}
\int^\beta_0
\dif \sigma\,
\e^{ \sigma \Delta}
\corr (- \sigma )
-
\gamma(0)
=
\se (- \Delta)
\quad
\longrightarrow
\quad
\Zdos =
-
\frac {1}{2}
\beta
\sum_{nm}
|F_{nm}|^2
\e^{-\beta \epsilon_n}
\se (-  \Delta_{nm})
\end{equation}
%__________________________
That is, the equilibrium quantities, as computed from the partition
function, will depend on the bath through the imaginary parts of the
coefficients $\R_{nn' mm'}$ [Eq.~(\ref{wuri})].
The real parts are associated to relaxation and decoherence times
(section \ref{sec:bloch}), that is, to how fast we reach the
stationary solution.
But now we are in the statics, and we do not care about the speed of
the approach, but about the final state instead; it then makes sense that the
equilibrium results do not depend on the real parts of $\R_{nn',mm'}$.

It is instructive to see that in the limit $\beta \hbar \to 0$, both
$\Zdos$ and the term $\req$ go to zero, recovering the classical result
of damping-independent equilibrium properties in a Gaussian bath.
Consistently, in the high temperature limit the quantum corrections to
the canonical result disappear too ($\ZTOT = \ZS \ZB$). %
\footnote {
In the high temperature limit, this is understandable, since letting
$T \to \infty$ we have $\ZTOT \to N_{\rm tot}$, with $N_{\rm tot}$ the
number of states.
But $N_{\rm tot} = N_{\rm s} N_{\rm b}$ so that $\ZTOT = \ZS \ZB$.
Physically, at very high $T$ the coupling potentials become negligible
compared to the kinetic energies.
}

We conclude with an analogy to a famous problem: 
the absence of orbital magnetism in classical statistical mechanics.
Let us start from the Hamiltonian of a charged particle in a magnetic field
%___________________________
\begin{equation}
\nonumber
{\mathcal H}
=
\frac{1}{2m} \left (\vec {p} - e {\vec A}
\right )^2 + V (\vec {q})
\end{equation}
%___________________________
and compute the partition function
$\mathcal {Z} 
= 
\int \dif \vec {q} \int \dif \vec {p} \, \e^{-\beta \mathcal {H}}$.
As the momentum integrals are Gaussian, on making the variable change
$\widetilde p_i = p_i - e A_i$, 
the vector potential disappears from the final result, producing a
zero magnetization
$\Mz \propto \partial \ln \Zr /\partial \Bz \equiv 0$.
This is the content of the Bohr-Van Leeuwen theorem.

In our case we have Gaussian integrals too, but it is $\cbath_\bindex$
what disappears from $\Zr$ (instead of $\vec A$), so that the equilibrium
quantities become independent of the system-bath couplings.
Quantum mechanically (we are thinking in the obtainment of  $\Zr$ from
the density matrix), the Gaussian argument does not hold.
Correspondingly, the equilibrium quantities do
depend on $\HI$, much as the charged particle developed a non-zero
magnetization (Landau).
%%

%%%%%%%%%%%%%%%%%%%%%%%%%%%%%%%%%%%%%%%%%%%%%%%%%%%%%%%%%%%%%%%%
\subsection{stationary solution of the master equation}

Let us address now the second question, that is, let us see if the
damping-dependent equilibrium distribution~(\ref{dmeq2nd}) is a
stationary solution of the quantum master equation~(\ref{dmebom}).
We would not hide that the answer is in the affirmative.
It should be so if we did things properly, since 
$[\e^{-\beta \HTOT}, \HTOT] = 0$,
while the master equations are just the Von Neumann evolution of the
total system with some tracing over the bath degrees of freedom
(remember the classical discussion in Sect.~\ref{sec:cllim}).
What we aim is to prove here is that the equality still holds when
$\e^{-\beta \HTOT}$ and $\HTOT$ are calculated order by order.
Let us see this.%
\footnote{
See \cite{gevrostan00} for a closely related calculation.
}

As all our calculations are perturbative (both in the dynamics and in
equilibrium), we will check 
${\cal L}_{\rm q} \dm_{\rm eq} \stackrel{?}{=} 0$
order by order.
Taking into account the decomposition~(\ref{dmeq2nd}) of the density
matrix
$\dm = \dm_{\rm eq}^{(0)} + \dm_{\rm eq}^{(2)}$
we obtain the following consistency conditions
%____________________________________
\begin{itemize}
\item[(i)]
\begin{equation}
\frac
{\dif  \dm^{(0)}_{\rm eq}}{\dif t} 
=
\frac{\iu}{\hbar}
[ \dm^{(0)}_{\rm eq}, \HS]
=
0
\end{equation}
\item[(ii)]
\begin{equation}
\frac
{\dif  \dm^{(2)}_{\rm eq}}{\dif t}
=
\frac{\iu}{\hbar}
[ \dm^{(2)}_{\rm eq}, \HS]
+
\R [\dm^{(0)}_{\rm eq}]
= 0
\end{equation}
\end{itemize}
%____________________________________
%%
Condition (i) is trivially fulfilled, being $\dm^{(0)}_{\rm eq}$ a
function of $\HS$.
In order to prove (ii), note first that 
$[\HS, (\Zdos/\ZS^2)\, \e^{-\beta \HS}] = 0$,
so that we need to check only if
%___________________________
\begin{equation}
\label{Hsreq}
[\HS, \req] + \R [\dm^{(0)}_{\rm eq}] \stackrel{?}{=} 0
\end{equation}
%___________________________
On writing 
$\req = \sum \req_{nm} \vl n \rangle \langle m \vl$
we have
$[\HS, \req] = \Delta_{nm} r_{nm} \vl n \rangle \langle m \vl$
with $\Delta_{nm}$ the energy difference between the $n$ and $m$
eigenstates of $\HS$.
Therefore, the equality~(\ref{Hsreq}) follows for the diagonal
elements ($n=m$), given that 
$\R [\dm^{(0)}_{nn}] \vl n \rangle \langle n \vl = 0$
(we are omitting the marker ``eq'' in $\dm$ to alleviate notation).
Finally, is not difficult to prove the relation
%_______________________
\begin{equation}
\frac{\iu}{\hbar}
\;
\req_{nm}
= 
\frac{1}{\Delta_{nm}}
\langle n \vl
\R [\dm^{(0)}]
\vl m
\rangle  
\end{equation}
%_______________________
showing that the second condition is obeyed too.

So we have just managed to prove the thermodynamic consistency of our
master equations.
This is also relevant for the linear-algebra eigenvalue problem
associated to the master equation.
Solving it, one of the eigenvalues will be zero, corresponding to the
stationary solution~(\ref{dmeq2nd}), as we have just checked.
As for the rest, we expect they having negative real parts, ensuring
convergence at long times to the stationary solution.
Positive real parts would imply a divergence with time which would not
be physically acceptable.

Well, at this point we have already answered our two questions.
That is, the stationary solution of the master equation~(\ref{dmebom})
is the equilibrium result (\ref{dmeq2nd}), and this does not agree in
general with the canonical distribution at zero coupling.
In other words, although by now it feels quite natural, the
equilibrium properties of quantum open systems do depend on the
damping strength.
When we started this chapter this would have struck us, due to the
intuition from the classical case and the Gibbsian prejudice.
Indeed, in many works it is still stated that the system would
approach the canonical $\e^{-\beta \HS}$.
Here we have argued that in general this is not the case.%
\footnote{
In the original Lindblad calculation, the convergence to
$\e^{-\beta\HS}$ is used as one of the constrains to fix the form of
the master equation from ``first principles''.
}

%%%%%%%%%%%%%%%%%%%%%%%%%%%%%%%%%%%%%%%%%%%%%%%%%%%%%%%%%%%%%%%%%%%%%%%%%%%
\section {computing thermodynamic quantities}
\label{sec:magter}

Once reached this point, we could use our formula (\ref{zgral}) for
$\Zr$ and compute thermodynamic quantities.
To this end, it will prove convenient to redefine the free energy as
follows (all other quantities will follow from this)
%________________________________________
\begin{equation}
\Free = \Free_{{\rm tot}} - \Free_{\rm b}
\;.
\end{equation}
%________________________________________
Here 
$\Free_{{\rm tot}} = - T \ln ( \ZTOT)$ 
with 
$\ZTOT = \Tr ( \e^{-\beta \HTOT} )$
the partition function of the whole system, while
$\Free_{\rm b} = - T \ln ( \ZB)$ with $\ZB$ would be the partition
function of the bath uncoupled from the system
$\ZB = \Tr_{\rm b} ( \e^{-\beta \HB} )$.
That is, we are redefining the zero of free energy, removing the
contribution of the bare bath.
Besides, in this way our definition above for $\Zr=\ZTOT/\ZB$ allows
obtaining this $\Free$ by the custom rules
%________________________________
\begin{equation}
\label{enlib}
\Free = -\kT \ln \Zr
\; ;
\qquad
\qquad
\Zr = {\ZTOT}/{\ZB}
\; .
\end{equation}
%________________________________
The idea will be to use now for $\Zr$ the perturbative result (\ref{zgral}).

From the definition of $\Free$ we can derive energies and entropies
in the usual way
%___________________________________
\begin{equation}
\label{enen}
%\Energy = \Free - T \frac{\partial \Free}{\partial T}
\Energy = \Free - T \partial_T \Free;
\qquad
\qquad
%\Entropy = - \frac{\partial \Free}{\partial T}
\Entropy = - \partial_T \Free
\; .
\end{equation}
%___________________________________
Thus, our convention gives 
$\Energy = \Energytot - \Energy_{\rm b}$
and 
$\Entropy = \Entropy_{\rm tot} - \Entropy_{\rm b}$ 
since all quantities follow from $\Free$ by linear operations.
Then, in processes not involving the bath, we are just redefining the
zero of all thermodynamical quantities.
But even in the bath is involved, as we have merely extracted its bare
contribution, it can be easily restored to the final results.

Let us illustrate what these definitions would give in a calculation
of the specific heat $\Cv$.
Imagine we want to find the $\Cv$ of our system, which as usual in
this work, is an open system and interacts, for example, with the
phonons of the lattice where it ``lives''.
To that end one applies a controlled heat pulse with a resistor, and
measures the temperature change $\Cv_{\rm tot} = \Delta Q/\Delta T$.
To isolate the contribution of the system from other sources
(phonons), one can repeat the measurement in the absence of ``system''
(for instance, if it consists of magnetic ions, we measure a compound
without spins, but having the same lattice). % \cite{carlin}).
This would give the bath specific heat $\Cv_{\rm b}$, which
experimentalists routinely substract from the original $\Cv$.
But the quantity $\Cv = \Cv_{\rm tot}-\Cv_{\rm b}$ is just what
directly comes from our redefinition of the energy, when computed in
the standard way $\Cv=-\partial_T \Energy$.

Instead of getting the thermodynamical energy as in (\ref{enen}), we
could have assumed that the relevant system energy is
$\langle \HS \rangle = \Tr ( \dm \HS )$.
But in general this would not agree with our $\Energy \neq \langle \HS
\rangle$. %
\footnote{
Such a way of computing the energy is possibly inherited from the
``reductionistic'' view of using the partial trace so extended in the
field of open systems.
}
The bare average of the system Hamiltonian $\HS$  misses the important
contribution of the system-bath coupling.
And we do not mean, as above, the additive part of the bath alone, but
the work done to ``insert'' the system in the bath.
Indeed, the use of $\langle \HS \rangle$ to discuss thermodynamical
principles instead of $\Energy$ can produce ``violations'' of the
second law (see the lucid discussion of Ford et al.~\cite{forlewoco85,
forlewoco88AP, foroco06, oco06}).

On the other hand, for observables not depending on $\HI$ ---we have in
mind the magnetization of a paramagnet where the coupling does not
depend on the magnetic field--- we have
%_________________________
\begin{equation}
\label{Mzfree}
\Mz = 
-
\frac {\partial \Free}{\partial B_z}
=
\langle S_z \rangle 
=
\Tr ( \dm S_z  )
\; .
\end{equation}
%________________________
That is, in this case both the definition from the free energy, or
directly from the reduced density matrix $\dm$, provide us with two
equivalent ways of computing the magnetization (that we will use in
chapter~\ref{chap:aplicationesII}).

%%%%%%%%%%%%%%%%%%%%%%%%%%%%%%%%%%%%%%%%%%%%%%%%%%%%%%%%%%%%%%%%%%%%%%%%%
\section{example: the harmonic oscillator}
\label{sec:HO}

Let us apply the thermodynamic formalism discussed to the example of a
harmonic oscillator $\HS = p^2/2m + \case m \who^2 \q^2$, bilinearly
coupled to the bath $F = -\q$.
This exactly solvable model has been thoroughly studied in the
literature \cite{weiss, haning05, foroco05}.
That allows us to compare our perturbative results with exact ones.
As $\F = - \q$ can be expressed as $\propto (a^+ +a)$ and the harmonic
oscillator's spectrum is equispaced 
$\Delta_{n n\pm 1} = \mp \hbar \who$, 
we have
%_______________________________
\begin{equation}
\Zdos =
-\beta \frac {\hbar}{2 m \who}
\bigg [
\se (- \hbar \who)
\sum_n \e^{-\beta \epsilon_n}
n
+
\se ( \hbar \who)
\sum_n \e^{-\beta \epsilon_n}
(n+1)
\bigg ]
\end{equation}
%_______________________________
%
with $\epsilon_n = \hbar \who (n +1/2)$.
Doing the geometric sums we arrive at
%____________________________________________________
\begin{equation}
\label{zdosho}
\frac {\Zdos}{\ZS}
=
-  \frac {\beta}{4 \who}
\Big [
\se_-
+
\coth \left (
\case \beta \hbar \who
\right )
\se_ +
\Big ]
\end{equation}
%____________________________________________________
with $\se_\pm \equiv \se (\hbar \who) \pm \se (-\hbar \who)$ and the
$\se$ function from (\ref{wuri}).

Now we can obtain $\Free$ directly via (\ref{enlib}), and by
derivation $\Energy$, $\Entropy$, $\Cv$, etc, as in (\ref{enen}).
To illustrate, we obtain the following specific heat in the
high-temperature range $ \hbar \who /\kT \ll 1$ 
%_________________________________
\begin{equation}
\frac {\Cv}{k_{\rm B}} \cong
1
-
\frac {\hbar \dampp}{2 \pi \kT }
\end{equation}
%________________________________ 
while in the regime of low temperatures $ \hbar \who /\kT \gg 1$:
%________________________________
\begin{equation}
\label{cvlt}
\frac {\Cv}{k_{\rm B}} \cong
\left ( \frac {\hbar \who}{\kT} \right )^2
\e^{-\beta \hbar \who}
+
\frac{\pi}{3}
\frac {\dampp\,\kT}{ \hbar \who^2}
\end{equation}
%________________________________
Both results agree with the expansion to first order in the damping of
the exact results by Ingold and Hanggi \cite{inghan06}.
%%%
%
\footnote{
We have used an Ohmic bath spectral distribution
$J (\omega) \sim m \dampp\,\omega$
regularized with an exponential cutoff $\cutoff$.
Then we know the explicit form of $\se$:
%_____________________
%_________________________________
\begin{equation}
\label{seohmic}
\se (\Delta) =
\dampp\,\frac {\Delta}{\pi}
\left [
2 \ln \left ( 
\frac {\Delta}{2 \pi \kT}
\right )
-
{\rm Re}\, \psi \left (1 + \iu  \frac {\Delta}{2 \pi \kT}   \right )
+
\ln \left ( \frac {\hbar \cutoff}{\Delta} \right )
\right ]
\end{equation}
%_________________________________
%_____________________ 
in terms of the $\psi$ function \cite{arfken} and the bath cutoff $\cutoff$.
%
%(This result will appear in $\equilibrio$, where we compute other
%quantities as well).
%
\label{foot:seoh}
} % END OF FOOTNOTE

Notice that in the low temperature regime, the dissipative corrections
give $\Cv \propto T$.
This is is contrast with the celebrated  $\dampp = 0$ result where $\Cv$
drops exponentially at low temperature, due to the gap between the
ground state and the first excited state.
To make sense of this, we repeat again that now the total system is
our system plus the bath, which removes the gap
(the reservoir has infinitely many oscillators with a quasi-continuous
distribution of $\omega_\bindex$'s).

\begin{figure}
\centerline{\resizebox{8.cm}{!}{%
\includegraphics[angle = -0]{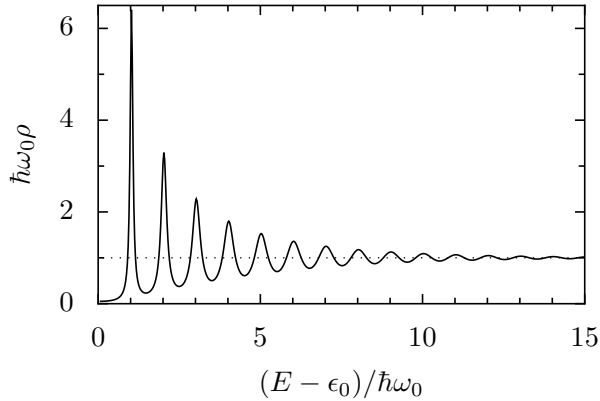}
}
}
\caption{
Density of states~(\ref{densitys}) for the harmonic oscillator (figure
extracted from \cite{haning05}).
Only the ground state contributes a delta to $\dens (\epsilon)$, the
other get ``blurred'' by the coupling with the bath, a feature often
used when reasoning about open systems.
}
\label{fig:density}
\end{figure}

Another way of looking at this is resorting to the definition of
density of states
%________________________________________
\begin{equation}
\label{densitys}
\dens (\epsilon)
=
\frac{\hbar}{2 \pi \iu}
\int_{-\iu \infty + c}^{+\iu \infty + c}
\dif \beta
\,
\Zr (\beta) 
\,
\e^{\beta \hbar \epsilon} 
\end{equation}
%________________________________________
That is, the inverse Laplace transform of the partition function.
For the central harmonic oscillator, Hanke and Zwerger \cite{hanzwe95}
obtained $\dens(\epsilon)$ exactly, finding that between the ground
state and the first excited state there exists a finite density of
states
(that is, there is no gap, see Fig.~\ref{fig:density}) 
%___________________________________________________
%___________________________________________________
%___________________________________________________
%___________________________________________________
%___________________________________________________
%___________________________________________________
%%

\newpage

%%%%%%%%%%%%%%%%%%%%%%%%%%%%%%%%%%%%%%%%%%%%%%%%%%%%%%%%%%%%%%%%%%%%%%
\section{final comments}

\label{sec:comenteq}

Performing the RWA in the master equation one gets Eq.~(\ref{dmeRWA}),
whose stationary solution coincides with the canonical.%
\footnote
{
This can be seen noting that the non-diagonal terms go to zero.
Inserting $\e^{-\beta \HS}$ in the diagonal part, and using detailed
balance (\ref{detailed}) one sees that $\e^{-\beta \HS}$ is indeed a
stationary solution.
}
That is, the equilibrium distribution at zero coupling.
Possibly this result has contributed to the ``belief'' that the system
always equilibrates towards $\e^{-\beta \HS}$ in the quantum case as
well.
But in general this is not so, and here we have studied the actual
form of the equilibrium distribution, and its relation with the
stationary solutions of the quantum master equations.

In the low damping range, the corrections computed here to the
thermodynamical quantities, though sensible, are probably small and not
easy to detect experimentally.
In a way, this contrasts with the situation in dynamics, where weak
coupling can produce sizable effects via relaxation and decoherence.
Anyway, from the conceptual point of view the rigorous treatment of
the equilibrium properties contributes to the understanding of those
dissipative contributions, absent in the classical case.
%%

%%%%%%%%%%%%%%%%%%%%%%%%%%%%%%%%%%%%%%%%%%%%%%%%%%%%%%%%%%%%
%%%%%%%%%%%%%%%%%%%%%%%%%%%%%%%%%%%%%%%%%%%%%%%%%%%%%%%%%%%%
%%%%%%%%%%%%%%%%%%%%%%%%%%%%%%%%%%%%%%%%%%%%%%%%%%%%%%%%%%%%
%%%%%%%%%%%%%%%%%%%%%%%%%%%%%%%%%%%%%%%%%%%%%%%%%%%%%%%%%%%%
%%%%%%%%%%%%%%%%%%  phase-space
%%%%%%%%%%%%%%%%%%%%%%%%%%%%%%%%%%%%%%%%%%%%%%%%%%%%%%%%%%%%
%%%%%%%%%%%%%%%%%%%%%%%%%%%%%%%%%%%%%%%%%%%%%%%%%%%%%%%%%%%%
%%%%%%%%%%%%%%%%%%%%%%%%%%%%%%%%%%%%%%%%%%%%%%%%%%%%%%%%%%%%

\chapter{working in phase space}
\label{chap:phase-space}

We conclude the general discussion of the formalism of open
systems with the reformulation of the master equations for $\dm$ in
phase space.
The phase space approach, together with canonical quantization and
path integral are three equivalent ways of quantizing \cite{zac02}.
In this formalism the theory ``lives'' in the space defined by the
canonical conjugate variables of Hamiltonian dynamics (phase-space).
Given that quantum and classical theory share now the same
mathematical space, the formalism provides an ideal framework to study
the transition from quantum to classical (or the other way around).
In this frame we will be able to relate explicitly the classical and
quantum theories of dissipation (chapters \ref{chap:clasico} and
\ref{chap:dinamica}).

We will motivate this chapter as Wigner did, in 1932, in his seminal
article \cite{wig32}.
We will introduce historically the development of the theory for
spinless particles.
Afterwards we will provide a more modern view, following the
postulates of Stratonovich to map Hilbert operators and functions in phase
space.
We then address the problem of systems with spin.
In both cases (particles and spins) we transform the master equations
into phase space.
Then the corresponding \FP\ equations are obtained in the
limit ``$\hbar \to 0$''.

Keeping in mind the spirit of Caldeira and Leggett, who posed the
theory of dissipation as a quantization problem \cite{calleg83}, this
chapter closes the circle, getting now the classical limit from the
quantum case.
The formalism discussed here will be very useful in
Chap.~\ref{chap:aplicationesI}, where we will solve quantum master
equations in phase space by adapting techniques developed to solve
classical \FP\ equations.

\newcommand{\Sdos}{{\mathbb{S}}^2}
%%%%%%%%%%%%%%%%%%%%%%%%%%%%%%%%%%%%%%%%%%%%%%%%%%%%%%%%%%%%%%%%%%%%%%
\section{the problem}

In classical mechanics the dynamics is given by the {\it Liouville
equation}:
%_______________________
\begin{equation}
\nonumber
\partial_t W
=
\{\HO, W \}
\end{equation}
%_______________________
where $W$ and $\HO$ are functions of the point $\pto$ in some
phase-space $\PS$.
To illustrate, $\pto = (\q,\p)$ in the one-dimensional particle case
($\PS = \mathbb R^2$) while $\pto = (\theta, \phi)$ for classical
``spins'' ($\PS = \Sdos$ with $\Sdos$ the sphere in three dimensions).

In quantum mechanics, the Liouville equation is substituted by {\it
  Von Neumann \/} equation:
%___________________
\begin{equation}
\nonumber
\frac
{ \dif \dm}{ \dif t} 
=
 -\frac{\iu}{\hbar} [\HO, \dm]
\end{equation}
%___________________
where now $\HO$ and $\dm$ are Hermitian operators in some Hilbert space.

In spite of the formal similarities 
$\{ \, , \, \} \leftrightarrow [\, ,\, ]$
these two equations seem difficult to relate.
We mean, it is not straightforward to take some appropriate limit in
Von Neumann equation and arrive at Liouville's.
They live in different mathematical spaces!
In this sense it turns to be involved to study the
classical-to-quantum transition, or the quantum-to-classical, in a
continuous way.

To bypass this problem we resort to the phase-space formalism, in
which both classical and quantum mechanics ``live'' in the same space.
Formulating the quantum theory as closely as possible to the classical
case would allow to study their crossover, and also see how the corrections
to the classical case are built;
as well as studying the problems from another point of view.
In particular, one can export notions and tools of the classical case
to the quantum world.
%%

%%%%%%%%%%%%%%%%%%%%%%%%%%%%%%%%%%%%%%%%%%%%%%%%%%%%%%%%%%%%%%%
\section{a constructive tale}

We are going to review the construction of the phase-space formulation
of quantum mechanics from a historical perspective.

%%%%%%%%%%%%%%%%%%%%%%%%%%%%%%%%%%%%%%%%%%%%%%%%%%%%%%%%%%%%%%%%
\subsection{Wigner's function (1932)}
\label{sub:1932}

In that year Wigner introduces the following ``distribution''
\cite{wig32}:
%________________________________
\begin{equation}
\label{wignf}
\W_\psi (\q, \p)
=
\frac{1}{2 \pi \hbar}
\int_{-\infty}^{\infty}
\dif y
\,
\e^{-\iu y \p/\hbar}
\psi^* (\q - \case y)
\psi (\q + \case y)
\; ,
\end{equation}
%________________________________
where $\psi$ is the quantum-mechanical wave function.
We have added the subindex to $\W_\psi$ in order to distinguish it
from the classical probability distributions 
(note that $\W_\psi = \W_\psi^*$).
Next introducing position ``eigenstates'' $\vl \q \rangle$, one defines
the transform of any operator $A$ as follows
%________________________________________________
\begin{eqnarray}
\label{WAwf}
\W_ {A}(\q,\p) = 
\int_{-\infty}^{\infty} \dif y \, \e^{\iu p y/\hbar} \langle \q -
\case y \vl A \vl \q + \case y \rangle
\end{eqnarray}
%____________________________
Note that $\W_A = \W_{A^+}^*$, that is, $\W_A$ is real for Hermitean $A = A^+$.
With these definitions the expectation value of $A$ in the state
$\vl\psi\rangle$ can be written as an ``average'' over phase space:
%_________________________________
\begin{equation}
\label{mvwf}
\langle \psi \vl A \vl \psi \rangle
=
\int \dif \p \int \dif \q \,
W_A(\q, \p) \W_\psi (\q, \p)
\end{equation}
%_________________________________
Thus we see that $\W_ \psi$ plays the role of a distribution.
Advancing results to fix ideas, $\W_A(\q,\p)$ is the classical
dynamical variable $A(\q,\p)$ corresponding to the operator $A$.
Putting $A = \mathbb I$ (the identity), we have the normalization
$ \int \dif \p \int \dif \q \W_\psi (\q, \p) =1 $.

The transformation properties of $\W_\psi$ are as follows \cite{hiletal84}.
Applying
 $\psi(\q) \to \psi (\q + r)$ 
we have
$\W_\psi (\q, \p) \to \W_\psi (\q + r, \p)$.
When
$\psi(\q) \to \psi (-\q)$, 
$\W_\psi$ transforms as 
$\W_\psi (\q, \p) \to \W_\psi (-\q , -\p)$.
Finally 
$\psi(\q) \to \psi^* (\q)$ 
gives
$\W_\psi (\q, \p) \to \W_\psi (\q , -\p)$.

We thus see that $\W_\psi$ is real, normalized, respects the
symmetries, and the averages are computed like in classical mechanics
(same for $\W_\dm$).\footnote
{
%\label{foot:wdm}
One can proceed analogously with $\dm$ (mixed states) introducing the
analogue to $\W_\psi$ (\ref{wignf}):
%_____________________________
\begin{equation}
\label{dmwf}
\W_\dm (\q, \p)
=
\frac{1}{2 \pi \hbar}
\int_{-\infty}^{\infty}
\dif y
\,
\e^{\iu y \p/\hbar}
\dm (\q - \case y,\q + \case y)
\end{equation}
%_____________________________
So the calculation of averages (\ref{mvwf}) now reads:
%____________________________
\begin{equation}
\Tr (A \dm )
=
\int \dif \p \int \dif \q \,
W_A(\q, \p) \W_\dm (\q, \p)
\; .
\end{equation}
%_____________________________
\label{foot:wdm}
} % END OF FOOTNOTE
But the analogy is not fully complete, some differences should show up!
$\W_\psi$ is not positive, and besides $\W_\psi \leq \pi/\hbar$.
That is, the value at any point $(\q, \p)$ of the quantum distribution
$\W_\psi$ is bounded.
This precludes, for instance, that one could write
$\W_\psi (\q, \p) = \delta(\q - \q_1) \delta(\p - \p_1)$
for a distribution localized at $(\q_1, \p_1)$.
As Moyal pointed out in 1949, this reflects the {\it uncertainty
principle}, forbidding well defined $\q$ and $\p$ ``simultaneously''
\cite{moy49}.

%%%%%%%%%%%%%%%%%%%%%%%%%%%%%%%%%%%%%%%%%%%%%%%%%%%%%%%%%%%%
\subsection 
{the 40s: Groenewold \& Moyal}
\label{sub:1946}

In 1946 Groenewold expresses the transform of the product of two
operators 
%______________________________
\begin{eqnarray}
\W_ {BA}(q,p) = 
\int_{-\infty}^{\infty} \e^{\iu p y/\hbar} \langle \q -
\case y \vl BA \vl \q + \case y \rangle \dif y
\end{eqnarray}
%______________________________
as follows
%______________________________
%_____________________________________
\begin{eqnarray}
\label{green}
\psF_{\opB\opA} (\q,\p)
=
\psF_{\opB} \,
\e^{\iu\hbar\Lambda/2}
\psF_{\opA}
%\vert_{(\q,\p)}
\qquad
\quad
\Lambda
:=
\frac{\overleftarrow {\partial}} {\partial \p}
\frac{\overrightarrow {\partial}}{\partial \q}
-
\frac{\overleftarrow {\partial}} {\partial \q}
\frac{\overrightarrow {\partial}}{\partial \p}
\end{eqnarray}
%_____________________________________
where the arrows indicate on which $\W$ the derivatives act upon.

In modern terminology the transform of the product defines an inner
product in $\PS$, called {\em star\/} product:
%_____________________-
\begin{equation}
\label{star}
\W_{BA}
=
\W_B
\star 
\W_A
\end{equation}
%______________________
Then the algebra of the product of operators (composition) is
transformed in the $\star$-product in $\PS$, inheriting its
properties.
Namely, it will be associative, $\W_{C(BA)} = \W_{(CB)A}$, but in
general non-commutative, $\W_{AB} \neq \W_{BA}$.

Von Neumann equation gets transformed in phase-space as
%__________________________
\begin{equation}
\label{vn}
\frac
{\dif \varrho}{\dif t}
=
-\frac{\iu}{\hbar} [H, \varrho]
\qquad
\longmapsto
\qquad
\partial_t \W_\dm
=
-\frac{\iu}{\hbar} ( \psF_{H} \star \W_\dm  - \W_\dm
 \star \psF_{H}
)
\end{equation}
%_______________________
%
Introducing now $\HS = \p^2/2m + V(\q)$ and using Groenewold's
result~(\ref{green}) we obtain Von Neumann dynamics in phase space
%_____________________________________
\begin{equation}
\label{moyal}
%%%%%%%%%%%%%%%%%%
\partial_{t}
\W (\q, \p, t)
=
\Big[
-\frac{\p}{m} \,
\partial_\q
+
V'\partial_\p
+
\sum_{s=1}^\infty
\frac{(\iu \hbar/2)^{2s}}{(2s +1)!}
V^{(2s +1)}
\partial_\p^{2s+1}
\Big ]
\W(\q, \p, t)
\; .
\end{equation}
%________________________________________
To abbreviate we removed the subindex $\dm$ for the density matrix
transform, and written $V^{(2s +1)} \equiv \partial_{\q}^{2s+1} V $.
Besides, we have extracted the $s=0$ term from the sum
($V'\partial_\p$) gathering all $\hbar$-dependent terms.
In this way the formal limit $\hbar \to 0$ recovers the classical
Poisson bracket:
%___________________________________
\begin{equation}
\partial_{t}
\W(\q, \p, t)
=
\left (
-\p /m \,
\partial_\q
+
V'\partial_\p
\right )
\W (\q, \p, t)
=
\{ \HO, \W \}
\; .
\end{equation}
%__________________________________
With the glasses of phase-space, the equation for the quantum dynamics
is built from the classical one simply adding more terms to the Poisson
bracket.%
\footnote
{
In this sense, the $\star$-product can be seen as a ``deformation'' of
the Poisson bracket.
This motivated the inverse problem, that is, given a Poisson bracket
how should we deform it to get quantum mechanics, {\it deformated
  quantization}; see \cite{hirhen02} for an introduction to this program.
} % END OF FOOTNOTE

Note that for the harmonic oscillator $\V = m \who^2 \q^2/2$ we have
$\V^{(2s +1)} = 0$ when $s \geq 1$, so Von Neumann's equation in phase
space is identical to the Liouville equation.
What is going on? Quantum and classical mechanics are the same for the
harmonic oscillator?
No, what happens is that in the classical and quantum cases the
admissible solutions are different.
Recall that $\W$ could take negative values and in addition is bounded
\cite{balvor90}.

In 1949 Moyal rounds up the work noting that Wigner's distribution
(\ref{WAwf}) is just the inverse of Weyl's prescription for symmetric
quantization ordering \cite{balescu}, that is:
%_____________________________
\begin{equation}
\label{weyl}
\q \p \stackrel {{\rm Weyl}}{\longrightarrow} 
\frac{1}{2}(\hat {\q} \hat {\p} + \hat {\p} \hat {\q} ) 
\stackrel{\rm Wigner}{\longmapsto}
\W_{(\hat \q \hat \p + \hat \p \hat \q )/2 } = \q \p 
\; .
\end{equation}
%______________________________
Besides, he computes the uncertainty relations in phase space
\cite{moy49}.
Therefore, reached this point ---calculation of observables,
$\star$-product, and dynamics--- the formalism in phase space was
completed, constituting an alternative to the standard canonical
quantization in Hilbert space.

From the 50s onwards many applications were developed of this
formalism.
Many of them exploited its analogy with the classical formalism,
adapting its schemes and tools to the quantum world \cite{hiletal84,
lee95, schleich, cofetal07}.

%%%%%%%%%%%%%%%%%%%%%%%%%%%%%%%%%%%%%%%%%%%%%%%%%%%%%%%%%%%%
\subsection {Bopp (1961)}
\label{sub:1961}

In 1961 Bopp introduces an operationally better way of computing the
star product
%__________________________________________
\begin{eqnarray}
%\nonumber
\psF_{\opB}
\star
\psF_{\opA}
=
\opB(\bopQ, \bopP) \, \psF_{\opA}
%\\
\; ;\qquad \qquad
\bopQ := \q +\frac {\iu\hbar}{2} \frac {\partial}{\partial \p}
,
\quad
\bopP := \p -\frac {\iu\hbar}{2} \frac {\partial}{\partial \q}
\label{bopppar}
\end{eqnarray}
%________________________________
where $\opB(\bopQ, \bopP)$ acts on the functions in $\PS$, with the
same functional form a
$\opB (\hat \q, \hat \p)$
but with
$\hat \q$ $\&$  $\hat \p$ 
being replaced by the differential operators $\bopQ$ and $\bopP$.
As an aside, note that these operators obey in $\PS$ the same
commutation rules as their Hilbert analogues  $\hat \q$ and $\hat \p$,
that is, 
$[\bopQ, \bopP] = \iu \hbar$. %
\footnote{
This can be checked directly from the definitions (\ref{bopppar}) of
the Bopp's or by simply noting that
 $\W_{[\hat \q, \hat \p] A} = \iu \hbar \W_{A}$.
} %END OF FOOTNOTE

Let us apply Bopp's result to Von Neumann dynamics.
To this end we use $\dm = \dm^+$ and $\HO = \HO^+$, whence 
$(\HO \dm)^+ = \dm \HO$,
together with $\W_A = \W_{A^+}^*$ for adjointing.
Then, $\W_{\HO \dm} = \W_{\dm \HO}^*$, and we obtain
%______________________________
\begin{eqnarray}
\label{vnpsp}
\partial_t \W
=
2
{\rm Im}
\Big (
 \HO(\bopQ,\bopP) \Big )  \W 
\; .
\end{eqnarray}
%_______________________________
The Hamiltonian $\HO$ is now a function of the Bopp operators $\bopQ$
and $\bopP$ which act on the functions in $\mathbb R^2$.
The equation above is a compact way of writing the
overwhelming~(\ref{moyal}).
%%

%%%%%%%%%%%%%%%%%%%%%%%%%%%%%%%%%%%%%%%%%%%%%%%%%%%%%%%%%%%%
\subsection {Caldeira--Leggett equation in phase space (1983)}
\label{sub:1983}

Let us apply the formalism to the dynamics of open systems.
This will prove useful in Chapter \ref{chap:aplicationesI}.
We want to transform into phase space the master
equation~(\ref{cleq}), which we briefly recall
%____________________________________
\begin{equation}
\nonumber
\frac {\dif \dm }{\dif t}
=
-\frac {\iu}{\hbar}
[\HS, \dm]
-\frac{m \dampp\,\kT}{\hbar^2}
[\q, [\q, \dm]] -
\iu \frac {\dampp}{2  \hbar}
[\q, [ \p, \dm ]_{+}]
\; ,
\end{equation}
%___________________________________
with $[\, , \,]_{+}$ the anticommutator.
To this end, the Bopp operators~(\ref{bopppar}) are very convenient,
since
%___________________________
\begin{equation}
\label{tr1}
[\q, [\q, \dm]]
=
\q^2 \dm - 2 \q \dm \q + \dm \q^2
\longmapsto
%4 {\rm Im}^2 (\bopQ) 
\Big ( \bopQ  - {\rm c.c.} \Big )^2
\W 
=
- \hbar^2
\frac{\partial^2}{\partial \p^2}
\W
\end{equation}
%__________________________
(c.c. $\equiv$ complex conjugate) and
%__________________________
\begin{equation}
\label{tr2}
[\q, [ \p, \dm ]_{+}]
=
\q \p \dm 
+
\q \dm \p
-
\p \dm \q
- \dm \p \q
\longmapsto
\Big (  \bopQ \bopP -  {\rm c.c.} \Big )\W
+
\Big (  \bopQ \bopP^* -  {\rm c.c.} \Big )\W
=
\iu
\hbar 
\frac{\partial}{\partial \p} \p \W 
\end{equation}
%__________________________
Introducing these two results in the Caldeira--Leggett equation, and
accounting for the transformation~(\ref{moyal}) of the unitary part
$-(\iu/\hbar)[\HS, \dm]$, one arrives at
%_____________________________________
\begin{equation}
\label{cleqps}
%%%%%%%%%%%%%%%%%%
\partial_{t}
\W  (\q, \p, t)
=
\Big[
-\frac{\p}{m} \,
\partial_\q
+
V'\partial_\p
+
\dampp
\partial_\p
\big(\p+ m\,\kT \partial_\p \big)
+
\sum_{s=1}^\infty
\frac{(\iu \hbar/2)^{2s}}{(2s +1)!}
V^{(2s +1)}
\partial_\p^{2s+1}
\Big ]
\W(\q, \p, t)
\; .
\end{equation}
%________________________________________
We have written the equation is such a way that the first three terms
correspond to the classical Klein--Kramers equation~(\ref{wkk}).
And the last term contains the purely quantum terms coming from the
closed evolution.
It is important to remark that the terms (\ref{tr1}) and (\ref{tr2})
are the diffusion and dissipation terms in the classical case.
That is, the Caldeira--Leggett equation is nothing but the Von
Neumann evolution augmented with the classical irreversible terms.
This identification has been possible thanks to the phase space
formalism.
Who could have guessed that 
$[\q, [\q, \dm]]$ and $[\q, [ \p, \dm ]_{+}]$
were indeed the dissipative terms of the classical Klein--Kramers
equation?
%%

%%%%%%%%%%%%%%%%%%%%%%%%%%%%%%%%%%%%%%%%%%%%%%%%%%%%%%%%%%%
\section{other symmetries and Stratonovich's postulates}

We have seen how the phase-space formalism developed from the
introduction of the Wigner distribution.
That is, everything followed from the map between operators and
dynamical variables in phase space [Eqs.~(\ref{WAwf}) and
(\ref{dmwf})].
But the map is restricted in the sense of ``only'' being applicable to
``particle systems'', we mean, to the algebra
$[\hat \q, \hat \p] = \iu \hbar$,
corresponding to the Poisson bracket $\{\q, \p\}=1$.
Then, how could we proceed if we want to tackle problems with other
commutation rules?
We have in mind a phase-space description for systems with
angular-momentum type algebra
$[S_i, S_j] = \iu \hbar \epsilon_{ijk} S_k$,
corresponding to the classical Poisson bracket 
$\{m_i, m_j\} = \gyr \epsilon_{ijk} m_k$
of Eq.~(\ref{poiss}) ($\gyr$ is the gyromagnetic ratio, and
  $\epsilon_{ijk}$ the completely anti-symmetric unit tensor).

The answer was given by Stratonovich \cite{str56}, completed by
Varilly \& Gracia-Bondia \cite{vargra89}, and later generalized by Briff
and Mann for any finite Lie group \cite{briman98}.
In brief, the idea is to find a map between Hilbert operators and
phase-space functions by imposing the ``physically'' reasonable
properties obeyed by the Wigner function (section \ref{sub:1932}).

%%%%%%%%%%%%%%%%%%%%%%%%%%%%%%%%%%%%%%%%%%%%%%%%%%%%%
\subsection{the Stratonovich postulates (1956)}
\label{sub:1956}

Mathematically one starts with a physical system having a given
``symmetry group'' 
($[\hat \q, \hat \p] = \iu \hbar$, 
or
$[S_i, S_j] = \iu \hbar \epsilon_{ijk} S_k$, etc)
corresponding to a classical phase space $\PS$. %
\footnote{
A somewhat more mathematical summary can be read in $\colibri$.
} % END OF FOOTNOTE
Then one seeks for a map between the Hilbert space and $\PS$ obeying the
following Stratonovich postulates:
\begin{itemize}
\item [(i)] linearity

\vskip0.2cm 
$\opA \mapsto \psF_{A}$ 
\quad
is linear and bijective.
\vskip0.2cm

\item [(ii)] ``reality''

\vskip0.2cm
$\psF_{A^+} = (\psF_{A})^{*}$
\quad
where $\opA^{+}$ is the adjoint of $\opA$.  
\vskip0.2cm

\item [(iii)] Normalization  ($\dif \mu$ is the area element in $\PS$):
 
$$\int_{\PS} \psF_{A}(\pto)\dif\mu(\pto) = \Tr (A).$$

\item [(iv)] trace property
$$\int_{\PS}\psF_{\opA}(\pto) \psF_{\opB}(\pto)\dif\mu(\pto) 
=
\Tr(\opA\opB).$$

\item [(v)] group covariance
$$\psF_{A^g}(\pto) = \psF_{A}(g\cdot\pto) \;, \quad \forall g\in G.$$
where  $G$ is our symmetry group and 
$A^g:={\cal U}(g^{-1})A \, {\cal U}(g)$ 
with ${\cal U}$ the group generators in Hilbert space.

\end{itemize}

In \cite{vargra89, briman98} it was  proved that these properties
determine uniquely the map.
The map is bijective (i) and hence invertible.
Condition (ii) ensures that $\W_A$ is real if $A = A^+$ [see after
(\ref{WAwf})].
Postulate (iii) is simply the normalization condition [see below
(\ref{mvwf})].
Linearity plus (iv) ensure that the averages are computed as in
Eq.~(\ref{mvwf}).
Finally (v) expresses respect to the basic symmetries of the system.
In the case of particles, these postulates produce the Wigner function
discussed above.

%%%%%%%%%%%%%%%%%%%%%%%%%%%%%%%%%%%%%%%%%%%%%%%%%%%%%%%%%%%%%%%%%%%%%%%%%%%
\section{the case of spin systems}

Let us apply the formalism to spin problems.
Here the dynamics is governed by a spin Hamiltonian 
$\HO (S_x, S_y, S_z)$ 
with
%_____________________________
\begin{equation}
[S_i, S_j]
= \iu \epsilon_{ijk} S_k
\; .
\end{equation}
%_____________________________
In the standard basis
$S_z \vl m \rangle = m \vl m \rangle$ 
and 
$S^2 \vl m \rangle = S(S+1) \vl m \rangle$, 
for
$m = -S, \dots, S$, the spin quantum number.%
\footnote{
Recall that $S = (\mu_{\rm B}/g)/\hbar$ with $\gyr$ the gyromagnetic
ratio and $\mu_{\rm B}$ Bohr's magnetron (section~\ref{subs:ejcl}).
In this section and in chapter~\ref{chap:aplicationesII} we will set
$\mu_{\rm B} = \hbar =1$ (as in $\colibri$).
Then the classical limit will correspond to $S \to \infty$.
\label{foot:units}
} % END OF FOOTNOTE
In this case, following Ref.~\cite{vargra89} we have the map
%__________________________________________
\begin{equation}
\label{swsspins}
\W_A(\theta, \phi)
=
\Tr (\sws (\theta, \phi) A)
\, ;
\qquad 
\sws (\theta, \phi)
=
\sqrt {\frac{4 \pi}{2S +1}}
\sum_{l=0}^{2S}
\sum_{m=-l}^{l}
T_{lm} Y_{lm}^{*} (\theta, \phi)
\; ,
\end{equation}
%__________________________________________
where 
$T_{lm} 
= 
\sqrt {(2l+1)/(2S+1)} 
\times 
\sum_{j, j'} 
\langle S, j; S, j' \vl S, j'\rangle 
\vl S, j'\rangle \langle S, j \vl$ 
are irreducible tensors, the
$ \langle S, j; S, j' \vl S, j'\rangle$ 
are Clebsch-Gordan coefficients and $Y_{lm}$ spherical harmonics
\cite{galindo, arfken}.

Klimov and Espinoza \cite{kliesp02} obtained a differential form for
the spin $\star$-product which reads:
%__________________________________________
\begin{equation}
\label{ksp}
\psF_{\opB}
\star
\psF_{\opA}
=
N_S
\sum_{j}
a_j
\widetilde {G}^{-1} (\cas^2)
\Big[\Big(
\opS^{+, (j)}
\widetilde {G} (\cas^2)
\psF_{\opB}\Big)
\otimes
\Big(
\opS^{-, (j)}
\widetilde {G} (\cas^2)
\psF_{\opA}\Big)
\Big]
\end{equation}
%_______________________________________
Here $N_S = \sqrt {2S +1}$ and $\cas^2$ is the angular momentum
operator on the sphere
%_______________________________________
\begin{equation}
\cas^2
= 
-
\left [
\frac {\partial^2}{\partial \theta^2}
+
\cot \theta
\frac {\partial}{\partial \theta}
+
\frac {1}{\sin^2 \theta}
\frac {\partial^2}{\partial \phi^2}
\right ]
\; ,
\qquad
\qquad
\cas^2 Y_{lm}
=
l (l+1) Y_{lm}
\; .
\end{equation}
%______________________________________
Besides
%_______________________________________
\begin{equation}
\widetilde {G} (\cas^2) Y_{lm}
= G (l) Y_{lm}
\; ,
\qquad
\qquad
\qquad
G(l) = \sqrt {(2S + l +1)!(2S- l)!}
\;,
\end{equation}
%_______________________________________
%____________________________________
\begin{equation}
\label{opS}
\opS^{\pm, (j)}
=
\prod_{k=0}^{j-1}
\left (
k \cot \theta
- 
\frac {\partial}{\partial \theta}
\mp
\frac {\iu}{\sin \theta}
\frac {\partial}{\partial \phi}
\right )
\; ,
\qquad
a_j
=
\frac {(-1)^j}{j! (2S + j +1)!}\; .
\end{equation}
%____________________________________
%
In formula~(\ref{ksp}) the operators $S^{\pm}$ and $\widetilde G$
inside the brackets act only inside the bracket, while $\widetilde
G^{-1}$ on the left acts on everything to its right.

Equations~(\ref{swsspins}) and (\ref{ksp}) are, respectively, the spin
analogues of the Wigner map and Groenewold $\star$-product for
particles.

%OJO
\vspace*{1.em}

%%%%%%%%%%%%%%%%%%%%%%%%%%%%%%%%%%%%%%%%%%%%%%%%%%%%%%%%%%%%%%%%%%%%%%%%%%
\subsection{Bopp operators for spins}

At first sight the $\star$-product (\ref{ksp}) looks of little use in
explicit calculations.
Then, in the spirit of Bopp, we would like to find a simpler
representation for $\W_B \star \W_A$.
To this end we explicitly calculated in $\colibri$ the transformation
$\W_{S_i A}$, $i = x, y, z$, finding
%_____________________________
\begin{equation}
\label{boppgral}
\psF_{S_i} \star \psF_{A} = \bopps_i \,
\psF_{A},\quad\quad i=x,y,z \; ,
\end{equation}
%_____________________________
where
%_______________________________
%_________________________________
\begin{equation}
\bopps_i = 
m_i \,
\tilde{\eta}_1(\cas^2) + \iu
({\bf m} \times {\bf L})_{i} \,
\tilde{\eta}_2(\cas^2) + 
\case L_{i}
\end{equation}
%_________________________________
Here
${\tilde\eta}_i (\cas^2)Y_{lm}={\eta_i}(l)Y_{lm}$ for $i=1,2$,
and the numerical coefficients are
%_________________________________
\begin{eqnarray}
\label{eta1}
\eta_1 (l)
&=&
\frac{1}{2 (2l+1)}
\left [
(l+1)
\sqrt {(2S+1)^2 - (l+1)^2}
+
l
\sqrt {(2S+1)^2 + l^2}
\right ]
\\
\label{eta2}
\eta_2(l) 
&=&
\frac{1}{2 (2l+1)}
\left [
\sqrt {(2S+1)^2 - (l+1)^2}
-
\sqrt {(2S+1)^2 + l^2}
\right ]
\end{eqnarray}
%_____________________________________
The vector
$ \bm (\theta,\phi) = ( \sin \theta \cos \phi, \sin \theta \sin
\phi,\cos \theta)$ 
was introduced in (\ref{poiss}) while ${\bf L}$ is
%_________________________________
\begin{equation}\label{commL}
{\bf L} = -\iu \left (
{\bf m} \times \frac {\partial}{\partial {\bf m}}
\right )
\end{equation}
%_________________________________
obeying
 $[L_i, L_j] = \iu \epsilon_{ijk} L_k$.
\footnote{ 
The following formulas are of assistance
%_________________________________
\begin{eqnarray}
\label{conmvec}
[m_i, L_j] = \iu \epsilon_{ijk} m_k
\; ;
\qquad
[({\bf m \times L})_i, L_j] 
= 
\iu \epsilon_{ijk} ({\bf m \times L})_k.
\end{eqnarray}
%_______________________________
%
} % END OF FOOTNOTE
Finally, invoking the associativity of the star product, for an
arbitrary operator $\opB(S_x, S_y,S_z)$ we have
%______________________________________
\begin{equation}
\label{mrb}
\psF_{\opB\opA}
=
\opB (\bopps_x, \bopps_y, \bopps_z) 
\psF_{\opA}
\end{equation}
%______________________________________
where 
$\opB (\bopps_x, \bopps_y, \bopps_z)$ 
is the same function of its arguments as $\opB(S_x, S_y,S_z)$, but now
the $\bopps_i$ are ``classical'' differential operators on the sphere.
To abbreviate, instead of writing
$\opB (\bopps_x, \bopps_y, \bopps_z)$ 
we simply write $\opB (\bopps)$.

Thus the $\bopps_i$ in the sphere play the same role as 
$\bopQ=\q + \iu(\hbar/2) \partial_{\p}$ 
and
$\bopP= \p - \iu(\hbar/2) \partial_{\q}$
in $\mathbb R^2$.
For example, in analogy with the $\bopQ$ and $\bopP$ commutation, we
have here
$[\bopps_i, \bopps_j] = \iu \epsilon_{ijk} \bopps_k$.
That is, the $\bopps_i$ obey the same commutation rules as their $S_i$
partners (for the same reason as discussed in Sec.~\ref{sub:1961}).
To conclude the analogy with the particle problem, the dynamics can
also be written in the following compact form [cf. Eq.~(\ref{vnpsp})]
%_______________________________________
\begin{equation}
\label{vnpss}
\partial_t \W
=
2
{\rm Im}
\Big (
 H(\bopps) 
\Big )
 \W
\end{equation}
%_______________________________________

%%%%%%%%%%%%%%%%%%%%%%%%%%%%%%%%%%%%%%%%%%%%%%%%%%%%%%%%%%%%
\subsection{an example}

Let us apply the formalism to the spin Hamiltonian studied in this work
%[see section \ref{sub:ourspin} and Eq.~(\ref{spinham})]:
%________________________
\begin{equation}
\label{spinhamex}
\HO = -\ani S_z^2 - \Bz S_z
\end{equation}
%________________________
%
The Poisson bracket (\ref{poiss}) produces in this case the
differential equation:
%_______________________________
\begin{equation}
\label{clex}
\partial_t \W
=
-
\big [
2 D S
\cos \theta
+
\Bz
\big ]
\frac {\partial\W}{\partial \phi}
\end{equation}
%_______________________________
In the quantum case we employ (\ref{mrb}) and (\ref{boppgral})
getting:
%_____________________________________
\begin{equation}
\partial_t \W
=
-
\bigg [
2 D
\Big (
{\cos \theta\,
\tilde\eta}_1
+
{\sin \theta\,
\frac {\partial}{\partial \theta}
\tilde \eta_2}
\Big )
+
\Bz
\bigg]
\frac {\partial\W}{\partial \phi}
\end{equation}
%_____________________________________
The first thing one notices is that for $D =0$ (isotropic spin) the
differential equation is the same in the classical and quantum cases:
%_____________________________________
\begin{equation}
\partial_t \W
=
-
\Bz
\frac {\partial\W}{\partial \phi}
\end{equation}
%_____________________________________
much as in the harmonic oscillator problem (section \ref{sub:1946}).
But as discussed there, the admissible solutions are different in the
classical and quantum problems.
Second, when $D\neq 0$ we can expand $\tilde {\eta}_1$ and
$\tilde{\eta}_2$ for large $S$ [Eqs.~(\ref{eta1}) and (\ref{eta2})] ,
obtaining
%_________________________________
\begin{eqnarray}
\label{asympfs}
&& 
2{\tilde\eta}_1 
= 
2S+1 - \frac{\cas^2 +1}{2(2S+1)}
+ {\rm O} \left (\frac {1}{S^2} \right )\;,
\cr
&& 
2{\tilde\eta}_2 
= 
-\frac {1}{2 (2S+1)} + {\rm O} \left (\frac {1}{S^2} \right ) 
\; .
\end{eqnarray}
%_________________________________
That is 
${\tilde\eta}_1 \cong 2S$ and ${\tilde\eta}_2 \cong 0$
as $S \to \infty$, recovering the classical equation~(\ref{clex}) for
the anisotropic spin.
Thus using the expansion~(\ref{asympfs}) one could get the
corresponding semiclassical evolution with $1/S$ corrections.%
\footnote { 
In $\colibri$ the dynamics for a general quadratic Hamiltonian was
obtained.
Besides, we took the classical limit for an arbitrary $\HS$, getting
the classical Poisson bracket as in the example discussed here.
} % END OF FOOTNOTE
%

%%%%%%%%%%%%%%%%%%%%%%%%%%%%%%%%%%%%%%%%%%%%%%%%%%%%%%%%%%%%%
\subsection{application to spin master equations}

\label{sub:2006}

As we did in the particle case (section \ref{sub:1983}) let us apply
the phase-space formalism to open systems.
We will consider the semiclassical regime, and eventually take the
limit $S \to \infty$.
Therefore, it will suffice to use the semiclassical master
equation~(\ref{RhT})
%______________________________
\begin{equation}
\label{dmehts}
\frac{\dif \varrho}{\dif t}
=
-\iu [\HS, \varrho]
-
\damps T
\Big (
[\F, \F \varrho] - \frac{1}{2T} [\F, [\HS, \F]\varrho] + {\rm h.c.}
\Big )
\, ,
\end{equation}
%______________________________
which has been written with the conventions used in this chapter
(recall footnote \ref{foot:units}).

In order to transform the master equation to phase space, we will
resort to our spin Bopp operators (\ref{boppgral}).
We need to compute [cf. (\ref{tr1}) and (\ref{tr2})]:
%_________________________________
\begin{equation}
[\F, \F \varrho] + {\rm h.c.} 
\quad
\mapsto
\quad
\Big (
\F (\bopps) - {\rm c.c.}
%\F^* (\bopps^{(\ord)})
\Big )^2 \W
\end{equation}
%_________________________________ 
%(c.c. $\equiv$ complejo conjugado) 
as well as
%__________________________________
\begin{equation}
%\hspace{-1.5cm}
[\F, [\HS, \F]\varrho]
+
{\rm h.c.}
\quad
\mapsto
\quad
\Big (
\F (\bopps) - {\rm c.c.}
%\F^* (\bopps^{(\ord)})
\Big )
\Big (
[\HS (\bopps), \F(\bopps)]
- {\rm c.c.}
%[\HS^* (\bopps^{(\ord)}), \F^*(\bopps^{(\ord)})]
\Big )
\W
\end{equation}
%__________________________________
Then the master equation (\ref{dmehts}) gets transformed into
%__________________________________
\begin{eqnarray}
\label{qfp}
\hspace{-2.ex}
\partial_t \W
=
\bigg \{
2{\rm Im}
\big (
\HS (\bopps)
\big )
+ 4\damps T
\Big [
{\rm Im}^2
\big (
F (\bopps)
\big )
%\nonumber\\[8pt]
-
\frac {1}{2T}
 {\rm Im}
\big (
F (\bopps)
\big )
 {\rm Im}
\big (
[\HS (\bopps), \F (\bopps)]
\big )
\Big ]
\bigg \}
\W
\end{eqnarray}
%__________________________________

As we did for the Caldeira--Leggett particle case, we will consider
{\em bilinear\/} coupling, that is $F= \sum \alpha_i S_i$, linear in
the system variables.
In contrast with the mechanical case, here $[\HS, F]$ depends always
of $\HS$, so that we cannot write the dissipative terms in a generic
way.%
\footnote {
Recall that in the generic case $\HS = \p^2/2m + V(\q)$ with $F =-\q$,
one has $[\HS, F] = \iu \hbar \p/m$, which does not depend on $V(\q)$.
} % END OF FOOTNOTE 
Considering the simplest case $\HS = \sum B_i S_i$ (isotropic spin),
we obtain:
%_________________________________________
\begin{eqnarray}
\label{qfpbil}
\hspace{-4.ex}
\partial_t \W
=
-\frac{1}{S}
\frac {\partial}{\partial {\bf m}}
\cdot
\bigg \{
({\bf m} \times {\bf B}_{\rm eff})
-
{\bf m}
\times
\hat \tL
\Big [
%\Big ( 
{\bf m}
%+
%{\bf M} \Big )
\times
\Big (
{\bf B}_{\rm eff}
-
T
\frac{\partial}{\partial {\bf m}}
\Big )
%\\ \nonumber
+
{\bf M}
\times 
{\bf B}_{\rm eff}
\Big ]
\bigg \}
\W
\end{eqnarray}
%_________________________________________
Here 
${\bf B}_{\rm eff} = -\partial \HS/\partial {\bf m}$
is the effective field of the classical equations, while we got an
extra term
${\bf M} = {\bf m}({\tilde\eta_1} - S)+\iu ({\bf m}\times {\bf
L}){\tilde\eta}_2$.
However, noting that 
$ {\tilde\eta_1} - S= {\rm O}(1/S)$
the term ${\bf M}$ disappears in the classical limit, recovering the
\FP\ equation~(\ref{clfp}). %
\footnote{
In $\colibri$ we took the classical limit for generic $\HS$ and $\F$,
also arriving at the \FP\ equation~(\ref{clfp}).
} % END OF FOOTNOTE

In this problem, even in the simplest case, the diffusion and
relaxation terms contain quantum corrections.
Recall that in the particle problem the diffusion and relaxation terms
where identical to the classical ones [Eq.~(\ref{cleqps})].

There exists an analogy with classical spin problems.
There a bilinear coupling induces multiplicative noise in the Langevin
equation, in contrast with the particle case, where the noise enters
additively for bilinear coupling.%
\footnote{
Compare the particle Langevin equation~(\ref{lanrig}) with the spin
Langevin equation~(\ref{lanspin}).
In the former, the noise $f(t)$ enters additively, whereas in the
latter $f(t)$ appears multiplying ${\bf m}$ and 
$\partial F / \partial {\bf m}$.
} % END OF FOOTNOTE
Formally, this is due to the angular Poisson bracket 
$\{m_i, m_j\} = \gyr \,\epsilon_{ijk} m_k$
which, even for $\F$ linear in $m_i$, gives a non-constant $\{A, \F\}$
in the Hamilton equation~(\ref{hama}) for any dynamical variable $A$
\cite{gar2000}.
Quantum mechanically, it is the angular-momentum commutation relations
which render $[\F, [\HS, \F]]$ non constant, which eventually
produces the term ${\bf M}$ in the master equation.
%% 

%%%%%%%%%%%%%%%%%%%%%%%%%%%%%%%%%%%%%%%%%%%%%%%%%%%%%%%%%%%%%%%%%%%%%%%%%%%
\section{other distributions}

What we have told here is not the whole story.
Canonical quantization comes associated with the ordering problem.
In path integrals, similarly, it shows up in the choice of the
evaluation point \cite{greifq}.
Analogously, in phase space this results in the ambiguity in the map
between Hilbert operators and $\PS$ functions --- we can construct other
distributions, each associated with a given ordering.
In the 60s, Cahill and Glauber \cite{cahgla69} constructed distributions
(called P and Q functions) in terms of the creation and destruction
operators of the harmonic oscillator.
Those distributions differ from the Wigner distribution~(\ref{wignf})
in that they correspond to normal and anti-normal ordering,
respectively \cite{hiletal84}.
For normal ordering (P-function) [cf. Eq.~(\ref{weyl})]:
%_________________________-
\begin{equation}
\frac{1}{2}(\hat {\q} \hat {\p} + \hat {\p} \hat {\q} ) 
\mapsto
P_{(\hat \q \hat \p + \hat \p \hat \q )/2 } = \q \p - 2 \iu \hbar
\end{equation}
%_________________________
while for the  Q-function 
$P_{(\hat \q \hat \p + \hat \p \hat \q )/2 } = \q \p + 2 \iu \hbar$.
That is, the phase-space function $\q \p$ corresponds to the operator
$\hat \q \hat \p$ in normal ordering and to   $\hat \p \hat \q$ for anti-normal.

These distributions can also be generated from the Stratonovich
postulates of Sec \ref{sub:1956}.
Then the map gets parametrized by a superindex $\ord$ and the
transformed function is denoted by $\W_A^{(\ord)}$; for $\ord = \pm 1$
one has the normal and anti-normal cases, while $\ord = 0$ corresponds
to the Weyl symmetric ordering discussed throughout.
Postulate (iv), the trace property, is then modified as
$$\int_{\PS}\psF_{\opA}^{(\ord)}(\pto)
\psF_{\opB}^{(-\ord)}(\pto)\dif\mu(\pto) = \Tr(\opA\opB).$$
Therefore the symmetric function ($\ord = 0$) is special, unique, in
the sense that to compute the averages we only resort to functions
with $\ord = 0$.
Any other $\ord$ requires transformed functions with $\ord$ and
$-\ord$.

In the time before Varilly $\&$ Gracia-Bondia, the spin distributions
followed the approach of Cahill and Glauber.
The construction was done, instead of using $a$ and $a^+$, in terms of
Bloch states or spin coherent states \cite{zhafengil90} (those were
the 70s).
With this approach Takahashi and Shibata found the spin Bopp operators
corresponding to the orderings $\ord = \pm 1$ \cite{takshi76}; an
important result they called the product theorem.
In $\colibri$, under the modern postulated-based approach, we
generalized the formalism to any ordering (here we have described the
$\ord=0$ case only).

%%%%%%%%%%%%%%%%%%%%%%%%%%%%%%%%%%%%%%%%%%%%%%%%%%%%%%%%%%%%%%%%%%%%%%%%%%%
\section{summary}

In this chapter we have surveyed the construction of the phase-space
approach to quantum mechanics.
The particle Bopp operators, and our generalization to spin problems,
have allowed us to transform easily the master equations into phase
space (sections \ref{sub:1983} and \ref{sub:2006}).
With the equations so transformed we have recovered, in the classical
limit, the corresponding \FP\ equations, and we have seen how the
first quantum corrections are generated.
In this way we have connected the classical and quantum theories of
dissipation of chapters \ref{chap:clasico} and \ref{chap:dinamica}.
On the other hand, as we will see in Chap.~\ref{chap:aplicationesI},
transforming the Caldeira--Leggett equation~(\ref{cleqps}) to phase
space with be instrumental for its numerical solution.

%%%%%%%%%%%%%%%%%%%%%%%%%%%%%%%%%%%%%%%%%%%%%%%%%%%%%%%%%%%%
%%%%%%%%%%%%%%%%%%%%%%%%%%%%%%%%%%%%%%%%%%%%%%%%%%%%%%%%%%%%
%%%%%%%%%%%%%%%%%%%%%%%%%%%%%%%%%%%%%%%%%%%%%%%%%%%%%%%%%%%%
%%%%%%%%%%%%%%%%%%%%%%%%%%%%%%%%%%%%%%%%%%%%%%%%%%%%%%%%%%%%
%%%%%%%%%%%%%%%%%%  METHODS
%%%%%%%%%%%%%%%%%%%%%%%%%%%%%%%%%%%%%%%%%%%%%%%%%%%%%%%%%%%%
%%%%%%%%%%%%%%%%%%%%%%%%%%%%%%%%%%%%%%%%%%%%%%%%%%%%%%%%%%%%
%%%%%%%%%%%%%%%%%%%%%%%%%%%%%%%%%%%%%%%%%%%%%%%%%%%%%%%%%%%%

\chapter{methods}
\label{chap:metodos}

This chapter is a transition between the general treatment hitherto
discussed and the chapters to come, devoted to specific applications.
Here we explain what we intend to calculate and how to carry it out
practically.

We present first the linear response theory for the obtainment of
dynamical susceptibilities.
Then we use the example of the two-level system to illustrate how
those susceptibilities can give us relaxation and decoherence times.
Finally, we introduce the continued-fraction method to solve master
equations and the practical computation of response functions.
%%

%%%%%%%%%%%%%%%%%%%%%%%%%%%%%%%%%%%%%%%%%%%%%%%%%%%%%%%%%%%%
\section{linear response theory (general formalism)}
\label{sec:LRTI}

We start recalling the protocol we discussed in
chapter~\ref{chap:dinamica} for the study of the dynamics using master
equations (section \ref{sec:dmeclmodel}, fig. \ref{fig:timeline}).
During $-\infty <  t < 0$ we let the system thermalize to the bath.
At $t=0$ we add some field, or change some parameter in the system
Hamiltonian $\HS$, and study the dynamics.
Thus master equations are a natural frame to address problems where
the system is perturbed, and study its adjustment to the new
conditions.

In order to not modifying the intrinsic nature of the system studied,
the probe can be made small enough \cite{dattagupta}.
But then the dynamics can be studied within perturbation theory.
Linear response theory (LRT), along with dealing with such
adjustments, provides the link with a typical experimental situation
--- the measurement of the dynamical susceptibilities.

So hands at work.
For $-\infty <t <0$ we can write the dynamics formally as
%___________________________
\begin{equation}
\partial_t
\dm
=
\Ldif_0
\dm
\;,
\end{equation}
%___________________________
where $\Ldif_0$ is a compact way of writing the master equation with
$\HS = \HO_0$ for $t<0$. %
\footnote{
Throughout this chapter we use the notation corresponding to the
dynamics of $\dm$.
The treatment, however, is general and hence valid for the examples of
classical \FP\ equations of chapter~\ref{chap:clasico} or their
quantum counterparts in phase space (chapter \ref{chap:phase-space}).
} % END OF FOOTNOTE
When $t=0$ the previous dynamics would have brought the system to the
stationary solution (recall chapter~\ref{chap:equilibrio})
%_______________________________
\begin{equation}
\label{staL0}
\Ldif_0
\dm_0
=
0
\;.
\end{equation}
%_______________________________
Then at $t=0$ we modify $\HS$ as $\HS = \HO_0 + \pert \cdot \HO_1$, with
$\pert$ accounting for the size of the perturbation.
To first order in $\pert$ we can write:
%_______________________
\begin{equation}
\label{L1}
\Ldif (t>0) = \Ldif_0 + \pert \cdot \Ldif_1
\;,
\end{equation}
%_______________________
and the formal solution of the master equation is:
%_________________________
\begin{equation}
\label{dm01}
\dm(t)
=
\dm_0
+
\pert \cdot
\dm_1(t)
\; ;
\qquad
\dm_1(t)
=
%\pert
%\cdot
\int_{-\infty}^{t}
\dif s
\,
\e^{(t-s) \Ldif_0}
\Ldif_1(s)
\dm_0
\end{equation}
%_________________________
%%

We now ask about the evolution of the average of some observable 
$\langle A \rangle = \Tr ( A \dm )$ 
and introduce:
%____________________________
\begin{equation}
\label{DeltaA}
\Delta A (t)
:=
\langle A \rangle (t)
-
\langle A \rangle_0
\end{equation}
%____________________________
where
$\langle A \rangle_0 = \Tr (A \dm_0)$ 
is the average in the absence of perturbation.
Then $\Delta A (t)$ measures how far we are from the reference
unperturbed evolution.

The generic form of the change is
$\HS \to \HO_0 + \pert \cdot \HO_1 f(t)$
with $f(t)$ some function of time.
Correspondingly
%_____________________________
%\begin{equation}
%\label{L1t}
$
\pert \cdot \Ldif_1 (t)
\to
\pert \cdot 
\Ldif_1
\,
f(t)
$,
%\end{equation}
%_____________________________
whence
%________________________
\begin{equation}
\Delta A (t)
=
\pert
\cdot
\int_{-\infty}^{\infty}
\dif s\,
\Resp (t-s)
f(s)
\end{equation}
%________________________
The response function $\Resp$ follows from (\ref{dm01}) in the form
%_______________________
\begin{eqnarray}
\Resp(t)
=
\left \{
\begin{array}{cc}
\Tr 
\left [
A 
\e^{(t-s) \Ldif_0}
\Ldif_1%(s)
\dm_0
\right ]
&
\qquad t >0
\\
0
&
\qquad t <0
\end{array}
\right .
\;.
\end{eqnarray}
%_______________________
Therefore our task will be to calculate $\Resp(t)$, which is independent
of $f(t)$ within linear response theory.
We will carry this out in two typical situations: the relaxation
experiment, and the AC response.
%%

%%%%%%%%%%%%%%%%%%%%%%%%%%%%%%%%%%%%%%%%%%%%%%%%%%%%%%%%%%%
\subsection{relaxation experiment}
\label{sub:relex}

\begin{figure}
\centerline{\resizebox{6.7cm}{!}{%
\includegraphics[angle = -0]{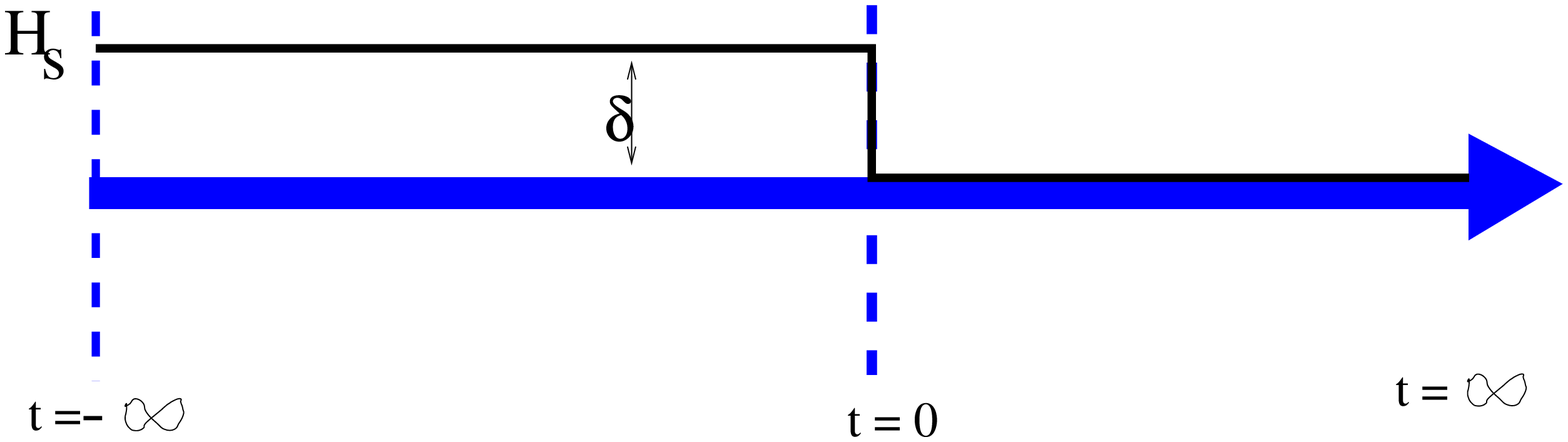}
}
\qquad
\quad
\resizebox{6.7cm}{!}{%
\includegraphics[angle = -0]{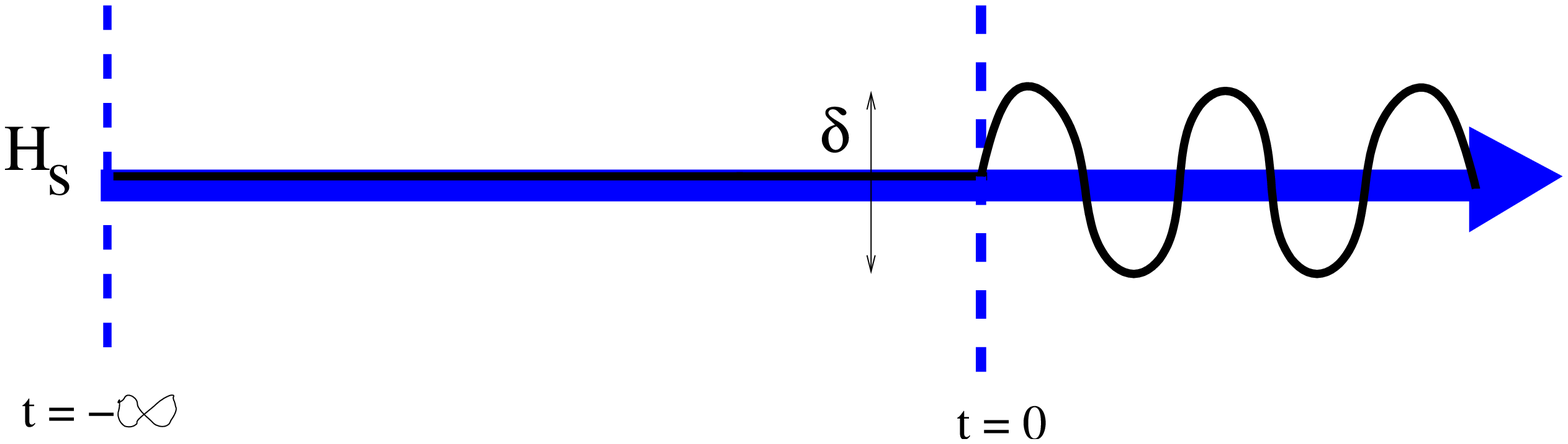}
}
}
\caption{
Left: scheme of the relaxation experiment.
We let the system equilibrate in the presence of a weak force $\pert$
and at $t=0$ we switch it off.
Right: AC experiment with a time periodic probe $\pert \cdot \sin
(\omega t)$ applied from $t=0$ on.
} 
\label{fig:relyAC}
\end{figure}

Let us consider a step function perturbation $f(t) = \Theta(-t)$, so
that $f(t)=1$ for $t<0$ while $f(t)=0$ when $t>0$.
Note that we have reversed the ``set up'', saying that in the interval
$-\infty < t <0$ the system was ``perturbed'' with a constant $\pert$,
and at $t=0$ the perturbation is removed (Fig.~\ref{fig:relyAC}). %
\footnote{
This is just a matter of terminology, as we could have said that $\HS$
is unperturbed in the interval $-\infty < t <0$, while at $t=0$ we
included the perturbation.
Either way the result is the same.
However, the language used in this section seems more natural, as we
will see.
} % END OF FOOTNOTE
%
%___________________________________________________
%___________________________________________________
%___________________________________________________
%___________________________________________________
%___________________________________________________
%___________________________________________________

In this case $\langle A \rangle_0 = \langle A \rangle (\infty)$,
since the ``unperturbed'' evolution takes place for $t>0$, and we know
that the system will eventually reach the stationary solution.
Then (\ref{DeltaA}) is written as
%____________________________
\begin{equation}
\label{relex}
\Delta_r A (t)
:=
\langle A \rangle (t)
-
\langle A \rangle (\infty)
\end{equation}
%____________________________
That is, $\Delta_r A(t)$ describes how the equilibrium is approached
as $t \to \infty$.
Hence the name ``relaxation experiment'' and the subindex ``$r$'' used.

%%%%%%%%%%%%%%%%%%%%%%%%%%%%%%%%%%%%%%%%%%%%%%%%%%%%%%%%%%%%
\subsection{AC response (frequency domain)}
\label{sub:ACgral}

According to the experimentalists, it is often easier to measure the
response to an oscillating probe than to conduct a relaxation
experiment.
In is convenient to introduce then the Fournier transform of the
response function $\Resp$:
%__________________________
\begin{equation}
\sus (\omega)
=
\int_{-\infty}^{\infty} \dif t\,
\e^{-\iu \omega t}
\Resp (t)
\; .
\end{equation}
%__________________________
At $t=0$ we switch on an oscillating probe $f(t) = \e^{\iu \omega t}$
(see Fig.~\ref{fig:relyAC}) and the response follows as
%_______________________________
\begin{equation}
\label{ACresp}
\Delta_{\rm {AC}} A(t) = 
\langle A \rangle (t)
-
\langle A \rangle (0)
=
\pert \cdot
\sus (\omega)
\e^{\iu \omega t}
\end{equation}
%_______________________________
Besides linear response theory teaches us that $\sus(\omega)$ of the
AC experiment is related with the response $\Delta A_r(t)$ in a
relaxation experiment~\cite[App.~C]{cofcrekal93}:
%___________________________
\begin{equation}
\label{fdt}
\sus(\omega)
=
\suseq 
\left [
1
-
\iu \omega
\int_0^\infty
\dif t
\,
\frac{\Delta A_r(t)}{\Delta A_r(0)}
\,
\e^{-\iu \omega t}
\right ]
\;.
\end{equation}
%___________________________
Here $\suseq = \sus (\omega=0)$ is the response to an infinitely slow
field, so that
%_________________________
\begin{equation}
\suseq = 
%\frac 
{\Delta A_r(0)}
/
{\pert}
\end{equation}
%________________________

%%%%%%%%%%%%%%%%%%%%%%%%%%%%%%%%%%%%%%%%%%%%%%%%%%%%%%%%%%%%
%%%%%%%%%%%%%%%%%%%%%%%%%%%%%%%%%%%%%%%%%%%%%%%%%%%%%%%%%%%%

\section{Bloch equations}
\label{sec:bloch}

To fix ideas and give content to the formalism, let us consider the
solvable problem of the two-state system.
In the basis of eigenstates $\vl 1 \rangle$ and $\vl 2 \rangle$ we can
always write
%________________________________
\begin{equation}
\HS = - \case \,\Bz\, \sigma_z
\end{equation}
%________________________________ 
%
where $\sigma_z$ is the Pauli  $z$ matrix and the energy levels are
$\epsilon_1 = - \Bz/2$ and $\epsilon_2 = \Bz/2$.
The notation reflects that we cannot help thinking of a $\case$ spin with a
Zeeman Hamiltonian.
In this way we bear in mind a physical realization of the two-level
model.

\newpage
The above $\HS$ is coupled to a bosonic bath as in the model~(\ref{clmodelct}).
The dynamics will be given in general by the master
equation~(\ref{dmebomnm}):
%
%_________________________________
\begin{eqnarray}
\label{dmebomnmbis}
\frac{\dif \dm_{nm}}{ \dif t}
&=&
- \frac{\iu}{\hbar} \Delta_{nm} \dm_{nm}
+
\sum_{n'm'}
\R_{nn'mm'}
\dm_{n'm'}
\end{eqnarray}
%___________________________________
As we set the problem ourselves, we make our life easier assuming
that the rotating wave approximation can be made (section
\ref{sub:RWA}).
Then the above system of ODEs simplifies; in particular the
off-diagonal elements decouple and evolve trivially.
For the two-state problem the remainder equations for the diagonal
[Eq.~(\ref{dmeRWA})] are called the Bloch equations \cite{blum}:
%______________________________
\begin{eqnarray}
\dif \dm_{11}/\dif t
&=&
-\Ptrans_{2 \vl 1}
\dm_{11}
+
\Ptrans_{1 \vl 2}
\dm_{22}
\\
\dif  \dm_{22}/ \dif t
&=&
%\quad 
+\Ptrans_{2 \vl 1}
\dm_{11}
-
\Ptrans_{1 \vl 2}
\dm_{22}
\end{eqnarray}
%______________________________
Here 
$P_{1 \vl 2} = W_{12} \vl F_{12} \vl^2$, 
with the matrix element of the coupling $F(\vec{\sigma})$ and the rate
$W$, providing detailed balance
$P_{1 \vl 2} = \e^{-\beta B} P_{2 \vl 1} $
[Eq.~(\ref{detailed})].

As mentioned the non-diagonal elements simply obey
($\dm_{12} = \dm_{21}^*$)
%______________________________
\begin{eqnarray}
\dif \dm_{12}/\dif t
&=&
+
 \frac{\iu}{\hbar} \Bz \dm_{12}
-
\gd
\dm_{12} 
\\
\dif \dm_{21}/\dif t
&=&
- \frac{\iu}{\hbar} \Bz \dm_{21}
-
\gd
\dm_{21} 
\end{eqnarray}
%______________________
with $\gd = -\Resp_{1122} = -\Resp_{2211} > 0$.
The solution~(\ref{nondiaRWA}) then reads
%_________________________________-
\begin{equation}
\label{nondiaRWA2}
\dm_{12}(t)
= 
\dm_{12}(0)
\e^{\left (\iu\,\Bz/\hbar - \gd\right ) t}
\; ,
\end{equation}
%______________________________
and its conjugate for $\dm_{21}(t)$.
Therefore one introduces the time constant $\Tdos \equiv 1/\gd$,
measuring the speed at which the off-diagonal elements go to zero ---
the {\em decoherence time}.

We are left with solving the Bloch equations for the diagonal
elements.
Using that
$1=\Tr (\dm) = \dm_{11} + \dm_{22}$ 
and introducing the population difference
$\Mz \equiv \dm_{11} - \dm_{22}$,
we have
%_____________________________
\begin{equation}
\label{rt}
%\langle r \rangle (t)
\Mz (t)
=
%\langle r \rangle (0) 
\Mz(0)
\cdot \e^{- \gr t}
+
\Mz (\infty)
\; ,
\end{equation}
%____________________________
%
with the relaxation constant
 $\gr = (P_{1\vl 2} + P_{2 \vl 1})/2$.
(The notation again tells that we have in mind the spin problem, with
the magnetization $\Mz\propto\langle \sigma_z\rangle$.)
Following the literature one defines the relaxation time
$\Tuno \equiv 1/\gr$,
which measures how fast the diagonal terms equilibrate to their
stationary values.
And these, as we pointed out in \ref{sec:comenteq}, agree with the
Boltzmann distribution
$\Mz (\infty) \propto \e^{-\beta \epsilon_1} - \e^{-\beta \epsilon_2}$.
\footnote{
Recall that under the RWA the ``pointer basis'' are the eigenstates of
$\HS$ (chapter \ref{chap:dinamica}).
}

This solvable example, plus LRT, will illustrate how one can obtain
$\Tuno$ and $\Tdos$ from the susceptibility curves.
Given that the susceptibility is an experimentally accessible
quantity, this provides a way of measuring relaxation and decoherence
times.
%%

%%%%%%%%%%%%%%%%%%%%%%%%%%%%%%%%%%%%%%%%%%%%%%%%%%%%%%%%%%%%%%%%%
\subsection{LRT and $\Tuno$}
\label{sub:tuno}

Let us reconsider the relaxation experiment of section \ref{sub:relex}
in a two-level system.
For $t<0$ the system was perturbed by an extra constant $\pert \Bz$:
%________________________________
\begin{equation}
\HS =
- \frac{1}{2}(\Bz + \pert \Bz) \sigma_z
\; .
\end{equation}
%________________________________
At $t=0$ we remove the probe, $\pert \Bz =0$, and follow the
evolution.

Immediately before $t=0$ the system would have reached the stationary
state in the presence of $\Bz + \pert \Bz$.
Under the RWA, we repeat again, the stationary solution is the
canonical one:
$\dm_{11}(0) = \e^{\beta (\Bz+ \pert\Bz) }$ , 
$\dm_{22}(0)= \e^{-\beta (\Bz +\pert \Bz)}$
and
$\dm_{12}(0)=\dm_{21}(0)=0$.
The same at $t =\infty$, but now with $\pert \Bz =0$.
As $\pert \Bz$ is suitably small we can expand to first order getting
$\Mz(0) = \Mz(\infty) + \pert \Bz \cdot \partial_B M_z$.
All the quantities here are equilibrium ones; by definition 
$\partial_B M_z = \suseq$
is the equilibrium susceptibility, so we can write
%_____________________________
\begin{equation}
\Delta \Mz (t)
=
\pert \Bz \cdot \suseq_{z}\,
\e^{-\gr t}
\end{equation}
%_____________________________ 
%%

\begin{figure}[t!]
\centerline{\resizebox{8.5cm}{!}{%
\includegraphics[angle = -90]{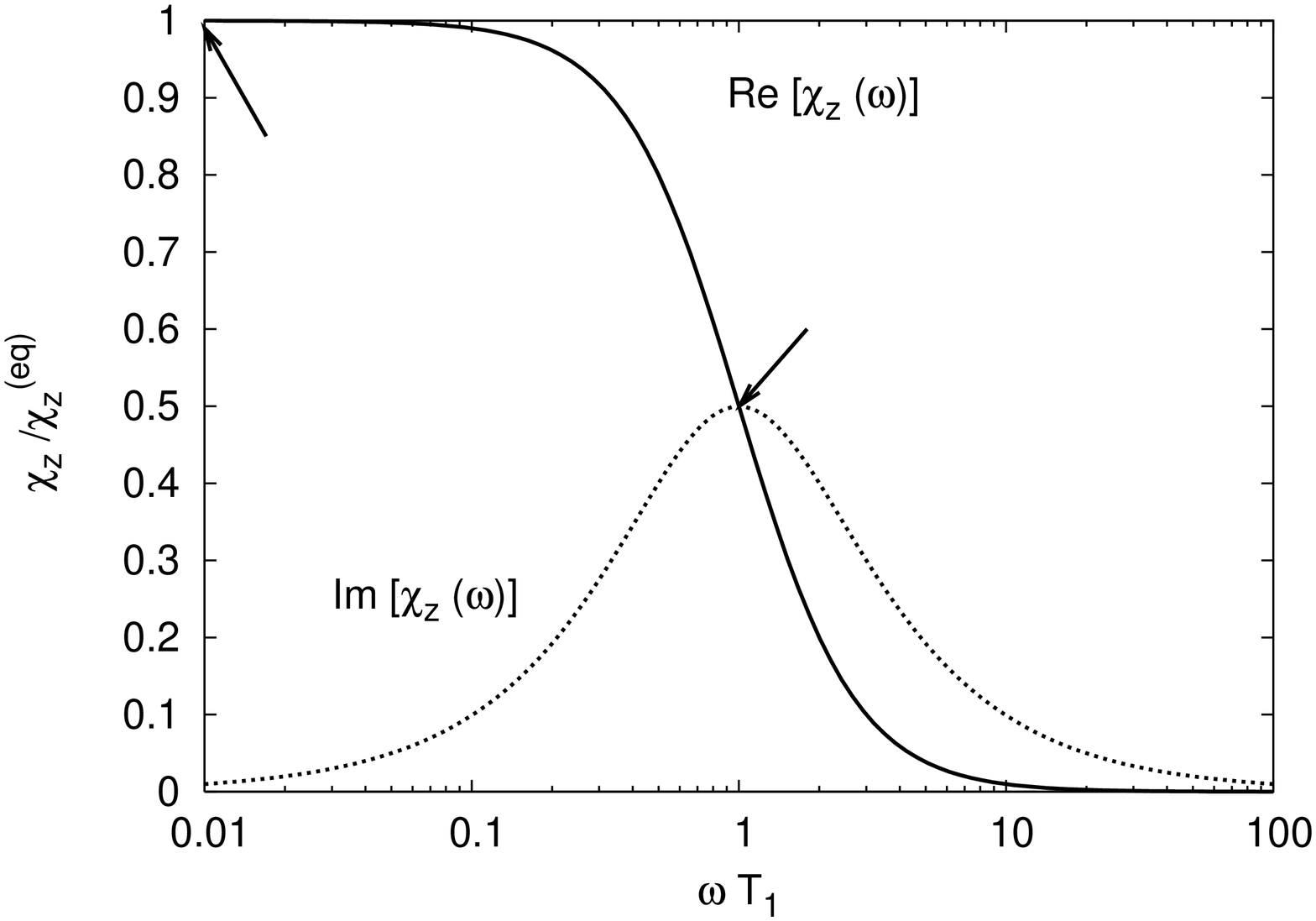}
}
\resizebox{8.5cm}{!}{%
\includegraphics[angle = -90]{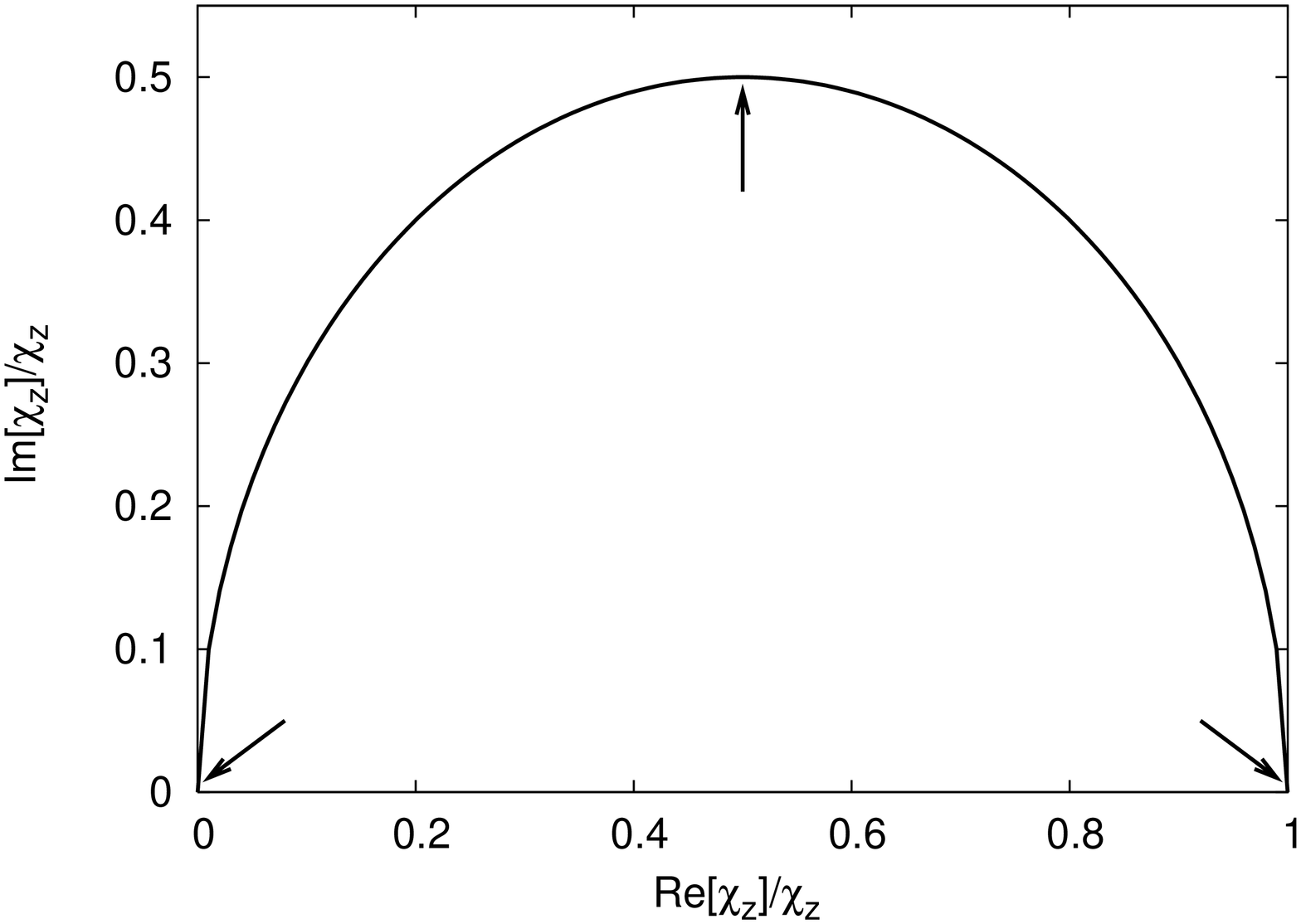}
}
}
\caption{
Left: real part (solid line) and imaginary part (dotted) of the Debye
curve~(\ref{debye}) vs.\ frequency.
The arrows mark special points in the spectrum, namely, the peak
location of the imaginary part, which gives $T_1$, and the equilibrium
susceptibility approached in the zero frequency limit.
Right: Cole-Cole representation (Im vs.\ Re).
}
\label{fig:LResp1}
\end{figure}

We can plug the above relaxation function into Eq.~(\ref{fdt}),
getting the dynamical susceptibility $\sus_z (\omega)$ as %
\footnote{
Here we use
%__________________________
\begin{equation}
\label{susint}
1
-
\iu \omega
\int_0^\infty
\dif t
\,
\e^{- (\Lambda + \iu \omega )t}
=
\frac {\Lambda}{\Lambda + \iu \omega}
\; .
\end{equation}
%_____________________________
\label{foot:susint}
} % END OF FOOTNOTE
%____________________________
\begin{eqnarray}
\label{debye}
\frac {\sus_z (\omega)}{\suseq_z}
=
\frac {1}{1 + \iu \omega \Tuno}
=
\frac {\gr}{\gr + \iu \omega }
\; .
\end{eqnarray}
%____________________________
We have written the result for $\sus_z(\omega)$ in two equivalent ways
found in the literature.
These functional forms are called after Debye's seminal work on
dielectric relaxation.
Let us split into the real and imaginary parts
%____________________________
\begin{equation}
\label{debyeRI}
\frac{\sus_z(\omega)}{\suseq_z}
=
\frac {1}{1 +  (\omega \Tuno)^2}
-
\iu 
\frac {\omega \Tuno}{1 +  (\omega \Tuno)^2}
\end{equation}
%____________________________
which have been plotted in Fig.~\ref{fig:LResp1}.
%
%___________________________________________________
%___________________________________________________
%___________________________________________________
%___________________________________________________
%___________________________________________________
%___________________________________________________
%
In the zero frequency limit the real part tends to the equilibrium
susceptibility.
Besides, the maximum in the imaginary part is located at $\omega =
1/T_1$.
In an ac experiment, where one applies a sinusoidal field and records
the response, one can directly obtain $\suseq$ and $\Tuno$.

Another usual form of representation is the Cole-Cole plot, with the
imaginary part plotted vs.\ the real part (Fig \ref{fig:LResp1},
right).
For Debye relaxation one has 
${\rm Im} = \sqrt { (1 - {\rm Re}) {\rm Re}}$,
in evident notations; that is, a semicircle with radius $1/2$ located
at $(0,1/2)$.
Note the marked points: the zeroes of the imaginary part are at
$\omega=0$ and $\infty$, corresponding to $1$ (maximum) and $0$
(minimum) of the real part.
The maximum of the imaginary part, ${\rm Im} = 1/2$, is reached at
$\omega = 1/T_1$, taking half the value of the maximum real part.
This plot is used as indication of non-Debye behavior if deviations
from a semicircle are found.
%%

%%%%%%%%%%%%%%%%%%%%%%%%%%%%%%%%%%%%%%%%%%%%%%%%%%%%%%%%%%%%%%%%%
\subsection{LRT and $\Tdos$}
\label{sub:tdos}

Let us now apply the LRT formalism to a problem with a transverse
probing field (see also $\jpaii$).
At negative times the Hamiltonian of our two-level system is then
%__________________________
\begin{equation}
\label{Hpertrans}
\HS = -\frac{1}{2} \Bz \sigma_z
+
\pert B_x \sigma_x
\end{equation}
%__________________________
where $\sigma_x$ is the corresponding Pauli matrix (we could use
$\sigma_y$ instead).
At $t=0$ we switch $\pert B_x$ off.
As before, for the stationary solutions we will use the canonical
form.
Then to first order in $\pert B_x$ we have
$\dm_{11}(0) = \dm_{11}(\infty) =\e^{\beta \Bz}$,
$\dm_{22}(0) =\dm_{22}(\infty) = \e^{-\beta \Bz}$,
for the diagonal elements, while the coherences (off-diagonals) read
$\dm_{12}(0) = \dm_{21}(0) = \delta B_x \Mz/\Bz$
and
$\dm_{12}(\infty) = \dm_{21}(\infty) = 0$.

\begin{figure}
\centerline{\resizebox{9.cm}{!}{%
\includegraphics[angle = -90]{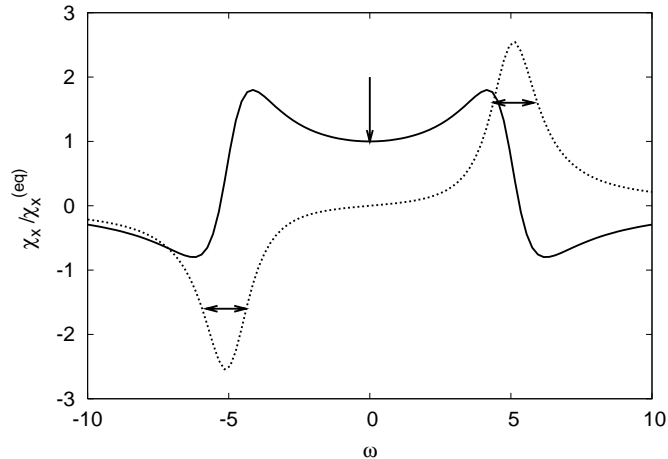}
}
}
\caption{
Real part (solid) and imaginary part (dotted) of the transverse
susceptibility $\sus_x (\omega)$ [Eq.~(\ref{transRI})].
The parameters are $\Bz = 5$ and $\gd = 1$.
The arrows mark the equilibrium susceptibility and the width of the
absorption peaks.
} 
\label{fig:TResp1}
\end{figure}

Now we monitorize $\sigma_x$ (the operator coupled to the
perturbation) through its average
$M_x \equiv \dm_{12} - \dm_{21}$.
This relaxes from its equilibrium value $\propto\suseq_x = \Mz/\Bz$ as
follows
%__________________________
\begin{equation}
\Delta M_x (t)
=
\pert B_x
\suseq_x
\left [
\e^{(\iu B - \gd)t}+
\e^{(-\iu B - \gd)t}
\right ]
\;.
\end{equation}
%__________________________
From this relaxation curve we can get the AC susceptibility
$\sus_x(\omega)$ [Eqs.~(\ref{fdt}) and (\ref{susint})]:
%______________________
\begin{equation}
\label{trans}
\frac
{\sus_{x} (\omega)}
{\suseq_x}
=
%\left [
\frac{ \iu B + \gd}{\iu (\omega + B) + \gd}
+
\frac{ -\iu B + \gd}{\iu (\omega - B) + \gd}
%\right ]
\; .
\end{equation}
%______________________
This is different from a simple Debye form, and we are going to see
that includes {\em absorption}.
Let us extract the real and imaginary parts 
%____________________________
\begin{eqnarray}
\label{transRI}
\frac
{\sus_{x} (\omega)}
{\suseq_x}
=&&
%\left [
\frac{ \gd^2 + (\omega + \Bz)\Bz}{ (\omega + B)^2 + \gd^2}
+
\frac{ \gd^2 + (\omega - \Bz)\Bz}{ (\omega - B)^2 + \gd^2}
\\ \nonumber
&+&
\iu
\left [
\frac{ \omega \gd}{ (\omega + B)^2 + \gd^2}
+
\frac{ \omega \gd}{ (\omega - B)^2 + \gd^2}
\right ]
\end{eqnarray}
%___________________________
These are plotted in Fig.~\ref{fig:TResp1}, where we see again that
the equilibrium result is recovered as $\omega \to 0$.
%
%___________________________________________________
%___________________________________________________
%___________________________________________________
%___________________________________________________
%___________________________________________________
%___________________________________________________
As for the imaginary part, it consists of two Lorentzian-type curves,
centred around the level difference $\pm \Bz$ and having width $\gd$.
In this case the perturbation did not commute with the base Hamiltonian
(\ref{Hpertrans}), inducing transitions between its levels.
Spectroscopists call $T_2$ ($=1/\gd$) the life-time of the levels; for
us it is the decoherence time (the time to erase the off-diagonals).
Letting $\gd \to 0$ (closed system) the absorption lines go over two
Dirac deltas at $\pm \Bz$.
Their broadening in an open system gives $\Tdos$ directly. %
\footnote{
The ``magnetic resonance'' experiment can be done alternatively by fixing
$\omega$ and varying $\Bz$ (the level splitting), looking for the
resonance.
} % END OF FOOTNOTE

%%%%%%%%%%%%%%%%%%%%%%%%%%%%%%%%%%%%%%%%%%%%%%%%%%%%%%%%%%%%%%%%%%%%%%%%
%%%%%%%%%%%%%%%%%%%%%%%%%%%%%%%%%%%%%%%%%%%%%%%%%%%%%%%%%%%%%%%%%%%%%%%%
\section{beyond the 2-state model}
\label{sec:LRTII}

The above example was a simple case where we could do everything
analytically.
In general, we will have to deal with quantum master equations
like~(\ref{dmebom}).
But we saw that if $\HS$ has a discrete and finite spectrum, we can
write a density-matrix equation for the $\dm_{nm}$ as
in~(\ref{dmebomnmbis}).
This is a linear system with $N \times N$ equations.
Diagonalizing the system's matrix (numerically most likely), we obtain
$N \times N$ eigenvalues and eigenvectors.
Then, with the appropriate initial conditions we can write
%____________________________________
\begin{equation}
\dm_{nm} (t)
=
\dm_{nm} (\infty)
+
\sum_{i=2}^{N \times N}
c_i^{(n,m)}
\e^{-\Lambda_i t}
\;
,
\end{equation}
%____________________________________
where we have split the contribution of the zero eigenvalue (giving
the stationary solution).

With these $\dm_{nm}(t)$ we can construct the relaxation curve $\Delta
A(t)$ for any observable $A$ of interest [Eq.~(\ref{relex})].
Plugging it into Eq.~(\ref{fdt}), and doing the integrals
with~(\ref{susint}), we obtain the corresponding dynamic
susceptibility
%________________________________
\begin{equation}
\label{eigensus}
\sus_A (\omega) 
=  
\sum_i a_i 
\frac {\Lambda_i}{\Lambda_i + \iu \omega}
\; ,
\end{equation}
%_____________________________
%
The
$a_i \propto \sum_{n,m} c_i^{(n,m)}\langle n \vl A \vl m \rangle$
involve the matrix elements of $A$ ($\prb$).
Real $\Lambda_i$ (the $\gr$ in the two-level system) contribute
Debye-type relaxation terms to the susceptibility.
On the other hand, complex eigenvalues (like $\iu \gd  \pm \Bz$ in the
2-level system) would contribute with absorption profiles.

\begin{figure}
\centerline{\resizebox{8.5cm}{!}{%
\includegraphics[angle = -90]{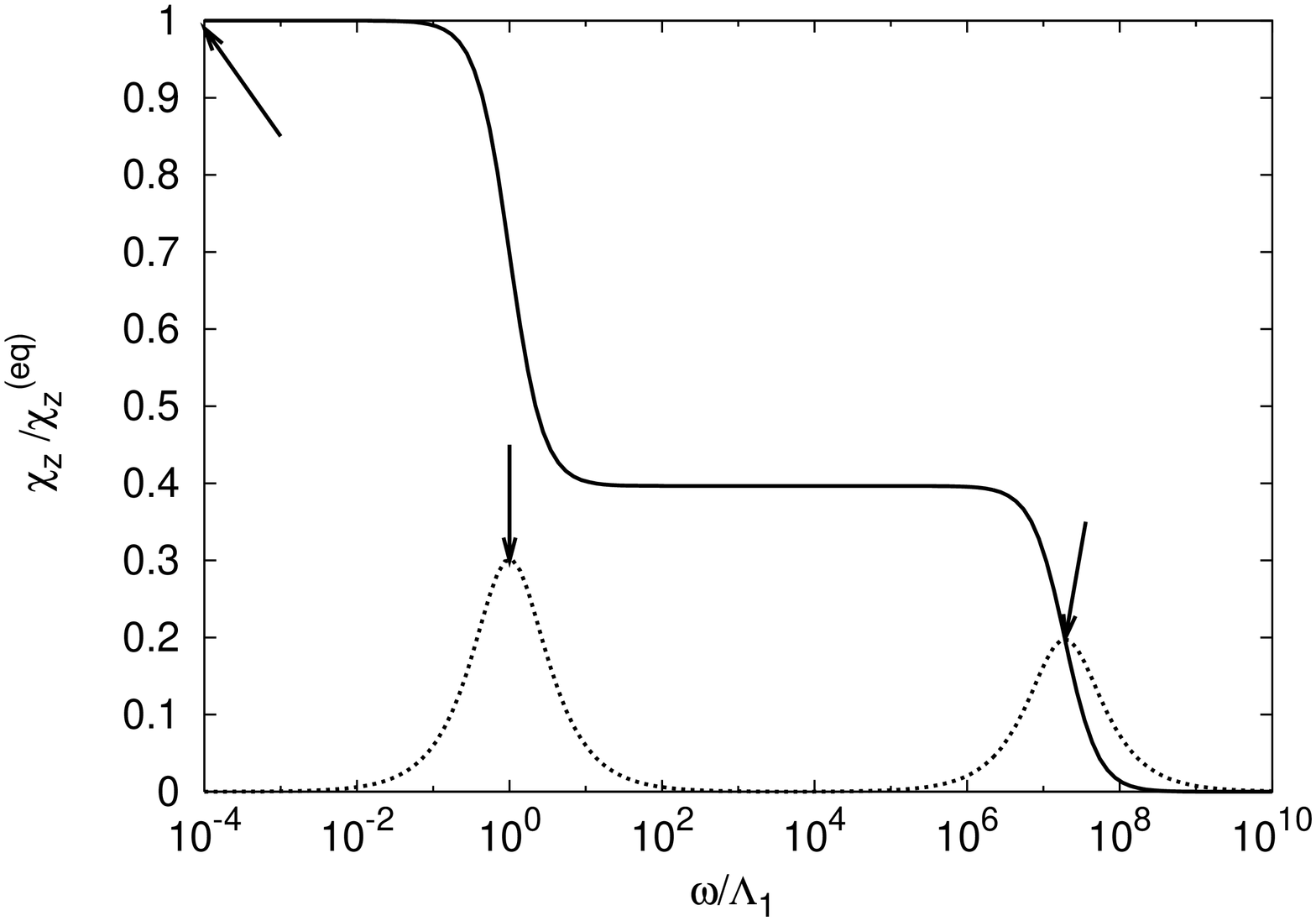}
}
\resizebox{8.5cm}{!}{%
\includegraphics[angle = -90]{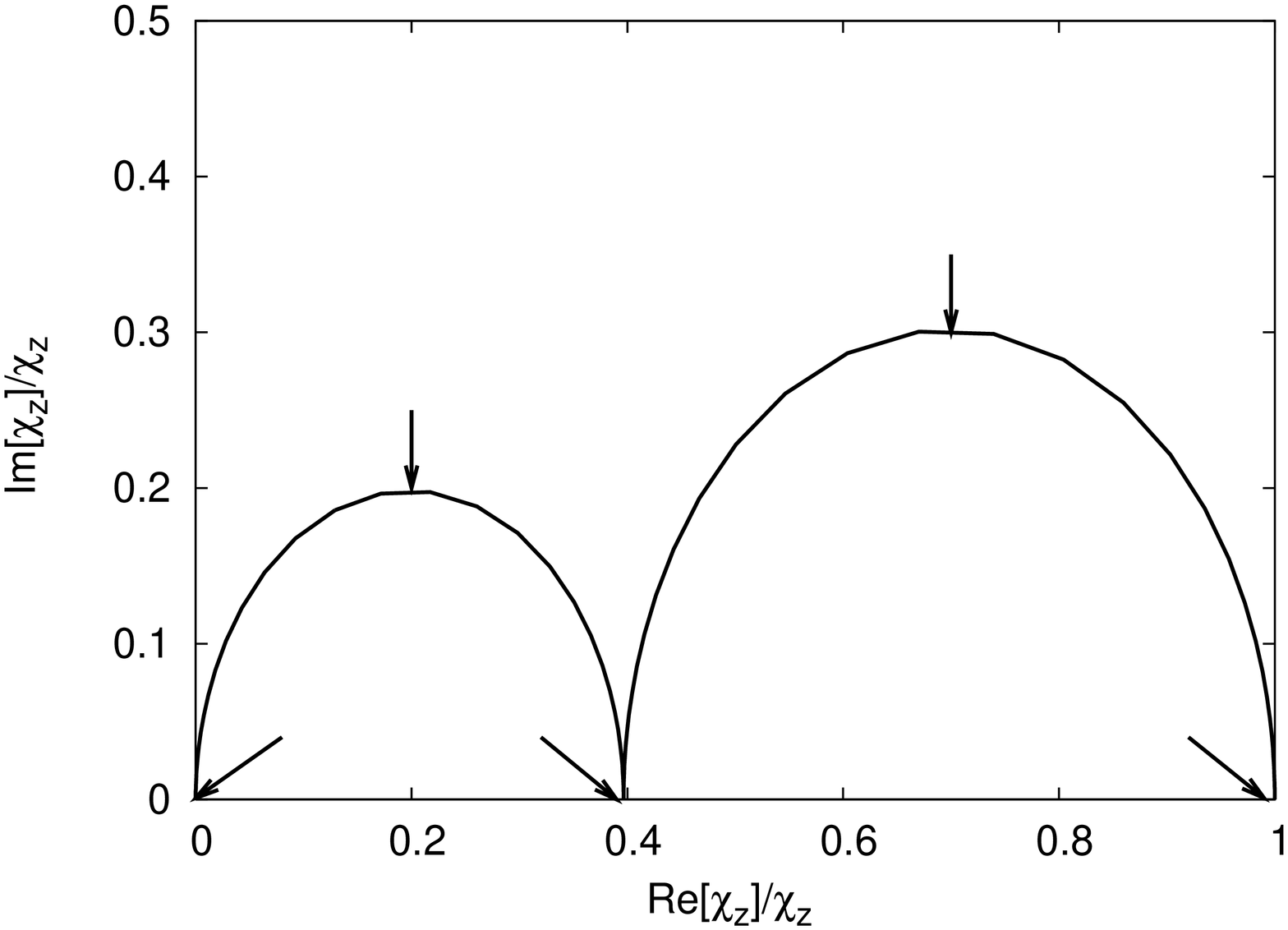}
}
}
\caption{
Left: real and imaginary parts of $\sus(\omega)$ given by the sum of
two Debye terms, with weights $a_i\propto0.6$ (left) and $0.4$
(right).
The arrows mark the relaxation times and the equilibrium limit.
Right: the Cole-Cole plot (cf. Fig.~\ref{fig:LResp1}) is approximately
given by two semicircles, as the two relaxation times are separated by
several orders of magnitude.
The zeros of the imaginary part correspond to the real parts $1$,
$0.4$ and $0$ (see left panel).
The maxima ${\rm Im} =0.2$ and $0.3$ correspond to ${\rm Re} = 0.2$
($=0.4/2$) and ${\rm Re} = 0.7$ ($=0.4+0.6/2$).
The arrows mark these special points.
} 
\label{fig:LResp2}
\end{figure}

Let us consider now an observable such that the $a_i$ of the complex
eigenvalues are all zero (the observable does not see those modes).
Then the {\em total\/} susceptibility is made up Debye's;
an example with two of them is plotted in Fig.~\ref{fig:LResp2}.
Each peak on the imaginary part is located at the corresponding
$\Lambda_i$.
%___________________________________________________
%___________________________________________________
%___________________________________________________
%___________________________________________________
%___________________________________________________
%___________________________________________________
%

In a multiexponential case like this, one tries to characterize the
response with a few effective parameters.
The integral relaxation time $\tint$ is defined as the area below the
relaxation curve:
%______________________________
\begin{equation}
\label{susloww}
\tint =
\int_{0}^{\infty} \dif t \, \frac{\Delta A(t)}{\Delta A(0)}
\; ;
\qquad
\quad
\sus (\omega) \cong \suseq ( 1 - \iu \omega \tint + \cdot \cdot \cdot)
\end{equation}
%______________________________
This quantity governs the low frequency behavior of the
susceptibility.
On the other hand, the high-frequency properties are better
characterized by the initial ($t=0$) slope of the relaxation curve.
This defines the effective time $\tef$
%_____________________________
\begin{equation}
\label{sushighw}
\tef^{-1}=
-
\frac{\dif}{\dif t}
\left (
\frac {\Delta A(t)}{\Delta A(0)}
\right )
\bigg \vl_{t=0}
\; ;
\qquad
\quad
\sus (\omega)
\cong
\suseq
\frac{\iu}{\omega \tef}
\end{equation}
%_____________________________
Both time constants coincide in a problem with a single exponential decay.
We will see examples with several relaxation/absorption peaks in
Chapters~\ref{chap:aplicationesI} and \ref{chap:aplicationesII}.

%%%%%%%%%%%%%%%%%%%%%%%%%%%%%%%%%%%%%%%%%%%%%%%%%%%%%%%%%%%%%%%%%%%%%%%%%%
\section{continued-fraction methods}
\label{sec:fc}

It is not always possible, or practical, to diagonalize our master
equations (large number of levels $N$, continuous spectrum, etc\dots).
In some cases we can resort to the continued-fraction method \cite{risken}.
This method is a relative of the solution of kinetic equations by
expansion into complete sets (\`a la Grad)  \cite{balescuII}.
It has been fruitfully exploited to solve classical \FP\ equations
\cite{risken, kalcof97} and quantum master equations
\cite{shi80,shiuch93,vogris88, vogris89}.

The ideas is to expand the unknown solution $\dm$ of our dynamical
equation
$\partial_t \dm = \Ldif \dm$
in some basis $\{u_j\}$ (all this applies to the classical or quantum
$W$)
%________________________________
\begin{equation}
\label{grad}
\dm =
\sum_i \Ccf_i u_i
\; .
\end{equation}
%________________________________
Plugging this into 
$\partial_t \dm = \Ldif \dm$
one gets a system of coupled differential equations for the expansion
coefficients, that is:
%________________________________
\begin{equation}
\label{gradsis}
\dot \Ccf_i
=
\sum_{j = -I}^I
Q_{i, i+j}
\Ccf_j
\end{equation}
%________________________________
Here $I$ is the range of the coupling.
If $I=0$, then
$\dot \Ccf_i = Q_{i, i} \Ccf_i$,
and 
$\Ccf_i (t) = \Ccf_i(0) \e^{Q_{i,i}t}$
solves the problem --- we had diagonalized the master equation!

When $I \neq 0$, on the other hand, one proceeds taking the Laplace
transform 
$\bar g[s] = \int_0^\infty \dif t \, \e^{-st} g(t)$
(it is after all a system of ODEs with constant coefficients).
Using the transform of the derivative
$\bar {\dot g} [s] = s \bar g[s] - g(0)$,
the system of differential equations is finally transformed into a set
of algebraic equations for the $\Ccf_i$, with recurrences between the
indexes
%__________________________
\begin{equation}
\label{rrgral}
-  \Ccf_i (0)
=
(Q_{i,i} -s) \bar \Ccf_i
+
\sum_{j \neq 0}
Q_{i, i+j}
\bar \Ccf_j
\; .
\end{equation}
%__________________________
Compared with (\ref{gradsis}), the coefficient of the central term was
simply augmented to $(Q_{i,i}-s)$, while $-\Ccf_i (0)$ appears as an
inhomogeneous term.

%%%%%%%%%%%%%%%%%%%%%%%%%%%%%%%%%%%%%%%%%%%%%%%%%%%%%%%%%%%%
\subsection{solving recurrence relations: case $I$=1}
\label{sub:rec1}

Let us assume that, with a smart basis choice, we have ended up with a
recurrence like~(\ref{rrgral}) with $I=1$ (first non-trivial case).
Then we introduce some simplified notation: 
$f_i \equiv \Ccf_i(0)$
for the source terms (``forcing''), and the constants
$Q^-_i \equiv Q_{i,i-1}$,
$Q^+_i \equiv Q_{i,i+1}$ 
and 
$Q_i \equiv Q_{i,i}-s$.
Then the recurrence~(\ref{rrgral}) would take the ``canonical'' form:
%______________________________
\begin{equation}
\label{3term}
Q_i^- C_{i-1}
+
Q_i C_i
+
Q_i^+ C_{i+1}
=
-f_i 
\; .
\end{equation}
%_____________________________
At this point we could diagonalize the matrix associated to the
system~(\ref{3term}).
Then, instead of diagonalizing $\cal L$ directly, we would have made
an analytical effort leading to a tri-diagonal matrix; the numerical
work would be more efficient.
But we can take an analytical step further.
If we introduce the following {\it ansatz} \cite{risken},
%____________________________
\begin{equation}
\label{ansatz}
 C_i =  S_i C_{i \mp 1} +  a_i
\end{equation}
%_____________________________
in equation (\ref{3term}), we easily get the following relations for
the ``ladder'' coefficients $S_i$ and the {\it shifts} $a_j$
(associated to the inhomogeneous term)
%_____________________________
\begin {eqnarray}
\label{s_i}
S_i =
-
\frac{Q_i^ {\mp}}{Q_i + Q_i^{\pm} S_{i \pm 1}}
\; ;
\qquad
%\label{a_i}
a_i 
=
-
\frac {f_i + Q_i^{\pm} a_{i \pm 1}}{Q_i + Q_i^{\pm} S_{i \pm 1}}
\end{eqnarray}
%______________________________
Now we can explain the name ``continued fraction'' given to this
method; iterating the denominators with the same formula one could
write:
%_____________________________
%
% CONTINUED FRACTION
%_____________________________
\begin{equation}
\label{sj}
 S_i
=
\cfrac{ Q^\mp_ i}{Q_{i,} + Q^\pm_i
 \cfrac
{ Q^\mp_{i-1}}{ Q_{i,-1} +  Q^\pm_{i-1}
  \cfrac
{Q^\mp_{i-2}}{Q_{i,-2} + Q^\pm_{i-2}
\dotsb
}}}
\end{equation}
%________________________________
%
This would provide a ``closed form'' for the coefficients $C_i[s]$.
We could say that this is a semi-analytical technique, where we end up
evaluating (numerically) continued fractions.
Eventually, inverse Laplace transformation would give the time
evolution of the coefficients $C_i$, and hence of the full distribution $\dm$.
\footnote {
This last step may be seen as a handicap of the method, as the inverse
Laplace transformation can be quite unstable \cite{arfken}, requiring
many $s$-points to attain accurate results.
As we will see later, in the cases treated here (statics and
stationary ac responses) we do not need to use Laplace transformation.
} %END OF FOOTNOTE

%%%%%%%%%%%%%%%%%%%%%%%%%%%%%%%%%%%%%%%%%%%%%%%%%%%%%%%%%%%%%%%%
\subsection{Brinkman hierarchy: matrix recurrence relations}
\label{sub:brickmanncl}

The best way to learn how to convert a kinetic equation into a suitable
recurrence relation is through an example.
Let us consider the Klein--Kramers equation~(\ref{wkk}), which in
appropriate dimless units reads
%___________________
%_____________________________
\begin{equation}
\nonumber
%%%%%%%%%%%%%%%%%%
\partial_{t}
\W (\q,\p)
=
\Big[
-\p\,
\partial_\q
+
V'\,
\partial_\p
+
\gamma \,
\partial_p \,
\big(\p+\partial_\p\big)
\Big ] 
\W (\q,\p)
%%%%%%%%%%%%%%%%%%
\;.
\end{equation}
%_____________________________
%_____________________________
%__________________
We follow Risken~\cite{risken} and begin expanding in a basis
[cf. Eq.~(\ref{grad})]
%_________________________
\begin{equation}
\label{expW}
\W (\q, \p)
= 
\w (\q, \p) \sum
c_n(\q,t) \psi_n(\p)
\end{equation}
%_________________________
with the pre-factor $\w$ extracted for later convenience.
Inserting this expansion in the \FP\ equation we get the following
``equations of motion'' for the coefficients $c_n$:
%__________________________________________
\begin{equation}
\label{Lbar}
\dot c_n 
=
\sum_{m}
Q_{n, n+m}
c_{n+m}
\; ;
\qquad
Q_{n,m}
=
\int 
\dif \p
\,
\psi_{n}
\bar {\cal L}
\psi_{m}
\; ;
\qquad
\bar {\cal L}
\equiv
\w^{-1}
{\cal L}
\w
\end{equation}
%__________________________________________
Here ${\cal L}$ is the differential operator on the right-hand side of
the Klein--Kramers equation, while $\bar{\cal L}$ is some similarity
transformed version of it:
$\bar{\cal L}\,f=\w^{-1}{\cal L}(\w\,f)$.

Following Risken we put
$\w = \e^{-(p^{2}/2 + V(\q))/2}$,
which results in
%____________________________-
\begin{equation}
\w^{-1}
\,
\partial_p \,
\big(\p+\partial_\p\big)
%\,
\w
\;
\psi_m
=
\left (
\partial^2_\p +
\frac{1}{4} \p^2
-
\frac{1}{2}
\right )
\psi_m
\;.
\end{equation}
%____________________________
That is, the dissipative part acts formally as the Schr\"odinger
Hamiltonian of a harmonic oscillator (on the $\p$ dependence).
This cries for introducing ``ladder'' differential operators
$a = \partial_\p + \p/2$
and
$a^+ = \partial_\p - \p/2$
obeying $[a, a^+]=1$ and giving
%__________________________
\begin{equation}
\w^{-1}
\,
\partial_p \,
\big(\p+\partial_\p\big)
%\,
\w
\;
\psi_m
=
a^+ a
\;
\psi_m
\;
.
\end{equation}
%__________________________
On the other hand, the closed part (Poisson bracket) gets transformed
as
%___________________________-
\begin{equation}
\w^{-1}
%\,
\left (
-\p\,
\partial_\q
+
V'\,
\partial_\p
\right )
\w
\;
\psi_m
=
(-D_{+}
a
+ 
D_{-}
a^+
)
\;
\psi_m
\end{equation}
%____________________________
with the shifted derivatives 
$D_{\pm} = \partial_\q \mp V'/2$.
Finally, if we choose for the $\psi_n$ the Hermite functions,
$\psi_n(\p) \propto H_n(p/\sqrt{2})$, 
the term $a^+ a$ becomes diagonal, giving $Q_{nn}$, while $a$ and $a^+$
produce $Q_{n n \pm 1}$ [Eq.~(\ref{Lbar})].
Putting all this together one gets the so-called Brickmann
hierarchy~\cite{bri56I, bri56II}
%______________________________
\begin{equation}
\label{brickcl}
\dot c_n
=
-
\left (
\sqrt n
D_-
c_{n-1}
+
\gamma
n
c_n
+
\sqrt {n+1}
D_+
c_{n+1}
\right )
\;
,
\end{equation}
%______________________________
having the form of a 3-term recurrence.

However, as the $D_{\pm}$ still contain operators on $\q$, we expand
on a $\q$-basis $\{ u_\alpha (\q) \}$, obtaining a recurrence
relation, now without operators, but involving two indexes:
%_________________________
$
\dot c_n^\alpha
=
Q_{n, n+m}^{\alpha, \alpha + \beta}
c_{n+m}^{\alpha + \beta}
$
.
%_________________________
In order to convert this into a one-index recurrence (the problem we
know how to solve), we introduce the following vectors and matrices:
%_________________________
\begin{eqnarray}
\label{matricial}
\mc_n
=
\left (
\begin{array}{c}
c_{n}^{-L}
\\
\vdots
\\
c_{n}^{L}
\end{array}
\right )
\; ;
\qquad
\qquad
\quad
\dot{\mc}_{\ix}
=
\mQ^-
%_{\ix,\ix-1} 
{\mc}_{\ix-1}
+
\mQ
\;
%_{\ix,\ix} 
{\mc}_{\ix}
+
\mQ^+
%_{\ix,\ix+1} 
{\mc}_{\ix+1}
\; .
\end{eqnarray}
%_____________________________
%%
Here $L$ is some truncation index for the $\q$-basis, and the
matrices $\mQ$ are
%_____________________________
\begin{eqnarray}
\label{matrixelements}
\nonumber
\mQ^-_{\alpha, \alpha + \beta} \equiv Q_{n, n-1}^{\alpha, \alpha +\beta} &=&
-\sqrt n \left [
(\partial_\q)_{\alpha, \alpha + \beta}
+
\case
V'_{\alpha, \alpha + \beta}
\right ] 
\\ \nonumber
\mQ_{\alpha, \alpha + \beta} \equiv Q_{n, n}^{\alpha, \alpha +\beta} &=&
-n \gamma \, \delta_{\alpha, \alpha + \beta}
\\
\mQ^+_{\alpha, \alpha + \beta} \equiv Q_{n, n+1}^{\alpha, \alpha +\beta} &=&
-\sqrt {n+1} \left [
(\partial_\q)_{\alpha, \alpha + \beta}
-
\case
V'_{\alpha, \alpha + \beta}
\right ] 
\end{eqnarray}
%_____________________________
with $\q$-matrix elements 
$A_{\alpha\beta}=\int{\rm d}x\,u_{\alpha}A\,u_{\beta}$.

Finally, we could Laplace transform as in (\ref{rrgral}), getting an
algebraic recurrence like~(\ref{3term}), with the $Q$'s and $\Ccf$'s
given by the vectors $\mc$ and matrices $\mQ$.
The resulting recurrence could be solved like the scalar case of
subsection~\ref{sub:rec1}.
The only difference is that the fraction bars in~(\ref{sj}) have to be
understood as matrix inversions ($A/B \to B^{-1} A$).
In the statics no Laplace transform is required, and Risken solved the
problem studying parameter ranges out of the reach of other techniques
\cite{risken}.

%%%%%%%%%%%%%%%%%%%%%%%%%%%%%%%%%%%%%%%%%%%%%%%%%%%%%%%%%%%%%%%%%%%%%%
\subsection{the case $I >1$}
\label{sub:Ip1}

What happens if the recurrence involves $5$ or more terms?
Then we build appropriate matrices and vectors (by mounting the ones
above) and the recurrence can always be cast into the ``canonical''
3-term form \cite{risken}.
The method will be useful as long as $I$ is not too large.
We will solve problems where $I = 1$ or $ 2$.

%%%%%%%%%%%%%%%%%%%%%%%%%%%%%%%%%%%%%%%%%%%%%%%%%%%%%%%%%%%%%%%%%%%%%%%%%
\section{stationary solutions, LRT \& continued fractions}

Let us see how to apply the continued-fraction method to obtain
stationary solutions and dynamical susceptibilities.
The approach will be used later in chapters \ref{chap:aplicationesI}
and \ref{chap:aplicationesII}.

%%%%%%%%%%%%%%%%%%%%%%%%%%%%%%%%%%%%%%%%%%%%%%%%%%%%%%%%%%%%%%%%%%%%%%%%%
\subsection{obtaining stationary solutions}
\label{sub:staFC}

When we want to obtain the stationary solution $\Ldif \dm =0$ we do not
need to resort to Laplace's transform.
Indeed, plugging the expansion $\dm =\sum_i \Ccf_i u_i$ in
$\Ldif\dm=0$ [cf.~Eq.~(\ref{staL0})], we get
%________________________________
\begin{equation}
\label{gradest}
\sum_{i = -I}^I
Q_{j, j+i}
\Ccf_i
=
0
\; ,
\end{equation}
%________________________________
for the coefficients of the stationary solution.
This is an algebraic recurrence relation like~(\ref{3term}), with
$f_i=0$, solvable simply as discussed in Sect.~\ref{sec:fc}.
%%

%%%%%%%%%%%%%%%%%%%%%%%%%%%%%%%%%%%%%%%%%%%%%%%%%%%%%%%%%%%%
\subsection{application to LRT}
\label{sub:LRTFC}

Let us have another look at the AC experiment of section
\ref{sub:ACgral}.
Recall that from $t=0$ on we apply a sinusoidal probe
$f(t)=\e^{\iu\omega t}$;
then the evolution operator becomes
$\Ldif
=
\Ldif_0
+
\pert \cdot \e^{\iu \omega t} \Ldif_1$,
while the solution can be written as
$\dm
=
\dm_0
+
\pert \cdot \e^{\iu \omega t} \dm_1$.
Compare with Eqs.~(\ref{L1}) and~(\ref{dm01});
here we have explicitly extracted the $t$ dependence $\e^{\iu \omega
  t}$ from $\dm$ and $\Ldif$.
Therefore, both $\dm_1$ and $\Ldif_1$ are time independent (of course,
$\dm_0$ and $\Ldif_0$ too).

Using now 
$\dot {\dm} = \iu \omega \dm_1 \e^{\iu \omega t}$
and equating terms with and without oscillating factors we get
%_________________________
\begin{eqnarray}
0
&=&
\Ldif_0 \dm_0
\\  
\iu \omega \dm_1
&=&
\Ldif_0 \dm_1 + \Ldif_1 \dm_0
\end{eqnarray}
%_________________________
Inserting the expansion $\dm_0 =\sum_i \Ccf_i u_i$ in the first
equation we get the recurrence~(\ref{gradest}), which we can solve by
continued fractions, getting the $\Ccf_i$.
Next, we insert the solution for $\dm_0$ in the second equation and
expand $\dm_1$ as in~(\ref{grad}), finding a recurrence
like~(\ref{gradest}), but now with $f_i \neq 0$.
The inhomogeneous part comes from $\Ldif_1 \dm_0$, thus justifying the
carrying along of the forcing term from the generic
recurrence~(\ref{3term}).
In summary, with two recurrence relations we can get $\dm$ to first
order in $\pert$.
With this $\dm$ we obtain the AC response of the averages of interest
[Eq.~(\ref{ACresp})].
%%

%%%%%%%%%%%%%%%%%%%%%%%%%%%%%%%%%%%%%%%%%%%%%%%%%%%%%%%%%%%%
\section {final comments}

In this chapter we have considered the system's response to first
order in the probing field/force.
To go beyond LRT, one would need the response to higher orders
[cf.~Eq.~(\ref{dm01})]
%____________________________
\begin{equation}
\dm =
\dm_0
+
\pert \cdot \dm_1
+
\pert^2 \cdot \dm_2
+
\cdots
\end{equation}
%____________________________
%
We will see in chapter \ref{chap:aplicationesII} how to proceed along
this line.

In the second part of the chapter we have briefly introduced the
continued-fraction method to solve master equations.
This is a semi-analytic technique, and for the cases we are interested
in (stationary and first few order responses) we will see that it can be very
efficient.
Besides, it can be used when the spectrum is continuous
(Chap.~\ref{chap:aplicationesI}).
In addition, one gets the full $\dm$, not only the observables.

It has some drawbacks too.
It is very specific of the problem addressed, requiring a specific
pre-analysis in each case (finding a good basis $\{ u_j\}$ to expand
$\dm$, obtainment of the required matrix elements, etc.).
Besides, it can become numerically unstable in certain parameter
ranges (typically very low damping and temperatures \cite{risken}).
%%

%%%%%%%%%%%%%%%%%%%%%%%%%%%%%%%%%%%%%%%%%%%%%%%%%%%%%%%%%%%%
%%%%%%%%%%%%%%%%%%%%%%%%%%%%%%%%%%%%%%%%%%%%%%%%%%%%%%%%%%%%
%%%%%%%%%%%%%%%%%%%%%%%%%%%%%%%%%%%%%%%%%%%%%%%%%%%%%%%%%%%%
%%%%%%%%%%%%%%%%%%%%%%%%%%%%%%%%%%%%%%%%%%%%%%%%%%%%%%%%%%%%
%%%%%%%%%%  APLICATIONES PARTICULAS %%%%%%%%%%%%%%%%%%%%%%%%
%%%%%%%%%%%%%%%%%%%%%%%%%%%%%%%%%%%%%%%%%%%%%%%%%%%%%%%%%%%%
%%%%%%%%%%%%%%%%%%%%%%%%%%%%%%%%%%%%%%%%%%%%%%%%%%%%%%%%%%%%
%%%%%%%%%%%%%%%%%%%%%%%%%%%%%%%%%%%%%%%%%%%%%%%%%%%%%%%%%%%%

\chapter[\!\!applications I:\! brownian\! particle\! in\! periodic\! potential]
{applications I:\\ brownian particle in \\ a periodic potential}
\label{chap:aplicationesI}

In this chapter we apply all the preceding open-system formalism to
the problem of a quantum particle in a periodic potential
$V(\q)=V(\q+L)$.
That is, we study the Bloch problem taking into account the coupling
to a dissipative bath.
Besides the substrate potential, the system can be acted upon by a
force $-\q\cdot f(t)$, allowing the study of transport properties (the
current).
The external force breaks periodicity and the band picture, so that
the problem becomes a delicate one (continuous unbounded spectrum,
Bloch oscillations, Wannier-Stark ladders \cite{glukolkor02} \dots).
On top of this the particle interacts with the bath, further
complicating matters.

The chapter begins with the solving of the Caldeira--Leggett
equation~(\ref{cleq}) in phase space [Eq.~(\ref{cleqps})] by means of
continued fractions.
In the classical limit this equation, which goes over the
Klein--Kramers equation~(\ref{wkk}), was solved in this way by Risken
(section \ref{sub:brickmanncl}).
Here we discuss the extension of that work to the quantum regime
($\jpai$).
In the paradigmatic example of a cosine potential $V(\q) = \cos (x)$
we will see how the classical and quantum distributions $\W(\q, \p)$
can differ.
After this we address transport problems by adding a driving
$-\q\cdot f(t)$.
We compute the current and the mobility in the cosine potential as
well as in potentials lacking space-inversion symmetry: ratchet
potentials of the type
$V(\q) = \sin(\q) + r/2\sin(2\q)$.
The phase-space approach will allow us to connect smoothly with the
classical results.

%%%%%%%%%%%%%%%%%%%%%%%%%%%%%%%%%%%%%%%%%%%%%%%%%%%%%%%%%%%%%%%%%%%%%%%%%%
%%%%%%%%%%%%%%%%%%%%%%%%%%%%%%%%%%%%%%%%%%%%%%%%%%%%%%%%%%%%%%%%%%%%%%%%%%
%%%%%%%%%%  

\section[solving the  Caldeira--Leggett master equation]
{solving the Caldeira--Leggett master\\ equation by
continued fractions}

We start writing the Caldeira--Leggett equation in phase space as
[cf. Eq.~(\ref{cleqps})]:
%_____________________________
%_____________________________
\begin{equation}
\label{cleqpsadim}
%%%%%%%%%%%%%%%%%%
\partial_{t}
\W (\q,\p)
=
\Big[
-\p\,
\partial_\q
+
V'\,
\partial_\p
+
\gammaT \,
\partial_p \,
\big(\p+\partial_\p\big)
+
\sum_{\iq=1}^{\infty}
\qcoef^{(\iq)}
\,
V^{(2\iq+1)}(\q)
\,
\partial_\p^{(2\iq+1)}
\Big]
\W(\q, \p)
%%%%%%%%%%%%%%%%%%
\; .
\end{equation}
%_____________________________
%_____________________________
The first two terms are simply the classical Poisson bracket.
The third one includes the dissipation and diffusion originating from
the coupling to the bath.
Being a high temperature semiclassical equation
(Sect.~\ref{sub:cldme}), those terms are equal to the classical ones.
The sum in the last term incorporates the quantum mechanical
``corrections'' to the closed Hamiltonian dynamics (Moyal terms,
section \ref{sub:1946}).

Equation~(\ref{cleqpsadim}) was made dimless by means of a
characteristic length $\q_0$ (e.g., the period of the potential) and a
characteristic energy $E_0$ (the potential amplitude).
Besides the parameters have been thermally rescaled introducing 
$\omega_T = (\kT /m \q_0)^{1/2}$.
Thus the momentum $\p$ in (\ref{cleqpsadim}) was made dimless and
thermalized (the potential $V \to V/\kT$ too), while the other
parameters read
%______________________________
\begin{equation}
\label{defadim}
\gammaT = \frac {\dampp}{\omega_T}
\; ;
\qquad
\qcoef^{(\iq)}
=
\frac{(-1)^\iq}{(2 \iq + 1)!}
\left (
\frac{\pi\omega_0}{\omega_T \kondobar}
\right )^{2 \iq}
\; ;
\qquad
\frac {\kondobar}{2 \pi}
=
\frac{S_0}{\hbar}
\; .
\end{equation}
%______________________________
%______________________________ 
Here $\omega_0$ and $S_0$ are the characteristic frequency and action
[$ \omega_0 = (E_0/m \q_0^2)^{1/2}$ and $S_0 = E_0/\omega_0$].
The temperature has been absorbed in the definition of $\gammaT$ and
the Moyal coefficients $\qcoef^{(\iq)}$.
In this way the form (\ref{defadim}) shows that the quantum terms
become less important the higher $\kondobar$ and/or the temperature
are.
Indeed the parameter $\kondobar$ tells how quantum the system is
($\kondobar \propto 1/\hbar$, Chap.~\ref{chap:clasico},
Sect.~\ref{sec:sis}).
Thus the limit $\kondobar \to \infty$ recovers the Klein--Kramers
equation from the Caldeira--Leggett master equation in phase space.
From now on we will simplify further the notation by omitting the
subindex $T$ en $\dampp$.
\footnote{
Note the relation between $\kondobar$ and the thermal de~Broglie
wavelength.
The definition 
$\lambda_{\rm dB} = \Delta \q = \hbar /\Delta \p$
with $\Delta \p = \sqrt{M\kT}$ (thermal uncertainty from
equipartition) gives
%_____________________
\begin{equation}
\lambda_{\rm dB}
=
\pi
\q_0
\sqrt {\frac {E_0}{\kT}}
%\frac {\omega_0}{\omega_T}
\frac{1}{\kondobar}
\end{equation}
%_____________________
} %ENDOF FOOTNOTE  
%

%%%%%%%%%%%%%%%%%%%%%%%%%%%%%%%%%%%%%%%%%%%%%%%%%%%%%%%%%%%%%%%%%%%%%%%%%%%%
%%%%%%%%%%%%%%%%%%%%%%%%%%%%%%%%%%%%%%%%%%%%%%%%%%%%%%%%%%%%%%%%%%%%%%%%%%%%
\subsection {the classical limit  ($\kondobar \to \infty$)}

As we stated in chapter~\ref{chap:phase-space}, one advantage of the
phase-space formalism is that by varying one parameter the master
equations go over their classical \FP\ counterparts.
In our problem, letting $\kondobar \to \infty$ in (\ref{cleqpsadim})
we recover the Klein--Kramers equation~(\ref{wkk}), in dimensionless form
%_____________________________
%_____________________________
\begin{equation}
\label{wkkadim}
%%%%%%%%%%%%%%%%%%
\partial_{t}
\W (\q,\p)
=
\Big[
-\p\,
\partial_\q
+
V'\,
\partial_\p
+
\dampp \,
\partial_p \,
\big(\p+\partial_\p\big)
\Big ] 
\W (\q,\p)
%%%%%%%%%%%%%%%%%%
%\;,
\quad
\end{equation}
%_____________________________
%_____________________________
In section \ref{sub:brickmanncl} we transformed this equation into a
recurrence form, the so called Brickmann hierarchy~(\ref{brickcl}).
Recall that what we did was to expand the distribution in Hermite
polynomials for $\p$ while the $\q$-basis was left unspecified, 
$\W (t,\q, \p)
\propto
\sum c_n^\alpha(t)u_\alpha (\q) \psi_n(\p)$.
This led to a $2$-index recurrence
$\dot c_n^\alpha
=
Q_{n, n+m}^{\alpha, \alpha + \beta}
c_{n+m}^{\alpha + \beta}$
which was converted into a recurrence in one index by defining the
vectors~(\ref{matricial}) and the matrix coefficients~(\ref{matrixelements})
%________________________________
\begin{equation}
\label{brickn}
\dot{\mc}_{\ix}
=
\mQ^-
%_{\ix,\ix-1} 
{\mc}_{\ix-1}
+
\mQ
\;
%_{\ix,\ix} 
{\mc}_{\ix}
+
\mQ^+
%_{\ix,\ix+1} 
{\mc}_{\ix+1}
\; .
\end{equation}
%________________________________
In this form, the recurrence could be solved by continued fractions
(Sect.~\ref{sec:fc}).
The $\q$-basis choice is usually dictated by the space symmetries of
the problem including boundary conditions.

%%%%%%%%%%%%%%%%%%%%%%%%%%%%%%%%%%%%%%%%%%%%%%%%%%%%%%%%%%%%%%%%%%%%%%%%%%%%
\subsubsection{application to periodic potentials}

If the substrate potential is periodic, one Fourier expands as 
$V'(\q) = \sum_q V'_q \e^{\iu q \q}$.
Then it is natural to choose plane waves for the $\q$ basis
%______________________________
\begin{equation}
\label{planewaves}
u_\alpha (x) 
=
\frac {\e^{\iu \alpha \q}}{\sqrt {2 \pi}}
\;
\qquad
\quad
\alpha = 0, \pm 1, \pm 2, \dots
\end{equation}
%______________________________
To get matrix elements 
$A_{\alpha\beta}=\int{\rm d}x\,u_{\alpha}A\,u_{\beta}$,
we use orthogonality
$\int_{0}^{2 \pi} \dif \q \,
\e^{\iu q \q}\e^{-\iu \alpha \q} = 2 \pi \delta_{q \alpha}$
obtaining
%_____________________________
\begin{equation}
(\partial_\q)_{\alpha ,\alpha + \beta}
= \iu \alpha \delta_{\alpha, \alpha + \beta}
\; ;
\qquad
\qquad
(V')_{\alpha, \alpha + \beta}
=
V'_{\beta}
\; .
\end{equation}
%_____________________________  
From these we can get the matrix elements of $\mQ$, which have the structure
$\mQ^{\pm}_{\alpha, \alpha + \beta} \propto [\iu \alpha\,
\delta_{\alpha, \alpha + \beta} \pm V'_{\beta}]$.
Hence, the number of diagonals with non-vanishing elements equals the
number of harmonics of $V$ (only one for the cosine potential).

Hitherto we always had in mind solving the recurrence~(\ref{brickn})
in the index $n$.
But we could also have chosen to transform the $2$-index recurrence 
$\dot c_n^\alpha
=
Q_{n, n+m}^{\alpha, \alpha + \beta}
c_{n+m}^{\alpha + \beta}$
by introducing the following vectors
%____________________________
\begin{eqnarray}
\label{matricialx}
\mc_\alpha
=
\left (
\begin{array}{c}
c_{\alpha}^{-N}
\\
.
\\
.
\\
c_{\alpha}^{N}
\end{array}
\right )
\; ;
\qquad
\qquad
\quad
\dot{\mc}_{\alpha}
=
\sum_{\beta = -I}^{I}
\mQ_{\alpha, \alpha + \beta}
%_{\ix,\ix-1} 
\mc_{\alpha+\beta}
\; .
\end{eqnarray}
%_____________________________
In this case $I$ equals the number of harmonics in $V$, since 
$\mQ_{\alpha, \alpha + \beta} = 0$
for $\beta$ larger than that number.
Both recurrences (in $\p$ or in $\q$) can be employed at convenience
as Risken already did in the classical case \cite{risken}.

%%%%%%%%%%%%%%%%%%%%%%%%%%%%%%%%%%%%%%%%%%%%%%%%%%%%%%%%%%%%%%%%%%%%%%%%%%%%%
\vspace{3.ex}
\qquad \quad {\it example: the cosine potential}
\vspace{3.ex}

\begin{figure}
\centerline{\resizebox{7.cm}{!}{%
\includegraphics[angle = -0]{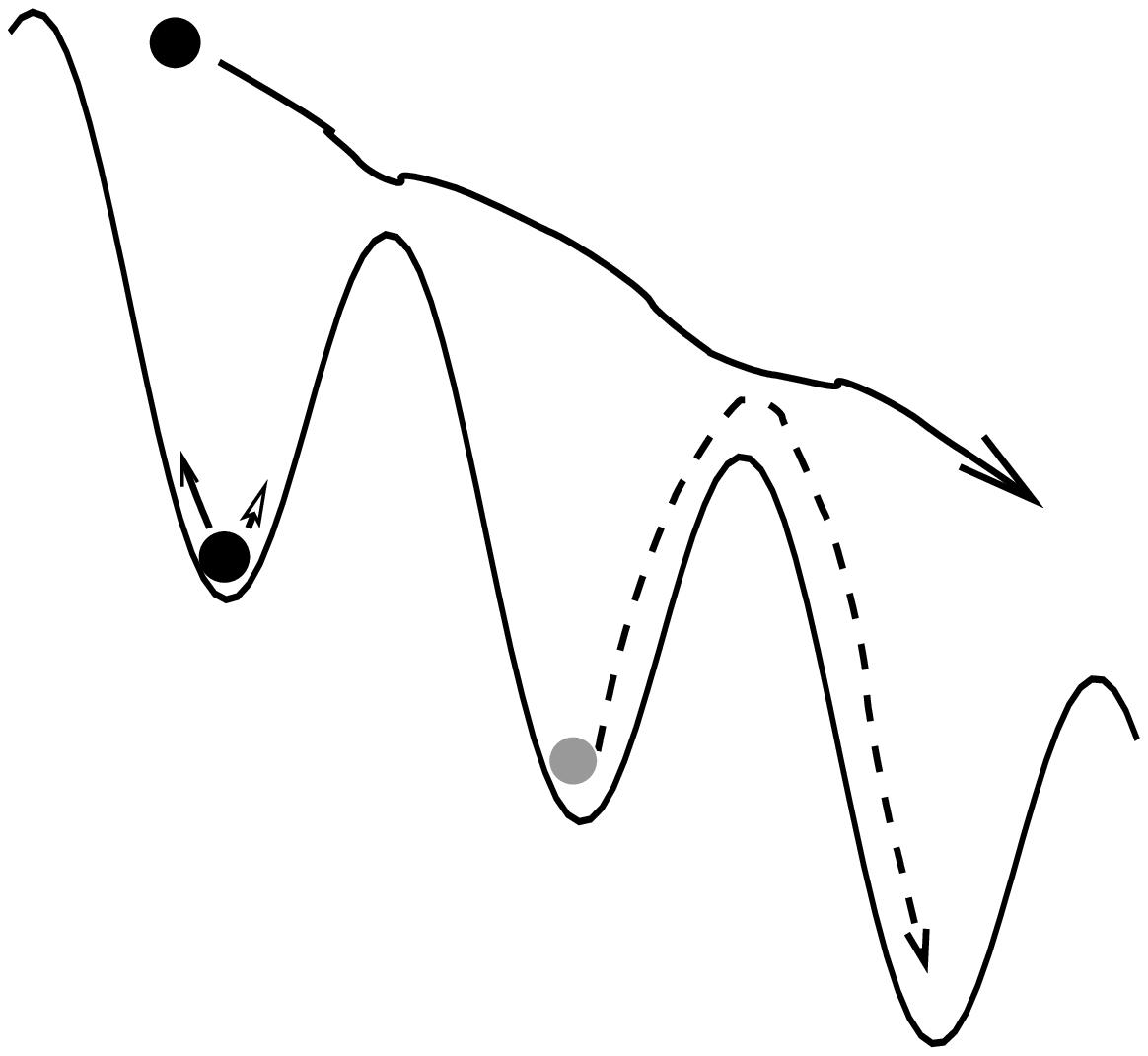}
}
}
\caption{
Potential profile $V(\q) = -\cos(\q) - F\q$.
In the Hamiltonian case ($\dampp = 0$) the particle can be ``locked'',
oscillating in one well, or ``running'' with $\langle p \rangle \neq
0$.
These two cases are depicted by solid trajectories.
When coupling the particle to a bath it experiences fluctuations and
dissipation.
These can lead to the eventual trapping of a running solution into one
well, or to the thermal launching of a locked particle over the
potential barrier (dashed line) possibly into a running solution.
} 
\label{fig:canaleta}
%\end{figure}
%___________________________________________________
%___________________________________________________
%___________________________________________________
%___________________________________________________
%___________________________________________________
%___________________________________________________
%\begin{figure}
\centerline{\resizebox{7.cm}{!}{%
\includegraphics[angle = -0]{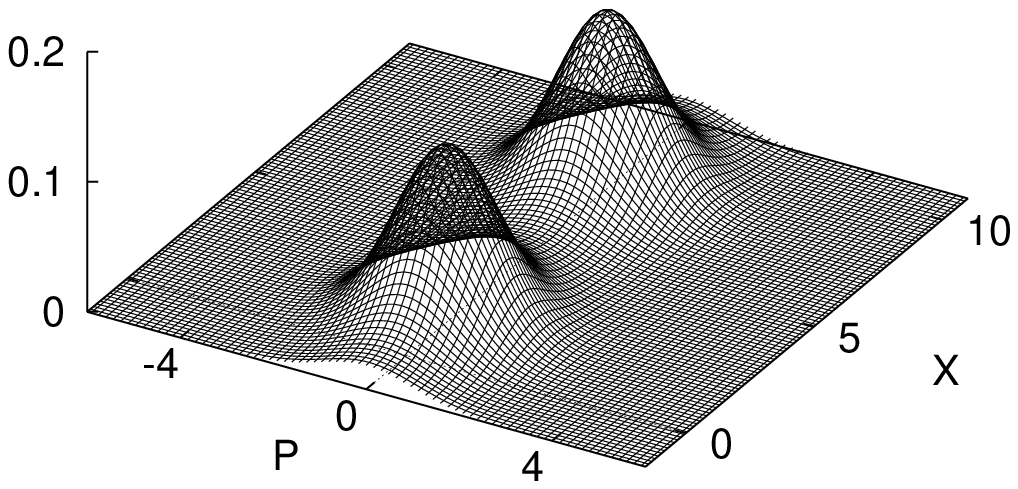}
}
\resizebox{7cm}{!}{%
\includegraphics[angle = -0]{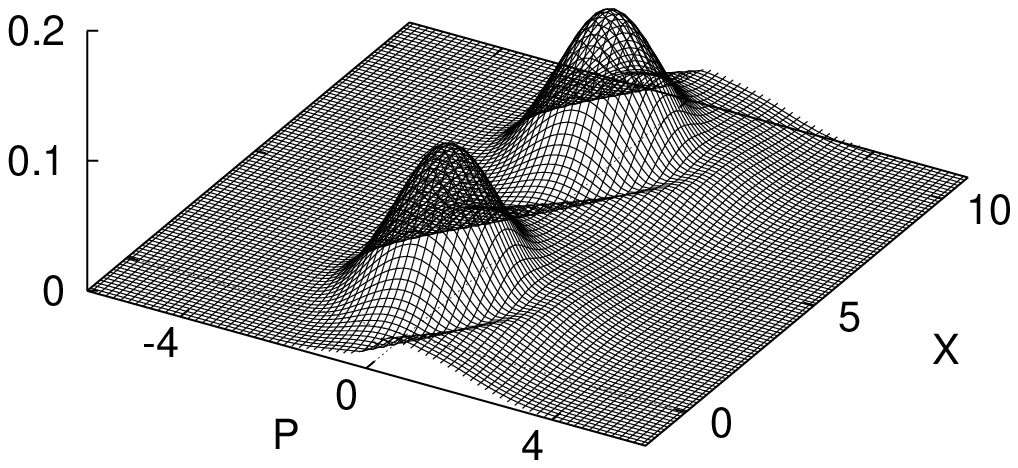}
}
\vspace*{-1.em}
}
\centerline{\resizebox{7.cm}{!}{%
\includegraphics[angle = -0]{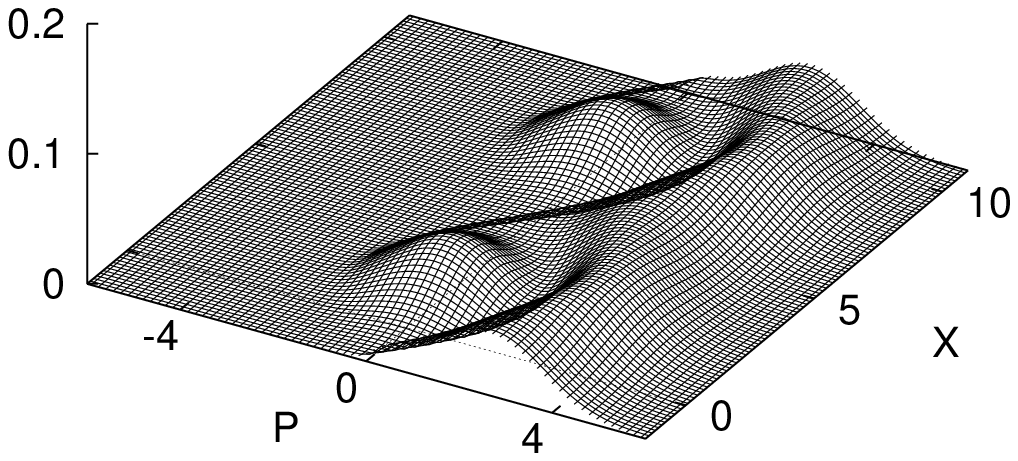}
}
\resizebox{7cm}{!}{%
\includegraphics[angle = -0]{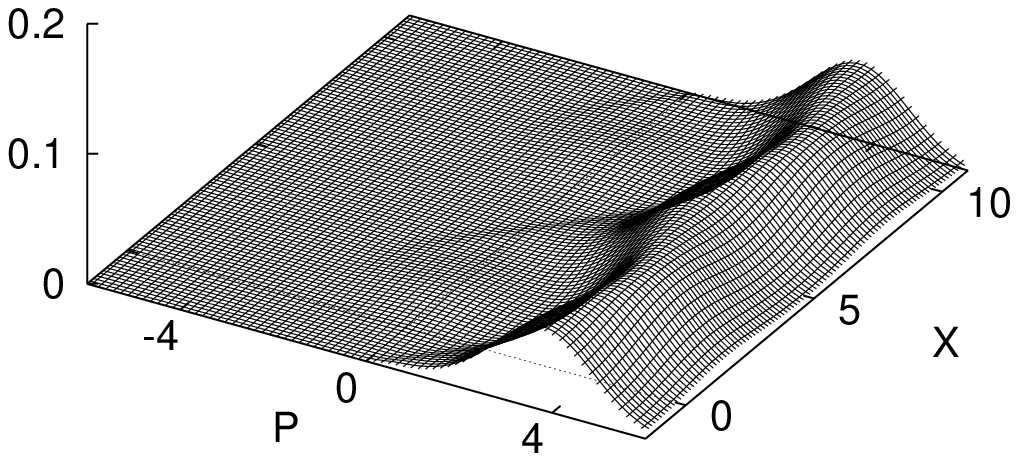}
}
}
\caption{
Risken's graphs: classical distributions for a particle in the tilted
periodic potential~(\ref{cosinetilt}).
The distribution is periodic in $\q$ so we have just plotted two
periods for illustration.
The damping is $\dampp = 0.05$ and the forces $F = 0$, $0.075$,
$0.15$, and $0.2$ (left to right, top to bottom).
Note the $\langle p \rangle = F/\dampp\sim4$ for the largest force,
and the bistability between running and locked solutions for
intermediate forces.
} 
\label{fig:risken}
\end{figure}

Let us consider the stationary solutions $\W(t\to \infty; \q, \p)$ for
a particle in a cosine potential, tilted by a constant force (see
fig.~\ref{fig:canaleta}):
%____________________________
\begin{equation}
\label{cosinetilt}
V(\q)
=
-V_0 \cos (\q) - \q \cdot F
\end{equation}
%____________________________
To this end we set $\dot \mc_{n}=0$ (or $\dot \mc_{\alpha}=0$) and
solve the vector recurrences.
With the $c$'s so obtained we reconstruct 
$\W (\q, \p)
\propto
\sum c_n^\alpha\,u_\alpha (\q) \psi_n(\p)$
[cf.~Eq.~(\ref{expW})], which is plotted in figure~\ref{fig:risken} (still
classical results).
%___________________________________________________
%___________________________________________________
%___________________________________________________
%___________________________________________________
%___________________________________________________
%___________________________________________________
%%

Let us follow Risken in the interpretation of these graphs.
In the absence of applied forces ($F = 0$) the stationary solution is
the equilibrium Boltzmann distribution $\propto \e^{-p^2/2 - V(\q)}$.
In the opposite case, when the force is very large, the substrate
potential matters only a little, and we can think of a damped particle
dragged by a constant force.
In this limit $\langle p \rangle = F/\dampp$.
Therefore, from small to large forces the distribution shifts in $\p$,
from the Boltzmann case with $\langle p \rangle=0$ towards the
stationary solution $\propto\e^{-\frac{1}{2}(\p - F/\dampp)^2}$.
In other words, a transition from ``locked'' to ``running'' solutions
as a function of the force (with bistability in between).
%

%%%%%%%%%%%%%%%%%%%%%%%%%%%%%%%%%%%%%%%%%%%%%%%%%%%%%%%%%%%%%%%%%%%%%%%%%
%%%%%%%%%%%%%%%%%%%%%%%%%%%%%%%%%%%%%%%%%%%%%%%%%%%%%%%%%%%%%%%%%%%%%%%%%
\subsection {the quantum regime ($\kondobar < \infty$)}

When $\kondobar$ is finite, we have to take into account the rest of
the terms in the Moyal series in equation (\ref{cleqpsadim}).
Here is when we start to appreciate the advantages of the phase-space
formalism:
one part of the equation is already solved --- Risken did it for us!
Now we are left ``only'' with handling the last sum, i.e.,
$\sum_{\iq=1}^{\infty}
\qcoef^{(\iq)}
\,
V^{(2\iq+1)}(\q)
\,
\partial_\p^{(2\iq+1)}
\W(\q, \p)$.
To this end, we make use of the result
%________________________
\begin{equation}
\w^{-1}
\partial_\p^{2 \iq +1}
\big( \w \; \psi_m \big)
=
(a^+ + a)^{2 \iq +1}
\psi_m
,
\end{equation}
%________________________
where recall that $a=\partial_\p + \p/2$ and $a^+ = \partial_p - p/2$.
Then, using the Hermite $\p$-basis each term in the sum will couple
$n$ with $n \pm (2 \iq + 1)$.
That is \cite{gar2004}, the Moyal term will in principle give coupling
to all indexes in the quantum generalization of the Brickmann
hierarchy~(\ref{brickcl})
%__________________________
\begin{equation}
\label{brickq}
\dot \mc_n
=
\sum_m
\mQ_{n, n+m} \mc_{n+m}
\; .
\end{equation}
%__________________________
%%
The explicit form of the matrices $\mQ$  can be found in $\jpai$,
sections 4 \&~5.
It is important to remark that for polynomial potentials
$V^{(2\iq+1)}=0$, after some $\iq$.
Then the recurrence would be finite, providing the finite $I$ we
required in the previous chapter to solve by continued fractions.
On the other hand, for non-polynomial potentials there exist
non-vanishing derivatives at all orders, like in a cosine potential.
We are back in the generic case of the quantum Brickmann
hierarchy~(\ref{brickq}) with infinite coupling range, making
continued fractions of no use.
But we still have the $\q$-recurrence, let us see if we are more lucky
with it \dots

%%%%%%%%%%%%%%%%%%%%%%%%%%%%%%%%%%%%%%%%%%%%%%%%%%%%%%%%%%%%%%%%%%%%%%%%%
\subsubsection{the case of periodic potentials}

In this case we proceed expanding the $\q$ dependence in plane
waves~(\ref{planewaves}).
The ``new part'' (the Moyal sum in powers of $\hbar$) brings in the
derivatives 
$V^{(\iq)} = \sum_q V^{(\iq)}_q \e^{\iu q \q}$.
To get their contribution to the matrix elements 
$A_{\alpha\beta}=\int{\rm d}x\,u_{\alpha}A\,u_{\beta}$,
we use the following result in the plane-wave basis:
%________________________________
\begin{equation}
[V^{(\iq)}(\q)]_{\alpha, \alpha + \beta} = V_\beta^{(\iq)}
\; .
\end{equation}
%________________________________
Therefore we have
$Q_{n, n+m}^{\alpha, \alpha + \beta} = 0$
whenever $\beta$ is larger than the number of harmonics in the
potential.
Then, using the vector recurrence~(\ref{matricialx}) in the $\alpha$
index, we have a finite coupling range, just as in the classical case.
The matrices $\mQ_{\alpha, \alpha + \beta}$ will contain more stuff,
$\hbar$-dependent terms, but what is important is that the index $I$
is finite and the same as in the classical problem: equal to the
number of harmonics.
In this case we could not chose between $\q$ or $\p$ recurrence (of no
use), but luckily one of them works.
%%

%%%%%%%%%%%%%%%%%%%%%%%%%%%%%%%%%%%%%%%%%%%%%%%%%%%%%%%%%%%%%%%%%%%%%%%%%%%%%
\vspace{3.ex}
\qquad \quad {\it example: cosine potential}
\vspace{3.ex}

\begin{figure}
\centerline{\resizebox{7.cm}{!}{%
\includegraphics[angle = -0]{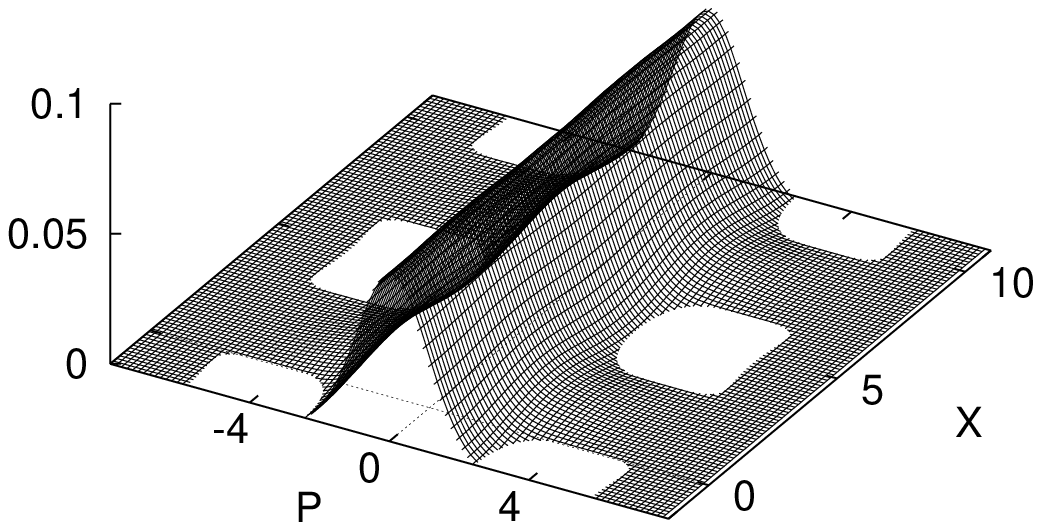}
}
\resizebox{7cm}{!}{%
\includegraphics[angle = -0]{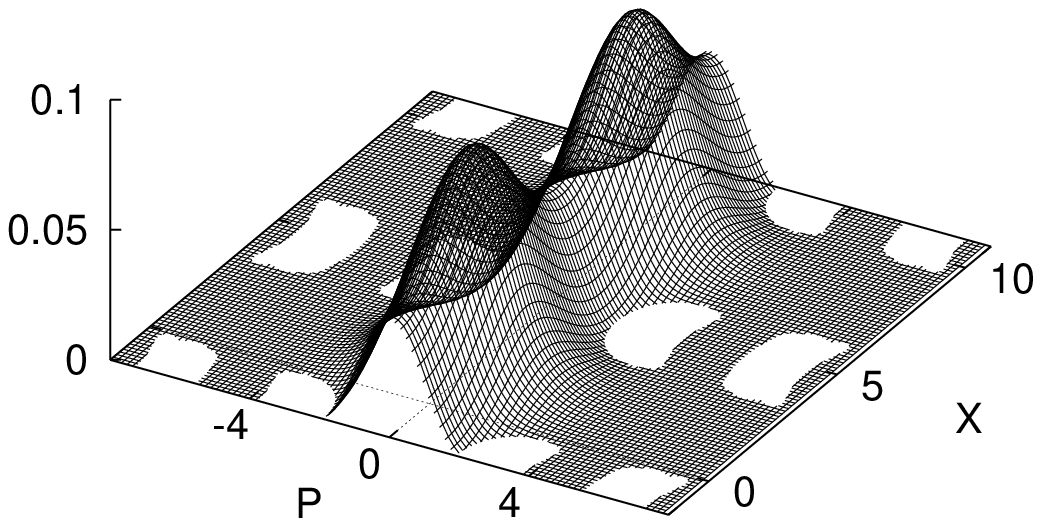}
}
}
\caption{ 
Wigner distributions in the deep quantum regime ($\kondobar=1$,
left, \&~ $\kondobar=2$, right).
For the temperature $T=1$ and damping $\dampp = 10^{-6}$ the
distributions are close to the canonical equilibrium (see text).
The white islands correspond to zones where $\W$ becomes negative
(slightly: $\W \sim -10^{-3}$ in the minima of the islands closer to
$\p=0$).
}
\label{fig:kandemir}
\end{figure}

As we did in the classical case, let us compute the stationary
solutions of the master equation in a cosine potential.
This potential has one harmonic ($I=1$) yielding a $3$-term
recurrence relation.
Setting $\dot \mc _\alpha = 0$ and solving, we get the distributions
of Fig.~\ref{fig:kandemir}.
Here we have set a ``very'' quantum regime, with $\kondobar = 1,2$
(recall $\kondobar\propto S_0/\hbar$).
In order not to leave the validity range $\hbar\dampp/\kT\ll1$ of the
Caldeira--Leggett master equation (Sect.~\ref{sub:cldme}), we have
used dimless $\dampp = 10^{-6}$ and $T=1$ (ensuring
$\dampp/T\kondobar\ll1$).
%___________________________________________________
%___________________________________________________
%___________________________________________________
%___________________________________________________
%___________________________________________________
%___________________________________________________
%%

The first striking feature (already announced in
chapter~\ref{chap:phase-space})) is the distribution becoming negative
in some regions (the white ``islands'' in figure~\ref{fig:kandemir}).
Besides, compared with the classical graphs~\ref{fig:risken}, the
distribution is quite {\em delocalized\/} in $\q$.
Specially for $\kondobar =1$; for $\kondobar =2$ we start to recognize
the first symptoms of localization.

To interpret with the hands the distributions of
figure~\ref{fig:kandemir} we recall first that in the range 
$\dampp/T = 10^{-6} \ll 1$
the stationary solutions should be well approximated by canonical
distributions (chapter~\ref{chap:equilibrio}).
Recalling also how we build the Wigner function from the density
matrix [Eq.~(\ref{dmwf})], the equilibrium $\W_{\rm eq}$ would be:
%__________________________________________-
\begin{equation}
\label{stacosine}
\W_{\rm eq} (\q, \p)
=
\frac{1}{2 \pi \hbar}
\int_{-\infty}^{\infty}
\dif y
\,
\e^{\iu y \p/\hbar}
\dm_{\rm eq} (\q - \case y,\q + \case y)
\; ;
\qquad
\dm_{\rm eq}
=
\int \dif q
\;
\e^{-\beta \epsilon_q} \vl q \rangle \langle q \vl
\end{equation}
%_____________________________
This will be approached by our $t \to \infty$ stationary solutions.
The Bloch waves $\vl q \rangle$, for small $\kondobar$, can be
constructed perturbatively (\cite{lanlif3}, section ``potential treated as a
perturbation'')
%_________________________
\begin{equation}
\label{bloch}
\vl q \rangle
=
\e^{\iu q \q}
(
1
+
\eta\, \e^{\iu \q}
+ 
\eta^*
\e^{-\iu \q}
)
\; ;
\qquad
\eta = \iu \left (\frac{\kondobar}{2 \pi} \right )^2
\; .
\end{equation}
%_________________________
Thus, when the characteristic action is ``small'' compared with
$\hbar$ (small $\kondobar/2\pi$), the Bloch particle behaves almost as
a free particle (see the energy bands in figure~\ref{fig:chenleb}).

Inserting eventually the above $\vl q \rangle$ in the
equilibrium~(\ref{stacosine}) we obtain:
%_________________________________________________
%_____________________________
%
% WIGNER FUNCTION: TURKEYS
%_____________________________
\begin{eqnarray}
\label{kandemir}
%%%%%%%%%%%%%%%%%%
\W(\q,\p)
&=&
\e^{-\frac{\p^2}{2 \vl \eta \vl^2}}
+
|\eta|^{2}
\left [
\e^{-\frac{(\p-1)^2}{2 \vl \eta \vl^2}}
+
\e^{-\frac{(\p+1)^2}{2 \vl \eta \vl^2}}
\right ]
\nonumber\\
%%%%%%%%%%%%%%%%%%
&-&
2|\eta|^{2}
\cos(2\q)\,
\e^{-\frac{\p^2}{2 \vl \eta \vl^2}}
+
2\iu\eta
\sin(\q)
\left[
\e^{-\frac{(\p-1/2)^2}{2 \vl \eta \vl^2}}
+
\e^{-\frac{(\p-1/2)^2}{2 \vl \eta \vl^2}}
\right]
%%%%%%%%%%%%%%%%%%
\; .
\end{eqnarray}
%_____________________________
%_____________________________
This remarkable form captures qualitatively the features of the
numerically obtained distributions.
The first term (a Gaussian in momentum) ``is'' the free particle
$\e^{\iu q \q} \to \exp (-\p^2/2 \vl \eta \vl^2) $, 
while the second comes from shifted plane waves 
$\e^{\iu (q \pm 1) \q}$.
They are responsible for the delocalization along the $\q$ axis.
The last two terms in~(\ref{kandemir}) follow from the cross products
between 
$\e^{\iu (q + 1) \q}$ \& $\e^{\iu (q - 1) \q}$
and
$\e^{\iu q  \q}$ $\&$ $\e^{\iu (q \pm 1) \q}$,
respectively.
Specifically, the third term incorporates a modulation due to the
potential profile (as it is of order $\vl \eta \vl^2$ is less
noticeable for $\kondobar=1$).
The fourth term, finally, produces negative regions.

The above terms, and hence the negative islands, can be seen as an
expression of the wave character of quantum mechanics.
By this we mean that $\W$ generates the averages and hence it is made
of squared wave functions [$\psi^*(\q - u/2) \psi (\q + u/2)$, section
\ref{sub:1932}].
Seen in this way, $\W$ is like an intensity (its Fourier transform)
and the last terms in~(\ref{kandemir}) depict a typical interference
pattern between sinusoidal waves.

%%%%%%%%%%%%%%%%%%%%%%%%%%%%%%%%%%%%%%%%%%%%%%%%%%%%%%%%%%%
%%%%%%%%%%%%%%%%%%%%%%%%%%%%%%%%%%%%%%%%%%%%%%%%%%%%%%%%%%%
%%%%%%%%%%%%%%%%%%%%%%%%%%%%%%%%%%%%%%%%%%%%%%%%%%%%%%%%%%%
\section{computation of observables: transport problems}

We did not mention it in this chapter, but we recall that once the
distribution $\W(\q, \p)$ is reconstructed from the expansion
coefficients $c$'s, we can compute any observable following
essentially the classical prescription [Eq.~(\ref{mvwf})]
%_________________________________
\begin{equation}
\nonumber
\langle A \rangle
=
\int \dif \p \int \dif \q\,
A (\q, \p) \W (\q, \p)
\; .
\end{equation}
%_________________________________
%
%
The ``classical observable'' $A(\q,\p)$ is actually the $\W_A$
transform of the operator $A$ in Hilbert space, Eq.~(\ref{WAwf}).

In some cases, it is not required to construct $\W$ explicitly, as
some observables of interest can be expressed directly in terms of
expansion coefficients.
For example, in transport problems the current $\langle \p \rangle$
can be written as
%____________________________________
\begin{equation}
\langle \p \rangle
=
\sum_{n}
K_{2n +1}^{(1)}
\left (
\sum_{\alpha}
c_{2n+1}^{\alpha}
I_{\alpha}
\right )
\end{equation}
%___________________________________
The coefficients $I_{\alpha}$ and $K_{2n +1}^{(1)}$ depend on the
prefactor $\w$ in Eq.~(\ref{expW}), and on properties of the integrals of
Hermite polynomials and plane waves [given explicitly in Eqs. (29) and
(30) of $\jpai$].

\section{example I: quantum transport in a cosine potential}

Let us focus now on transport problems, and the effects of the
dissipative bath.
We consider first that the substrate potential is the simplest
possible, the cosine, which is however tilted by a constant force $-\q
\cdot F$ (see figure~\ref{fig:canaleta}).
For this problem we solve the Caldeira--Leggett equation and compute the
current in stationary conditions $\langle \p \rangle (t\to\infty)$.

\begin{figure}[h]
\centerline{\resizebox{7.5cm}{!}{%
\includegraphics[angle = -90]{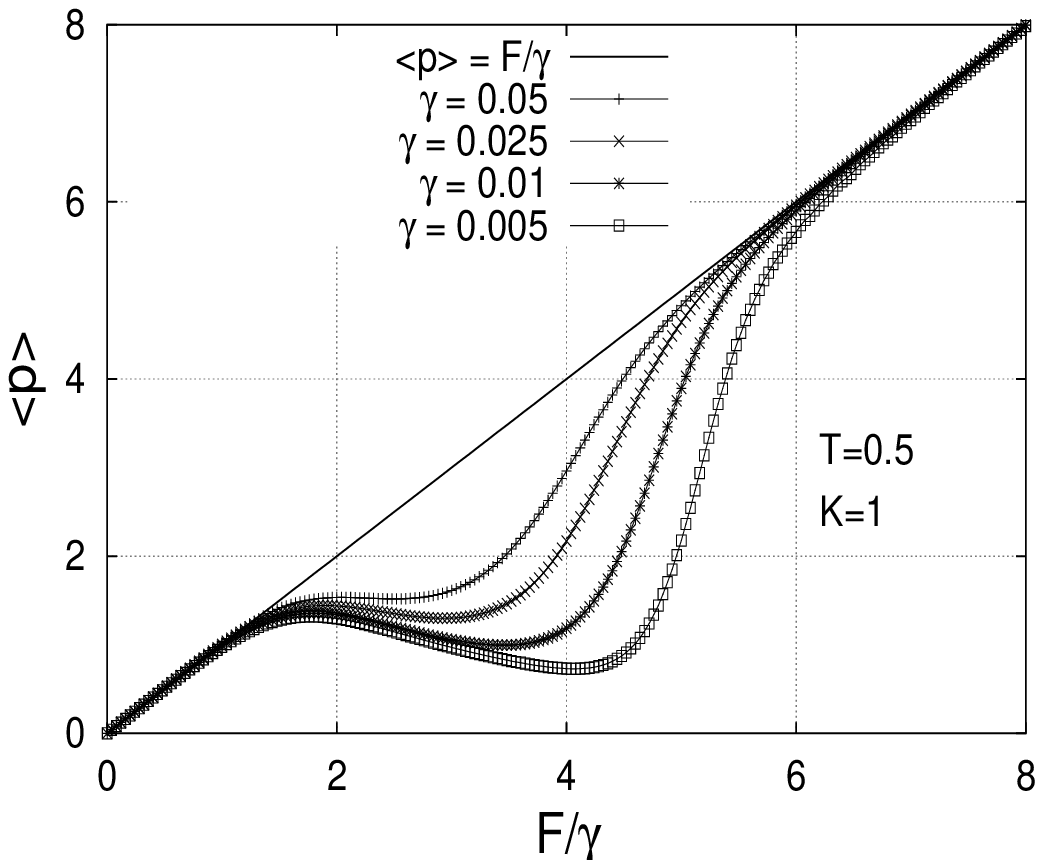}
}
  \resizebox{8.cm}{!}{%
\includegraphics[angle = -90]{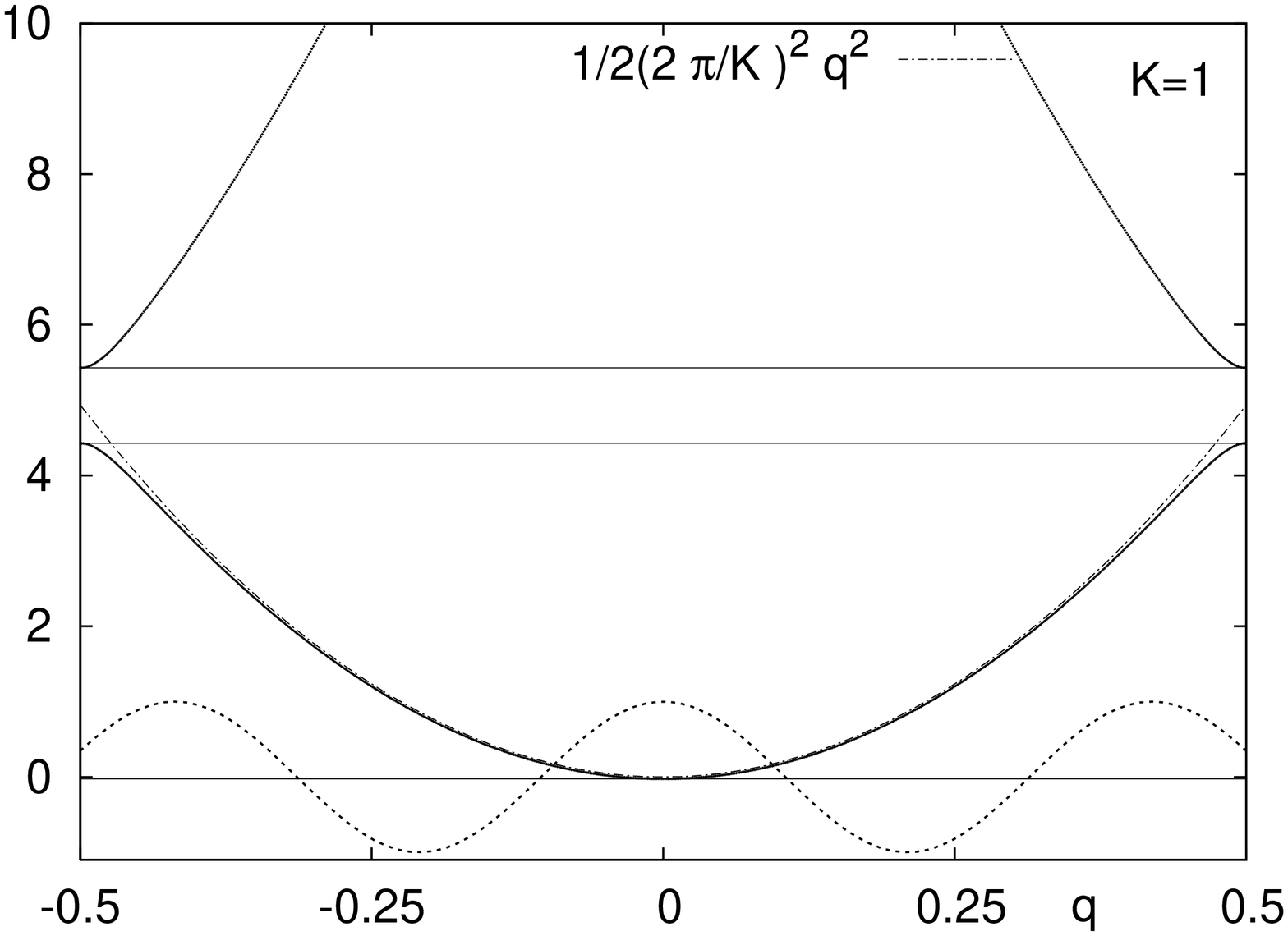}
}
}
\caption{
Left: average velocity $\langle p \rangle$ as a function of the
applied force, for various $\dampp$ in the weak damping regime.
Right: energy bands of the (untilted) cosine potential $V(\q) = -V_0
\cos (\q)$ for $\kondobar = 1$.
}
\label{fig:chenleb}
\end{figure}

Figure~\ref{fig:chenleb} shows the results for several values of the
damping.
The phenomenology is that studied by Chen and Lebowitz, which did a
difficult path-integral calculation treating the potential
perturbatively \cite{cheleb92a, cheleb92b}.
%_________________________________________________
%___________________________________________________
%___________________________________________________
%___________________________________________________
%___________________________________________________
%___________________________________________________
%___________________________________________________
%
The curves can be understood as follows.
When the force is small the particle moves on the band bottom, which
for $\kondobar=1$ has a near free particle form
(figure~\ref{fig:chenleb}, right).
Then $\langle\p\rangle=F/\dampp$ as in a damped and dragged classical
free particle.
Increasing the force, we approach the band limit.
There Bragg scattering reduces the group velocity $\propto \partial E
/\partial q$, and hence $\langle \p \rangle$ decreases (see any solid
state book, e.g. \cite{ashcroft}).
But increasing further the force, we go into the next band,
progressively recovering the free particle behavior
$\langle\p\rangle=F/\dampp$.
Figure \ref{fig:chenleb} also shows how increasing the damping the
curves approach this free particle current.
This is understood in terms of the folk argument of bath as blurring
the energy levels (recall figure~\ref{fig:density}), which in our case
amounts to bridge the band gap, responsible for the decrease.

%%%%%%%%%%%%%%%%%%%%%%%%%%%%%%%%%%%%%%%%%%%%%%%%%%%%%%%%%%%%%%%%%%%%%%%%%%
\section{example II: directional motion in a ratchet potential}

\begin{figure}
\centerline{\resizebox{8.cm}{!}{%
\includegraphics[angle = -90]{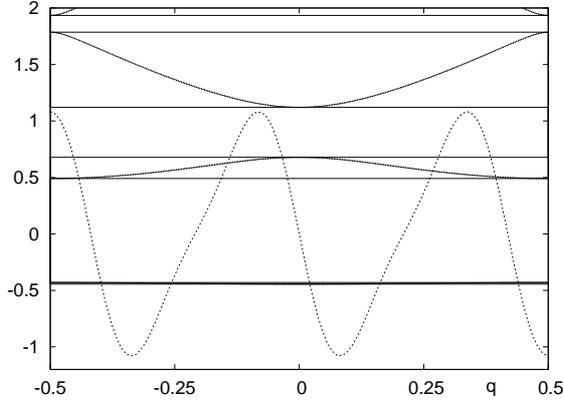}
}
}
\caption{
Ratchet potential profile~(\ref{Vrat2}) for $r=0.44$, and the
associated band structure for $\kondobar=5$, having only two bands
below the barrier top.
}
\label{fig:Vrat}
\end{figure}

Up to this point we have treated the cosine potential.
In this section we consider the minimal extension of adding a second
harmonic
%______________________________
\begin{equation}
\label{Vrat2}
V(\q)
=
-
V_0
\left [
\sin \q
+
(\rat/2) \sin (2 \q)
\right ]
\end{equation}
%______________________________
with $r$ the ratchet parameter ($r=0$ brings us back to the one-harmonic
case).
This potential is asymmetric, as seen in Figs.~\ref{fig:bands}
and~\ref{fig:Vrat}.
Looking at the potential profile we realize that the slope to the
right is less steep (easy side) than to the left (hard side).

An important property of ratchet potentials is that they can
accommodate directional motion.
By this we mean that the system can have a net current even if the
average of the external forces is zero (on average, the force does not
do work).
One also speaks of {\em rectifying\/} zero-mean perturbations
(generating a net response).
How is this possible?
First, if the system is out of equilibrium, the second law of
thermodynamics does not hold.
Besides, we need the potential being asymmetric, as otherwise there
would not exist a privileged direction \cite{rei02, asthan02}.
Then a system modelized by an asymmetric potential like~(\ref{Vrat2}),
and coupled to a dissipative bath, can show such a phenomenology if a
driving force $\q \cdot f(t)$ keeps it out of equilibrium.
%%
%___________________________________________________
%___________________________________________________
%___________________________________________________
%___________________________________________________
%___________________________________________________
%___________________________________________________
%___________________________________________________
%___________________________________________________
%___________________________________________________
\begin{figure}
\centerline{
\resizebox{8.cm}{!}{%
\includegraphics[angle = -90]{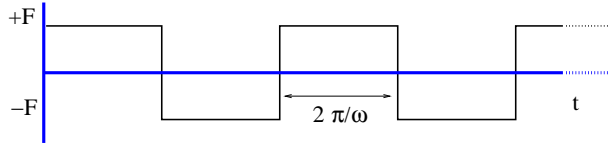}
}
}
\caption{
Scheme of the square-wave force profile applied in the problem of
rectification in a ratchet potential.
} 
\label{fig:adiabatic}
\end{figure}
%___________________________________________________
%___________________________________________________
%___________________________________________________

For the numerical solution we are going to assume, for simplicity,
that $f(t)$ has a square wave profile, which changes between positive
and negative values $\pm F$ with a period $2 \pi/\omega$ (see
Fig.~\ref{fig:adiabatic}).
Besides, we use the adiabatic trick: the period is long enough
(compared with the rest of time scales of the problem), so that most
of the time inside each step $+F$ or $-F$ the current is the
stationary one at the corresponding force.
Then the net current (averaged over time) would be:
%_____________________________
\begin{equation}
\dampp\,\langle p \rangle_{\rm r}
=
\dampp\,\langle p \rangle_{+F}
+
\dampp\,\langle p \rangle_{-F}
\; ,
\end{equation}
%_____________________________
with $\langle p \rangle_{\pm F}$ the corresponding stationary
solutions.%
\footnote{
Note that for the cosine problem we have
$\langle p \rangle_{+F}
= - \langle p \rangle_{-F}$
and hence no rectification $\langle p \rangle_{\rm r} = 0$.
} %ENDOF FOOTNOTE

Working in the stationary regime, we set $\dot \mc_{\alpha} = 0$ in
the recurrence (\ref{matricial}) and we are back in the case discussed
in section \ref{sub:staFC}.
As for the continued-fraction solution, note that having two harmonics
in the potential, the associated recurrences now have $I=2$ as coupling
range (section \ref{sub:Ip1}).

%%%%%%%%%%%%%%%%%%%%%%%%%%%%%%%%%%%%%%%%%%%%%%%%%%%%%%%%%%%%%%%%
\subsection{directed motion in the classical limit}

\begin{figure}
\centerline{\resizebox{9cm}{!}{%
\includegraphics[angle = -0]{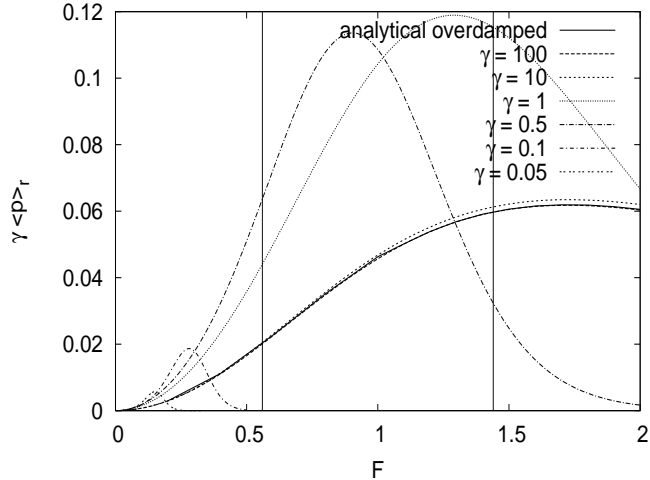}
}
}
\caption{
Rectified current in the classical limit for the ratchet
potential~(\ref{Vrat}) driven by the square wave of
figure~\ref{fig:adiabatic}.
Results shown for various values of the damping at $T=1$.
The thicker line is the analytical result \cite{risken} in the
overdamped regime $\dampp \to \infty$.
The vertical lines mark the two critical
forces~(\ref{criticalforces}).
}
\label{fig:ratchet}
\end{figure}

Let us try to understand first the classical problem in order to
address later the effects of adding quantum effects.
The classical rectified current in shown in figure~\ref{fig:ratchet}.
The main feature is that the net current is positive, meaning that the
particle moves on average to the right, the easy way.
%%
%___________________________________________________
%___________________________________________________
%___________________________________________________
%___________________________________________________
%___________________________________________________
%___________________________________________________
A way of understanding this is introducing the critical forces for
barrier disappearance, to the left and to the right:
%_____________________________
\begin{equation}
\label{criticalforces}
\vl F_c^+ \vl/V_0  = 1- r
\; ;
\qquad
\qquad
\vl  F_c^- \vl/V_0  = 1+ r
\; .
\end{equation}
%_____________________________
Thus, the barrier felt by the particle when tilting the potential with
$+F$ is lower than that when we tilt with $-F$, so it is easier to move to
the right.

To understand the other features, we will think first on the $T=0$
limit; that is, only dissipation, no fluctuations.
The associated deterministic system has two more characteristic forces
$F_1$ and $F_2$, obeying $F_1 < F_2 < F_c$.
When  $F <F_1$ the dynamically stable solution (the attractor) is the
locked solution, while for  $F>F_2$ the running solution is the stable
one.
Well, if the system is locked close to a minimum ($F < F_1$), or if it
is in a wild running mode, in both cases the potential asymmetry (and
other details) become less relevant.
Therefore, we understand that the rectification will be more
noticeable in some intermediate force range, some ``window'', as seen in
figure~\ref{fig:ratchet}.
Besides, the characteristic forces $F_1$ and $F_2$ are damping
dependent, and happen to decrease with $\dampp$ \cite{risken}.
Thus the window of maximum rectification shifts down along the force
axis when we decrease $\dampp$ \cite{borcosmar02}.
Finally, to restore $T$ in the picture, we can think of its effect as
allowing jumping between attractors, smoothing the sharp ranges, and
in particular allowing rectification below $F_c^+$ ($\jpai$ and
$\pe$).
\enlargethispage{0.5cm}

%%%%%%%%%%%%%%%%%%%%%%%%%%%%%%%%%%%%%%%%%%%%%%%%%%%%%%%%%%%%%%%%
\subsection{quantum corrections}

Our goal here is to study how the rectified currents of the previous
section are modified when the quantum properties of the system are
taken into account.
To this end we focus on one of the curves (we fix one $\dampp$) and
progressively decrease the quantumness parameter $\kondobar \propto
S_0/\hbar$.
This makes quantum effects more and more important.
%%
%___________________________________________________
%___________________________________________________
%___________________________________________________
\begin{figure}
\centerline{
\resizebox{9.cm}{!}{%
\includegraphics[angle = -90]{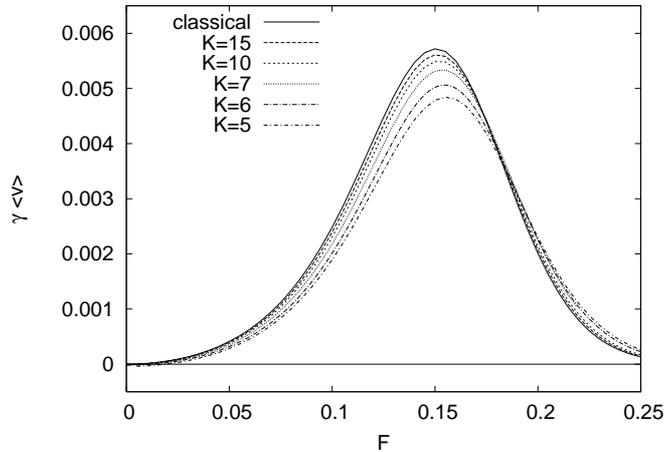}
}
}
\vspace{2.ex}
\caption{
Rectified current in the quantum problem, for several values of
$\kondobar \propto S_0/\hbar$.
We fixed the damping to $\dampp = 0.05$ and plotted the classical
curve ($\kondobar\to\infty$) as reference.
}
\label{fig:ratchetq}
\end{figure}
%___________________________________________________
%___________________________________________________
%___________________________________________________
%___________________________________________________
%___________________________________________________
%___________________________________________________
\begin{figure}
\centerline{
\resizebox{8cm}{!}{%
\includegraphics[angle = -0]{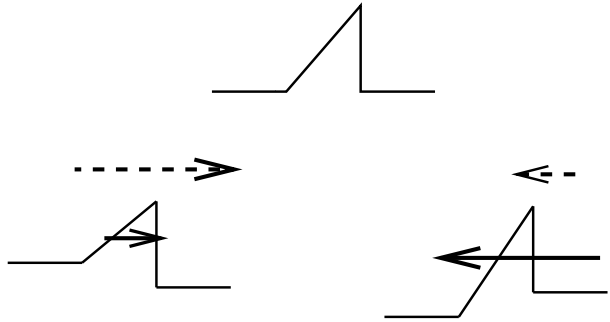}
}
}
\vspace{2.ex}
\caption{
Saw-tooth potential in the absence of force, and as deformed by $\pm
F$.
The barrier is lower to the easy side than to the hard side.
The solid arrows indicate the transmission for energies smaller than
the barrier (tunnelling) and the broken arrows the transmission for
energies larger than the barrier (affected by overbarrier wave
reflection).
The length of the arrows indicates qualitatively the probability of
those processes.
}
\label{fig:saw}
\end{figure}
%___________________________________________________
%___________________________________________________
%___________________________________________________

In figure~\ref{fig:ratchetq} we see that for large amplitudes of the
force wave $\vl F \vl$ the quantum corrections increase the rectified
current
$\langle p \rangle_{+F}+\langle p \rangle_{-F}$,
whereas the rectification is decreased at low forces with respect to
the classical case.

To understand the reduction/increase, in $\jpai$ and $\pe$ we resorted
to the semiclassical calculation of the transmission through a
saw-tooth potential.
This is an asymmetric potential, much as the ratchet potential, with
easy and hard sides (figure~\ref{fig:saw}), but where analytical
calculations are possible.
We studied the transmission coefficient when deforming the potential
with $+F$ and $-F$ (appendix F, $\jpai$).
We found that for energies over the barrier the transmission is higher
to the easy side, because the wave-reflection is less intense than to
the hard side (schematically indicated by the broken arrows in
figure~\ref{fig:saw}). 
On the other hand, when the particle has an energy lower than the
barrier, the transmission is higher to the hard side, as tunneling is
more probable towards that direction (solid arrows in
figure~\ref{fig:saw}). %
\footnote{
Recall that in quantum mechanics, much as particles with energy lower
than the barrier can be transmitted to the other side, by {\em
tunneling}, the complementary effect, equally counter-intuitive, can
also take place: particles with energies larger than the barrier can
bounce back: {\em overbarrier wave-reflection}.
} %ENDOF FOOTNOTE

Then, for large enough forces, particles are more energetic and there
is more probability of being launched over the barrier by thermal
fluctuations; then the wave-reflection dominates and we see an
amplification of the rectification by quantum effects.
On the other hand, for small forces/energies, it is more probable to
find particles with energies lower than the barrier.
Here tunnel events win, and the rectified current is reduced.

In $\jpai$ we presented results for fixed amplitude $F$, studying the
dependence of the current on the temperature.

%%%%%%%%%%%%%%%%%%%%%%%%%%%%%%%%%%%%%%%%%%%%%%%%%%%%%%%%%%%%%%%%%%%%%%%
\section{example III: AC response in periodic potentials (LRT)}
\label{sub:LRTcosine}

In the previous examples we have addressed stationary transport in a
cosine potential, subjected to a constant force, and then rectified
currents in a ratchet potential, with a driving force in the adiabatic
regime.
Thus, we could reduce the studies to the computation of stationary
solutions, and combinations of them.
Now we are going to address a genuine AC problem, with the driving out
of the adiabatic regime.

Let us return to the cosine potential~(\ref{cosinetilt}), and let 
$f(t)=\delta \cdot \e^{\iu \omega t}$.
If the amplitude is small $\delta \ll 1$, we can use linear response
as in section~\ref{sec:LRTI}.
The observable of interest now is the AC current
[cf. Eq.~(\ref{ACresp})]
%______________________________
\begin{equation}
\label{perLRT}
  \langle p \rangle = \delta  \cdot \sus (\omega) \; \e^{\iu \omega t}
\; .
\end{equation}
%______________________________
To compute the amplitude $\sus (\omega)$ we can use the
continued-fraction method as explained in section~\ref{sub:LRTFC}.
Note that, being in linear response, $\langle \p \rangle$ is
proportional to the force, so that $\langle f(t) \rangle  = 0$ gives a
zero time average of the current, irrespective the potential.
Then, the asymmetry of the potential does not bring much new, and for
simplicity we return to the cosine potential.
%___________________________________________________
%___________________________________________________
%___________________________________________________
\begin{figure}
\centerline{\resizebox{8.cm}{!}{%
    \includegraphics[angle = -0]{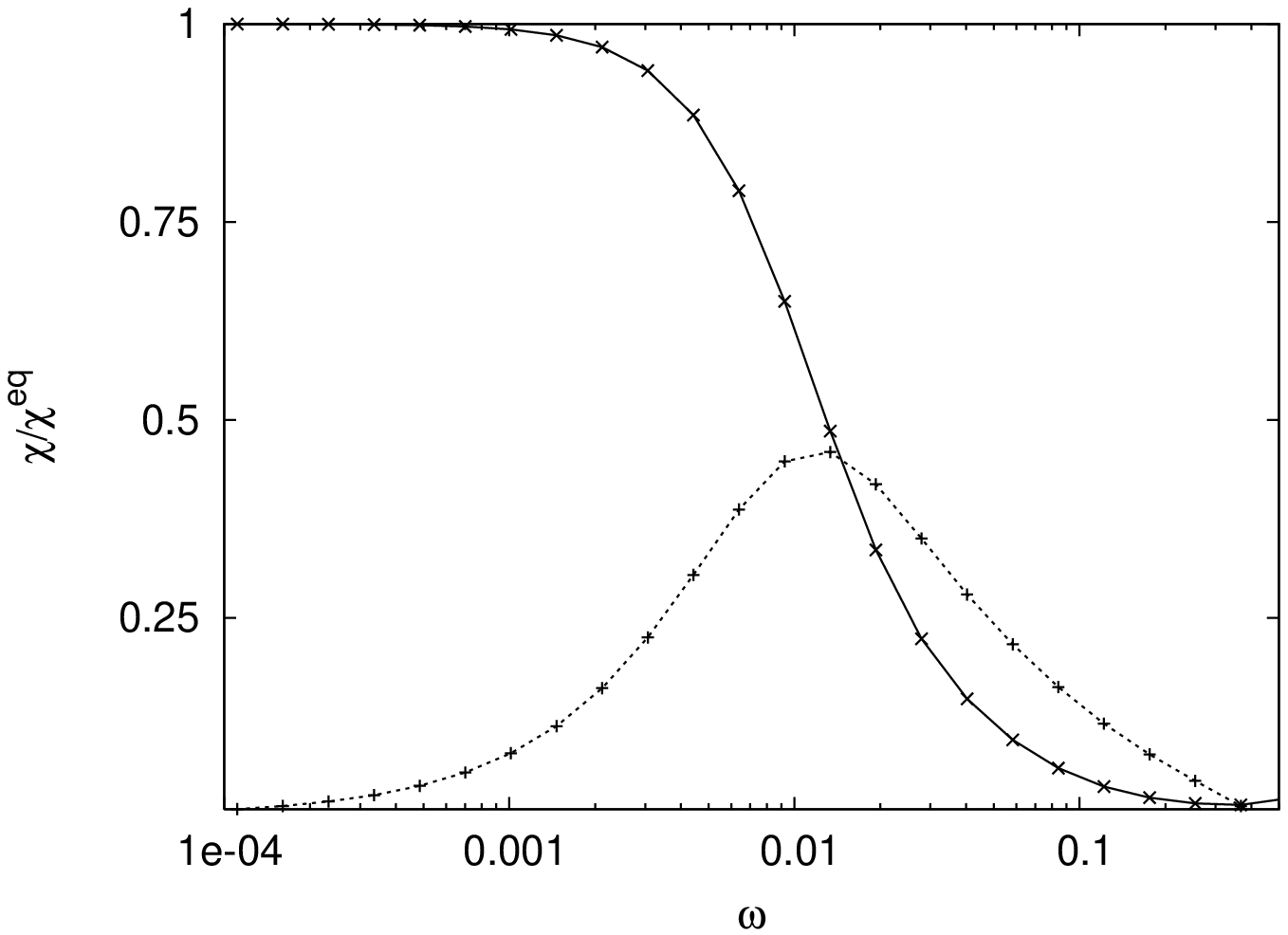} 
}
\resizebox{7.9cm}{!}{%
\includegraphics[angle = -0]{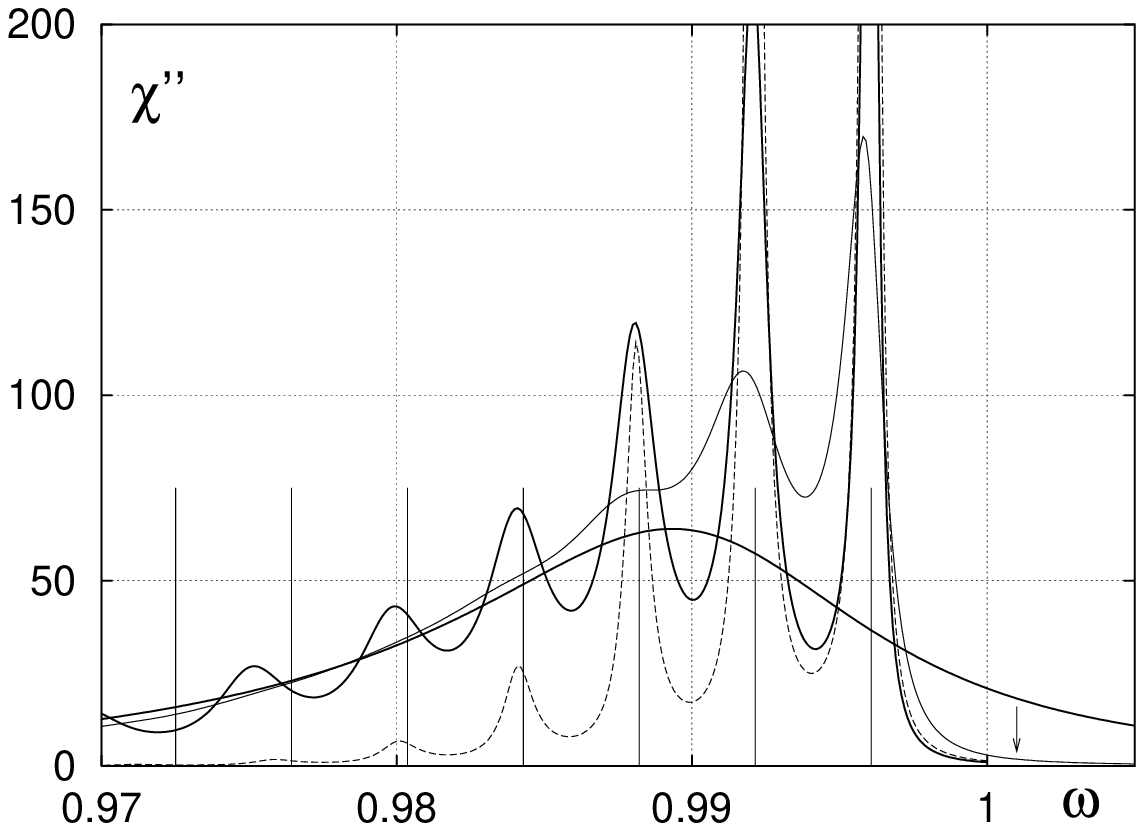}
}
}
\caption{
Left: low frequency range of the susceptibility $\sus(\omega)$ (cf.
figure~\ref{fig:LResp1}).
The real and imaginary parts are plotted for $K=1000$ (classical),
$T=0.5$, $\dampp = 0.1$ and $F= 0.1$.
Right: $\sus (\omega)$ in the high frequency range (imaginary part
only), for $K=200$, $T=0.05$, and $F=0$ (as in $\jpai$).
The $\dampp=0.01$ curve is essentially the classical result, and for
$\dampp=0.0003$ and $\dampp=0.0001$ the peaks become narrower.
The vertical lines mark the transition frequencies of the associated
approximate quartic oscillator (see the text).
}
\label{fig:lrtcosine}
\end{figure}
%___________________________________________________
%___________________________________________________
%___________________________________________________

Figure \ref{fig:lrtcosine} shows details of the response in the low
frequency range (left panel) and at high frequencies (right).
The low frequency spectrum shows a Debye profile, while at high
frequencies we see a pattern of absorption lines (we just plotted the
imaginary parts).
These figures can be understood qualitatively recalling the discussion
of section~\ref{sec:LRTII}.
There we argued that in general the response would be a combination of
Debye's and absorption curves, pretty much as what we see in
figure~\ref{fig:lrtcosine}.
The absorption peaks are Lorentzian curves centered on the energy
differences, and with a width $\propto \dampp$ (or inversely
proportional to the decoherence times, if you prefer).
On the other hand, the broad relaxation maximum of the low frequency
Debye curve (note the log scale), would give the relaxation time.

The reason of the Debye part is the following.
Along with the erasing of the non-diagonal terms in the density matrix
(decoherence), the system has to readjust the populations of the wells
(the cosine potential is now tilted by the AC force).
This leads to the relaxation time (slow dynamics) we find in the low
frequency range.

As for the high frequency curves, the widths of the absorption
peaks can be calculated semiclassically ($\jpai$, appendix F).
We performed a WKB calculation of the widths $W_b$ of the lower bands
for the example displayed ($\kondobar = 200$), finding
 $W_b = (4/\kondobar )\exp [ -K (1 - \epsilon)/2]$
with $\epsilon = E/V_0$.
Therefore the widths are exponentially small, and for $\kondobar=200$
they can be considered as discrete levels (recall Fig. \ref{fig:bands}).
Besides, their position can be estimated from the energy levels of the
anharmonic oscillator obtained expanding the cosine to fourth order 
$\cos(x) \cong a \q^2 + b \q^4$.
Indeed, plotting the associated energy differences, we find that the
absorption peaks are centered around them, as expected.
Now, if we increase the damping, we see how the absorption curves
broaden, with a width proportional to $\dampp\,\propto 1/T_2$.
Eventually, as the damping increasingly blurs the energy levels, the
curves approach a smooth peakless ``classical'' profile.
%%

%%%%%%%%%%%%%%%%%%%%%%%%%%%%%%%%%%%%%%%%%%%%%%%%%%%%%%%%%%%%
%%%%%%%%%%%%%%%%%%%%%%%%%%%%%%%%%%%%%%%%%%%%%%%%%%%%%%%%%%%%
%%%%%%%%%%%%%%%%%%%%%%%%%%%%%%%%%%%%%%%%%%%%%%%%%%%%%%%%%%%%
%%%%%%%%%%%%%%%%%%%%%%%%%%%%%%%%%%%%%%%%%%%%%%%%%%%%%%%%%%%%
\section {summary}

The example of the particle in a periodic potential has allowed us to
review many of the concepts and tools introduced in the preceding chapters:
Fokker--Planck equations, quantum master equations, phase-space
methods, relaxation and decoherence times, \dots

The continued-fraction technique, plus phase space, has made possible
to solve the problem as an extension of Risken approach to the
classical problem.
In this way we have bypassed the problem of the underlying 
spectrum (demanding, as is continuous and non-bounded).
During the whole chapter we had the classical limit as reference, in
order to understand our results as corrections (recall that the
Caldeira--Leggett equation is semiclassical).

We finally remark the differences between the classical and quantum
distributions (recall the wave-mechanical interpretation).
We have discussed how the interplay of overbarrier wave reflection and
tunneling can increase or decrease the rectified current in an
asymmetric potential, depending on the force/energy range.
We have also seen how the damping can make the curves to approach the
classical counterparts.
We closed the chapter with the study of the dynamical susceptibility
spectra and characteristic times.

%%%%%%%%%%%%%%%%%%%%%%%%%%%%%%%%%%%%%%%%%%%%%%%%%%%%%%%%%%%%
%%%%%%%%%%%%%%%%%%%%%%%%%%%%%%%%%%%%%%%%%%%%%%%%%%%%%%%%%%%%
%%%%%%%%%%%%%%%%%%%%%%%%%%%%%%%%%%%%%%%%%%%%%%%%%%%%%%%%%%%%
%%%%%%%%%%%%%%%%%% SUPERPARAMAGNETS %%%%%%%%%%%%%%%%%%%%%%%%
%%%%%%%%%%%%%%%%%%%%%%%%%%%%%%%%%%%%%%%%%%%%%%%%%%%%%%%%%%%%
%%%%%%%%%%%%%%%%%%%%%%%%%%%%%%%%%%%%%%%%%%%%%%%%%%%%%%%%%%%%
%%%%%%%%%%%%%%%%%%%%%%%%%%%%%%%%%%%%%%%%%%%%%%%%%%%%%%%%%%%%

\chapter [applications II: superparamagnets] 
{applications II:\\ statics \& dynamics of superparamagnets}
\label{chap:aplicationesII}

Quantum nanomagnets are an example of mesoscopic objects described
effectively by ``one'' degree of freedom (their net spin).
In this chapter we are going to study the statics and dynamics of these
systems, where the interaction with the environment is important.
We start reintroducing the model and investigating its equilibrium
properties.
Next we specify the master equation to be solved to study
non-equilibrium behavior.
As examples we will address the relaxation mechanisms (analogous to
the calculation of $T_1$ with Bloch's equations), by analyzing the
longitudinal susceptibilities (linear and non-linear).
We will conclude with the calculation of the transverse susceptibility
(study of $T_2$).
%%

%%%%%%%%%%%%%%%%%%%%%%%%%%%%%%%%%%%%%%%%%%%%%%%%%%%%%%%%%%%%%%%%%%%%%%%%%%
%%%%%%%%%%%%%%%%%%%%%%%%%%%%%%%%%%%%%%%%%%%%%%%%%%%%%%%%%%%%%%%%%%%%%%%%%%
\section{spin Hamiltonian}
\label{sec:hamspin}

Superparamagnets are solids or clusters of nanometric size where each
molecule can be described by a spin Hamiltonian \cite{white}: 
%_________________________________
\begin{equation}
\label{superham}
\HS = -\ani S_z^2 -\Bz S_z
\; .
\end{equation}
%_________________________________
Here $\ani$ is the anisotropy parameter, originating from the
spin-orbit coupling, while $-\Bz S_z$ is the Zeeman term accounting
for the coupling to external fields.
As a first approximation the magnetic molecules can be considered as
independent of one another, so the total Hamiltonian is a sum of
contributions like~(\ref{superham}).
Its eigenvalues are $\epsilon_n = -\ani m^2 - \Bz m$. %
\footnote{
In this chapter we set $\hbar = \mu_{\rm B} = k_{\rm B} =1$; remember
the footnote~\ref{foot:units} in chapter \ref{chap:phase-space}.
} %ENDOF FOOTNOTE

For $\ani \geq 0$ the anisotropy term in~(\ref{superham}) yields a
bistable structure with minima at $m = \pm S$, while the Zeeman term
tilts this ``double well'' (figure \ref{fig:els},
chapter~\ref{chap:clasico}).
In equilibrium the spin orientations are like those of a paramagnet
(the magnetization and susceptibility are given by Brillouin and Curie
type laws), but due to the large $S$ they are called {\em
  superparamagnets}.
Physical realizations of this model are the ${\rm Mn}_{12}$ or
${\rm Fe}_8$ clusters, among others \cite{blupra2004}, which have a moderately
large spin $S \sim 10$, and where quantum effects of the spectrum's
discreteness can be studied \cite{mildri01}.

It is convenient to introduce scaled variables ($\beta = 1/T$):
%_____________________________________
\begin{equation}
\label{scaledxisigma}
\sigma
\equiv
\beta D S^{2}
\;,
\qquad
\xi
\equiv
\beta\Bz S
\;,
\qquad
\hef 
\equiv
\frac {\xi}{\left(2-\frac{1}{S}\right)\sigma}
\;.
\end{equation}
%_____________________________________
%
The parameter $\hef$ is called the reduced field.
When there is no external field obviously $\hef = 0$, whereas $\hef=1$
just when the barrier disappears.
In these variables, the thermal Hamiltonian reads 
$\beta \HS = - \sigma (m/S)^2 - \xi (m/S)$
and we will take the classical limit $S \to \infty$ keeping $\sigma$
and $\xi$ fixed (note $|m/S|\leq1$).
In this way the energies are kept constant, while introducing more and
more levels towards a continuum.
Thus the spin number $S$ plays the same role as $\kondobar$ in the
particle problems (characteristic action over $\hbar$).
In that case when $\kondobar \to \infty$ the number of bands below the
barrier was increased, but the potential height was kept fixed.

From a formal point of view, varying parameters allows the study of
problems with equispaced spectrum ($D =0$), or problems with more
featured spectra ($D \neq 0$).
In the case $D >0$, we have mentioned that the spectrum has a
double-well structure, while $\ani <0$ has a single well
(figure~\ref{fig:els}).
The level spacings are
%__________________________________
\begin{equation}
\label{deltam}
\epsilon_{m} - \epsilon_{ m\pm 1}
=
\Delta_{m m\pm 1}
=
\pm \left [
\ani (2m \pm 1) + B 
\right ]
\end{equation}
%__________________________________
so that, when $\Bz = 0$, the levels $m$ and $-m$ become degenerate.

%%%%%%%%%%%%%%%%%%%%%%%%%%%%%%%%%%%%%%%%%%%%%%%%%%%%%%%%%%%%%%%%%
\section {Hamiltonian coupling to the environment}
\label {sub:coups}

As these spins ``live'' in a crystal, they interact with the lattice
phonons, providing an example where the coupling to a bath of
oscillators is well justified.
Thus we write [cf.~(\ref{clmodelre})]:
%______________________________________
\begin{eqnarray}
\HTOT
=
\widetilde {\HS}
+
\underbrace {\sum_{\bi}
F_{\bi}({\bf S})
\otimes
c_\bi
( b_{\bi}^+
+
b_{-{\bi}} )}_{\substack{\cal H_{\rm s-ph}}}
+
\underbrace{
\sum_{\bi} \omega_{\bi} b_{\bi}^+ b_{\bi}}
_
{\substack{
\mathcal H_{\rm ph}
}
}
\end{eqnarray}
%__________________________________________
where $\widetilde {\HS}$ is the Hamiltonian accounting for
renormalizations (recall sect.~\ref{sec:dmeclmodel}).
As we see, these systems can be studied within the open-system
formalism of the previous chapters.

The coupling between the spin and the phonons includes
%_____________________________
%
% F: LINEAR
%_____________________________
\begin{equation}
\label{F:lin}
%%%%%%%%%%%%%%%%%%
F({\bf S})
=
\eta_{+}
\big [ v(S_z),\,S_+\big ]_+
+
\eta_{-}
\big [ v(S_z),\,S_-\big ]_+
%%%%%%%%%%%%%%%%%%
\;,
\end{equation}
%_____________________________
where the complex conjugate constants $\eta_- =\eta_+^*$ ensure
hermiticity, the anticommutator is represented as $[\, , \,]_+$, and
$v(S_z)$ is any function of the operator $S_z$.
Note that the coupling is realized through the ladder operators
$S_\pm = S_x \pm \iu S_y$,
which obey 
$S_\pm \vl m \rangle = \lf_{m ,m \pm 1} \vl m \pm 1 \rangle$
with the custom factors
$\lf_{m, m\pm 1}=\sqrt {S(S+1) - m(m \pm 1)}$.

In the case $v(S_z) = {\rm cte}$, the coupling 
$F \propto \eta_+ S_- + \eta_- S_+$
is linear in the system variables.
This is analogous to the coupling used in the particle chapter, and
here will serve us as control model.
On the other hand, $v(S_z) \propto S_z$ for spin-phonon
(magneto-elastic) coupling \cite{harpolvil96, garchu97}, and we have
to use the more realistic $F\sim[S_z,S_\pm]_+$.
In the former, linear coupling case we use an Ohmic distribution of
bath oscillators, $J(\omega) = \damps\,\omega$, as we did previously.
But when studying the magneto-elastic model we will use a super-Ohmic
bath with a spectral distribution $J(\omega) = \damps\,\omega^3$
appropriate for phonons (or photons) in three dimensions \cite{weiss}.

Both in the equilibrium properties and in the dynamics we will need
the explicit form for the matrix elements 
$F_{nm} \equiv \langle n \vl F \vl m \rangle$:
%_____________________________
%
% F: MATRIX ELEMENTS
%_____________________________
\begin{equation}
\label{Fnm}
%%%%%%%%%%%%%%%%%%
F_{nm}
=
L_{m,m-1}
\delta_{n,m-1}
+
L_{m+1,m}^{*}
\delta_{n,m+1}
%%%%%%%%%%%%%%%%%%
\;,
\quad
%%%%%%%%%%%%%%%%%%
L_{m,m'}
=
\eta_{+}
[v(m)+v(m')]
\lf_{m,m'}
%%%%%%%%%%%%%%%%%%
\; .
\end{equation}
%_____________________________
Note that when $v(m) \neq {\rm cte}$ (spin-phonon) the $L_{m,m'}$ add
an extra dependence on $m$ in the coupling/damping, compared with the
linear case; if you wish, the spin analogue of position-dependent
damping in particle problems \cite{risken}.

%%%%%%%%%%%%%%%%%%%%%%%%%%%%%%%%%%%%%%%%%%%%%%%%%%%%%%%%%%%%%%%%%%%
\section{equilibrium properties}

In chapter  \ref{chap:equilibrio} we discussed the equilibrium
properties of open systems.
In the previous chapter we already showed stationary Wigner
distributions for a particle in a periodic cosine potential.
There, as we worked with $\dampp/T \ll 1$, the stationary solution was
essentially the canonical distribution at zero damping
(figure~\ref{fig:kandemir}).
Now we want to study explicitly corrections to the thermodynamical
properties when $\damps \neq 0$ in spin problems.

Let us recall that to obtain the damping-dependent corrections we had
to compute certain $\Zdos$.
Then 
$\ZTOT/\ZB \cong \ZS + \Zdos$
with $\ZTOT$, $\ZB$ and $\ZS$ the partition functions of the whole
system, the bath and the closed system, respectively, Eq.~(\ref{zgral}).
From ${\cal Z}$ we can obtain the free energy, and hence all other
thermodynamical quantities, as in section \ref{sec:magter}.

For a spin-bath coupling linear in $S_{\pm}$ as in (\ref{Fnm}), we
have
$\vl F_{nm} \vl^2 = \lf_{n+1, n}^2 \delta_{m n+1}^2 + \lf_{n, n-1}^2
\delta_{m ,n-1}$.
Plugging this $\vl F_{nm} \vl^2$ in expression~(\ref{zse}) for $\Zdos$
we have
%______________________________________
 \begin{equation}
\label{Z2ani}
\Zdos =
 -\frac{\beta}{2}
\sum_{m}
\e^{\beta (\ani m^2 + \Bz m)}
\bigg [
\lf^2_{m+1, m}
\se \left (\Delta_{m+1, m} \right )
+
\lf^2_{m, m-1}
\se \left (\Delta_{m-1,m} \right)
\bigg ]
\end{equation}
%______________________________________
where we have used the energy levels 
$\epsilon_m = -\ani m^2 - \Bz m$,
while
$\Delta_{m, m\pm 1} = \pm \left [ \ani (2m \pm 1) + \Bz \right ]$
are their differences (``transition'' frequencies).

%%%%%%%%%%%%%%%%%%%%%%%%%%%%%%%%%%%%%%%%%%%%%%%%%%%%%%%%%%%%%%%%%%%%
\section{example 0: equilibrium of the isotropic spin}

\begin{figure}
\centerline{\resizebox{8.cm}{!}{%
    \includegraphics[angle = -90]{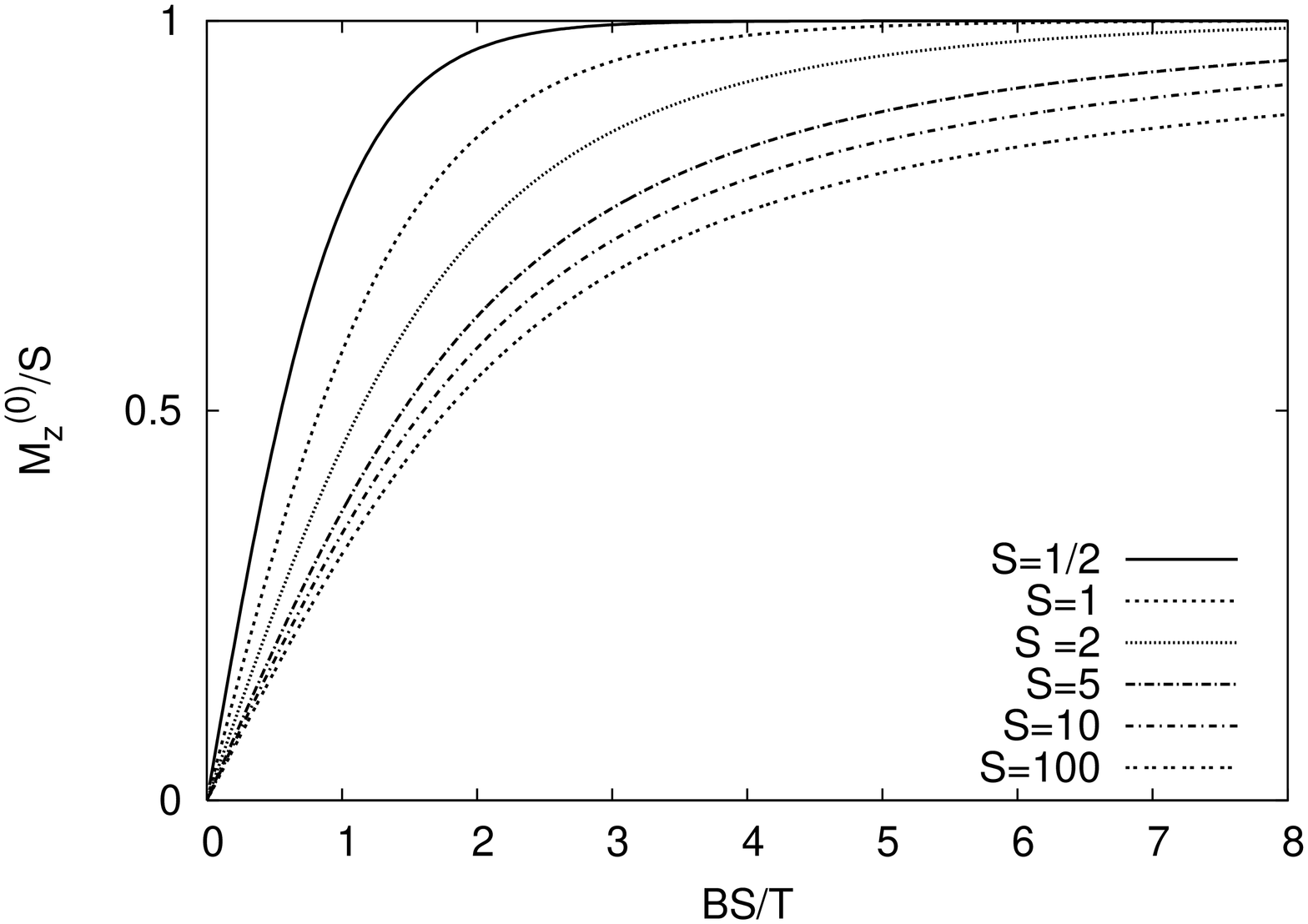} 
}
\resizebox{8.cm}{!}{%
 \includegraphics[angle = -90]{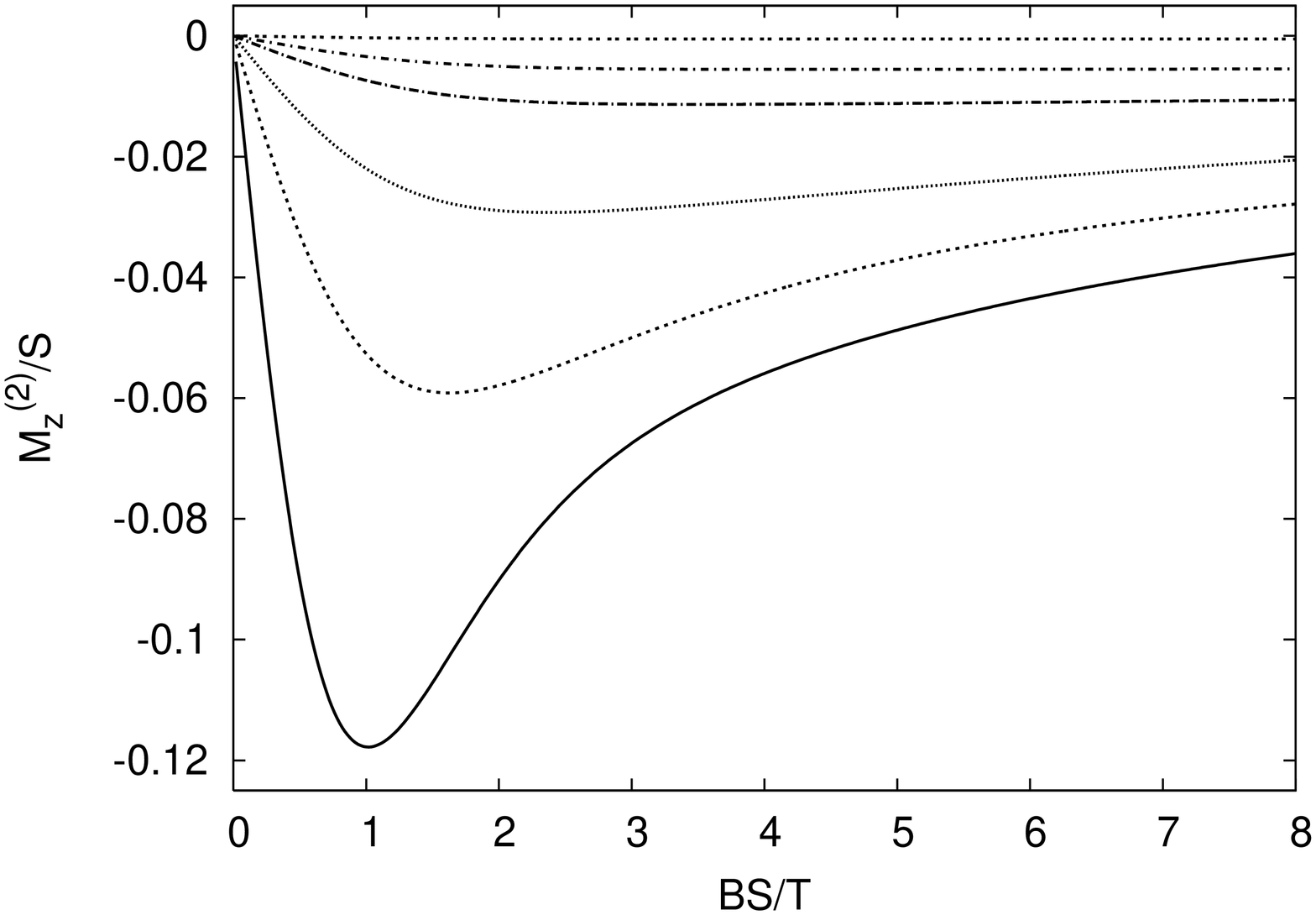} 
}
}
\centerline{\resizebox{8.cm}{!}{%
    \includegraphics[angle = -90]{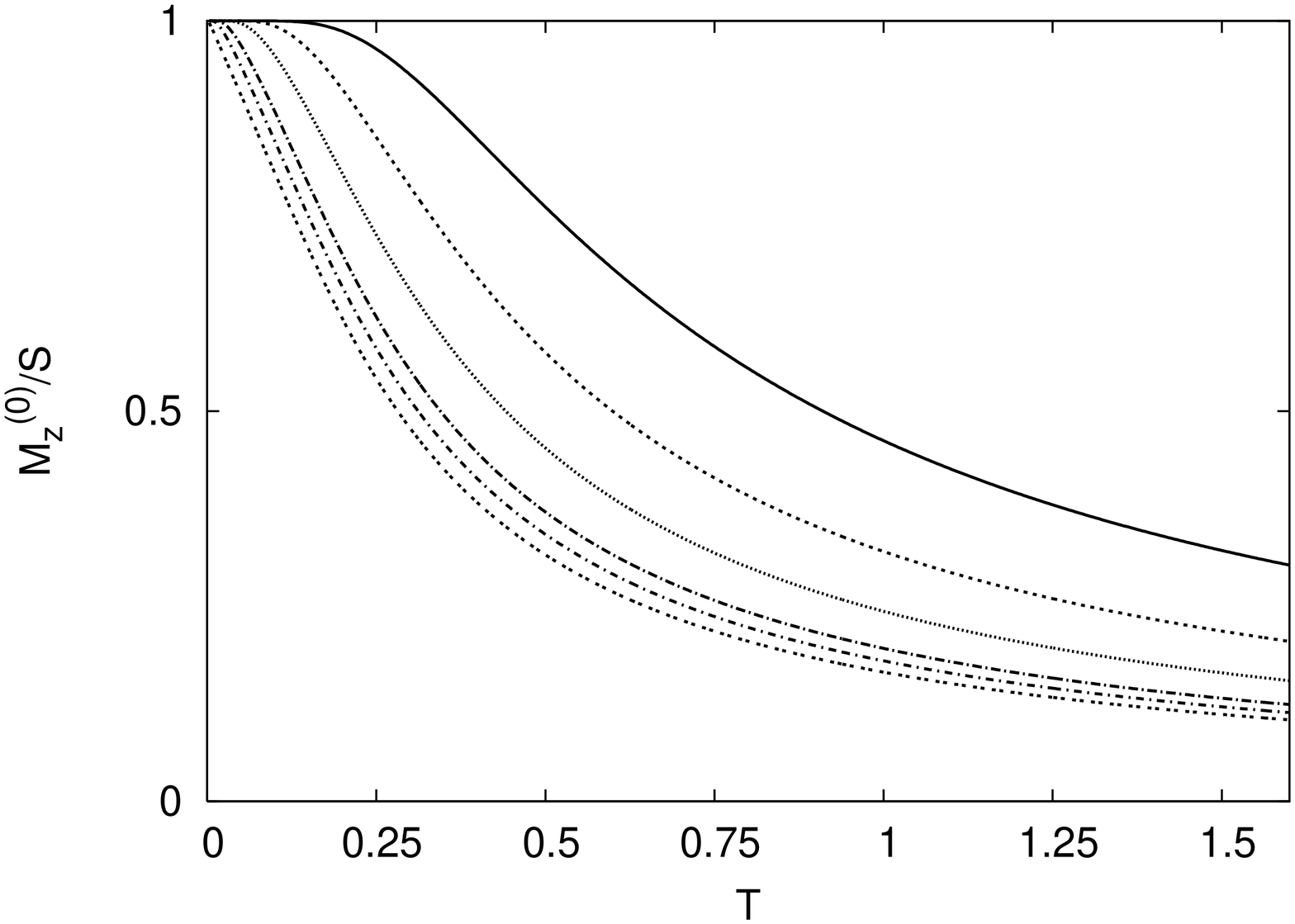}
}
\resizebox{8.cm}{!}{%
\includegraphics[angle = -90]{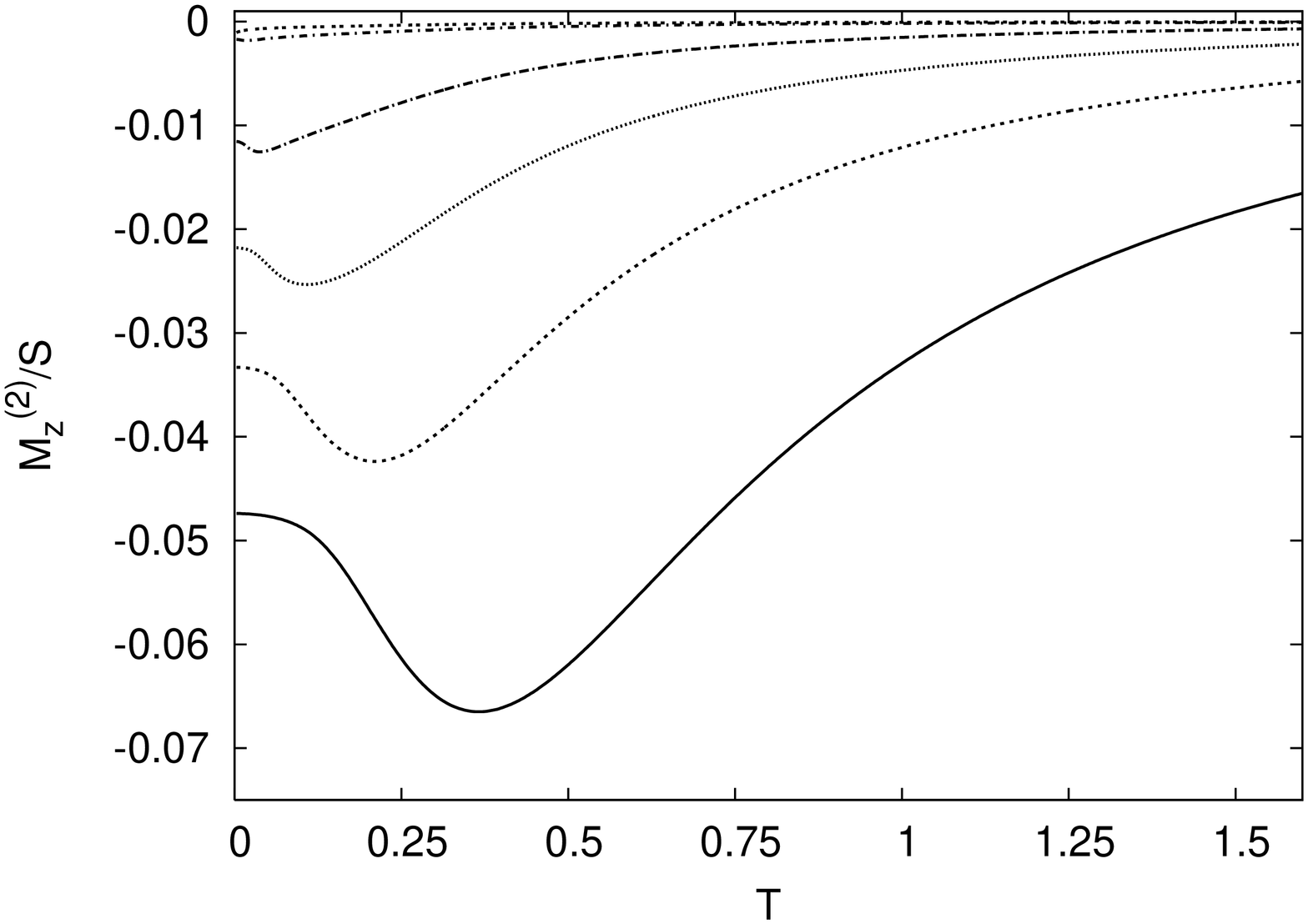}
}
}
\caption{ 
Top panels: Brillouin function (left) and its dissipative corrections
(right) for several $S$ as a function of $\xi=\Bz S/T$.
For the corrections we use $\damps S = 0.05$ (corresponding to the
 classical damping, see footnote~\ref{foot:cls} below) and $J(\omega)
 = \damps\,\omega$, in the model linear in ${\bf S}$ (section
 \ref{sub:coups}).
Bottom panels: the same quantities (left Brillouin, right dissipative
corrections), but plotted vs the temperature (complementing the top
plots vs.~$1/T$).
}
\label{fig:brillo}
\end{figure}

If we consider $\ani =0$ (isotropic spin) the spectrum becomes
equispaced and the arguments of the $\se$ in (\ref{Z2ani}) are simply
$\pm B_z$ (the constant level difference).
As these do not depend on the index $m$, they can be taken out of the
sum, as in the example of the harmonic oscillator in section
\ref{sec:HO}.
Then  $\Zdos$ can be explicitly calculated [cf. Eq.~(\ref{zdosho})]:
%__________________________________________
\begin{equation}
\frac {\Zdos}{\ZS} =
- \frac {1}
%{\etados}
{2T}
\Mcero
\left [
\se_{-} 
+
\coth \left (
\frac {B}{2T}
\right )
\se_{+}
\right ]
\; ,
\qquad
\se_{\pm} = \se (\Bz) \pm \se(-\Bz)
\end{equation}
%____________________________________________
Here $\ZS$ is the partition function of an isotropic spin
$\ZS = \sinh[ (S + 1/2) \Bz/T] /\sinh [ \Bz /2T]$
and $\Mcero$ the corresponding magnetization (at zero coupling), which
is nothing but the {\em Brillouin function}:
%___________________________________
\begin{equation}
\label{brillo}
\Mcero = \left (S + \frac{1}{2} \right )\coth 
\left[ \left ( S + \frac{1}{2} \right ) \frac{\Bz}{T} \right ] 
- 
\frac{1}{2}
 \coth \left [\frac{1}{2} \frac {\Bz}{T} \right ] 
\; .
\end{equation}
%____________________________________
%
Now, with our $\Zdos$ we can compute the magnetization including the
corrections due to the bath coupling:
%___________________________
\begin{equation}
\Mz \cong
\Mcero 
+
\underbrace{T \frac{\partial}{\partial \Bz} \frac {\Zdos}{\ZS}}_ {\substack
  {\Mdos}}
\; .
\end{equation}
%___________________________
Now we take the derivative and plot the results in
figure~\ref{fig:brillo}.
These are the corrections to Brillouin's law for several spin values,
and as a function of the field and the temperature.
On the left panels we plot the bare Brillouin curves as reference.
%
%___________________________________________________
%___________________________________________________
%___________________________________________________
%___________________________________________________
%___________________________________________________
%___________________________________________________

But let us first discuss the isolated spin limit.
At zero temperature the system is in the ground state $m = S$ and
hence the magnetization is ``saturated'' $\Mz = S$.
Increasing $T$ the levels $S-1$, $S-2$, \dots become populated,
and $\Mz$ starts to decrease.
Eventually, at very high temperatures all levels are equally
populated, and the magnetization goes to zero, say, by thermal
misalignment from the field direction ($\Mz = \sum m \dm_{mm}$, and
the $T \to\infty$ population of the levels $m$ and $-m$ is the same).

Keeping these $\damps \to 0$ results in mind, let us turn to the
corrections due to the bath.
First, we see that the finite-coupling corrections vanish in the
$S\to\infty$ classical limit, and in the high temperature limit too,
as we could have expected (section \ref{sec:cuaneq}).
Besides, the corrections are negative, and they are more noticeable at
intermediate values of $\Bz/T$.  %
\footnote{
We obtained results qualitatively similar for the anisotropic spin
with $\ani > 0$.
} %END OF FOOTNOTE

It is worth remarking that at $T \to 0$ the correction remains finite,
$\Mdos <0$, and hence $\Mz = \Mcero + \Mdos < S$, contrasting the
saturated $\Mz(\damps=0)=S$.
In other words, switching on $\damps \neq 0$ the excited levels became
populated.
What should we think about this?
Well, we have to take into account that the system proper is the total one,
spin plus bath.
Besides, $[\HS + \HB, \HI] \neq 0$.
Then the ground state is not 
$\vl S \rangle \otimes \vl {\rm vacuum} \rangle$
(with $\vl {\rm vacuum} \rangle$ the bath's ground state), but spin
and bath are entangled.
Seen in other way, the reduced density matrix is not diagonal in the
$S_z$ basis, which makes  $\dm_{mm} \neq 0$ for excited states even at
$T=0$.

We already mentioned in chapter~\ref{chap:equilibrio} that in the
systems we have in mind this type of corrections are probably small,
and we do not expect they being very relevant in a ${\rm Mn}_{12}$ or
${\rm Fe}_8$, for example.
Maybe very low temperature measurements, to check $\Mz < S$.
This would give a quantitative measure of the entanglement between
spin and bath (see Jordan and Buttikker \cite{jorbut04, butjor05} for
a similar discussion in the harmonic oscillator). 
Anyway, though the equilibrium formalism helped us to understand
and correct some misconceptions, in what follows (dynamics), we will
consider for all practical purposes that the equilibrium state is the
closed canonical one.

%%%%%%%%%%%%%%%%%%%%%%%%%%%%%%%%%%%%%%%%%%%%%%%%%%%%%%%%%%%%%
\section {dynamics}

It will not surprise the reader if to study the non-equilibrium
properties we use the quantum master equations of chapter
\ref{chap:dinamica}.
Specifically, we will use the second order master equation~(\ref{RWAmej}), in
what we called the improved rotating-wave approximation (RWA)
%________________________________
\begin{equation}
\label{RWAmej2}
\dif \dm_{nm}/\dif t
=
-\iu \Delta_{nm} \dm_{nm}
+
\R_{nn+1, m m+1} \dm_{n+1 m+1}
+
\R_{nn, m m} \dm_{n m}
+
\R_{nn-1, m m-1} \dm_{n-1 m-1}
\; .
\end{equation}
%________________________________
At the end of this chapter we will briefly touch the comparison
between crude RWA, and its improved version.

At some point we would like to add a small transverse field, e.g.,
$B_x$, to study the transverse response $\sus_x(\omega)$.
Then the system Hamiltonian $\HS$ becomes
%_________________________________
\begin{equation}
\label{superhamtran}
\HS = -\ani S_z^2 -\Bz S_z
- \case
( B_+ S_- + B_- S_+)
\end{equation}
%_________________________________
with $B_\pm  = B_x \pm \iu B_y$. 
But for arbitrary $B_\pm$ we do not know the eigenvalues and
eigenvectors of $\HS$ analytically, so we cannot calculate the
unperturbed evolution of the coupling term
$F_{nm} (\tau) = \F_{nm} \e^{-\iu \Delta_{nm} \tau}$
(which was required to calculate the relaxation term $\R$ in section
\ref{sub:conocemos}).
However, restricting ourselves to $B_{\pm} \ll 1$ we can approximate
the required evolution by the diagonal part.
Naturally, we can and will fully keep the transverse field in the
unitary evolution.
All things considered, we will have to solve
%_______________________________
\begin{eqnarray}
\label{RWAmej3}
\nonumber
\dif \dm_{nm}/\dif t
&=&
-\iu \Delta_{nm} \dm_{nm}
\\
\nonumber
&+&
\tfrac{\iu}{2} B_+
(\lf_{n n+1}
\dm_{n m+1}
+
\lf_{mm-1}
\dm_{n-1 m}
)
+
\tfrac{\iu}{2} B_-
(\lf_{n n-1}
\dm_{n m-1}
+
\lf_{mm+1}
\dm_{n-1 m}
)
\\ 
&+&
\R_{nn+1, m m+1} \dm_{n+1 m+1}
+
\R_{nn, m m} \dm_{n m}
+
\R_{nn-1, m m-1} \dm_{n-1 m-1}
\; ,
\end{eqnarray}
%________________________________
where $\Delta_{nm}$ are the energy differences for the longitudinal
part (\ref{deltam}).
For the explicit form of the $\R$'s, see $\jpaii$.

A final comment on the resolution method.
Note that the master equation~(\ref{RWAmej3}), can be formally written
as
$\dot\dm_{nm}
= 
\sum_{\alpha,\beta}Q_{nn+\alpha}^{mm+\beta}\dm_{n+\alpha,m +\beta}$.
Then, introducing the matrices
$[\mQ_{n,n+\alpha}]_{m, m+\beta} = Q_{n, n + \alpha}^{m, m + \beta}$
and the vectors 
$[\mc_n]_m = \dm_{nm}$, 
we have
%_____________________________  
\begin{eqnarray}
\label{matricials}
\dot{\mc}_{\ix}
=
\mQ%_{-1}
_{\ix,\ix-1} 
{\mc}_{\ix-1}
+
\mQ%_{0}
_{\ix,\ix} 
{\mc}_{\ix}
+
\mQ%_{+1}
_{\ix,\ix+1} 
{\mc}_{\ix+1}
\; .
\end{eqnarray}
%_____________________________
That is, we have rewritten (\ref{RWAmej3}) in a form suitable for the
use of continued fractions (chapter~\ref{chap:metodos}, $\jpaii$).
%%

%%%%%%%%%%%%%%%%%%%%%%%%%%%%%%%%%%%%%%%%%%%%%%%%%%%%%%%%%%%%%%%%%%%%%%
\section[example I: longitudinal linear susceptibility] 
{example\! I: longitudinal relaxation \& linear susceptibility}

Our first example of dynamics is a study of the relaxation mechanisms
by means of the linear susceptibility.
We will consider that no transverse fields are applied ($B_\pm =0$),
and analyze the time evolution of $\Mz= \langle S_z \rangle$:
%_____________________-
\begin{equation}
M_z(t) 
= 
{\rm Tr} ( S_z \dm) 
= 
\sum_{m = -S}^S m N_m 
\; ; 
\qquad
\qquad
N_m \equiv \dm_{mm}
\; .
\end{equation}
%_______________________
In this problem we just need the diagonal density-matrix elements to
construct the response $\Mz$.
Besides, for $B_\pm =0$ the diagonal and off-diagonal elements
in~(\ref{RWAmej3}) get decoupled.
Then it suffices to solve the system of $2S+1$ coupled differential
equations for the populations $N_m$ ({\em balance equations}):
%___________________________
%\vspace{-1.ex}
\begin{eqnarray}
\label{balance}
\dot N_{m}
=
\big (\Pmmplus N_{m+1}
-
\Pmplusm N_{m}
\big )
+
\big (
\Pmmminus N_{m-1}
-
\Pmminusm N_{m}
\big )
\end{eqnarray}
%_________________________________
where the transition probability is $P_{m \vl m'} = \R_{m m' m m'}$. %
\footnote{
As we work here with the diagonal elements only, while $\dm_{mm}$ is
the probability of having the spin in the state $m$, we can take the
classical limit directly.
On so doing, we get the longitudinal \FP\ equation~(\ref{clfptheta}).
In the general case (with non-diagonal terms), we have to resort to
phase-space methods, as in chapter~\ref{chap:phase-space}.
Taking the classical limit allows us to relate both theories,
classical and quantal.
In particular, for Ohmic baths, we get the relation
$\damps S = \lambda_{\rm LL}$
where $\damps$ is the damping used here, whereas $\lambda_{\rm LL}$ is
the phenomenological damping in the Landau--Lifshitz classical equation
(see $\prb$).
\label{foot:cls}
}

We will calculate the {\em linear\/} response of $\Mz$ to an
ac perturbation $\delta \Bz \e^{-\iu \omega t}$.
To this end, recall that $\susz (\omega)$ could be constructed as in
Eq.~(\ref{eigensus}), using the eigenvalues and eigenvectors of the
balance equation without perturbation; and that equation~(\ref{fdt})
relates the response in the time and frequency domains.

We can compute $\susz (\omega)$ directly with continued fractions too,
as in section \ref{sub:LRTFC}.
However, the eigenroute not only gives $\susz (\omega)$, but allows
to analyze and characterize its contributions (while continued
fractions are a kind of a black box).

%%%%%%%%%%%%%%%%%%%%%%%%%%%%%%%%%%%%%%%%%%%%%%%%%%%%%%%%%%%%%%%%%%%%%
\subsection{eigenvalues and eigenvectors of the relaxation matrix}

\begin{figure}
\centerline{\resizebox{10.cm}{!}{%
    \includegraphics[angle = -90]{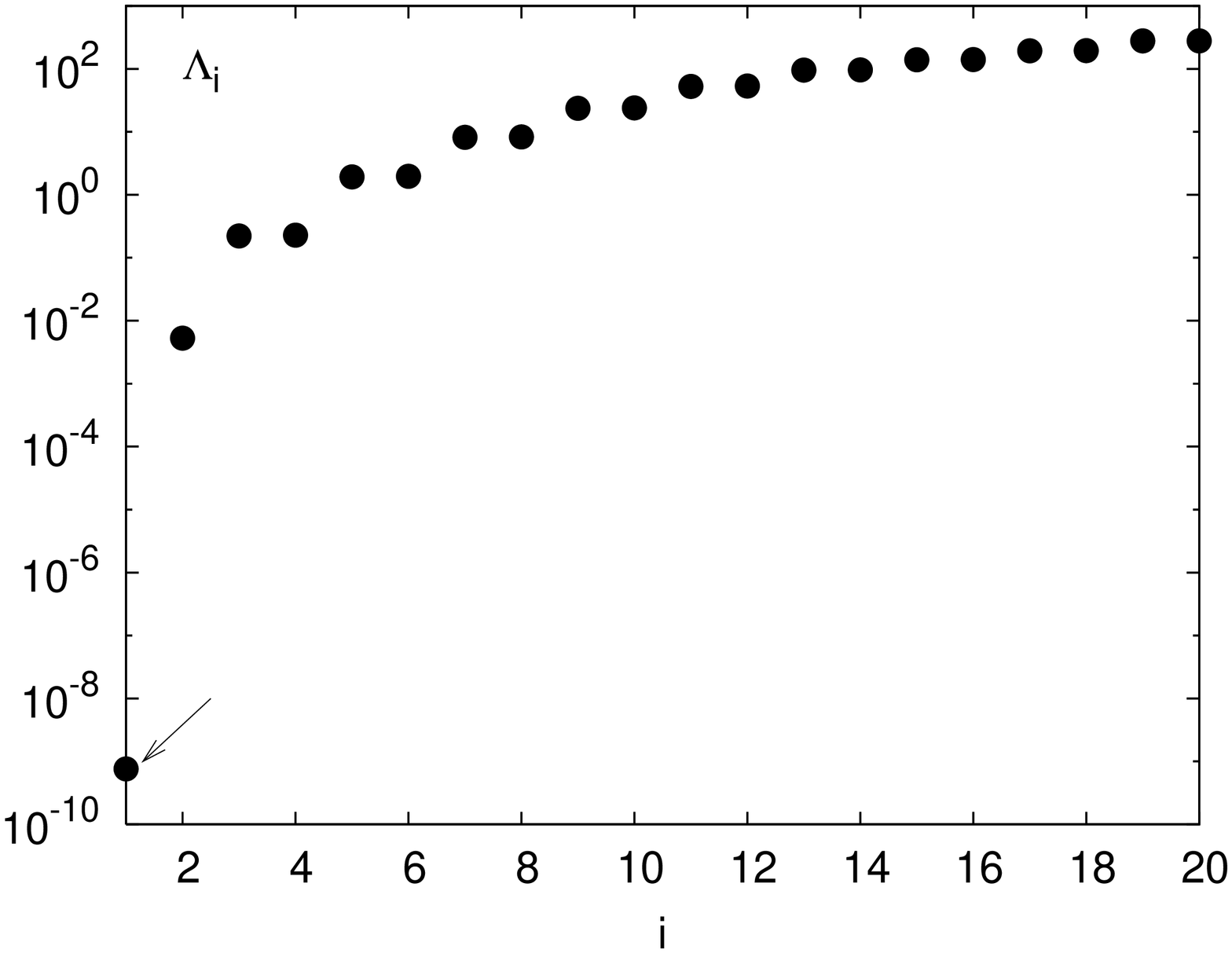} 
}
}
\centerline{\resizebox{8.cm}{!}{%
    \includegraphics[angle = -90]{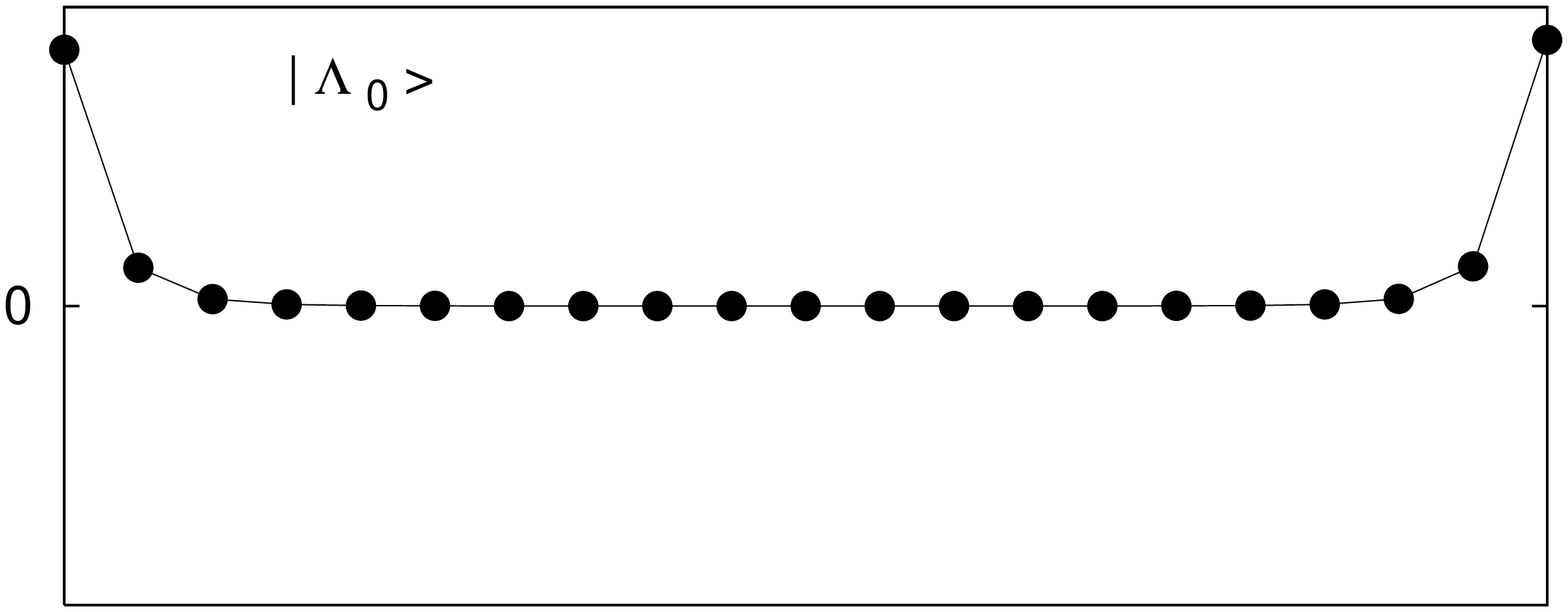} 
}
}
\centerline{\resizebox{8.cm}{!}{%
    \includegraphics[angle = -90]{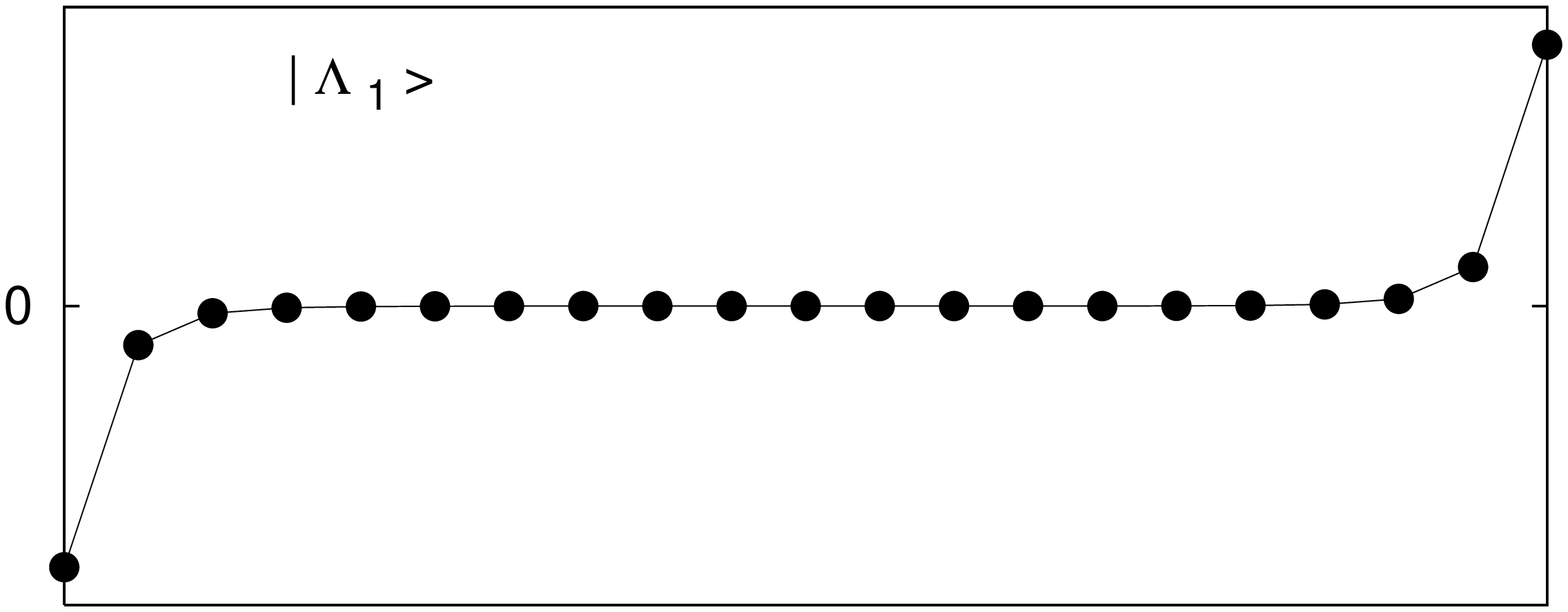} 
}
}
\centerline{\resizebox{8.cm}{!}{%
    \includegraphics[angle = -90]{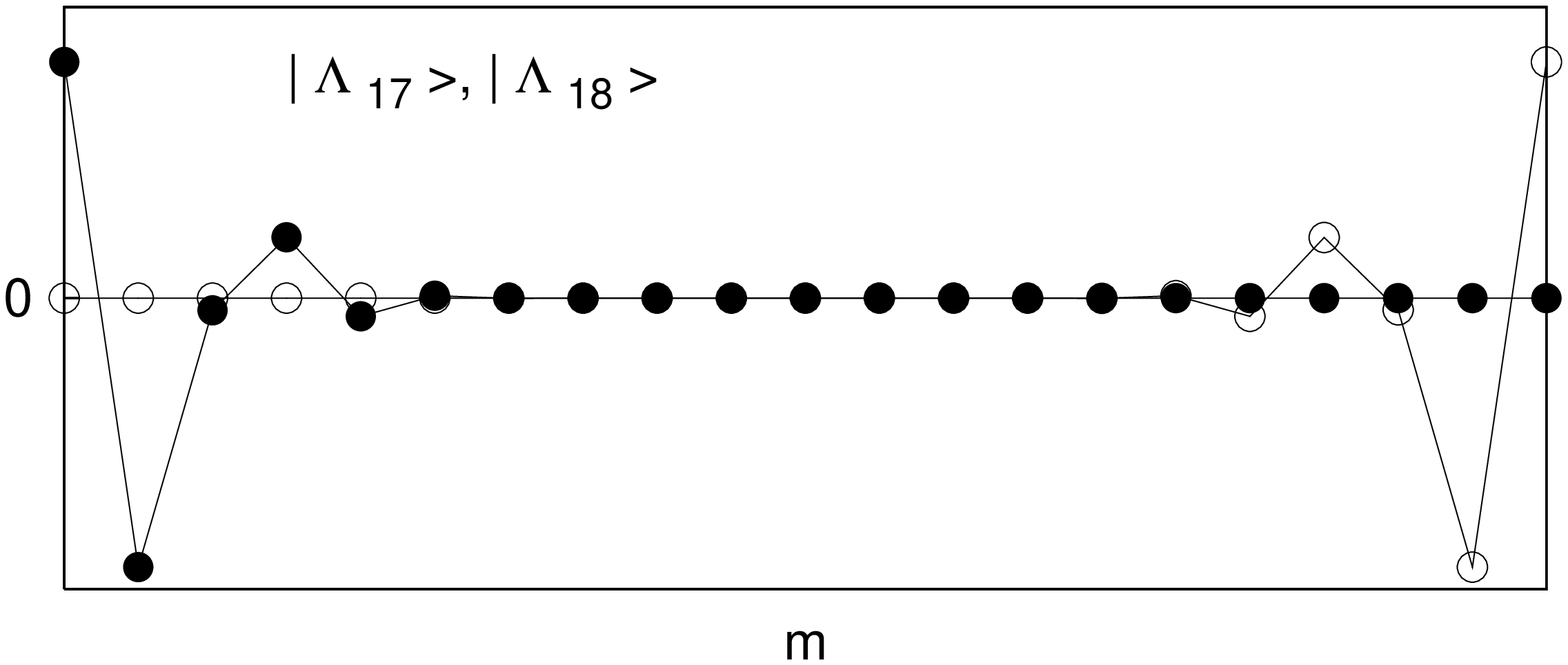} 
}
}
\caption{
Eigenvalues and eigenvectors of the matrix associated to the
system~(\ref{balance}).
The parameters are  $\sigma = 15$, $\Bz =0$ and $S =10$ .
The coupling is magneto-elastic and $J(\omega) = \damps\,\omega^3$,
with $\damps = 10^{-9}$.
Top panel: non-zero eigenvalues.
Lower panels: a few representative eigenvectors:
$\vl \Lambda_0 \rangle$ (stationary solution); 
$\vl \Lambda_1 \rangle$, determining the overbarrier dynamics, and two
fast modes
$\vl \Lambda_{17} \rangle$ $\&$ $\vl\Lambda_{18}\rangle$, associated
to the dynamics in the right and left wells, respectively.
}
\label{fig:autos}
\end{figure}

Let us first write the balance equation~(\ref{balance}) in a matrix form, with
the vector of populations $[{\bf N}]_{m}=N_{m}$
%_______________________________
\begin{equation}
\label{balmat} 
\dif {\bf N} \big/ \dif t = - \R\, {\bf N}
\; .
\end{equation}
%_______________________________
The eigenvalues $\Lambda_i$ and eigenvectors $\vl \Lambda_i \rangle$
of the $(2S+1)\times(2S+1)$ matrix $\R$ can be obtained by numerical
diagonalization.
We have plotted them in figure~\ref{fig:autos}.
One of the $2S + 1$ eigenvalues is zero, corresponding to the
stationary solution.
Then we can write the solution of (\ref{balmat}) as
${\bf N}=\sum_{i=1}^{2S}
c_{i}
\,
\e^{-\Lambda_{i} t}
\,
\vl\Lambda_{i}\rangle 
+
\vl\Lambda_{0}\rangle$
with $\vl\Lambda_{0}\rangle_m \propto \exp(-\beta \epsilon_m)$, that
is, the canonical distribution (second panel of
figure~\ref{fig:autos}).
%%
%___________________________________________________
%___________________________________________________
%___________________________________________________
%___________________________________________________
%___________________________________________________
%___________________________________________________

The remainder eigenvalues are positive (ensuring convergence to the
stationary solution as $t \to \infty$) and real (the density matrix is
Hermitian).
Figure~\ref{fig:autos} also shows that the eigenvalues are organized
in two ``groups''.
Far away from the others we find $\Lambda_1$, which generates the slow
long-time dynamics ($\tau_1 = \Lambda_1^{-1}$).
Indeed the structure of the eigenmode $\vl \Lambda_1 \rangle$
(figure~\ref{fig:autos}, 3rd panel), with population decrease in one
side and increase in the other, corresponds to transfer from one well
to the other.
As tunnel is not present (no transverse field), this population
transfer should be channelled over the barrier top, and hence we call
it the {\it overbarrier} process.
On the other hand, the rest of eigenvalues give population
readjustments inside each well (fast dynamics); we call these the {\it
intrawell} processes.
The former, $\Lambda_1$, depends exponentially on the barrier height,
whereas the intrawell modes depend only polynomially.
Whence the large separation of the scales of these two processes seen
in figure~\ref{fig:autos} (note the logarithmic scale).

\begin{figure}
\centerline{\resizebox{9.cm}{!}{%
    \includegraphics[angle = -90]{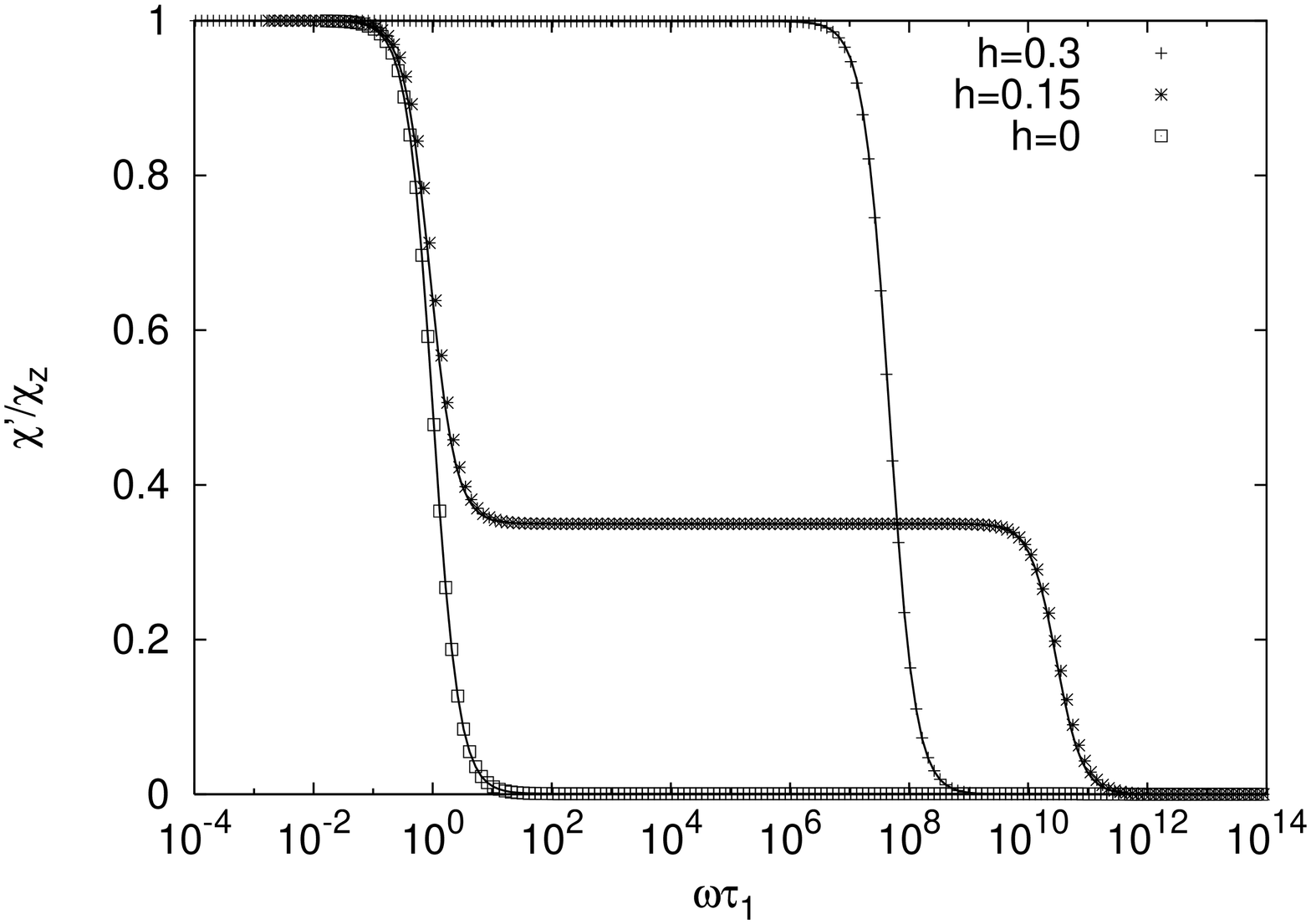} 
}
}
\centerline{\resizebox{9.cm}{!}{%
    \includegraphics[angle = -90]{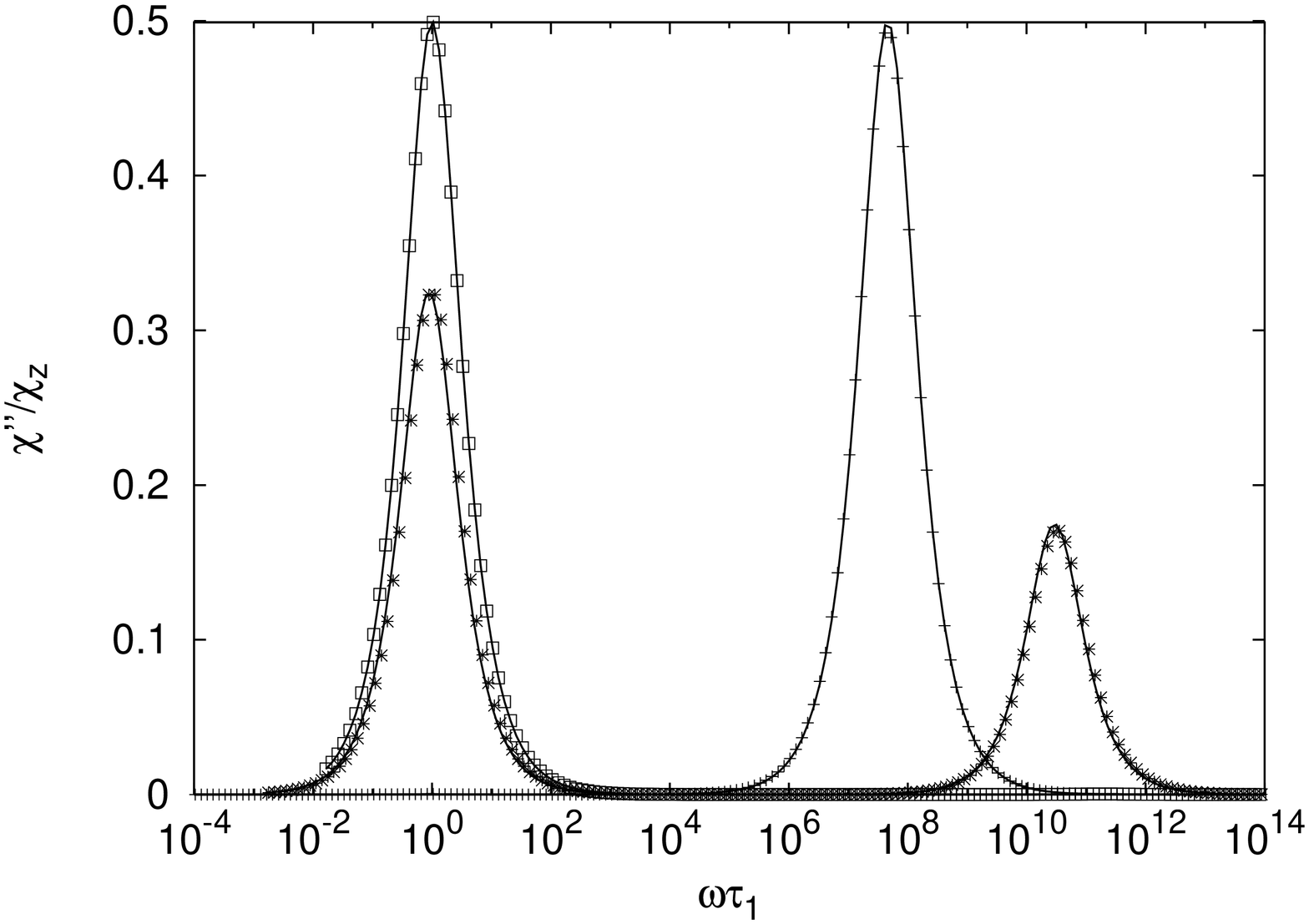}
}
}
\caption{
Real and imaginary parts of $\sus_z (\omega)$ for a spin $S=10$ in
various fields $\hef$.
The parameters are $\sigma = 15$, $T=0.1$ and $\damps = 10^{-9}$
(magneto-elastic model with $J(\omega) = \damps\,\omega^3$).
Note that the frequency is scaled by $\tauuno$.
The symbols are the numerically exact results and the lines the
bimodal approximation (\ref{aprbim}).
}
\label{fig:suslin}
\end{figure}

Well, having the eigenvalues and eigenvectors,%
\footnote{
To avoid possible misunderstandings, let us make clear that these
eigenvectors and not vectors of the Hilbert space of the problem, but
they are rather $n$-tuples containing the diagonal elements of $\dm$.
} %ENDOF FOOTNOTE
we can construct now the dynamical susceptibility.
The eigenvalues, being real, would produce a susceptibility in the
form of a sum of {\it Debye\/} profiles (section \ref{sec:LRTII})
%__________________________
\begin{equation}
\sus_z
(\omega) \sim  \sum_i  \frac{1}{1 + \iu \omega/\Lambda_i}
\; .
\nonumber
\end{equation}
%____________________________ 
%
We have plotted a few susceptibility curves in
figure~\ref{fig:suslin}.
%___________________________________________________
%___________________________________________________
%___________________________________________________
%___________________________________________________
%___________________________________________________
%___________________________________________________
The response is given, at most, by two main contributions.
We associate the low frequency peak with the slow overbarrier process.
While the peak at high frequency, when present, is to be attributed to
the fast intrawell dynamics.

%%%%%%%%%%%%%%%%%%%%%%%%%%%%%%%%%%%%%%%%%%%%%%%%%%%%%%%%%%%%%%%%%
\subsection{bimodal approximation (analytical)}

To analyze the susceptibility curves, given the mode's structure
above, we tried a relaxation profile consisting of two effective modes
(as Kalmykov {\it et al.} did in the classical case
\cite{kalcoftit2003})
%____________________________
\begin{eqnarray}
\label{aprbim}
\susz (\omega)
\cong
\suseq
\left [
\frac{a_1}{1+\iu\,\omega\tauuno}
+
\frac {1-a_1}{1+\iu\,\omega\tauW}
\right ]
\; .
\end{eqnarray}
%___________________________________
Here $\suseq$ is the equilibrium susceptibility and $\tauuno =
\Lambda_1^{-1}$, which follows an Arrhenius-Kramers law:
%_______________________________
\begin{equation}
\label{tauuno}
\tauuno
\propto%\cong\tau 
\; \e^{\beta \Delta U}
\end{equation}
%______________________________ 
This exponential dependence on the barrier height $\Delta U$ is
typical of problems of escape out of a metastable minimum
\cite{hantalbor90, mel91}.
The specific calculation for our problem is given in $\prb$.

The susceptibility above has two more parameters, $a_1$ and $\tauW$,
which can be determined following the idea proposed by Kalmykov {\it
et al.} in the classical problem \cite{kalcoftit2003}.
They argued that we have two parameters to determine, while
$\susz(\omega)$ can be written in closed form in both the limits of
low and high frequencies; in terms of $\tint$ for $\omega\to0$
[$\tint$ is the area below the relaxation curve, Eq.~(\ref{susloww})]
and in terms of $\tef$ at high $\omega$ [the initial slope of the
decay~(\ref{sushighw})].
The advantage is that both  $\tint$ and $\tef$ can be computed
analytically.
Then, these {\em two\/} limits would provide the {\em two\/} unknown
parameters $a_1$ and $\tauW$, so yielding a closed formula without
fitting parameters ($\prb$).
The results of the analytical ansatz are the lines in
figure~\ref{fig:suslin}, which account quite well for the symbols,
obtained constructing $\susz(\omega)$ from the numerical eigenstuff.

Let us comment on the curves.
The relative importance of the two effective modes is governed by the
parameter $a_1$: when $a_1 =1$ there is only overbarrier response,
while switching to $a_1=0$, the whole weight is shifted to the
intrawell response.
%%
%___________________________________________________
%___________________________________________________
%___________________________________________________
\begin{figure}
\centerline{\resizebox{8.cm}{!}{%
\includegraphics[angle = -90]{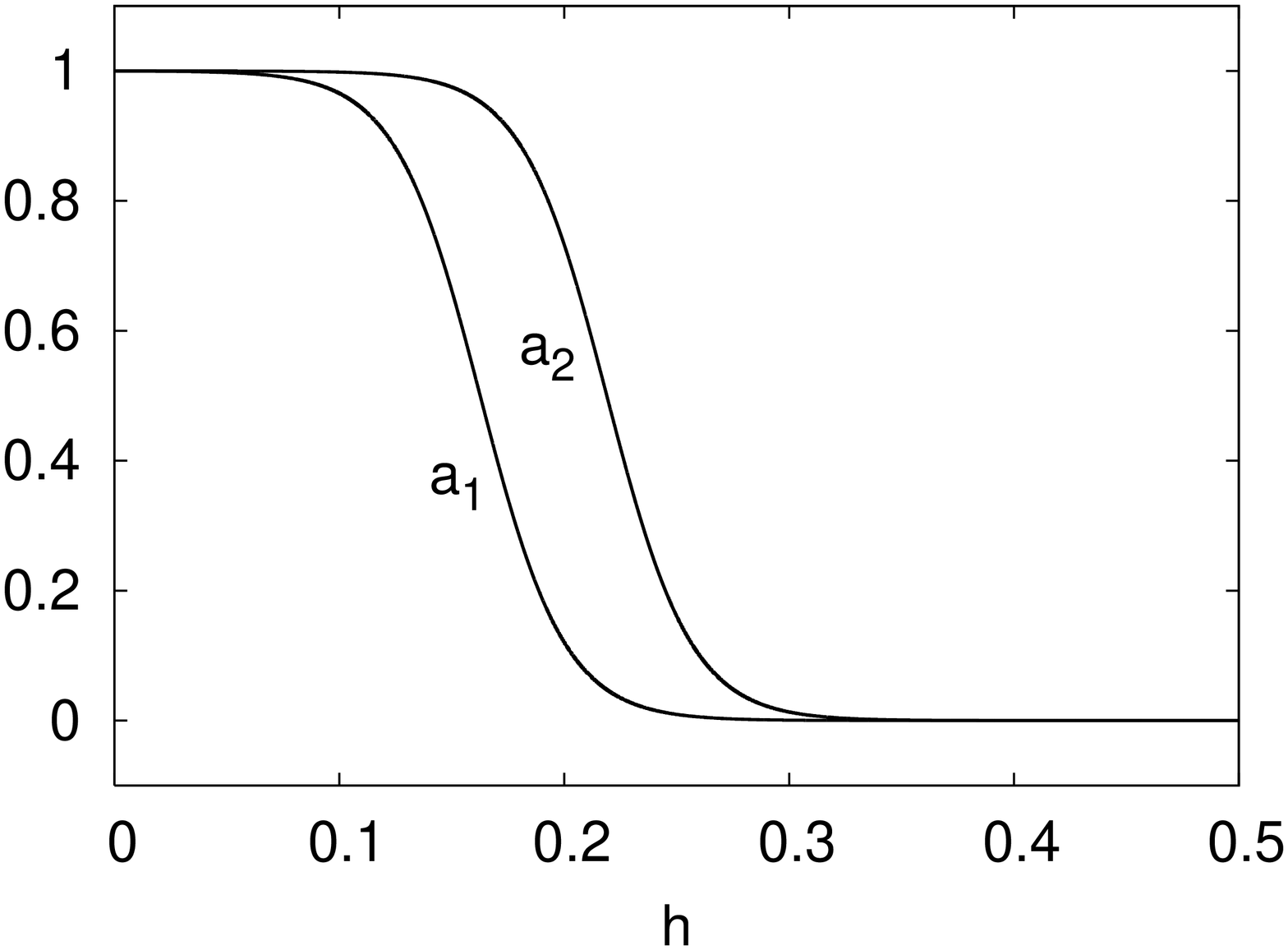}
}}
\caption[]{
Coefficients  $a_k$  ($k=1,2$) plotted vs.~$h$.
The $a_k$ are approximately independent of the spin-bath coupling
considered (they depend on the parameters of the spin Hamiltonian and
$T$).
Here we used $S=10$ at $\sigma =14$.
Notice that while the barrier disappears for $\hef=1$
(Sec.~\ref{sec:hamspin}) we have $a_k \cong 0$ quite before that
value.
}
\label{fig:aks}
\end{figure}
%___________________________________________________
%___________________________________________________
%___________________________________________________
Figure~\ref{fig:aks} shows how this is realized by changing the field
$\Bz$, with $a_1=1$ at zero field and $a_1$ decreasing towards zero at
large fields.
Indeed, at large $h$ there is no barrier, so the overbarrier
contribution is naturally expected to vanish.
However, figure~\ref{fig:aks} shows that $a_1 \cong 0$ much before the
barrier has disappeared.

Indeed, figure~\ref{fig:suslin} shows that at $\hef =0$ there was
only one peak, associated to the overbarrier mode.
Physically, at zero field the fast dynamics of the two wells is
equivalent (see figure~\ref{fig:autos}, lowest panel), and it averages
to zero when looking at $\langle S_z \rangle$.
At intermediate fields $\hef = 0.15$ the response clearly exhibits two
modes.
Increasing $\hef$ further only the intrawell peak is left (though
there is still barrier at $\hef=0.3$, the upper well is thermally
depopulated).

We close with the isotropic spin.
If we set $\ani =0$ , we always have a single peak (see
\cite{gar91llb} and $\jpaii$).
In a sense, this is clear now for us; in the isotropic spin there is
no barrier imparting a large time scale separation, and this kind of
relaxation measurement does not resolve the differences between the
fast modes.

%%%%%%%%%%%%%%%%%%%%%%%%%%%%%%%%%%%%%%%%%%%%%%%%%%%%%%%%%%%
\section {example II: non-linear susceptibility}

Let us illustrate, with the example of the longitudinal relaxation,
how to proceed if we want to go beyond the linear response regime.
In this and the next section we will see how going to higher orders in
the response, one can obtain more information on the system
\cite{lopetal05, gargar04}.
In the nonlinear case, we do not have anymore an equation
like~(\ref{fdt}) relating the behaviour in the time and frequency
domains.
Here we are going to study directly the response as a function of the
frequency of the driving field.

To get the dynamical susceptibilities, we excite our system with
$\delta B_z \e^{\iu \omega t}$ and proceed as we did in section
\ref{sub:LRTFC}.
That is, we start from the master equation written formally as:
%______________________
\begin{equation}
\partial_t \dm 
=
{\cal L} \dm
\end{equation}
%______________________
and we Fourier expand the evolution operator
$\Ldif
=
\Ldif_0
+
\pert \cdot \e^{\iu \omega t} \Ldif_1
+
\pert^2 \cdot \e^{\iu 2\omega t} \Ldif_2
+ \cdots$
as well as the solution 
$\dm
=
\dm_0
+
\pert \cdot \e^{\iu \omega t} \dm_1
+
\pert^2 \cdot \e^{\iu 2\omega t} \dm_2
+ \dots$
Here we have explicitly extracted the temporal dependencies $\e^{k\iu
\omega t}$ from $\dm$ and $\Ldif$.
Therefore, $\dm_k$ and $\Ldif_k$  ($k=1,2,\dots$) are time independent
(of course, $\dm_0$ and $\Ldif_0$ too).
Now, equating order by order we have
%_________________________
\begin{eqnarray}
\label{chain1}
0
&=&
\Ldif_0 \dm_0
\\  
\label{chain2}
\iu \omega \dm_1
&=&
\Ldif_0 \dm_1 + \Ldif_1 \dm_0
\\
\label{chain3}
2 \iu \omega \dm_2
&=&
\Ldif_0 \dm_2 + \Ldif_1 \dm_1 + \Ldif_2 \dm_0
\\[-1.ex]
\nonumber
& \vdots&
\end{eqnarray}
%_________________________
(we have written to second order only, but the perturbative structure
is clear).
Once again, this chain of equations can be solved by continued
fractions, as in \ref{sub:LRTFC}.
First we solve for $\dm_0$; this is then inserted in the second
equation, and we solve for $\dm_1$; knowing $\dm_0$ and $\dm_1$ we can
solve for $\dm_2$ in the third one.

In some cases this set can be solved analytically.
For example, for spin $\case$ we had the equation of motion for the
magnetization given by Bloch equations (section \ref{sec:bloch})
%__________________________
\begin{equation}
\dot \Mz = -\Gamma_r \Mz + \Mz(\infty)
\end{equation}
%__________________________
Then $\Ldif = \Gamma_r$ (a number) and operating like in the
chain~(\ref{chain1})-(\ref{chain3}) we find simple expressions.
To first order we recover the Debye~(\ref{debye}), while for the first
nonlinear susceptibility we have
%_____________________________________
\begin{eqnarray}
\label{susdosone}
\susdos =
\suseqdos 
\frac {1}
{1+\iu\,2 \omega \tauuno}
-
\susequno \frac {\iu \, \omega \tauuno '}
{(1+\iu\,\omega \tauuno)(1+\iu\,2 \omega\tauuno)}
\; ;
\qquad
\tauuno' \equiv \frac{\partial \tauuno}{\partial B_z}
\;.
\end{eqnarray}
%_____________________________________
Here we have written $\tauuno$ for the relaxation time, while 
$\sus^{\rm eq}_{k} = \partial^k \Mz /\partial \Bz^k$, $k=1,2$
are the equilibrium linear and nonlinear susceptibilities.
As we see, the nonlinear response gives information, not only of
$\tauuno$, but of its $\Bz$-derivative as well, $\tauuno'$.
As mentioned above $\susdos$ seems more sensitive to some parameters
of the problem.

Well, this was for $S=\case$, but we want to get the response for any $S$.
In general, the response will include relaxation ``blocks'' of the
type~(\ref{susdosone}), and maybe cross-terms.
As we did for the linear susceptibility, we will try to find an
approximate analytical modelization.
%___________________________________________________
%___________________________________________________
%___________________________________________________
\begin{figure}[h!]
\centerline{\resizebox{9.cm}{!}{%
\includegraphics[angle = -90]{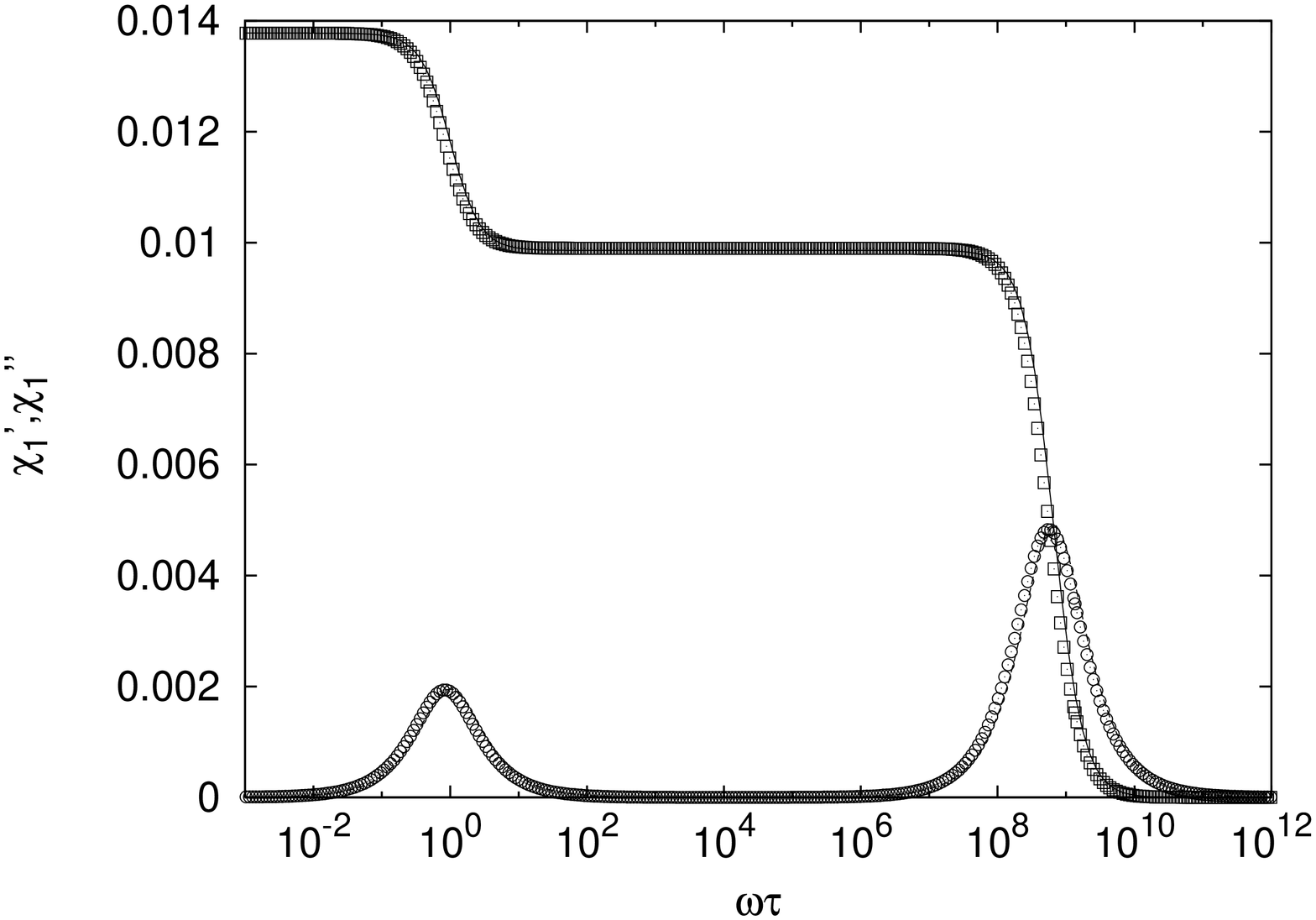}
}}
\centerline{\resizebox{9.cm}{!}{%
\includegraphics[angle = -90]{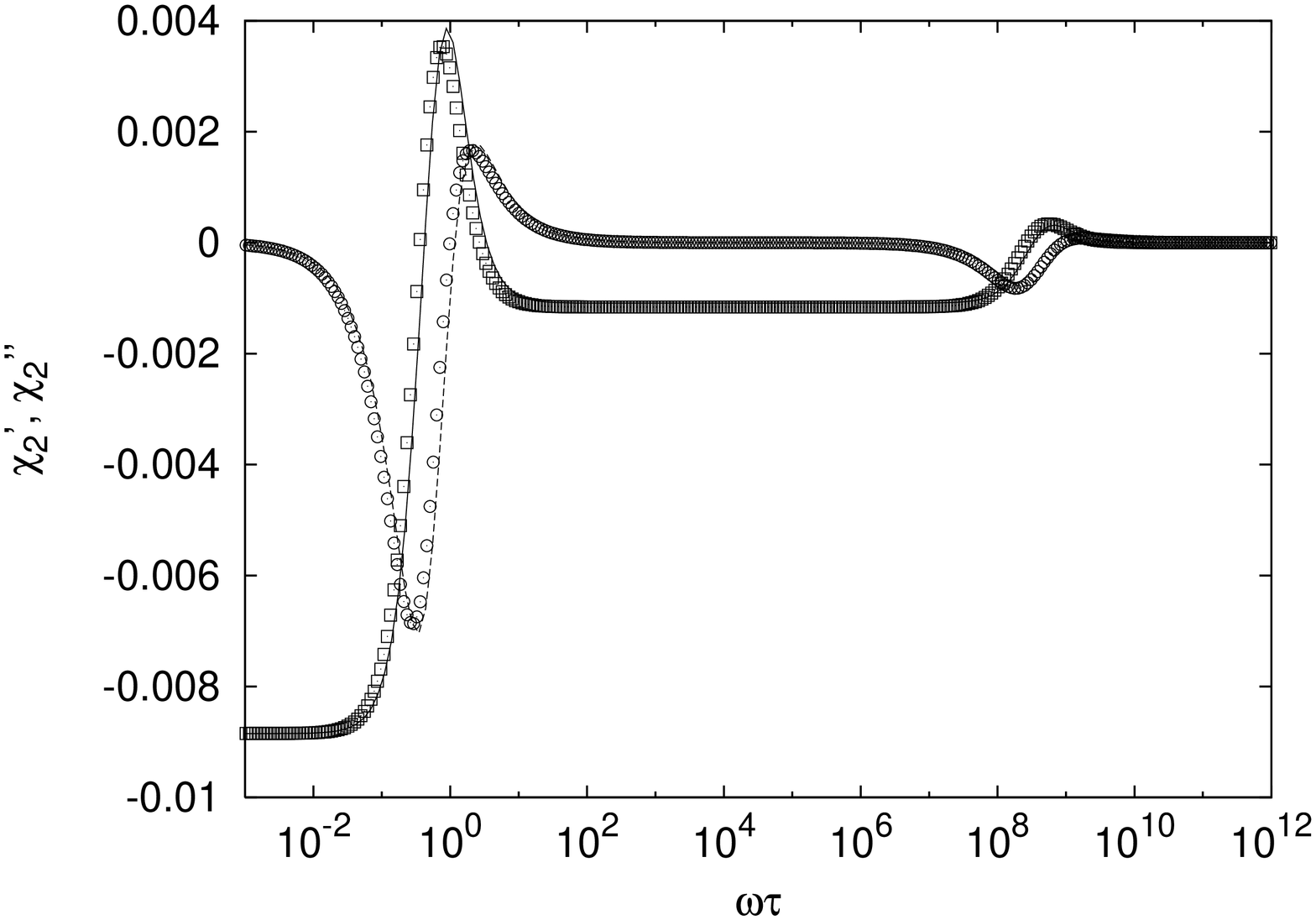}
}}
\caption{
Real part (squares) and imaginary part (circles) of the linear (top)
and nonlinear susceptibility (bottom).
The symbols are the numerically exact results and the lines the
bimodal approximations~(\ref{aprbim}) and~(\ref{susdosone}).
We have set $S=10$, and the parameters $\sigma = 14$ and $\xi = 5$ in
reduced units~(\ref{scaledxisigma}).
The coupling is magneto-elastic with $\F \sim S_z S_{\pm}$.
}
\label{fig:susdos}
\end{figure}
%_____________________________________________________
%_____________________________________________________
%_____________________________________________________
%

%%%%%%%%%%%%%%%%%%%%%%%%%%%%%%%%%%%%%%%%%%%%%%%%%%%%%%%%%%%%%%%%%%%%
\subsection {generalization of the bimodal approximation to nonlinear response}

The reader could rightly guess that we are going to try a similar
reduction of the $2S$ modes to $2$ effective ones (intrawell and
overbarrier) to account for the nonlinear response.
Thus, we construct an effective model with dynamics governed by the
two relevant relaxation times of our problem: $\tauuno$ and $\tauW$.
Although now we need to use $2 \times 2$ matrices in
(\ref{chain1})-(\ref{chain3}), one readily gets:
%_____________________
\begin{eqnarray}
\label{susdostwo}
\nonumber
\susdos &\cong&
 \suseqdos  
\frac { a_2}
{1+\iu\,2 \omega \tauuno}
-
 \susequno  
\frac { a_1  \iu \, \omega \tauuno '}
{(1+\iu\,\omega \tauuno)(1+\iu\,2 \omega\tauuno)}
\\ 
&+&
\suseqdos  
\frac { (1-a_2) }
{1+\iu\,2 \omega \tauW}
-
 \susequno 
\frac { (1-a_1)  \iu \, \omega \tauW '}
{(1+\iu\,\omega \tauW)(1+\iu\,2 \omega\tauW)}
\end{eqnarray}
%______________________________________________-
where
%______________________________________________
\begin{eqnarray}
%\tau_1' \equiv \frac{\partial \tau_1}{\partial B_z}
\tau_1' \equiv \partial \tau_1/\partial B_z
\; \qquad
%\tauW' \equiv \frac{\partial \tauW}{\partial B_z}
\tauW' \equiv \partial \tauW/\partial B_z
\end{eqnarray}
%___________________________________________________
This is a generalized bimodal formula for the nonlinear response,
without any free parameter, as before.
The new coefficient, $a_2$, tells when the intrawell modes enter the
scene (decreasing from 1), and was already plotted in
figure~\ref{fig:aks}.
We see that for a given field $a_2 > a_1$, so that the fast modes
cannot be detected at lower fields using $\susdos$.
Had we obtained $a_2 < a_1$, we would have used $\susdos$ to detect
the intrawell modes without needing as large fields as when using
$\sus_1$.
%
%Por otro lado, al ser $a_2$ siempre mayor que $a_1$ la $\susdos$ nos sera util
%para caracterizar el modo {\it overbarrier} a campos intermedios donde $a_1$
%sea ya casi cero and $a_2$ sea todavia apreciable.
%%

In figure~\ref{fig:susdos} we compare the new formulas with the
numerical results obtained by solving the
chain~(\ref{chain1})-(\ref{chain3}) with continued fractions.
We see how the bimodal ansatz works very well in the nonlinear case too.
We also see how the intrawell modes as less noticeable in the
nonlinear susceptibility $\susdos$, as we anticipated from $a_2(\Bz)$.
\enlargethispage*{0.75cm}

%%%%%%%%%%%%%%%%%%%%%%%%%%%%%%%%%%%%%%%%%%%%%%%%%%%%%%%%%%%%%%%%%%%%%%
\section {example III: $\tauuno$ and tunneling -- experiments in ${\rm Mn}_{12}$}

In this section we will use the nonlinear formalism for the analysis
of experiments in the superparamagnetic molecular cluster ${\rm
Mn}_{12}$.
In particular, we will see how to get more information on the relaxation
time  $\tauuno$ from the $\susdos$ vs.~$\omega$ spectra.

Well, in {\em actual\/} superparamagnets the Hamiltonian is a bit more
complicated than (\ref{superham}) \cite{mildri01}:
%___________________________________
\begin{eqnarray}
{\mathcal H}_{\rm S}  = -DS_z^2 -B_z S_z  
 - E (S_x^2 -  S_y^2) - B_xS_x + \cdots
\end{eqnarray}
%_______________________________
The term $- E (S_x^2 -  S_y^2)$ is a consequence of the distortion of
the anisotropy axis, while $B_x$ accounts for possible transverse
fields.
But when these terms are small they can be treated perturbatively.

\begin{figure}
\centerline{\resizebox{8.cm}{!}{%
\includegraphics[angle = -0]{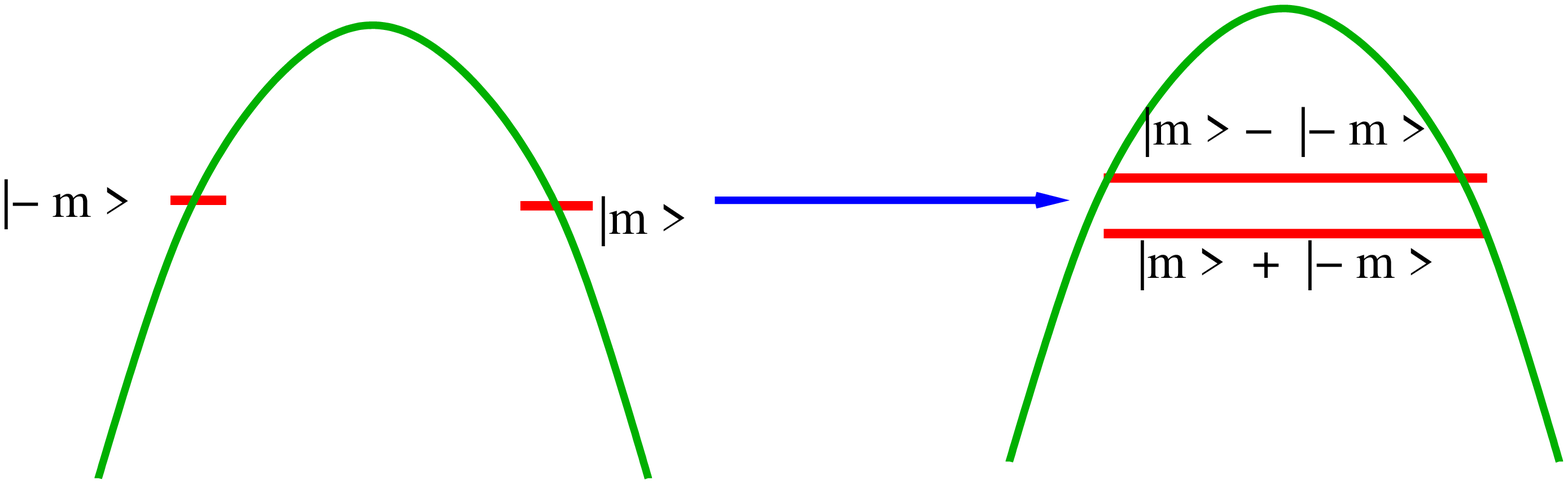}
}}
\centerline{\resizebox{8.cm}{!}{%
\includegraphics[angle = -0]{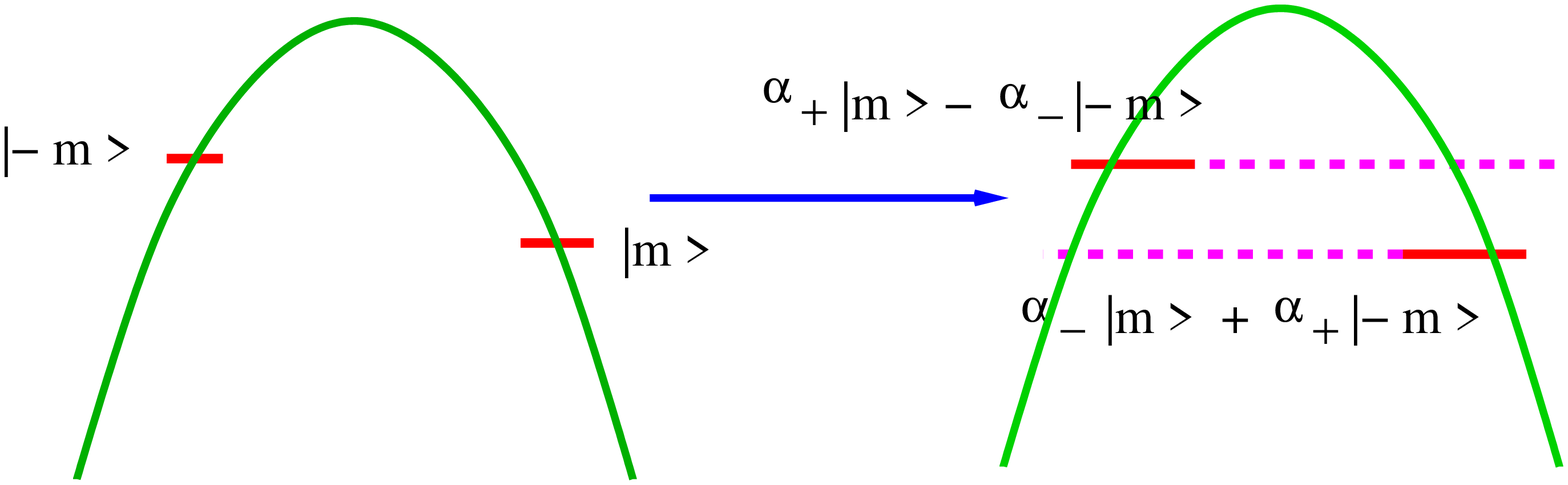}
}}
\caption{
Schematic representation of the splitting of the levels 
$\vl m \rangle$ and $\vl {-m} \rangle$
under resonant (top) and non-resonant conditions (bottom) as discussed
in the text.
}
\label{fig:hibri}
\end{figure}

Let us first discuss (with the hands) how these extra terms, which do not
commute with $S_z$, affect the relaxation time $\tauuno$
(\ref{tauuno}).
In the absence of transverse terms, the states 
$\vl m \rangle$ and $\vl {-m} \rangle$
are degenerated at $\Bz =0$.
Now, the terms $E$ and $B_x$ mix 
$\vl m \rangle$ and $\vl {-m} \rangle$ (figure~\ref{fig:hibri}).
The new eigenstates can be formed from their symmetric and
antisymmetric combinations
$\vl s \rangle$ and $\vl a \rangle$
%___________________________
\begin{eqnarray}
\nonumber
\vl s \rangle &\cong&
\frac{1}{\sqrt {2}}
\left (
\vl m \rangle + 
\vl {-m} \rangle
\right )
%%%
\\
\vl a \rangle &\cong&
\frac{1}{\sqrt {2}}
\left (
\vl m \rangle - 
\vl {-m} \rangle
\right )
\end{eqnarray}
%___________________________
with $\epsilon_{\vl a \rangle} - \epsilon_{\vl s \rangle} = \spl$ the
tunnel splitting, which can be calculated perturbatively
\cite{garchu97, luibarfer98}.
These states are delocalized (in $m$), and the spin can ``tunnel''.
%%
%___________________________________________________
%___________________________________________________
%___________________________________________________
%_____________________________________________________
%_____________________________________________________
%_____________________________________________________
%

If we now apply a field $\Bz$ the degeneration is lifted and we block
the tunnel channel.
This can be seen from the perturbed eigenstates:
%___________________________
\begin{eqnarray}
\begin{array}{c}
\vl s \rangle \cong
\alpha_+
\vl m \rangle + 
\alpha_-
\vl -m \rangle
%%%
\\
\vl a \rangle \cong
\alpha_-
\vl m \rangle - 
\alpha_+
\vl -m \rangle
\end{array}
\qquad
\alpha_\pm
=
\left [
\frac{1}{2}
\left (
1 \pm \frac {\Bz}{\sqrt{\Bz^2 + \spl^2}}
\right )
\right]^{1/2}
\; .
\end{eqnarray}
%___________________________
The transverse terms are assumed small ($\spl \ll 1$), so we have
$\alpha_+ \cong 1$ and $\alpha_- \cong 0$ at not very large fields
$\Bz$.
Therefore, out of resonance we have
$\vl s \rangle \cong \vl m \rangle$
and
$\vl a \rangle \cong \vl -m \rangle$,
and we are back in the behavior with no transverse terms.

However, increasing the field up to $\Bz = \ani$, the levels
$\vl {-m} \rangle$ and $\vl m -1 \rangle$
enter in resonance, making possible tunneling through these states.
This is repeated at all {\em crossing fields}, $\Bz = k \ani$, where
the states $\vl {-m} \rangle$ and $\vl m -k \rangle$ become degenerate
(figure~\ref{fig:els}).
Therefore, by changing the field $\Bz$, we ``open'' and ``close''
tunnel channels.

The tunnel splitting decreases exponentially from the states near the
barrier top to the lower laying states near the bottom of the wells.
This can be understood because the terms $S_\pm^2$ and $S_x$ couple
states with $m \pm 2$ and $m \pm 1$ in each perturbative order: then,
to connect $S$ all the way to $-S$ we need to go to higher orders in
the perturbation.
%
%Por lo tanto cualquier campo presente en el laboratorio rompe la degeneration
%haciendo muy poco probable el tunel entre el fondo de los pozos
\footnote{
This agrees with the semiclassical picture where tunnel is smaller the
higher the barrier over the tunnel channel.
} %ENDOF FOOTNOTE

\begin{figure}
\centerline{
\resizebox{6.8cm}{!}{%
    \includegraphics[angle = -0]{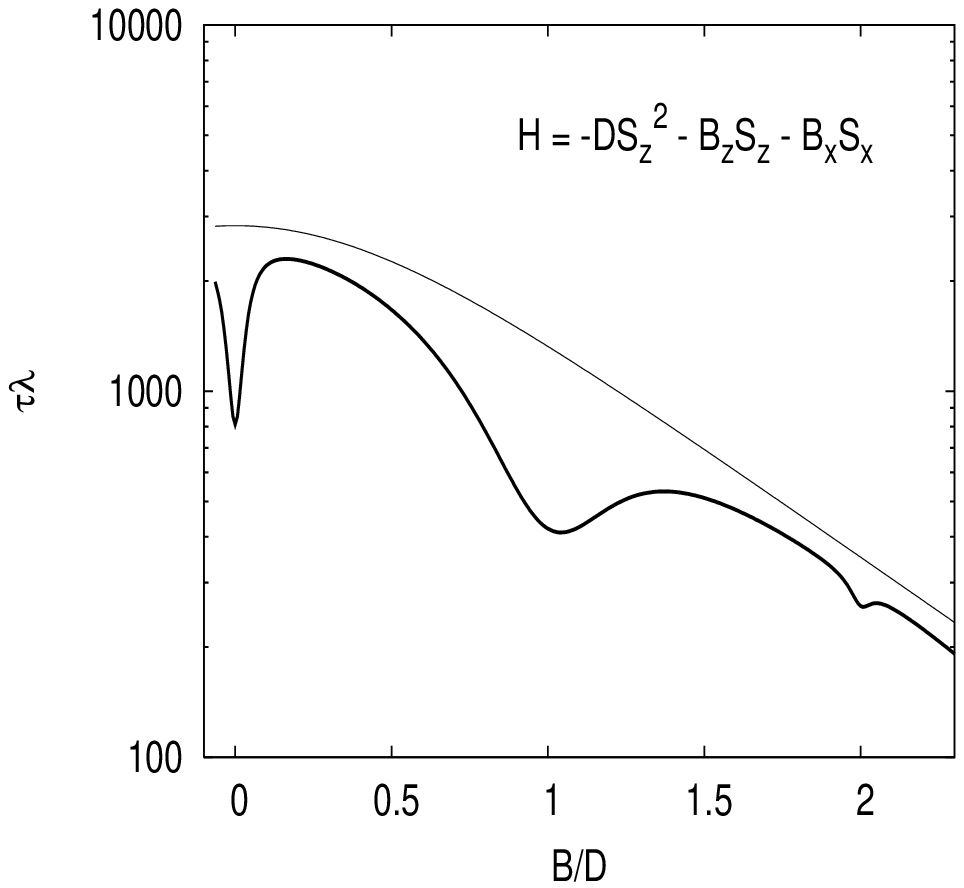} 
}
\resizebox{8.cm}{!}{%
\includegraphics[angle = -0]{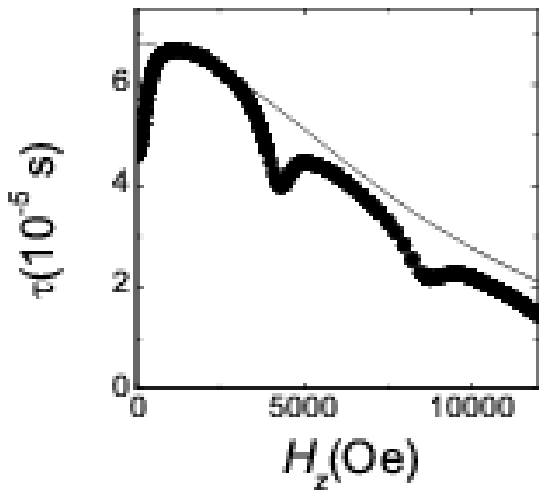}
}
}
\caption{ 
Left: the relaxation time $\tauuno$ from Eq.~(\ref{tauunoT}).
The curve corresponds to including a transverse field in
$\HS = -DS_z^2 - \Bz S_z - B_x S_x$
with $B_x = 10^{-3}$, at $\sigma = 15$ for a spin $S=10$.
We have also plotted for reference $\tauuno$ for $B_x =0$ (thin line),
where there is no tunneling, and the relaxation time does not display
dips at the resonance fields.
Right: $\tauuno$ in ${\rm Mn}_{12}$ at $T = 8K$ from susceptibility
measurements ($\prbbc$).
}
\label{fig:LT}
\end{figure}

On the other hand, in the temperature ranges we are going to address
($T > 2$\,K) tunnel proceeds incoherently.
That is, the decoherence time is so short, that coherent oscillations,
left-right, are never observed \cite{lopetal05}.
In this incoherent regime, the relaxation time $\tauuno$ can be
written as \cite{mildri01}:
%______________________________
\begin{equation}
\label{tauunoT}
\tauuno^{-1}
= 
\sum_m \tau_{m , m'} \e^{-\beta (\epsilon_{-m} - \epsilon_{-S})}
\end{equation}
%______________________________
where the sum extends to the resonant pairs $m$ and $m'$.
As we said, for the lowest laying pairs $\tau_{m ,m'} \cong 0$.
As a matter of fact, the barrier that the spin has to overcome
decreases in the crossing fields (as the tunnel channels get open) and
we have
$\tauuno \propto \e^{\beta \Delta U_{\rm ef}}$
with $\Delta U_{\rm ef} < \Delta U$. %in (\ref{tauuno}).
In figure~\ref{fig:LT} we have plotted the formula~(\ref{tauunoT}) for
a simple model with transverse field and displayed experimental data
for ${\rm Mn}_{12}$.
At the crossing fields, where the levels enter into resonance, the
relaxation time shows dips, as a consequence of the decrease of the
effective barrier \cite{luibarfer98}.
%%
%___________________________________________________
%___________________________________________________
%___________________________________________________
%___________________________________________________
%___________________________________________________
%___________________________________________________
%___________________________________________________
%___________________________________________________
%___________________________________________________
\begin{figure}
\centerline{\resizebox{6.8cm}{!}{%
    \includegraphics[angle = -0]{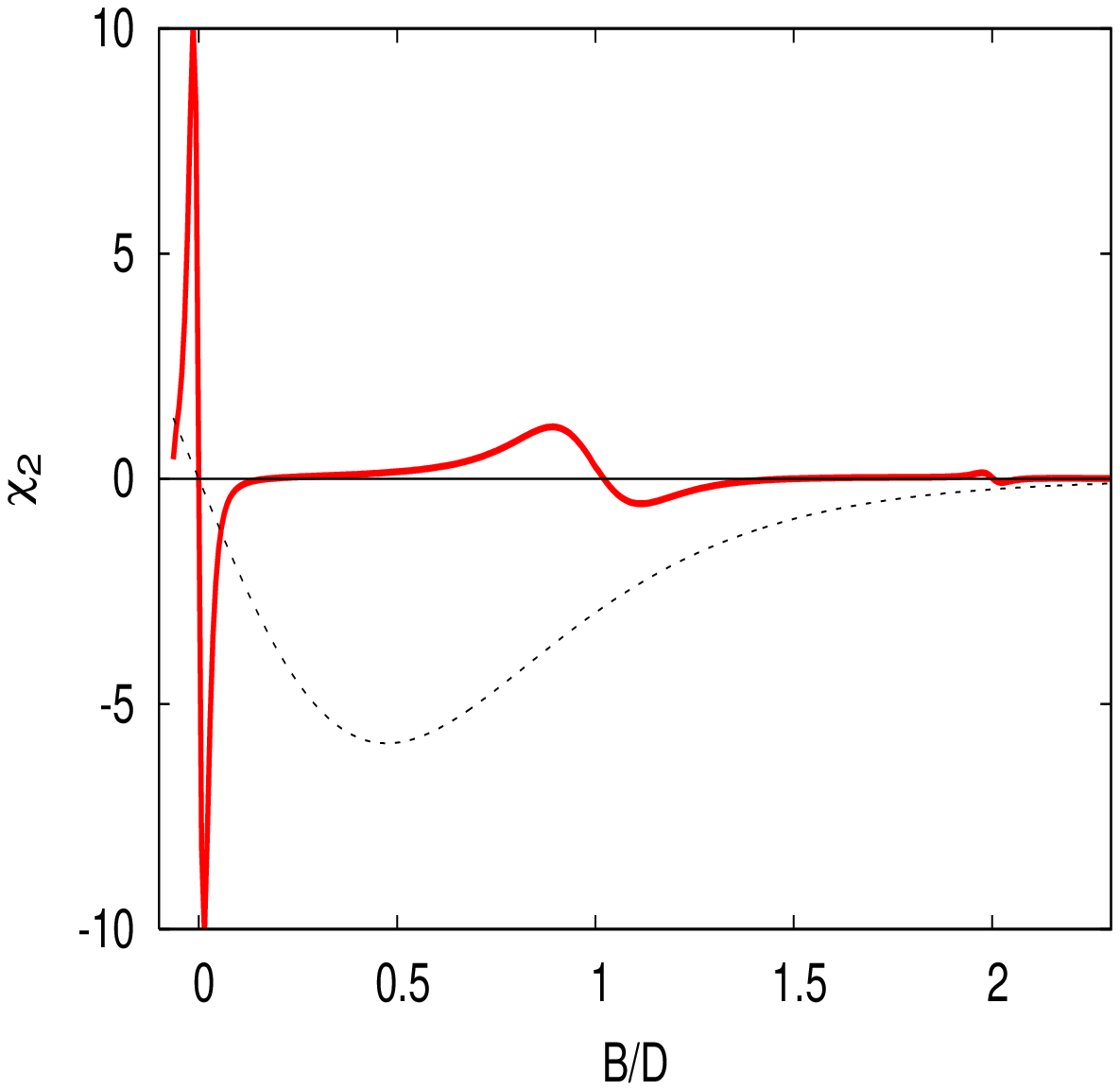} 
}
\resizebox{8.2cm}{!}{%
\includegraphics[angle = -0]{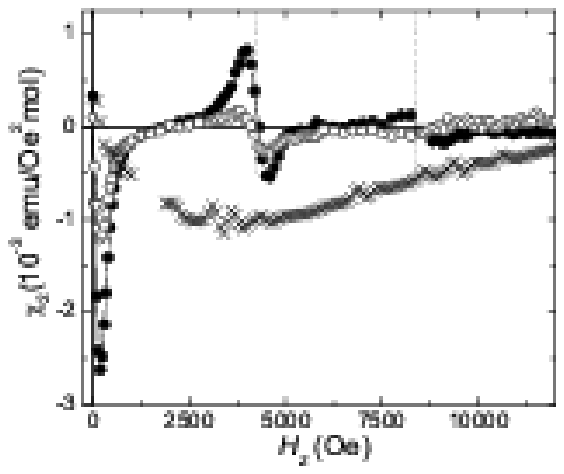}
}
}
\caption{
Left: prediction for $\susdos$ (imaginary part) with the theoretical
$\tauuno$ from figure~\ref{fig:LT}, left, plugged into the
analytical~(\ref{susdosone}).
Right: experimental $\susdos$ curves corresponding to the $\tauuno$
shown in figure~\ref{fig:LT}, right.
}
\label{fig:X2T}
\end{figure}
%___________________________________________________
%___________________________________________________
%___________________________________________________

How could we see this behavior studying $\susdos$ curves?
Well, if we work with not too strong static fields, we have 
$a_1 \cong a_2 \cong 1$.
Besides, if we probe the system with frequencies of the order of
$\tauuno^{-1}$, the fast intrawell contribution to~(\ref{susdostwo}) is
negligible ($\omega \tauW \ll 1$, since $\tauW\ll\tauuno$).
Then we can work with an approximate $\susdos$ dominated by the
overbarrier contribution
%_______________________________
\begin{eqnarray}
\label{susdostwoaprox}
\nonumber
\susdos &\cong&
 \suseqdos  
\frac { 1}
{1+\iu\,2 \omega \tauuno}
-
 \susequno  
\frac { \iu \, \omega \tauuno '}
{(1+\iu\,\omega \tauuno)(1+\iu\,2 \omega\tauuno)}
\end{eqnarray}
%____________________________
%
At fields $\Bz$ close to the crossing fields, the derivative
$\tauuno'$ becomes large (we are inside the dips).
Therefore, we expect that the term with $\tauuno'$ would dominate in
$\susdos$ in those ranges.
On the other hand, the derivative changes sign, and so would do
$\susdos$ at the crossing fields.

We have plotted $\susdos$ as a function of $\Bz$ in figure~\ref{fig:X2T}.
The experimental data shows the same qualitative features as those
just discussed on the basis of the formula~(\ref{susdostwoaprox}).
We thus see that $\susdos$ is a good tool to detect tunnel in these
systems, as it provides resonant features (peaks and sign changes) not
easily mistaken.
And all this because $\susdos$ is more sensitive to the variations in
$\tauuno$ that the linear susceptibility ($\prbbc$).

%%%%%%%%%%%%%%%%%%%%%%%%%%%%%%%%%%%%%%%%%%%%%%%%%%%%%%%%%%%%%%%%%%%%%%%%
\section {example IV:  transverse response}

\begin{figure}
\centerline{\resizebox{9.cm}{!}{%
    \includegraphics[angle = -90]{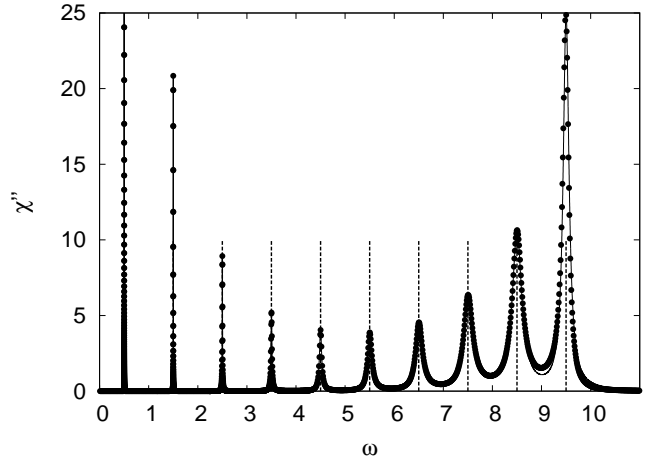} 
}
}
\caption{
Imaginary part of the transverse susceptibility $\sus_x'' (\omega)$
for a spin $S=10$, at $\sigma = 5$ with a weak damping $\damps =
3\cdot10^{-8}$ (magneto-elastic coupling).
Only the positive frequency is shown, as
$\sus_x''(\omega)=-\sus_x''(-\omega)$.
The symbols are the exact continued-fraction results ($\jpaii$), and the
continuous line is Eq.~(\ref{susx}).
We have also marked the level differences with vertical dashed lines.
}
\label{fig:FMR}
\end{figure}

The linear susceptibility is still useful, as it highlights some other
aspects of the problem.
In this last example we are going to compute the transverse
susceptibility, as we already did for the spin $\case$ case.
To this end, we excite with a transverse perturbation
$\pert \cdot B_x \e^{-\iu \omega t}$
and we monitor the response of $M_x=\langle S_x \rangle$
%_______________________________________________
\begin{equation}
M_x = \frac{1}{2} \langle S_+ + S_- \rangle 
=
\frac{1}{2} 
%\left [ 
\sum_m
\left (
\lf_{m-1, m}
\dm_{m-1,m}
+
\lf_{m+1, m}
\dm_{m+1, m}
\right )
%\right ]
\end{equation}
%_______________________________________________
In this problem, the unitary part in the master
equation~(\ref{RWAmej3}) mixes off-diagonal and diagonal elements,
incorporating the ladder factors 
$\lf_{m, m\pm 1}=\sqrt {S(S+1) - m(m \pm 1)}$.
Therefore, we have to consider all the density-matrix elements.
This can be done solving with the continued fraction method (section
\ref{sub:LRTFC}).
The response obtained is displayed in figure~\ref{fig:FMR}.

Based on the results of the spin-1/2 in \ref{sub:tdos}, one could
expect some absorption spectrum.
The imaginary part is indeed made of Lorentzians centered at the level
differences.
But due to the anisotropy, the $\HS$ spectrum is not equispaced
now, and we can find up to $2S+ 1$ peaks (cf. the cosine problem in
figure~\ref{fig:lrtcosine}).
%%
%___________________________________________________
%___________________________________________________
%___________________________________________________
%___________________________________________________
%___________________________________________________
%___________________________________________________

Following our custom, we will try to draw an approximate formula over
the exact curves.
Recall that under the crude rotating-wave approximation
Eq.~(\ref{nondiaRWA}), the time evolution of the non-diagonal
elements is simply given by
$\dm_{m\pm1, m} 
\propto 
\exp[-\iu (\Delta_{m_\pm 1 m} + \R_{m\pm1 m \pm 1, mm})t]$.
Then the response in the frequency domain follows from Eq.~(\ref{fdt}) as
%_______________________________________
\begin{eqnarray}
\label{susx}
\sus_x (\omega)
=
\frac {1}{2}
\suseq_x
\left
\{
\sum_m
\left [
a^+_m
\frac {\iu \Delta_{m-1 m} + \R_{m-1, m}}{\iu (\omega + \Delta_{m-1m}) 
+  \R_{m-1, m}}
+
a^-_m
\frac {\iu \Delta_{m+1 m} + \R_{m+1, m}}{\iu (\omega + \Delta_{m+1m}) + 
 \R_{m+1,m}}
\right ]
\right \}
\; \;
\end{eqnarray}
%_______________________________________
Here $\suseq_x$ is the equilibrium susceptibility, while $\R_{m\pm1,m}
\equiv \R_{m\pm1 m\pm1, mm}$ are relaxation coefficients and the
amplitudes $a_m^{\pm}$ depend essentially on the population difference
between the states $m$ and $m \mp 1$.
%___________________________________--
\begin{equation}
a_m^{\pm} = \frac{1}{2 \suseq_x \ZS} \frac{\e^{-\beta \epsilon_m} - \e^{-\beta
    \epsilon_{m \mp 1}}}{\Delta_{m \mp 1 m}} \lf_{m, m \mp 1}^2
\; .
\end{equation}
%_______________________________
This constitutes a generalization of equation (\ref{trans}) for
$S>\case$.
This formula is plotted as the continuous line in
figure~\ref{fig:FMR}, where we see that it fits quite well the
numerics (which correspond to the improved RWA, Eq.~(\ref{RWAmej3}), symbols).

If we now increase $S$ or the coupling $\damps$ (that is, decreasing
the quotient $\Delta_{m m \pm1}/\damps)$, we have observed that the
crude RWA (with which we got the formula) starts to fail
(see also \cite{raujohshn04} for more RWA failures).
We could still try to get the classical limit of the susceptibility
\cite{gardat05}.
But there the formula would be valid only in the limit $\damps =0$,
anyhow recovering Gekht's formula for the absorption line of a
classical superparamagnet \cite{gek83, garishpan90, gardat05}.
In any case, for the couplings and spin values in ordinary quantum
superparamagnets, equation~(\ref{susx}) could be a valuable tool to
fit experimental absorption curves.

\section {summary}

In this last chapter, we have sought to apply all the formalism
discussed along the thesis to the example of quantum superparamagnets.
What we learnt in chapter~\ref{chap:equilibrio} has been used to
calculate dissipative corrections to the celebrated Brillouin law of
quantum paramagnets.
With the master equation of chapter~\ref{chap:dinamica} together with
the methods presented in chapter~\ref{chap:metodos}, we have studied
relaxation mechanisms in superparamagnets.
We could contribute to the study of the phenomenon of tunnel of the
magnetic moment, studying the nonlinear susceptibility.
Finally, we derived approximate expressions for several quantities;
the formula for the dynamical transverse susceptibility can be used to
estimate the decoherence times in superparamagnets.

%%%%%%%%%%%%%%%%%%%%%%%%%%%%%%%%%%%%%%%%%%%%%%%%%%%%%%%%%%%%
%%%%%%%%%%%%%%%%%%%%%%%%%%%%%%%%%%%%%%%%%%%%%%%%%%%%%%%%%%%%
%%%%%%%%%%%%%  CONCLUSIONES %%%%%%%%%%%%%%%%%%%%%%%%%%%%%%%%
%%%%%%%%%%%%%%%%%%%%%%%%%%%%%%%%%%%%%%%%%%%%%%%%%%%%%%%%%%%%
%%%%%%%%%%%%%%%%%%%%%%%%%%%%%%%%%%%%%%%%%%%%%%%%%%%%%%%%%%%%

\chapter *{summary \& conclusions}
 \addcontentsline{toc}{chapter}{summary \& conclusions}
 \chaptermark{summary \& conclusions}

\label{chap:conclusiones}

As it is the custom, let us conclude remarking the most important
results presented in this thesis.

\begin{itemize}
  
\item 
In the classical limit we worked with the bath-of-oscillator
Hamiltonian~(\ref{clmodel}) to describe  open systems, obtaining the
corresponding Langevin and Fokker--Planck equations.
The model allows to understand the paradoxes of {\em
  irreversibility\/} from a Hamiltonian point of view.

\item
The dissipation theory is now formulated with a Hamiltonian: we can
quantize it.
We discussed the concept of quantum decoherence and derived quantum
master equations for the reduced density matrix.
The master equations used here are valid in the regimen of
weak system-bath coupling, compared with the self-correlation times of
the bath.

\item
We showed that the master equations are thermodynamically consistent,
that is, that they have as stationary solution the reduced distribution
obtained with quantum statistical mechanics.
We used the opportunity to discuss the differences between equilibrium
in the classical and quantum cases.
In particular, in contrast to the classical distribution, the quantum
reduced distribution depends on the damping.
We concluded the equilibrium properties discussing how to properly
define and compute thermodynamical quantities.

\item
With the phase-space formalism we closed the circle.
We transformed the equations for the density matrix $\dm$, into
equations for the Wigner distribution.
Working in phase space and taking the limit $\hbar \to 0$, the quantum
equations reduce to their classical counterparts, explicitly relating
the classical and quantum theories of dissipation.

\item 
We discussed the continued-fraction technique to solve master equations
(obtainment of stationary solutions and the linear and nonlinear
responses).
In spite of the demanding pre-treatment and specificity, when it does
work it happens to be a quite useful method.
In particular, it very suited to solve the Caldeira--Leggett master
equation in phase space for nonlinear potentials.
It also allows to study spin problems with $S \gg 1$, a limit very
demanding with other techniques.

\end{itemize}

\newpage

\begin{itemize}

\item
We applied the formalisms and the methods of the first part of the
thesis to the problem of a damped particle in a periodic potential.
The main results are:
\begin{itemize}
\item
We obtain the Wigner stationary solutions, and we see how the wave
nature of quantum mechanics takes reflection in the negative zones in
the ``distribution'', persisting at long times.

\item
Computing the currents in a cosine potential, we see how increasing the
damping we get closer and closer to the classical limit.

\item
We tackled the problem of directional motion in potentials without
space-reflection symmetry.
We computed the first quantum corrections to the rectified current.
The deviations from the classical limit can be understood appealing to
two complementary quantum phenomena: tunnel and wave reflection.

\item
We obtained a estimation of the decoherence times and the inter-well
jumps from the calculation of the linear susceptibility.
\end{itemize}

\item
Finally, we addressed the equilibrium and dynamical properties of
quantum superparamagnets.
The main results are:

\begin{itemize}

\item
We obtained dissipative corrections to the Brillouin magnetization law.
In particular, we could address the entanglement between system and bath
at very low temperatures from the equilibrium magnetization.

\item
We fully characterized the relaxation mechanisms in these systems:
overbarrier dynamics, and intrawell processes in the potential wells.
This allowed us to obtain a simple formula for the longitudinal linear
response and closed form expressions for the relaxation times for any
value of the spin $S$.

\item
We generalized our formalism to the nonlinear response.
We showed how to detect tunnel by means of the nonlinear susceptibility.

\item
We produced a simple expression for the transverse susceptibility,
which accounts for the absorption spectra in the ranges of parameters
relevant for ordinary experiments.

\end{itemize}

\end{itemize}

%%%%%%%%%%%%%%%%%%%%%%%%%%%%%%%%%%%%%%%%%%%%%%%%%%%%%%%%%%%%
%%%%%%%%%%%%%%%%%%%%%%   BIBLIOGRAPHY   %%%%%%%%%%%%%%%%%%%%
%%%%%%%%%%%%%%%%%%%%%%   BIBLIOGRAPHY   %%%%%%%%%%%%%%%%%%%%
%%%%%%%%%%%%%%%%%%%%%%   BIBLIOGRAPHY   %%%%%%%%%%%%%%%%%%%%
%%%%%%%%%%%%%%%%%%%%%%   BIBLIOGRAPHY   %%%%%%%%%%%%%%%%%%%%
%%%%%%%%%%%%%%%%%%%%%%   BIBLIOGRAPHY   %%%%%%%%%%%%%%%%%%%%
%%%%%%%%%%%%%%%%%%%%%%%%%%%%%%%%%%%%%%%%%%%%%%%%%%%%%%%%%%%%

  \addcontentsline{toc}{chapter}{bibliography}
%  \bibliography{/home/david/notas_articulos/cosas_tex/david}
%\bibliography{/home/jose/Desktop/COMMANDS/jlgarcia}
  \bibliography{david}
  \bibliographystyle{unsrt}

%%%%%%%%%%%%%%%%%%%%%%%%%%%%%%%%%%%%%%%%%%%%%%%%%%%%%%%%%%%%
%%%%%%%%%%%%%%%%%%%%%   END  DOCUMENT   %%%%%%%%%%%%%%%%%%%%
%%%%%%%%%%%%%%%%%%%%%%%%%%%%%%%%%%%%%%%%%%%%%%%%%%%%%%%%%%%%
%%%%%%%%%%%%%%%%%%%%%   END  DOCUMENT   %%%%%%%%%%%%%%%%%%%%
%%%%%%%%%%%%%%%%%%%%%%%%%%%%%%%%%%%%%%%%%%%%%%%%%%%%%%%%%%%%
%%%%%%%%%%%%%%%%%%%%%   END  DOCUMENT   %%%%%%%%%%%%%%%%%%%%
%%%%%%%%%%%%%%%%%%%%%%%%%%%%%%%%%%%%%%%%%%%%%%%%%%%%%%%%%%%%
%%%%%%%%%%%%%%%%%%%%%   END  DOCUMENT   %%%%%%%%%%%%%%%%%%%%
%%%%%%%%%%%%%%%%%%%%%%%%%%%%%%%%%%%%%%%%%%%%%%%%%%%%%%%%%%%%

\end{document}